\documentclass[10pt,twocolumn,letterpaper]{article}
\usepackage[T1]{fontenc}
\usepackage[pagenumbers]{cvpr}
\usepackage{amsthm,bm,array,multirow}
\usepackage[pass]{geometry}
\usepackage{iftex}
\definecolor{cvprblue}{rgb}{0.21,0.49,0.74}
\usepackage[pagebackref,breaklinks,colorlinks,allcolors=cvprblue]{hyperref}
\hypersetup{bookmarksnumbered=true,bookmarksopen=true,bookmarksopenlevel=1,bookmarksdepth=4,
  pdftitle={Depth from Differential Defocus beyond Geometric Optics},pdfauthor={Junjie Luo}}
\graphicspath{{figures/}}

\renewcommand{\paragraph}[1]{\par\vspace{0.4ex plus 0.3ex}\noindent\textbf{#1.}\enspace\ignorespaces}
\newtheorem{proposition}{Proposition}
\newtheorem{lemma}{Lemma}
\theoremstyle{definition}

\newcommand{\vx}{\bm{x}}
\newcommand{\vu}{\bm{u}}
\newcommand{\va}{\bm{a}}
\newcommand{\vf}{\bm{f}}
\newcommand{\ve}{\bm{e}}

\newcommand{\dd}{\,\mathrm{d}}

\newcommand{\Lap}{\nabla^2}

\title{Depth from Differential Defocus beyond Geometric Optics}
\author{Junjie Luo\\
Elmore Family School of Electrical and Computer Engineering, Purdue University\\
{\small luo330@purdue.edu}}

\begin{document}
\twocolumn[{%
\renewcommand\twocolumn[1][]{#1}%
\maketitle
\vspace{-1.6em}
\begin{center}
\captionsetup{type=figure}
\includegraphics[width=\textwidth]{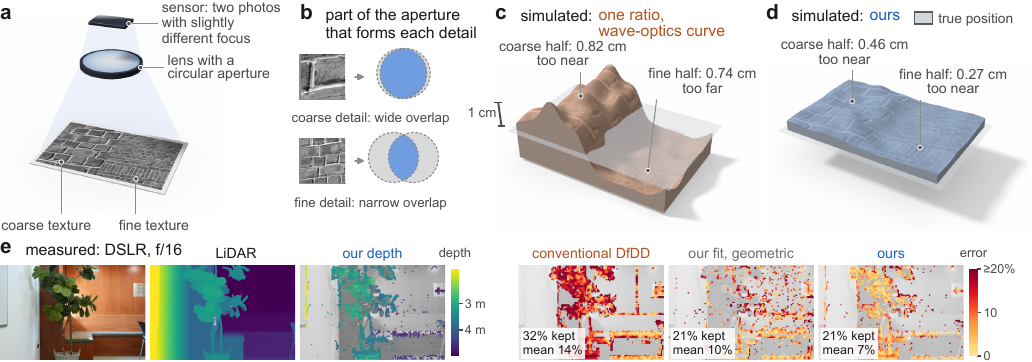}
\captionof{figure}{\textbf{In wave optics a circular aperture changes fine detail more slowly with focus than DfDD assumes, and a joint fit
of narrow bands extends the range of one image pair.} (a,~b)~Fine detail is formed by a narrow overlap (blue) of two copies of the aperture. (c,~d)~A prediction beyond our data, simulated for f/6, $2\,\mu$m pixels and bands up to $0.62f_c$
(Sec.~\ref{S-sec:teasersim}): on a flat print, one ratio over all frequencies places the two textures 1.56\,cm apart,
ours 0.19\,cm. (e)~Measured, one f/16 image pair of a DSLR focused at 7.8\,m, whose bands reach $0.23f_c$,
where this texture bias is too small to resolve (Sec.~\ref{sec:range}). Each estimator discarded its least
confident 75\% of windows over all scenes, as DfDD cameras do. Our depth and relative depth errors against
LiDAR on ZEDD scene 20~\cite{zuo2026}, held out, with the share kept here and its mean error. Over
all maps the mean errors are 20\%, 15\% and 8\%, and ours measures to 10\% over more distances
(Fig.~\ref{fig:depth}).}
\label{fig:teaser}
\end{center}
\vspace{0.4em}
}]

\begin{abstract}
Depth from differential defocus (DfDD) computes depth from two images with slightly different focus by modeling
defocus as diffusion, which in geometric optics is exact only for a Gaussian aperture. Real lenses have circular
apertures, and their blur contains diffraction. In wave optics, by Hopkins' small-aberration formula, the diffusion coefficient of a
circular aperture near focus is the second moment of the overlap of two displaced pupils. It falls with
spatial frequency $f$, to first order as $1-\frac{20}{3\pi}f/f_c$ with $f_c$ the diffraction cutoff, and rises near
the zeros of the defocused optical transfer function. On released data of a DfDD prototype the measured
coefficient falls with frequency at three iris diameters, along one curve in $f/f_c$. A test fixed in advance
confirmed a fall on new smartphone sweeps, at 95\,cycles/mm only and steeper than a clear disc predicts. Because one
ratio of the image change to the Laplacian, as in conventional DfDD, averages this coefficient over the texture, we
fit narrow frequency bands jointly to through-focus curves of a clear disc computed in wave optics and take a
confidence from the residual of the fit. On the prototype the fit measures distance to 15\,mm, against 22\,mm
with geometric curves. On a DSLR at f/16, with the least confident 75\% of windows
discarded, one image pair has a mean relative depth error of 8\%, less than half the 20\% of conventional DfDD, and
a test fixed in advance confirms the gap on 50 further scenes.
\end{abstract}

\section{Introduction}
\label{sec:intro}

Depth from differential defocus (DfDD) measures depth from two or three images whose focus, aperture or sensor position
differs only slightly~\cite{alexander2019,alexander2018,guo2017}. Because the images differ so little, depth follows in
closed form from a few image derivatives at each pixel, with no search over blur kernels, so DfDD cameras run on a
battery-powered embedded computer~\cite{luo2025fs} or need only 36 operations per pixel~\cite{luo2025cod}.

When the blur of an image grows slightly, the image changes in proportion to
its own Laplacian, $\Delta I\approx\sigma\Delta\sigma\,\Lap I$, where $\sigma$ is the size of the blur. This is the heat
equation, with the blur variance in the role of time~\cite{favaro2007}. Dividing the image change by the Laplacian gives
$\sigma$, and the thin-lens law turns $\sigma$ into depth,
\begin{equation}
  Z=\frac{\alpha}{\beta-I_\rho/\Lap I},
  \label{eq:depth_intro}
\end{equation}
where $I_\rho$ is the derivative of the image with respect to the optical power $\rho$ of the lens and the constants
$\alpha$ and $\beta$ depend on the optics (Sec.~\ref{sec:background}). The heat equation describes an idealized camera,
with geometric optics and a Gaussian aperture~\cite{alexander2019}. Its diffusion coefficient, the factor between
image change and Laplacian, is the same at every frequency. Real lenses have
circular apertures, and their blur contains diffraction. Current DfDD systems calibrate $\alpha$ and $\beta$ and treat
what remains as noise.

\begin{figure*}[t]
\centering
\includegraphics[width=\textwidth]{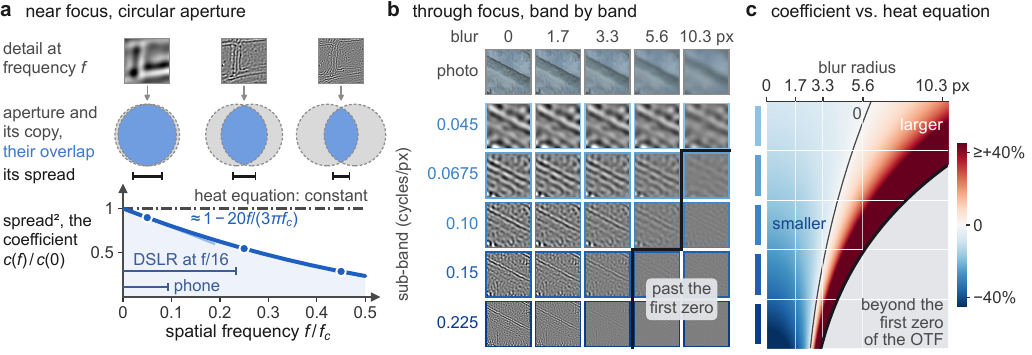}
\caption{\textbf{In wave optics the diffusion coefficient of a circular aperture depends on the spatial frequency.}
(a)~Detail at $f$ is formed by the overlap (blue) of the aperture and its copy displaced by $\lambda sf$.
Near focus the coefficient is the second moment of the overlap (bar: its spread), $1-\frac{20}{3\pi}f/f_c$ to first
order; the brackets end at our highest band on the phone and on the DSLR. (b)~The window of Fig.~\ref{fig:method} in
five f/16 photos, by blur radius, and in our five sub-bands, one gray scale per band; right of the stepped
line a clear disc is past the first zero of its OTF. (c)~The coefficient of a clear disc relative to the constant
$R^2/4$ of the heat equation, on the rows of (b) (f/16, exact OTF).}
\label{fig:overlap}
\end{figure*}

We analyze differential defocus in wave optics. The optical transfer function (OTF) at spatial frequency $\vf$ is an
integral over the overlap of two copies of the pupil displaced by $\lambda s\vf$, with $\lambda$ the wavelength and $s$
the sensor distance~\cite{goodman2005} (Fig.~\ref{fig:teaser}b). Near focus, the diffusion coefficient at $\vf$ is the
second moment of this overlap, which is Hopkins' small-aberration formula~\cite{hopkins1957} read for DfDD. A Gaussian
pupil gives an overlap of the same shape at every frequency, so the heat equation stays exact. For a clear disc, a uniform circular pupil without aberration,
the overlap narrows as the frequency grows, and to first order the coefficient falls as $1-2.12\,f/f_c$, with $f_c$
the diffraction cutoff; away from focus, the zeros of the OTF make it rise (Fig.~\ref{fig:overlap}). A phone camera
at f/1.8 with $1.4\,\mu$m pixels samples up to $0.35f_c$ (Table~\ref{S-tab:ls}), and on the DfDD
prototype of~\cite{luo2025cod}, at f/16--f/37, our bands reach 0.15--0.34$f_c$.

A coefficient that varies with frequency has three consequences for conventional DfDD, which takes a single ratio of
the image change to the Laplacian over all frequencies of a window; we call this estimate \emph{one ratio}. Its
geometric constants have the wrong scale. On a DSLR at f/16, the through-focus curve (the ratio against defocus) that the nominal
aperture predicts fits the data only after multiplication by a gain, 0.52 in geometric optics and 0.82 in wave
optics; the 0.82 is consistent with a pupil 8\% smaller than nominal and the camera noise (Sec.~\ref{sec:zedd}). The ratio also averages the coefficient over the frequencies of a window, weighted by the texture
spectrum, so the estimate depends on the texture (Fig.~\ref{fig:teaser}c,~d). And the ratio turns over as its dominant frequencies approach an OTF zero, which limits the
working range of one image pair (Figs.~\ref{fig:method}b,~\ref{fig:depth}a). We instead measure the ratio in narrow frequency bands, where
the coefficient is nearly constant, and fit all bands at once against through-focus curves computed from the pupil
(Fig.~\ref{fig:method}).

We test the law on the released data of a DfDD prototype whose iris changes the cutoff by a factor of
2.3~\cite{luo2025cod} (Fig.~\ref{fig:proto}). At three iris diameters the measured coefficient falls as a clear
disc predicts, along one curve in $f/f_c$. On a second
file of the prototype the joint fit measures distance with a mean absolute error of 15\,mm, against 22\,mm with
geometric curves and 38\,mm for conventional DfDD. On public focal stacks of a DSLR with LiDAR
depth~\cite{zuo2026}, curves of a clear disc with one gain per aperture match curves calibrated for every image
pair and band. With the least confident 75\% of windows discarded, as in DfDD cameras, one image pair there has a
mean relative depth error of 8\%, against 20\% for conventional DfDD (Fig.~\ref{fig:depth}), and read from its
posterior, in an exploratory analysis, the fit errs 31\% less within 0.35\,D of focus. A test fixed in advance confirmed
a fall on new smartphone sweeps~\cite{herrmann2020}; on the test scenes of the DSLR the turnovers and sign
changes of narrow bands, which no gain can move, exclude a disc of any one radius in geometric optics, and the sign changes
place a clear disc at 0.92 of the nominal radius.

Filter banks~\cite{guo2019}, texture-invariant filters~\cite{watanabe1998} and spectral
ratios fitted to a model of the OTF~\cite{pentland1987,morganmar2014} are not new; new here are the curves of the ratio computed
in wave optics, which save the calibration per image pair and band. Our contributions are:
\begin{itemize}[leftmargin=*,itemsep=1pt,topsep=2pt]
\item a wave-optics analysis of DfDD with a circular aperture: the coefficient as the second moment of the pupil
overlap, in closed form, and the texture bias of one ratio (Sec.~\ref{sec:theory});
\item a sub-band estimator that fits the band ratios jointly against through-focus curves computed from the pupil, with
one gain per aperture, a confidence from its residual and a posterior for the depth and its expected error
(Sec.~\ref{sec:bw});
\item tests on a DfDD prototype with three iris diameters, a DSLR and a smartphone, with rules fixed in advance
(Sec.~\ref{sec:real}).
\end{itemize}
Numbers with an S refer to the supplementary material.

\section{Related work}
\label{sec:related}

\paragraph{Depth from defocus}
Depth from defocus compares images with different amounts of
blur~\cite{pentland1987,subbarao1994,schechner2000,favaro2007}; Subbarao and Surya~\cite{subbarao1994} already divide
an image difference by a Laplacian, and Favaro et al.~\cite{favaro2008} treat defocus as diffusion. Frequency-domain
methods compare the spectra against a model of the OTF, frequency by frequency~\cite{pentland1987,xiong1993,morganmar2014}
or in local frequency bands~\cite{gokstorp1994,rajagopalan1997}; others calibrate the kernels~\cite{ens1993} or
engineer the optics~\cite{nayar1996,zhou2009}. Watanabe and
Nayar~\cite{watanabe1998} note that broadband operators such as the Laplacian of a Gaussian depend on the texture
spectrum and design rational filters that do not, and Schechner and Kiryati~\cite{schechner1999} relate the robustness
of depth from defocus to the frequency content of the scene and to the zeros of the OTF.
Tang et al.~\cite{tang2017} recover depth and flow from two handheld phone frames with a blur radius under three
pixels.
Most methods model the point-spread function (PSF) as a pillbox or a Gaussian~\cite{mannan2016}, adequate for most devices~\cite{saadat2026};
DfDD uses the focus derivative of the OTF, which differs from a Gaussian's near focus even where the
PSFs are close. Learned operators~\cite{favaro2002learning} and
networks~\cite{hazirbas2018,maximov2020,wang2021,yang2022,zuo2026}, some with a defocus model~\cite{fujimura2024} or a
single-image prior~\cite{ganj2025}, need no model of the pupil.

\paragraph{Depth from differential defocus}
DfDD replaces the large blur difference by a derivative in an optical parameter: the aperture
mask~\cite{farid1998}, the focus for a moving target~\cite{alexander2018} or of a deformable lens~\cite{guo2017}, two
simultaneous focus states~\cite{guo2019,luo2025fs}, coupled focus and aperture derivatives~\cite{luo2025cod}, or
differential defocus with stereo~\cite{luo2026d3s}. Guo et
al.~\cite{guo2019} fuse nine filters by their confidence across texture frequencies.
Alexander~\cite{alexander2019} shows that, in geometric optics, a texture-independent local constraint of this form
requires a Gaussian aperture, and Alexander et al.~\cite{alexander2021} relate depth from defocus to the
transport-of-intensity equation~\cite{streibl1984}. All use geometric optics.

\paragraph{Wave optics of defocus}
The defocused OTF of a disc is classical~\cite{hopkins1955,stokseth1969,lee1990,goodman2005,fitzgerrell1997}. For small
aberrations, Hopkins~\cite{hopkins1957} expressed the loss of the OTF through the variance of the wave-aberration
difference over the pupil overlap, and the OTF at any defocus is a slice of the pupil's ambiguity
function~\cite{brenner1983}. Learned optics for depth model the PSF in wave optics~\cite{chang2019,wu2019,ikoma2021};
phase diversity~\cite{gonsalves1982,paxman1992} and curvature sensing~\cite{roddier1988} infer the wavefront, not
depth, from two images.

\paragraph{Companion paper}
A companion paper~\cite{luo2026theory} derives exact constraints
for Gaussian pupils, proves that only they make defocus an exact diffusion, and applies Lemmas~\ref{lem:small}
and~\ref{lem:disc} to coupled optical differentiation. We cite its Gaussian-pupil result
(Sec.~\ref{sec:gauss}); circular apertures, the texture bias, the estimator and all
experiments appear only here.

\section{Differential defocus in geometric optics}
\label{sec:background}

\begin{figure*}[t]
\centering
\includegraphics[width=\textwidth]{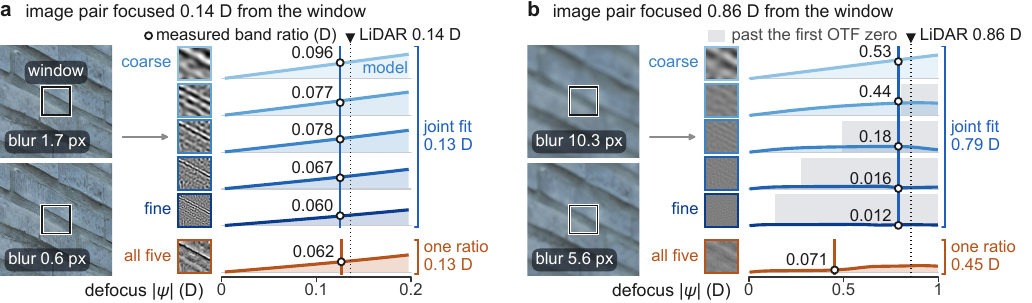}
\caption{\textbf{Sub-band estimation on one window, seen by two image pairs.} Rows: the window in five
sub-bands, 0.045 to 0.225\,cycles/px, and in their sum, each with its through-focus curve and its measured ratio
(circle; in diopters by~\eqref{eq:psiapp}); one defocus (blue line) must bring every curve to its ratio.
(a)~Focused 0.14\,D from the window: the ratios differ, where the heat equation predicts one value. (b)~Focused
0.86\,D from it: the finer bands are past their first OTF zero (gray) and the coarsest still fixes the defocus; one
ratio of the five (orange) reads 0.45\,D. ZEDD scene 38~\cite{zuo2026}, f/16, held out; of 556 windows the one whose
relative errors lie closest to their medians.}
\label{fig:method}
\end{figure*}

Following~\cite{luo2025fs,luo2025cod}, consider a fronto-parallel plane with texture $T$ at depth $Z$, imaged by a thin
lens of optical power $\rho$ onto a sensor at distance $s$. The image $I=k*P$ is the pinhole image
$P(\vx)=T(-Z\vx/s)$ convolved with the PSF $k$. In geometric optics the PSF is a scaled copy of
the aperture transmittance profile $K$,
\begin{equation}
  k(\vx)=\frac1{\sigma^2}\,K\Big(\frac{\vx}{\sigma}\Big),\qquad \sigma=As\psi,
  \label{eq:sigma}
\end{equation}
where $A$ is the aperture scale, $\sigma$ the defocus level and $\psi=1/Z-\rho+1/s$ the defocus in diopters (1\,D\,$=$\,1\,m$^{-1}$). For a
Gaussian profile the image obeys the heat equation $\partial I/\partial\sigma=\sigma\Lap I$~\cite{favaro2007}
(Fig.~\ref{S-fig:dbc}), and since $\partial\sigma/\partial\rho=-As$,
\begin{equation}
  I_\rho=-A^2s^2\psi\,\Lap I .
  \label{eq:dbc}
\end{equation}
Most DfDD methods rest on this relation. It gives the depth
equation~\eqref{eq:depth_intro} with $\alpha=A^2s^2$ and $\beta=A^2s^2(\rho-1/s)$; $I_\rho$ is the difference of two
images and $\Lap I$ is taken from their mean, both summed over a window. In the Fourier domain,~\eqref{eq:dbc}
multiplies the spectrum at every frequency $\vf$ by a factor proportional to $|\vf|^2$, with the same coefficient $c$,
\begin{equation}
  \widehat{I_\rho}(\vf)=4\pi^2s^2\psi\,c\,|\vf|^2\,\hat I(\vf).
  \label{eq:dbc_fourier}
\end{equation}
Here $c$ is the diffusion coefficient of the heat equation; the Gaussian profile gives $c=A^2$. A circular
aperture of radius $R$ gives a pillbox PSF, for which~\eqref{eq:dbc} holds only approximately, with $c=R^2/4$, and
less well as the blur grows~\cite{luo2025fs}; we call this model the \emph{geometric disc}. Wave optics adds an error that dominates near focus.

\section{Defocus in wave optics}
\label{sec:theory}

\subsection{The OTF as an overlap of two pupils}
\label{sec:model}
We use scalar, paraxial wave optics with quasi-monochromatic light of wavelength $\lambda$ and an incoherently lit
target. For a complex pupil function $p(\vu)$, the OTF $H$ is
\begin{equation}
\begin{aligned}
  H(\vf;\psi)&=\int O_{\va}(\vu)\,e^{-i2\pi\psi\,\vu\cdot\va/\lambda}\dd\vu,\\
  O_{\va}(\vu)&=\frac{p(\vu-\tfrac{\va}{2})\,p^*(\vu+\tfrac{\va}{2})}{\|p\|^2},\qquad \va=\lambda s\vf .
\end{aligned}
  \label{eq:otf}
\end{equation}
Here $O_{\va}$ is the normalized product of the pupil at $\vu-\va/2$ and its conjugate at $\vu+\va/2$ (Fig.~\ref{fig:overlap}a);
its integral is the in-focus OTF, the autocorrelation of the pupil~\cite{goodman2005}. Defocus changes the path
difference of these two points by $\psi\,\vu\cdot\va$, linearly across the overlap, which makes $H$ the Fourier transform
of $O_{\va}$ at $\psi\va/\lambda$. For a static scene $\hat I=H\hat P$, so the derivative of the image with respect to
an optical parameter $t$ that leaves $P$ fixed is a filtered copy of it,
\begin{equation}
  \widehat{I_t}(\vf)=\partial_t\log H(\vf)\;\hat I(\vf).
  \label{eq:spectral}
\end{equation}
A constraint that relates $I_t$ to the image and its derivatives up to second order, as in DfDD, thus holds for
every texture if and only if $\partial_t\log H$ is a polynomial of degree at most two in $\vf$~\cite{luo2026theory};
the heat equation~\eqref{eq:dbc} has $\partial_\rho\log H\propto|\vf|^2$.

\paragraph{Gaussian pupils}
\label{sec:gauss}
For a pupil with Gaussian amplitude, \mbox{$p(\vu)=\exp(-|\vu|^2/4A^2)$}, the overlap keeps its Gaussian shape at every
displacement, and the OTF is~\cite{luo2026theory}
\begin{equation}
  H(\vf;\psi)=\exp\!\big(-2\pi^2(\sigma^2+\sigma_d^2)|\vf|^2\big),\quad \sigma_d=\frac{\lambda s}{4\pi A},
  \label{eq:gauss_otf}
\end{equation}
with $\sigma=As\psi$ as in~\eqref{eq:sigma}. Defocus adds the geometric variance $\sigma^2$ to a diffraction variance
$\sigma_d^2$ that does not depend on the focus, as for a Gaussian beam~\cite{siegman1986}, so the focus derivative
removes $\sigma_d^2$ and~\eqref{eq:dbc} holds exactly. Only pupils with Gaussian amplitude make defocus an exact
diffusion~\cite{luo2026theory}.

\subsection{Circular apertures: the coefficient falls with frequency}
\label{sec:hard}

We define the diffusion coefficient at each frequency as
\begin{equation}
  c(\vf;\psi)=-\frac{\partial_\psi\log H(\vf;\psi)}{4\pi^2s^2\psi|\vf|^2}.
  \label{eq:cdef}
\end{equation}
It is defined where $H\neq0$, and at $\psi=0$ or $\vf=0$ by its limit. A Gaussian pupil gives $c=A^2$ at every
frequency.

\begin{lemma}[Small defocus]
\label{lem:small}
For a nonnegative, centrally symmetric pupil, $c(\vf;0^+)=M(\va)$, the second moment along $\vf/|\vf|$ of the
overlap $O_{\va}$ normalized to unit mass.
\end{lemma}

\noindent This is Hopkins' small-aberration formula~\cite{hopkins1957} for defocus, whose aberration
difference is linear across the overlap; it describes $c$ well inside the first OTF zero. In geometric optics the
overlap is the whole pupil at every frequency and $c(\vf;0^+)$ is constant; in wave optics the overlap of a disc
narrows along $\vf$ as $|\vf|$ grows, and $c$ falls.

\begin{lemma}[Circular aperture]
\label{lem:disc}
For a clear disc ($p=1$ for $|\vu|\le R$) and $q=|\vf|/f_c$, with $f_c=2R/(\lambda s)$,
\begin{align}
  \frac{c(\vf;0^+)}{c(0;0^+)}&=\frac{(1+4q^2)\arccos q-\tfrac{q}{3}(13+2q^2)\sqrt{1-q^2}}{\arccos q-q\sqrt{1-q^2}}\notag\\
  &=1-\frac{20}{3\pi}\,q+O(q^2),
  \label{eq:t4}
\end{align}
where $f_c=1/(\lambda N)$ is the diffraction cutoff for the working f-number $N=s/2R$.
\end{lemma}

\noindent Sec.~\ref{S-sec:hard} proves Lemmas~\ref{lem:small} and~\ref{lem:disc}.
Stokseth's approximation of the defocused OTF~\cite{stokseth1969} implies $(1-q)^2$, within 0.02 of the exact
ratio; the exact slope $20/(3\pi)\approx2.12$ is steep (Fig.~\ref{fig:overlap}a). At $0.1f_c$ the coefficient has
dropped by 20\%, at $0.25f_c$ by 45\%.

\subsection{Consequences for conventional DfDD}
\label{sec:errmap}
\label{sec:bias}

Away from focus the coefficient of a disc also rises above $R^2/4$ as the frequency approaches the first zero of the OTF
(Fig.~\ref{fig:overlap}c; Fig.~\ref{S-fig:errmap}). A depth estimate fits~\eqref{eq:dbc} over a window
and so averages the coefficient over its spectrum.

\begin{proposition}[Least-squares estimate]
\label{prop:ls}
Fit~\eqref{eq:dbc} with a constant coefficient $M_0$ by least squares over a large window, with texture power spectrum
$S_T$ and pre-filter $G$. Then
\begin{equation}
  \frac{\hat\psi}{\psi}=\frac{\sum_{\vf}|\vf|^2\,\partial_\rho H\,H\,S_T\,|G|^2}
  {4\pi^2s^2\psi M_0\sum_{\vf}|\vf|^4H^2S_T\,|G|^2},
\end{equation}
which equals 1 for every texture for a Gaussian pupil with $M_0=A^2$. For a circular aperture it depends on $S_T$.
\end{proposition}

\noindent Textures rich in fine detail put more weight on high frequencies, where the coefficient is small near focus,
so their defocus is underestimated more; a calibration corrects the average error but not its variation between
textures. Since the weights grow as $|\vf|^4$, the highest frequencies of a natural texture dominate, and the variation
is large only near an OTF zero or close to the diffraction cutoff. In a simulated f/6 camera with its cutoff near
Nyquist, one ratio with the exact curve of one texture is biased by $-13\%$ to $+14\%$ on others, and the joint fit
of Sec.~\ref{sec:bw} by about $\pm6\%$ (Fig.~\ref{S-fig:sim}).

\section{Depth estimation in sub-bands}
\label{sec:bw}

One ratio mixes frequencies with different diffusion coefficients. Within a narrow band the coefficient varies
little, so we measure the ratio in several bands, predict each from the pupil, and fit all bands at once
(Fig.~\ref{fig:method}).

\paragraph{Measurement in each band}
We filter each of two images $I_\pm$, captured at defocus $\psi\pm\Delta\psi/2$, with a radial log-Gabor filter $W_b$
centered at frequency $f_b$ and sum two band energies over a window, $e_b^\pm=\sum(W_b*I_\pm)\,\Lap(W_b*I_\pm)$ and
$E_b^\pm=\sum(\Lap W_b*I_\pm)^2$. The band ratio and its weight are
\begin{equation}
  r_b=\frac{e_b^+-e_b^-}{\Delta\psi\,(E_b^++E_b^-)},\qquad \omega_b=\tfrac12\big(E_b^++E_b^-\big).
  \label{eq:bw_ratio}
\end{equation}
For a small step, $r_b$ is the ratio of the heat equation~\eqref{eq:dbc} in one band, so $r_b=A^2s^2\psi$ in
every band for a Gaussian pupil. With the geometric coefficient $R^2/4$ of a disc it becomes an \emph{apparent
defocus},
\begin{equation}
  \psi_{\mathrm{app},b}=\frac{4\,r_b}{(Rs)^2},
  \label{eq:psiapp}
\end{equation}
equal to $\psi$ in every band if the heat equation held. When the signal dominates, $\omega_b$ is the inverse variance of $r_b$ for white
image noise. Each energy comes from one image, so the pair needs no pixel-accurate
registration (ZEDD leaves shifts of 0.5--1\,px; Sec.~\ref{S-sec:zedd}). Noise adds to
$E_b^\pm$, so a band of pure noise pulls the fit toward focus (Table~\ref{S-tab:bw_summary}); the confidence
and the posterior below account for it.

\paragraph{Through-focus curves}
For a texture with power spectrum $S_T$, the band ratio of a large window tends to
\begin{equation}
  r_b=\frac{\sum_{\vf}|W_b|^2\,L\,\big(|H_+|^2-|H_-|^2\big)\,S_T}{\Delta\psi\sum_{\vf}|W_b|^2\,L^2\,\big(|H_+|^2+|H_-|^2\big)\,S_T},
  \label{eq:bw_response}
\end{equation}
where $L(\vf)=-4\pi^2|\vf|^2$ and $H_\pm$ are the OTFs at $\psi\pm\Delta\psi/2$. Evaluating~\eqref{eq:bw_response} with
the exact OTF of the pupil, the pixel aperture and a nominal spectrum $S_0\propto|\vf|^{-2}$ gives the
model \emph{through-focus curve} $g_b(\psi)$ of band $b$. It contains
diffraction, the pupil shape and the finite focus step without linearization, so depth does not rest on
Lemmas~\ref{lem:small} and~\ref{lem:disc}, which say where such curves leave geometric optics. A curve can also be
calibrated on targets
at known depths, which absorbs unknown aberrations. Far from focus each band turns over near its first OTF zero, so lower
bands remain usable longer (Fig.~\ref{fig:method}b; Fig.~\ref{S-fig:response}).

\paragraph{Estimate}
We take the defocus that fits all bands best,
\begin{equation}
  \hat\psi=\arg\min_\psi\ \sum_b\omega_b\big(r_b-\gamma\,g_b(\psi)\big)^2 ,
  \label{eq:bw_estimate}
\end{equation}
a 1-D search per pixel. The gain $\gamma$, one per aperture, is chosen on training scenes for the smallest
median depth error; it absorbs the scales of the aperture and the reference depth. The share of
$\sum_b\omega_b r_b^2$ that the fit explains is our \emph{confidence}; we discard the least confident
estimates~\cite{luo2025fs} (Sec.~\ref{S-sec:confidence}).

\paragraph{Posterior}
\label{sec:posterior}
The minimum of~\eqref{eq:bw_estimate} ignores how flat the cost is. Let $\eta_b$ be the share of camera noise in
$\omega_b$, from one constant per band measured on defocused windows without depth. The band ratio then has mean
$\gamma(1-\eta_b)g_b$, and integrating out the unknown scale of the residuals of the $n$ bands gives the
posterior $p(\psi\,|\,r)\propto C(\psi)^{-n/2}$ over the search, with
$C=\sum_b\omega_b\,(r_b-\gamma(1-\eta_b)g_b(\psi))^2/(1-\eta_b/2)$. We report the distance that minimizes the
expected relative error under $p$; that expected error serves as its confidence
(Sec.~\ref{S-sec:posterior}).

\paragraph{Why it works}
The measured ratio and the through-focus curve average the same per-frequency response over the band, with weights
that differ only by the spectra $S_T$ and $S_0$, so away from OTF zeros the texture error of a band is at most the
variation of this response within an octave of the band center (Proposition~\ref{S-prop:bw}). To first order it is
1--4\% per unit of spectral exponent at 0.225\,cycles/px within 0.2\,D at f/16. In a narrow band $r_b$ is the
spectral ratio of depth from defocus~\cite{pentland1987,morganmar2014} in other units; in a band of finite width
that ratio also carries the mean square frequency of the texture, which $r_b$ divides out (Sec.~\ref{S-sec:bw}).

\section{Experiments}
\label{sec:real}

\begin{figure*}[t]
\centering
\includegraphics[width=\textwidth]{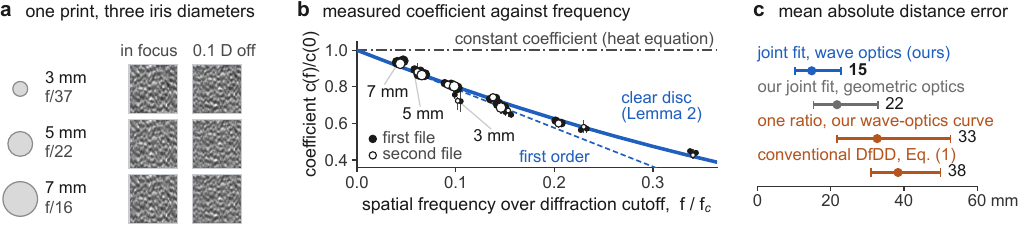}
\caption{\textbf{On a DfDD prototype~\cite{luo2025cod} the measured diffusion coefficient falls with spatial
frequency as a clear disc predicts, at three iris diameters.} (a)~One print through three irises.
(b)~Slope of the apparent defocus~\eqref{eq:psiapp} against $1/Z$ near focus, per band and
iris of two files, divided by one gain per iris. Lines: a constant coefficient and Lemma~\ref{lem:disc} with the
nominal cutoff; only the gains are fitted. (c)~Distance error on the second file
within 9.5\,px of blur, from the focus change alone, without the aperture change of~\cite{luo2025cod}. Bars: 95\% intervals.}
\label{fig:proto}
\end{figure*}

\begin{table}[t]
\centering
\caption{\textbf{Depth from one image pair on the DSLR at f/16}, held out, textured windows; prior methods reimplemented. Mean relative depth
error (MRE) [\%] within $\pm0.35$\,D of focus; its paired ratio to ours [95\% interval]; the MRE over the
9.5\,px search with the least confident 75\% discarded (post hoc), and on the windows ours keeps. Confidence: the
residual for joint fits with model curves, the expected error for the posterior (exploratory; f/16 foci), an image
energy otherwise. Bold: ours; underlined: best.}
\label{tab:est}
\resizebox{\columnwidth}{!}{%
\setlength{\tabcolsep}{3pt}
\begin{tabular}{@{}c@{\hspace{4pt}}lcccc@{}}
\toprule
 & Estimator and curve & MRE & Ratio to ours & 75\% disc. & Our windows \\
\midrule
\multirow{2}{*}{\rotatebox{90}{prior}}
 & conventional DfDD~\cite{guo2017,luo2025fs}: one ratio, $\alpha,\beta$ & 21.9 & 1.40 [1.29, 1.56] & 18.2 & 19.4 \\
 & spectral ratio~\cite{pentland1987,morganmar2014}, calibrated & 17.1 & 1.09 [1.02, 1.19] & 15.2 & 8.7 \\
\midrule
\multirow{2}{*}{\rotatebox{90}{\textbf{ours}}}
 & \textbf{joint fit, wave} & \textbf{15.6} & -- & \textbf{6.0} & \underline{\textbf{6.0}} \\
 & \textbf{\quad read from the posterior} (Sec.~\ref{sec:bw}) & \underline{\textbf{10.8}} & 0.69 [0.62, 0.77] & \underline{\textbf{4.0}} & -- \\
\midrule
\multirow{7}{*}{\rotatebox{90}{our variants}}
 & one ratio, wave & 15.3 & 0.98 [0.95, 1.00] & 11.6 & 11.9 \\
 & joint fit, geometric disc & 19.0 & 1.22 [1.16, 1.28] & 10.3 & 11.3 \\
 & \quad fitted radius and gain & 16.6 & 1.06 [1.03, 1.10] & 11.2 & 23.3 \\
 & \quad radius $\times(1-f_b/f_c)$, wave approx. & 15.7 & 1.01 [1.00, 1.02] & 6.5 & 6.6 \\
 & joint fit, calibrated per pair and band & 15.7 & 1.01 [0.95, 1.08] & 12.3 & 6.8 \\
 & \quad with a gain & 15.1 & 0.97 [0.91, 1.02] & 9.9 & 6.2 \\
 & perceptron on the band statistics & 11.7 & 0.75 [0.63, 0.92] & 11.6 & 6.1 \\
\bottomrule
\end{tabular}}
\end{table}

Does the diffusion coefficient fall with frequency as Lemma~\ref{lem:disc} predicts (\textbf{Q1})? Are curves computed in wave optics as accurate as
calibrated ones, and would geometric curves do as well (\textbf{Q2})? Over what distances does one image pair measure
depth (\textbf{Q3})? Does conventional DfDD depend on the texture (\textbf{Q4})? Depth errors are mean absolute
(MAE) and mean relative (MRE) errors of distance~\cite{luo2025fs,luo2025cod}.
\emph{Pre-specified} tests had their rules fixed before the analysis, in dated local records without a third-party
time stamp (Table~\ref{tab:tests}); every other number is exploratory (Sec.~\ref{S-sec:protocol}).

\subsection{The coefficient on a DfDD prototype (Q1)}
\label{sec:proto}

\paragraph{Data}
The released data of the coupled-differentiation prototype~\cite{luo2025cod} show a printed texture on a motorized
slide through an iris of 3, 5 or 7\,mm (f/37, f/22, f/16) and a tunable lens at three optical powers 0.06\,D apart
(Fig.~\ref{fig:proto}a). The iris moves the cutoff by a factor of 2.3 with all else unchanged, so a fall due to
aberrations or the sensor would not follow it. One file holds 97 distances from 0.45 to 1.41\,m; a second,
with another print, 0.91 to 1.87\,m. Our bands at 0.03--0.10\,cycles/px lie at 0.04--0.34 of the nominal cutoff.

\paragraph{The fall}
\label{sec:stats}
For two images 0.06\,D apart, the slope of the apparent defocus~\eqref{eq:psiapp} against $1/Z$ near
focus measures the coefficient of band $b$ in units of $R^2/4$, up to one gain per iris for the
errors of the nominal geometry. The slopes fall with frequency at all three diameters (Fig.~\ref{fig:proto}b).
At 3, 5 and 7\,mm they fall between 0.03 and 0.10\,cycles/px to 0.60, 0.68 and 0.74 of their value, where clear-disc
curves predict 0.52, 0.67 and 0.73 and geometric curves 1.02, 0.92 and 0.89. Bands of different irises at the same
$f/f_c$ agree to 3--5\% rms, and a constant coefficient misses the slopes by 6--21\%. With one gain per iris
(0.74--0.85) and the nominal cutoff, the 36 slopes lie within 4\% rms of
Lemma~\ref{lem:disc}. Post hoc, a sensor distance of 97\,mm, where the released code has 110\,mm, gives these
gains and scales the cutoff by 1.14, inside both intervals of Table~\ref{tab:tests} (Sec.~\ref{S-sec:cod}).

\paragraph{Depth on the second file}
With the cutoff scale, gains and foci of the first file and one focus shift from the second, the joint fit of
Sec.~\ref{sec:bw} measures distance within 9.5\,px of geometric blur radius, 0.96--1.53\,m, with an MAE of 15\,mm,
1.2\%. The same fit with geometric curves errs by 22\,mm, one ratio with our curve by 33\,mm and
conventional DfDD~\eqref{eq:depth_intro} by 38\,mm (Fig.~\ref{fig:proto}c). The three depth statements
pre-specified before the second file was analyzed, in defocus error, are confirmed, as are those of two controls fixed later
(Table~\ref{tab:tests}; Sec.~\ref{S-sec:codcontrols}); with windows of $64\times64$\,px the MAE is 17\,mm.

\subsection{Depth from one image pair on a DSLR (Q2)}
\label{sec:depth}

\paragraph{Data and setup}
We use the 50 validation scenes of ZEDD~\cite{zuo2026}: a Sony $\alpha$1\,II, a mirrorless camera that we call a
DSLR as the dataset does, with a 50\,mm lens, nine focus distances (0.82--8.10\,m), six apertures and LiDAR depth.
Two consecutive focus distances form an image pair; five of the eight pairs have small focus steps (0.02--0.11\,D)
and three large ones (0.17--0.48\,D). At f/16 our bands at 0.045--0.225\,cycles/px reach $0.23f_c$; at f/2.8--f/5.6
they reach at most $0.08f_c$, a control. We estimate the defocus of $64\times64$\,px
windows from one pair with five-fold cross-validation over scenes and turn it into a distance with the focus of
the pair. One ratio sums the five bands and uses $\alpha$
and $\beta$ calibrated in its linear range, or a through-focus curve. Model curves have one gain per aperture;
calibrated curves are fitted per pair and band. Every calibration, the focus included, uses only the training
scenes of the fold; intervals are over 2\,000 scene resamples (Sec.~\ref{S-sec:zedd}; depth maps in
Figs.~\ref{S-fig:supp_maps}--\ref{S-fig:maps_more}).

\paragraph{Accuracy}
Table~\ref{tab:est} takes the windows with uniform LiDAR depth in the more textured half of their scene, 45\% of
those with valid depth. Within 0.35\,D of focus the curve sets the accuracy. Curves computed in
wave optics with one gain err as much as curves calibrated for every pair and band, and within 3\% once those are
given a gain. Curves of a geometric disc err 22\% more than ours, and still 8\% more with a gain
for every band. The joint fit adds nothing near focus, where one ratio with our curve errs as little;
conventional DfDD errs 40\% more. Read from the posterior, the
same fit errs 31\% less [23, 38], as little as a perceptron on the same band statistics
(10.8\% against 11.7\%); the foci are then measured at f/16 without depth, and calibrated foci give 11.2\%
(Sec.~\ref{S-sec:posterior}). At f/2.8--f/5.6
wave-optics and geometric curves agree within 4\% (Table~\ref{S-tab:estfull}).

\paragraph{What needs wave optics}
A geometric disc with a fitted radius and gain ($0.63R$, 1.25) errs 6\% more than ours near focus
(Table~\ref{tab:tests}). Over the 9.5\,px search, filtered by the confidence of its own fit, it errs 1.9 times ours
[1.6, 2.2], as one radius cannot place the turnover of every band. Only a radius that shrinks as
$1-f/f_c$, Stokseth's approximation of wave optics~\cite{stokseth1969}, stays within 8\% of ours
(Table~\ref{S-tab:attrctl}).

\paragraph{Calibration}
With every method refitted on fewer training scenes, which also calibrate the nine focus settings, our curves fitted on
two scenes, 90 images at five apertures with LiDAR depth, err 5\% more than with all 40 and 33\% more in the worst of
20 such fits (Fig.~\ref{S-fig:learning}). Calibrated curves need 5--10 scenes to match this and the perceptron
10--20.

\emph{Answer to Q2:} nearly, at f/16 on textured windows; a clear disc with one gain per aperture matches a
calibration of every pair and band and beats geometric curves even with a gain per band. Near focus a
geometric disc with a fitted radius and gain comes within 6\%; over the range of one pair it takes
an overlap that narrows with frequency.

\begin{figure*}[t]
\centering
\includegraphics[width=\textwidth]{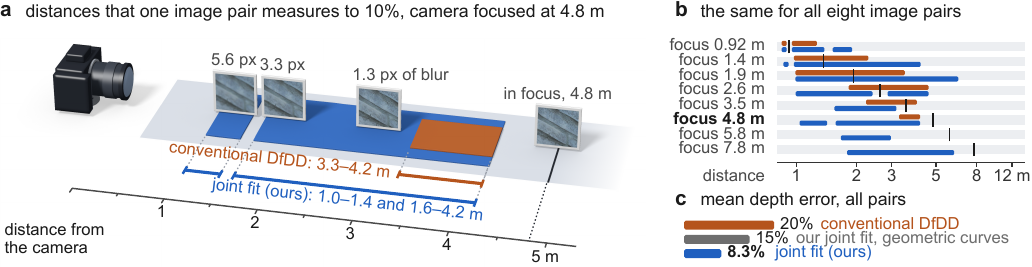}
\caption{\textbf{With the least confident 75\% of windows discarded, the joint fit measures distance to 10\% over
more distances than conventional DfDD} (post hoc). DSLR at f/16, held out, every window with LiDAR depth. (a)~Distances at which the mean relative depth error of one image pair stays below 10\%, for the pair whose two spans lie closest to the medians of the eight; cards: one window at the blur
it has there; at the focus neither reaches 10\%. (b)~The same for every pair (tick: its focus). (c)~The mean
error over all pairs and distances.}
\label{fig:depth}
\end{figure*}

\begin{table}[t]
\centering
\def\tyes{\textcolor{ForestGreen}{$\checkmark$}}\def\tund{$\circ$}\def\tno{\textcolor{BrickRed}{$\times$}}%
\caption{\textbf{Pre-specified tests that the text discusses}; Sec.~\ref{S-sec:protocol} lists every one with
its rule. Intervals are 95\%.}
\label{tab:tests}
\resizebox{\columnwidth}{!}{%
\setlength{\tabcolsep}{4pt}
\begin{tabular}{@{}lllll@{}}
\toprule
Data & Quantity & Criterion & Estimate & \\
\midrule
\multicolumn{5}{@{}l}{\emph{Does the coefficient fall as a clear disc predicts (Q1)?}} \\
prototype & cutoff scale, first file & 0.80--1.25 & 1.20 [1.13, 1.27] & \tund \\
prototype & cutoff scale, second file & 0.80--1.25 & 1.16 [1.07, 1.36] & \tund \\
phone & cells with a fall at 95; 65\,cycles/mm & 6 of 6 & 6; 5 & \tyes~at 95 \\
DSLR, test & band slope at 0.225\,cycles/px & 0.67, not 0.84 & 0.82 [0.73, 0.94] & \tno \\
DSLR, test & turnover ratio of the finest bands & above 0.69 & 0.72 [0.70, 0.76] & \tyes \\
DSLR, test & sign changes: clear-disc radius & 0.88--0.94, flat & 0.920 [0.912, 0.923] & \tund \\
\midrule
\multicolumn{5}{@{}l}{\emph{Does ours err less? Error of X over ours, in the measure of each rule (Q2, Q3)}} \\
prototype & X: geometric curves; one ratio & above 1 & $\ge$1.57 [1.19, 2.02] & \tyes \\
prototype & error of ours at 1.5--3.5; 3.5--9.5\,px & below 10\% & 4.7\%; 6.1\% & \tyes \\
prototype & X: geometric disc fitted per iris$^{\dagger}$ & above 1.05 & 3.03 [2.49, 3.77] & \tyes \\
prototype & X: geometric; conventional ($64\times64$\,px) & above 1.05 & $\ge$1.41 [1.28, 1.55] & \tyes \\
DSLR & X: disc of fitted radius & above 1.05 & 1.14 [1.09, 1.18] & \tyes \\
DSLR & X: fitted radius and gain & above 1.05 & 1.06 [1.03, 1.09] & \tund \\
DSLR, test & X: conventional DfDD & above 1.05 & 2.49 [1.97, 3.12] & \tyes \\
DSLR, test & X: geometric curves & above 1.05 & 1.77 [1.64, 1.92] & \tyes \\
\bottomrule
\multicolumn{5}{@{}l@{}}{\rule{0pt}{2.6ex}\tyes~confirmed, \tund~undecided, \tno~refuted. $^{\dagger}$The fitted radius ran to the bound of its search.} \\
\end{tabular}}
\end{table}

\subsection{Working range of one image pair (Q3, Q4)}
\label{sec:range}
DfDD cameras discard their least confident pixels and report the mean error of the rest; their working range is
where it stays below 5\%~\cite{luo2025fs} or 10\%~\cite{luo2025cod} of the depth. We do the same on every window
with LiDAR depth, with no selection by texture and the search extended to 9.5\,px of blur.
The joint fit uses the confidence of Sec.~\ref{sec:bw} and conventional DfDD the best of four measures of
image energy, chosen on the training folds (Sec.~\ref{S-sec:confidence}; post hoc).

\paragraph{Mean error with a confidence filter}
With nothing discarded no estimator is usable, at an MRE of 74--86\% (Table~\ref{S-tab:confpooled}). With the
least confident 75\% discarded, the joint fit has an MRE of 8.3\% [7.3, 9.5] and an MAE of 0.33\,m, conventional
DfDD 20.1\% and 0.98\,m, and the joint fit with geometric curves 14.6\% (Fig.~\ref{fig:depth}c). The gap is not a
choice of easy windows. On
the windows that our confidence keeps, conventional DfDD errs by 20.6\%, and on those that its own keeps, ours errs
by 13.7\% against 20.1\%. Most of the gap comes with the residual of the fit: with an image-energy
confidence the joint fit has 17.3\% [11.7, 24.3] and one ratio with our curve 14.7\%. The mean error stays below
10\% over a wider total span, in diopters, in all eight pairs, in seven with a paired interval above zero
(Fig.~\ref{fig:depth}a,~b). A confirmation on the 50 test scenes,
pre-specified and scored against depth from focus at f/2.8, gives 4.9\% for ours, 12.2\% for conventional DfDD
and 8.7\% with geometric curves (Table~\ref{tab:tests}).

\emph{Answer to Q3:} at f/16, with the least confident 75\% of windows discarded, one image pair measures distance
with a mean error of 8\%, against 20\% for conventional DfDD, a gap that a pre-specified test on the test scenes
confirms, and to 10\% over a wider span in seven of eight pairs. The range needs the confidence
(Sec.~\ref{sec:discussion}).

\paragraph{Texture dependence (Q4)}
We split windows into terciles by spectral slope. At f/2.8--f/5.6, near the OTF zeros, the relative bias of one
ratio with a calibrated curve differs by 22--62 points between the extreme terciles, against $-2$ to $+20$ for joint
fits with model curves; at f/16 we measure 0--2 points for one ratio and $-5$ [$-8$, $-1$] for ours
(Tables~\ref{S-tab:texpred},~\ref{S-tab:basetex}). On the prototype one ratio depends on the print. With the gains
of the first file it errs by 11.3\% in defocus on the second, with gains refitted there by 7.3\%.
\emph{Answer to Q4:} yes near the OTF zeros, where the joint fit removes most of the dependence, and between the
two prints of the prototype; at f/16 on the DSLR it is too small to resolve.

\subsection{Phone and DSLR (Q1)}
\label{sec:pixel}
\paragraph{Phone}
On 47 focal sweeps of a Pixel~3 (f/1.8)~\cite{herrmann2020}, dual-pixel views
summed~\cite{garg2019,punnappurath2020}, with bands that reach $0.093f_c$, the slope ratio of a fine to a
reference band falls below that of a constant coefficient
(Fig.~\ref{S-fig:evidence}). This was exploratory, so we pre-specified a confirmation on 351 sweeps in
other rooms before obtaining them (Sec.~\ref{S-sec:pixelconfirm}). It confirms the fall at 95\,cycles/mm, not at 65
(Table~\ref{tab:tests}).

\paragraph{DSLR}
\label{sec:zedd}
Where a band turns over, and where its energy is equal in two images so that their difference changes sign, do
not depend on a gain, a tone curve or a noise floor. A test of the turnovers, pre-specified on the 50 test scenes
with depth from focus at f/2.8 as the reference, is confirmed: the ratio of the turnovers of the two finest
bands, 0.72 [0.70, 0.76], lies above the 0.69 that a geometric disc of any radius can reach; a clear disc
gives 0.74--0.78. The sign changes
need no depth, since windows equally sharp in two near-focused images lie midway in diopters between the two foci
(Sec.~\ref{S-sec:signs}). In a pre-specified test on the test scenes, 14 sign changes at
0.15--0.24\,cycles/px give a clear disc of radius $0.920R$ [0.912, 0.923] to 1\%, where a geometric disc needs
a radius that falls from 0.76 to $0.63R$. The test is undecided (Table~\ref{tab:tests}), as its second criterion of
three, a clear-disc radius constant in frequency, is missed by 0.002. On the
validation scenes, with the camera noise removed, curves of a clear $0.92R$ disc fit the turnovers with a gain of
1.01, not 0.82.

\emph{Answer to Q1:} yes on a DfDD prototype, where the fall follows the cutoff of the iris; on the phone at
95\,cycles/mm only; on the DSLR, turnovers and sign changes exclude a geometric disc of one radius and the sign
changes give a clear disc of $0.92R$, while a test of the fall in the band slopes was refuted
(Sec.~\ref{sec:discussion}).

\section{Discussion and limitations}
\label{sec:discussion}

\looseness=-1 The evidence for the law comes from one
prototype, read from released files that do not document its light source, camera settings or distance origin; by
our rule its cutoff is undecided (Table~\ref{tab:tests}), and the sensor distance that accounts for it and for
the gains is inferred. On the
phone the fall appears at 95\,cycles/mm only, far steeper than for a clear disc and of unknown origin, and our
curves err 25\% more there than one calibrated ratio (Table~\ref{S-tab:real_est}). On the DSLR test scenes a pre-specified
test of the fall was refuted by a band slope of 0.82 [0.73, 0.94] at 0.225\,cycles/px against 0.62 on the
validation scenes and 0.84 for a constant coefficient, for a cause we have not found
(Sec.~\ref{S-sec:zedd}). A disc of $1.25R$, not $0.92R$, fits the in-focus loss, so our curves are a
calibrated approximation, not an identified pupil. The test scenes served six tests in sequence
(Sec.~\ref{S-sec:protocol}). The depth results come from one lens at f/16 and from JPEGs;
at f/2.8--f/5.6 our curves bring nothing over geometric ones. The range comes from the low bands and needs
the confidence: with nothing discarded the minimum of the fit gives no wider a span than conventional DfDD, and
with 75\% discarded the span includes the focus in three of eight pairs, in five for conventional DfDD
(Sec.~\ref{S-sec:conditions}). A perceptron errs 25\% less than the minimum of the fit near focus. The
confidence was designed
post hoc on the validation scenes and is confirmed on the test scenes only where depth from focus gives a
reference (53\% of their windows, within 7.4\,m); the posterior is exploratory, cross-validated over
scenes. The joint fit costs 2\,900 operations per pixel, against 36
in~\cite{luo2025cod} (Sec.~\ref{S-sec:flops}).

\section{Conclusion}
\label{sec:conclusion}

On a DfDD prototype the diffusion coefficient falls with frequency at three iris diameters,
along one curve in $f/f_c$; a smartphone confirms a fall at 95\,cycles/mm only, and on a DSLR, turnovers and
sign changes exclude a geometric disc of one radius. Narrow bands fitted jointly against clear-disc
curves measure distance from one image pair to 15\,mm on the prototype and, with the least confident 75\%
discarded, to 8\% on the DSLR, against 20\% for conventional DfDD. Code and test records will be
released.\label{sec:endtext}

\section*{LLM Usage Statement}
The author used Claude (Anthropic) to help with the research in this
project. The author and Claude both worked on the derivations, the search of related work, the code and the runs of
the simulations and of the experiments on public data, the figures, and the text of the paper and of the supplement,
and Claude and GPT (OpenAI) reviewed drafts in the manner of referees. The author directed the work, verified every
derivation, number and figure, and takes full responsibility for the content.

\clearpage
\newgeometry{margin=1in,headheight=12pt,headsep=25pt,footskip=30pt}
\onecolumn
\makeatletter
\iftutex
  \renewcommand{\encodingdefault}{TU}
  \renewcommand{\rmdefault}{lmr}
  \renewcommand{\sfdefault}{lmss}
  \renewcommand{\ttdefault}{lmtt}
\else
  \renewcommand{\encodingdefault}{T1}
  \renewcommand{\rmdefault}{lmr}
  \renewcommand{\sfdefault}{lmss}
  \renewcommand{\ttdefault}{lmtt}
\fi
\renewcommand{\familydefault}{\rmdefault}
\renewcommand{\bfdefault}{bx}
\DeclareRobustCommand\normalsize{%
   \@setfontsize\normalsize\@xipt{13.6}%
   \abovedisplayskip 11\p@ \@plus3\p@ \@minus6\p@
   \abovedisplayshortskip \z@ \@plus3\p@
   \belowdisplayshortskip 6.5\p@ \@plus3.5\p@ \@minus3\p@
   \belowdisplayskip \abovedisplayskip
   \let\@listi\@listI}
\DeclareRobustCommand\small{%
   \@setfontsize\small\@xpt\@xiipt
   \abovedisplayskip 10\p@ \@plus2\p@ \@minus5\p@
   \abovedisplayshortskip \z@ \@plus3\p@
   \belowdisplayshortskip 6\p@ \@plus3\p@ \@minus3\p@
   \def\@listi{\leftmargin\leftmargini
               \topsep 6\p@ \@plus2\p@ \@minus2\p@
               \parsep 3\p@ \@plus2\p@ \@minus\p@
               \itemsep \parsep}%
   \belowdisplayskip \abovedisplayskip}
\DeclareRobustCommand\footnotesize{%
   \@setfontsize\footnotesize\@ixpt{11}%
   \abovedisplayskip 8\p@ \@plus2\p@ \@minus4\p@
   \abovedisplayshortskip \z@ \@plus\p@
   \belowdisplayshortskip 4\p@ \@plus2\p@ \@minus2\p@
   \def\@listi{\leftmargin\leftmargini
               \topsep 4\p@ \@plus2\p@ \@minus2\p@
               \parsep 2\p@ \@plus\p@ \@minus\p@
               \itemsep \parsep}%
   \belowdisplayskip \abovedisplayskip}
\DeclareRobustCommand\scriptsize{\@setfontsize\scriptsize\@viiipt{9.5}}
\DeclareRobustCommand\tiny{\@setfontsize\tiny\@vipt\@viipt}
\DeclareRobustCommand\large{\@setfontsize\large\@xiipt{14}}
\DeclareRobustCommand\Large{\@setfontsize\Large\@xivpt{18}}
\DeclareRobustCommand\LARGE{\@setfontsize\LARGE\@xviipt{22}}
\DeclareRobustCommand\huge{\@setfontsize\huge\@xxpt{25}}
\DeclareRobustCommand\Huge{\@setfontsize\Huge\@xxvpt{30}}
\def\@listi{\leftmargin\leftmargini
            \parsep 4.5\p@ \@plus2\p@ \@minus\p@
            \topsep 9\p@   \@plus3\p@ \@minus5\p@
            \itemsep4.5\p@ \@plus2\p@ \@minus\p@}
\let\@listI\@listi
\def\@listii{\leftmargin\leftmarginii
             \labelwidth\leftmarginii
             \advance\labelwidth-\labelsep
             \topsep    4.5\p@ \@plus2\p@ \@minus\p@
             \parsep    2\p@   \@plus\p@  \@minus\p@
             \itemsep   \parsep}
\def\@listiii{\leftmargin\leftmarginiii
              \labelwidth\leftmarginiii
              \advance\labelwidth-\labelsep
              \topsep    2\p@ \@plus\p@\@minus\p@
              \parsep    \z@
              \partopsep \p@ \@plus\z@ \@minus\p@
              \itemsep   \topsep}
\normalfont\normalsize
\setlength\heavyrulewidth{.08em}
\setlength\lightrulewidth{.05em}
\setlength\cmidrulewidth{.03em}
\setlength\belowrulesep{.65ex}
\setlength\aboverulesep{.4ex}
\setlength\cmidrulekern{.5em}
\setlength\defaultaddspace{.5em}
\setlength\leftmargini{2.5em}
\setlength\leftmarginii{2.2em}
\setlength\leftmarginiii{1.87em}
\setlength\leftmarginiv{1.7em}
\setlength\leftmarginv{1em}
\setlength\leftmarginvi{1em}
\setlength\labelsep{.5em}
\setlength\labelwidth{\leftmargini}
\addtolength\labelwidth{-\labelsep}
\leftmargin\leftmargini
\setlength\parindent{17\p@}
\setlength\topskip{11\p@}
\setlength\maxdepth{.5\topskip}
\setlength\footnotesep{7.7\p@}
\setlength{\skip\footins}{10\p@ \@plus 4\p@ \@minus 2\p@}
\setlength\floatsep{12\p@ \@plus 2\p@ \@minus 2\p@}
\setlength\textfloatsep{20\p@ \@plus 2\p@ \@minus 4\p@}
\setlength\intextsep{12\p@ \@plus 2\p@ \@minus 2\p@}
\setlength\partopsep{3\p@ \@plus 1\p@ \@minus 1\p@}
\renewcommand{\topfraction}{0.85}\renewcommand{\bottomfraction}{0.5}\renewcommand{\textfraction}{0.1}
\renewcommand{\floatpagefraction}{0.7}\setcounter{topnumber}{3}\setcounter{totalnumber}{5}
\setcounter{bottomnumber}{1}
\fussy
\setlength\abovecaptionskip{10\p@}
\setlength\belowcaptionskip{6pt}
\long\def\@makecaption#1#2{%
  \vskip\abovecaptionskip
  \sbox\@tempboxa{#1: #2}%
  \ifdim \wd\@tempboxa >\hsize
    #1: #2\par
  \else
    \global \@minipagefalse
    \hb@xt@\hsize{\hfil\box\@tempboxa\hfil}%
  \fi
  \vskip\belowcaptionskip}
\setlist[itemize]{}
\setlist[enumerate]{}
\backrefparscanfalse
\renewcommand\section{\@startsection{section}{1}{\z@}%
                                   {-3.5ex \@plus -1ex \@minus -.2ex}%
                                   {2.3ex \@plus.2ex}%
                                   {\normalfont\Large\bfseries}}
\renewcommand\subsection{\@startsection{subsection}{2}{\z@}%
                                     {-3.25ex\@plus -1ex \@minus -.2ex}%
                                     {1.5ex \@plus .2ex}%
                                     {\normalfont\large\bfseries}}
\renewcommand\subsubsection{\@startsection{subsubsection}{3}{\z@}%
                                     {-3.25ex\@plus -1ex \@minus -.2ex}%
                                     {1.5ex \@plus .2ex}%
                                     {\normalfont\normalsize\bfseries}}
\renewcommand\paragraph{\@startsection{paragraph}{4}{\z@}%
                                    {3.25ex \@plus1ex \@minus.2ex}%
                                    {-1em}%
                                    {\normalfont\normalsize\bfseries}}
\setcounter{section}{0}
\setcounter{figure}{0}
\setcounter{table}{0}
\setcounter{equation}{0}
\setcounter{proposition}{0}
\renewcommand{\thesection}{S\arabic{section}}
\renewcommand{\thetable}{S\arabic{table}}
\renewcommand{\thefigure}{S\arabic{figure}}
\renewcommand{\theequation}{S\arabic{equation}}
\renewcommand{\theproposition}{S\arabic{proposition}}
\def\theHsection{supp.\arabic{section}}
\def\theHtable{supp.\arabic{table}}
\def\theHfigure{supp.\arabic{figure}}
\def\theHequation{supp.\arabic{equation}}
\def\theHproposition{supp.\arabic{proposition}}
\newcommand{\Para}[2][]{\paragraph*{#2.}\phantomsection\addcontentsline{toc}{paragraph}{#2}%
  \ifx\relax#1\relax\else\label{#1}\fi\ignorespaces}
\newcommand\SUPPstarttoc{}
\renewcommand\tableofcontents{\begingroup\hypersetup{linkcolor=black}%
  \section*{Contents}\hypertarget{supp-toc}{}\pdfbookmark[1]{Contents}{supp-contents}%
  \small\setlength{\parskip}{0pt}\raggedright
  \let\SUPP@contentsline\contentsline\def\contentsline##1##2##3##4{}%
  \def\SUPPstarttoc{\let\contentsline\SUPP@contentsline}%
  \@starttoc{toc}\par\endgroup}
\renewcommand*\l@section[2]{\par\leftskip=0pt\relax\addvspace{4pt}%
  \noindent{\normalsize\bfseries\setlength\@tempdima{2.2em}#1\nobreak\hfill\nobreak#2}\par\nobreak
  \leftskip=2.2em\relax}
\renewcommand*\l@paragraph[2]{\ifvmode\noindent\fi\mbox{#1\nobreak\enspace{\color{gray}\scriptsize#2}}%
  \hskip 0.9em plus 0.5em minus 0.2em\relax}
\def\ps@supp{\let\@mkboth\@gobbletwo\let\@oddhead\@empty\let\@evenhead\@empty
  \def\@oddfoot{\reset@font\footnotesize\hyperlink{supp-toc}{Contents}\hfil\normalsize\thepage\hfil
    \footnotesize\phantom{Contents}}\let\@evenfoot\@oddfoot}
\global\@topnum\z@
\makeatother
\pagestyle{supp}
\pdfbookmark[0]{Supplementary material}{supp-title}
\null
\vskip 2em
\begin{center}
{\LARGE Depth from Differential Defocus beyond Geometric Optics\\[4pt]\large Supplementary material\par}
\vskip 1.5em
{\large\lineskip .5em
\begin{tabular}[t]{c}
Junjie Luo
\end{tabular}\par}
\vskip 1em
{\large}
\end{center}
\par
\vskip 1.5em
\thispagestyle{plain}
\addtocontents{toc}{\protect\SUPPstarttoc}
\tableofcontents

\section{Overview}
\label{S-sec:overview}

Numbers with the prefix S (Sec.~S1, Eq.~(S1), Table~S1, Fig.~S1) refer to this document; all others refer to the
main paper. The contents above list every paragraph of this document. Each entry there, each reference to a section,
table, figure or equation, and the word ``Contents'' at the foot of every page are links, and the bookmarks of a PDF
reader show the same outline. The companion paper~\cite{S-luo2026theory}
derives the OTF of Gaussian pupils and the other DfDD cues in wave optics, and proves that only Gaussian pupils make
defocus an exact diffusion. The guide below maps each part of the main paper to the material that supports it.

\begingroup
\small
\renewcommand{\arraystretch}{1.18}
\noindent\begin{tabular}{@{}>{\raggedright\arraybackslash}p{3.9cm}>{\raggedright\arraybackslash}p{12.0cm}@{}}
\toprule
main paper & where to find the supporting material \\
\midrule
Secs.~\ref{sec:background} and~\ref{sec:theory}, Fig.~\ref{fig:overlap}: heat equation, OTF as an overlap of
two pupils, Lemmas~\ref{lem:small} and~\ref{lem:disc}, Proposition~\ref{prop:ls}
& Sec.~\ref{S-sec:gauss}, defocus as diffusion on a real texture (Fig.~\ref{S-fig:dbc}). Sec.~\ref{S-sec:hard}: the
\hyperref[S-par:proofsmall]{proofs}, exact values of the coefficient (Tables~\ref{S-tab:M} and~\ref{S-tab:map}), the
least-squares bias (Table~\ref{S-tab:ls}), \hyperref[S-par:errs]{where conventional DfDD errs} (Fig.~\ref{S-fig:errmap};
main-paper Fig.~\ref{fig:overlap}c shows its panel b on our five bands) and \hyperref[S-par:aberr]{aberrations}. Gaussian pupils and the other cues are in the companion paper. \\
Sec.~\ref{sec:bw}, Fig.~\ref{fig:method}: depth in sub-bands
& Sec.~\ref{S-sec:bw}: \hyperref[S-par:curves]{through-focus curves} (Fig.~\ref{S-fig:response}),
Proposition~\ref{S-prop:bw} and \hyperref[S-par:proofbw]{its proof}, the \hyperref[S-par:specratio]{relation to the
spectral ratio}, the \hyperref[S-par:simsetup]{simulation}
(Fig.~\ref{S-fig:sim}, Tables~\ref{S-tab:bw_summary} and~\ref{S-tab:bw_depth}), \hyperref[S-sec:flops]{computational cost}
(Table~\ref{S-tab:flops}), band ablation (Table~\ref{S-tab:bandabl}) and \hyperref[S-par:cascade]{a cascade of two cheap
ratios} (Table~\ref{S-tab:cascade}); the window of Fig.~\ref{fig:method} is described in
\hyperref[S-par:methodfig]{Sec.~\ref{S-sec:zedd}}. Sec.~\ref{S-sec:posterior}: the posterior
(Table~\ref{S-tab:posterior}). \\
Sec.~\ref{sec:proto}: Q1 on a DfDD prototype, Fig.~\ref{fig:proto}
& Sec.~\ref{S-sec:cod}: the record, its \hyperref[S-par:codoutcome]{outcome} (Table~\ref{S-tab:cod}),
\hyperref[S-par:codposthoc]{why the scale comes out high}, the \hyperref[S-par:codamend]{second file} and
\hyperref[S-par:codamendout]{its outcome} (Table~\ref{S-tab:coddepth}), the \hyperref[S-par:codfig]{slopes against the
closed form} (Table~\ref{S-tab:codcells}), \hyperref[S-par:coddist]{depth in units of distance}
(Tables~\ref{S-tab:coddist} and~\ref{S-tab:codmae}) and the \hyperref[S-sec:codcontrols]{controls with an effective
aperture and with small windows} (Tables~\ref{S-tab:codctlA} and~\ref{S-tab:codctlB}, Fig.~\ref{S-fig:protomaps}). \\
Sec.~\ref{sec:depth}: Q2, Table~\ref{tab:est}
& The full table in diopters (Table~\ref{S-tab:estfull}) and in distance (Tables~\ref{S-tab:conft1} and~\ref{S-tab:conft1b}), the
\hyperref[S-sec:attrctl]{attribution controls}
(Table~\ref{S-tab:attrctl}), \hyperref[S-par:range]{error against defocus and training scenes}
(Fig.~\ref{S-fig:depth_curves}),
\hyperref[S-par:depth]{depth from one image pair} (Table~\ref{S-tab:est_bins}, Fig.~\ref{S-fig:zedd_est}),
\hyperref[S-par:baselines]{baselines and ablations} (Tables~\ref{S-tab:baselines} and~\ref{S-tab:basetex}), the
\hyperref[S-par:srcontrol]{spectral ratio with the same curves} (Table~\ref{S-tab:srcontrol}),
\hyperref[S-par:attribution]{where each improvement comes from} (Table~\ref{S-tab:attribution}),
\hyperref[S-par:perpair]{errors per pair} (Table~\ref{S-tab:perpair}), \hyperref[S-par:budget]{what a calibration needs}
(Table~\ref{S-tab:budget}), \hyperref[S-par:training]{training data}
(Fig.~\ref{S-fig:learning}, Table~\ref{S-tab:learning}), \hyperref[S-par:weights]{band weights} (Table~\ref{S-tab:weights}),
\hyperref[S-par:texpred]{texture dependence} (Table~\ref{S-tab:texpred}), \hyperref[S-par:transfer]{depth on the test
scenes} (Tables~\ref{S-tab:transfer} and~\ref{S-tab:transfer_range}), \hyperref[S-par:learned]{physical and learned depth}
and \hyperref[S-par:correction]{the correction of learned depth} (Tables~\ref{S-tab:learned} and~\ref{S-tab:sameinput}). \\
Sec.~\ref{sec:range}: Q3 and Q4, the working range of one image pair, Fig.~\ref{fig:depth}
& \hyperref[S-sec:confidence]{Depth with a confidence} (Tables~\ref{S-tab:confpooled}--\ref{S-tab:confpair}) and its
\hyperref[S-sec:conftest]{confirmation on the test scenes} (Table~\ref{S-tab:conftest}), the
\hyperref[S-sec:depthunits]{working range by the median error} (Fig.~\ref{S-fig:rangemedian}, Tables~\ref{S-tab:depthrange}
and~\ref{S-tab:depthunits}), the \hyperref[S-par:range]{working range in defocus}, its
\hyperref[S-par:depthold]{drawing as distances} (Fig.~\ref{S-fig:depthold}) and
\hyperref[S-par:rangepair]{its value per pair}
(Tables~\ref{S-tab:range_pairs}, \ref{S-tab:range_energy}, \ref{S-tab:texcrit} and~\ref{S-tab:range}),
\hyperref[S-par:gainfar]{a gain chosen far from focus}, \hyperref[S-par:coverage]{error against coverage}
(Fig.~\ref{S-fig:coverage}, Table~\ref{S-tab:coverage}) and \hyperref[S-par:lowband]{conventional DfDD on low bands}
(Table~\ref{S-tab:lowband}). \\
Sec.~\ref{sec:pixel}: Q1 on the phone, Fig.~\ref{S-fig:evidence}a
& Sec.~\ref{S-sec:pixel}: the \hyperref[S-par:pixref]{reference depth}, \hyperref[S-par:pixcontrols]{controls}
(Table~\ref{S-tab:alias}), \hyperref[S-par:pixratios]{band ratios} (Table~\ref{S-tab:real_xb}, Fig.~\ref{S-fig:crossband}),
the \hyperref[S-par:pixpupil]{pupil fit} (Tables~\ref{S-tab:aberr} and~\ref{S-tab:aberr_par}, Fig.~\ref{S-fig:aberr}), the
\hyperref[S-sec:pixelconfirm]{pre-registered confirmation on new rooms} and \hyperref[S-par:pixoutcome]{its outcome}
(Tables~\ref{S-tab:pixelconfirm} and~\ref{S-tab:pixelconfirm_all}), the \hyperref[S-par:pixdup]{sweeps that the archives
hold twice}, the \hyperref[S-par:pixviews]{pre-specified test on the two dual-pixel views} and
\hyperref[S-par:pixviewsout]{its outcome} (Tables~\ref{S-tab:viewsA}, \ref{S-tab:viewsB} and~\ref{S-tab:viewsC}), and
\hyperref[S-par:pixdepth]{depth on the phone} (Table~\ref{S-tab:real_est}, Fig.~\ref{S-fig:realerr}). \\
Sec.~\ref{sec:zedd}: Q1 on the DSLR, Fig.~\ref{S-fig:evidence}b and c
& Sec.~\ref{S-sec:zedd}: \hyperref[S-par:focuscal]{focus calibration} (Table~\ref{S-tab:calib}),
\hyperref[S-par:robust]{robustness} (Table~\ref{S-tab:robust}), the \hyperref[S-par:chain]{measurement chain}
(Tables~\ref{S-tab:pipesim} and~\ref{S-tab:nonlin}), \hyperref[S-par:allpairs]{all image pairs} (Table~\ref{S-tab:allpairs}),
the \hyperref[S-par:nuis]{nuisances} and the \hyperref[S-par:radius]{pupil radius} (Tables~\ref{S-tab:spread},
\ref{S-tab:fig3b} and~\ref{S-tab:turnover}), the \hyperref[S-par:dslrconfirm]{pre-registered confirmation},
\hyperref[S-par:dslroutcome]{its outcome}, \hyperref[S-par:twosets]{how far the two sets differ} and the
\hyperref[S-par:posthoc]{post hoc diagnostics} (Table~\ref{S-tab:confirm}), the \hyperref[S-par:tone]{tone curve} and
\hyperref[S-par:tonebracket]{a registered bracket of tone curves} (Table~\ref{S-tab:tonecurve}),
\hyperref[S-par:pair4]{where the difference sits}, the
\hyperref[S-par:controls]{controls at f/2.8--f/5.6} (Table~\ref{S-tab:controls}), \hyperref[S-par:infocus]{in focus}
(Table~\ref{S-tab:infocus}), the \hyperref[S-par:scale]{absolute scale} (Table~\ref{S-tab:eiv}) and
\hyperref[S-par:onepupil]{one effective pupil}. Sec.~\ref{S-sec:signs}: turnovers and sign changes
(Table~\ref{S-tab:signs}). \\
Fig.~\ref{fig:teaser} and the depth maps
& \hyperref[S-par:maps]{Depth maps} (Tables~\ref{S-tab:maps} and~\ref{S-tab:maps_vergence}, Fig.~\ref{S-fig:supp_maps}),
\hyperref[S-par:pairmaps]{maps from one image pair} (Figs.~\ref{S-fig:maps}, \ref{S-fig:maps_conf},
Table~\ref{S-tab:pairmaps}) and
\hyperref[S-par:mapsmore]{more estimators on the same maps} (Fig.~\ref{S-fig:maps_more}). \\
Sec.~\ref{sec:discussion}: limits, cost, transfer, the phone
& Sec.~\ref{S-sec:conditions}: the limits in full. \hyperref[S-sec:flops]{Computational cost},
\hyperref[S-par:transfer]{transfer to the test scenes} and
\hyperref[S-par:pixdepth]{depth on the phone} (Table~\ref{S-tab:real_est}). \\
Tests fixed in advance & Sec.~\ref{S-sec:protocol} (Table~\ref{S-tab:protocol}); three further tests on the DSLR in
Sec.~\ref{S-sec:furthertests}. \\
Code and data & Sec.~\ref{S-sec:repro}. \\
\bottomrule
\end{tabular}
\par\endgroup
\medskip

Units: defocus $\psi$ in diopters (D), spatial frequency in cycles/mm on the sensor or cycles/px, lengths in mm
unless stated. ``Broadband'' estimates are conventional DfDD: one ratio of the image change to the Laplacian over a wide
band, as in Eq.~\eqref{eq:depth_intro} of the main paper. As in the main paper, a clear disc is a uniform circular
pupil without aberration, and the geometric disc is the same aperture in geometric optics, whose point-spread function
is a pillbox. Table~\ref{S-tab:regimes} summarizes what the three models of
defocus imply for DfDD.

\begin{table}[htbp]
\centering
\caption{\textbf{What three models of defocus imply for DfDD, near focus.} $f_c$ is the diffraction cutoff.}
\label{S-tab:regimes}
\begin{tabular}{@{}>{\raggedright\arraybackslash}p{2.4cm}>{\raggedright\arraybackslash}p{3.3cm}%
>{\raggedright\arraybackslash}p{3.0cm}>{\raggedright\arraybackslash}p{4.6cm}@{}}
\toprule
& geometric & \multicolumn{2}{c}{wave optics} \\
\cmidrule(l){3-4}
& optics & Gaussian pupil & circular aperture \\
\midrule
heat equation exact & Gaussian aperture only$^{\ast}$ & yes (Eq.~\eqref{eq:gauss_otf}) & no~\cite{S-luo2026theory} \\[2pt]
coefficient & constant & constant & falls, $\approx1-\frac{20}{3\pi}\frac{f}{f_c}$ (Lemma~\ref{lem:disc}) \\[2pt]
texture bias & none & none & yes (Proposition~\ref{prop:ls}) \\[2pt]
calibration & constants $\alpha$, $\beta$ & constants $\alpha$, $\beta$ & a curve per band (main-paper Sec.~\ref{sec:bw}) \\
\bottomrule
\end{tabular}

\vspace{2pt}
{\footnotesize $^{\ast}$For other apertures the heat equation holds only in the limit of small blur; its error grows
with the blur.}
\end{table}

\section{Tests fixed in advance and their records}
\label{S-sec:protocol}

The main paper calls a test \emph{pre-specified} if its data, statistic and decision rule were written down before
the data it decides on were analyzed. Each record is a dated text file on our own machine, kept with a list of the
SHA-256 of the scripts and inputs it freezes; every list was written before the run it covers and checked after it. None carries
a time stamp from a third party, so the records show what was fixed and cannot prove to a reader when
(\hyperref[S-par:localrecords]{Sec.~\ref{S-sec:pixel}}). Two of the records are titled preregistrations, and the sections
on the phone and on the DSLR call their tests registered; the word means nothing more there. Table~\ref{S-tab:protocol}
lists every test, what we had seen when each rule was fixed, and the outcomes. Main-paper Table~\ref{tab:tests}
gives the outcomes of the tests that bear on the main text. Of the others, the tests on the two dual-pixel views and
on all eight pairs at f/16 are described in Secs.~\ref{S-sec:pixel} and~\ref{S-sec:zedd}, and three more in
Sec.~\ref{S-sec:furthertests}. The records and hash lists will be released with the code (Sec.~\ref{S-sec:repro}).

\begin{table}[htbp]
\centering
\caption{\textbf{The tests whose rules were fixed in advance.} Dates are those of the records (local files, 2026).
``Seen before'' names the data that had been analyzed when the rule was written.}
\label{S-tab:protocol}
\small
\renewcommand{\arraystretch}{1.15}
\begin{tabular}{@{}>{\raggedright\arraybackslash}p{3.3cm}>{\raggedright\arraybackslash}p{1.3cm}%
>{\raggedright\arraybackslash}p{4.3cm}>{\raggedright\arraybackslash}p{5.9cm}@{}}
\toprule
test & record & seen before & outcome by the rule \\
\midrule
Prototype, cutoff of Lemma~\ref{lem:disc}, first file (Sec.~\ref{S-sec:cod}) & 13 Nov.
& the file structure, stored settings, the released code and two contact sheets; no spatial-frequency statistic
& undecided: cutoff scale 1.20 [1.13, 1.27], tolerance 0.80--1.25 \\
Prototype, second file (amendment, \hyperref[S-par:codamend]{Sec.~\ref{S-sec:cod}}) & 13 Nov.
& the first file in full; of the second, its settings, image means and one contact sheet
& cutoff undecided (1.16 [1.07, 1.36]); depth from one pair: the three rules confirmed
(Table~\ref{S-tab:coddepth}) \\
Prototype, an effective aperture (\hyperref[S-sec:codcontrols]{Sec.~\ref{S-sec:cod}}) & 14 Nov.
& both files in full with whole crops; no estimate with a geometric disc of another radius
& confirmed by the rule (3.03 [2.49, 3.77]), with the fitted radius at the end of its search; radius alone 1.55
[1.03, 2.51], undecided by the same margin (Table~\ref{S-tab:codctlA}) \\
Prototype, windows of $64\times64$ pixels (\hyperref[S-sec:codcontrols]{Sec.~\ref{S-sec:cod}}) & 14 Nov.
& the same; no estimate from a window
& confirmed against geometric curves (1.41 [1.28, 1.55]) and against conventional DfDD (3.06 [2.71, 3.48];
Table~\ref{S-tab:codctlB}) \\
DSLR depth at f/16, an effective aperture? (\hyperref[S-sec:attrctl]{Sec.~\ref{S-sec:zedd}}) & 14 Nov.
& all results of this document on the validation scenes; no estimate with a fitted radius
& fitted radius: confirmed (1.14 [1.09, 1.18]); fitted radius and gain: undecided (1.06 [1.03, 1.09], margin 1.05) \\
DSLR depth at f/16 with a confidence filter, test scenes (\hyperref[S-sec:conftest]{Sec.~\ref{S-sec:zedd}}) & 14 Nov.
& the validation scenes in full, with the confidence designed on them; of the test scenes the coefficient test and
defocus errors of the joint fits, with a gate on the band energy; no distance error, no estimate of conventional DfDD
& confirmed against conventional DfDD (2.49 [1.97, 3.12]) and against geometric curves (1.77 [1.64, 1.92];
Table~\ref{S-tab:conftest}) \\
Phone, fall on new rooms (\hyperref[S-sec:pixelconfirm]{Sec.~\ref{S-sec:pixel}}) & 8 Nov.
& the 47 test captures; no image of the training split
& confirmed at 95 cycles/mm in all six cells, not at 65 cycles/mm in one cell \\
Phone, the two dual-pixel views (\hyperref[S-par:pixviews]{Sec.~\ref{S-sec:pixel}}) & 12 Nov.
& the confirmation above, on the sum of the two views; the views apart on no training capture
& disparity: holds in all 18 cells; law of total variance: undecided (\hyperref[S-par:pixviewsout]{outcome}) \\
DSLR, all eight pairs at f/16 (\hyperref[S-par:allpairs]{Sec.~\ref{S-sec:zedd}}) & --
& the validation scenes on other statistics; fixed before this test was run on them
& undecided: $T=+18$ [$-13$, $+75$] (Table~\ref{S-tab:allpairs}) \\
DSLR, small steps on the test scenes (\hyperref[S-par:dslrconfirm]{Sec.~\ref{S-sec:zedd}}) & 7--10 Nov.
& the 50 validation scenes; no test scene (four amendments before the run, a fifth after it for a post hoc bracket)
& refuted: 0.82 [0.73, 0.94] against 0.84 for a constant coefficient and 0.67 for wave optics \\
\bottomrule
\end{tabular}
\end{table}

\begin{table}[htbp]
\addtocounter{table}{-1}
\renewcommand{\theHtable}{\thetable.cont}
\centering
\caption{\textbf{The tests whose rules were fixed in advance} (continued): five more tests on the test scenes of the DSLR,
fixed on 14--16 Nov.}
\small
\renewcommand{\arraystretch}{1.15}
\begin{tabular}{@{}>{\raggedright\arraybackslash}p{3.3cm}>{\raggedright\arraybackslash}p{1.3cm}%
>{\raggedright\arraybackslash}p{4.3cm}>{\raggedright\arraybackslash}p{5.9cm}@{}}
\toprule
test & record & seen before & outcome by the rule \\
\midrule
DSLR, turnover ratio of the two finest bands, test scenes (Sec.~\ref{S-sec:signs}) & 14 Nov.
& the validation scenes in full, with this ratio on them; of the test scenes the refuted test above and its
diagnostics near focus; no turnover of a band
& confirmed by a margin of 0.007: 0.719 [0.697, 0.757], geometric discs at most 0.690, wave optics from 0.738 \\
DSLR, two constants per pair, test scenes (\hyperref[S-par:twocalib]{Sec.~\ref{S-sec:furthertests}}) & 14 Nov.
& the validation scenes in full; of the test scenes the tests above and the errors of the uncalibrated fits; no
estimate with a calibrated constant
& confirmed: 0.782 [0.657, 0.891], rule: interval below 1 \\
DSLR, astigmatism without depth, test scenes (\hyperref[S-par:astig]{Sec.~\ref{S-sec:furthertests}}) & 15 Nov.
& the validation scenes in full, with the profile measured on them; of the test scenes the tests above; no
orientation statistic
& both criteria met: 1.026 [0.992, 1.054] in 0.85--1.15 and 0.725 [0.657, 0.792] in 0.50--1.10 \\
DSLR, noise floor from camera constants, test scenes (\hyperref[S-par:camerafloor]{Sec.~\ref{S-sec:furthertests}}) & 15 Nov.
& the validation scenes in full; of the test scenes the tests above
& fixed, not run \\
DSLR, sign changes without depth, test scenes (Sec.~\ref{S-sec:signs}) & 16 Nov.
& the validation scenes in full, with a dry run of the test and its reproduction by a second implementation; of the
test scenes the tests above; no narrow-band energy
& undecided: radius 0.920 [0.912, 0.923] in 0.88--0.94; slope of the clear disc $-0.047$ [$-0.062$, $-0.021$]
outside $\pm0.06$ by 0.002; slope of the geometric disc $-0.252$ [$-0.272$, $-0.224$] below $-0.10$
(Table~\ref{S-tab:signs}) \\
\bottomrule
\end{tabular}
\end{table}

\Para{What was not fixed in advance}
The choice of the bands, of the window size and of the texture selection on the DSLR, the first analysis of the phone
(47 captures), the small-step subset on the validation scenes, every row of main-paper Table~\ref{tab:est} other
than the attribution controls, the working range in defocus and in distance, the confidence of the joint fit and every error
filtered by it on the validation scenes (\hyperref[S-sec:confidence]{Depth with a confidence}), the mean errors on the
prototype (Table~\ref{S-tab:codmae}), and all diagnostics of this document that are marked post hoc. Main-paper Table~\ref{tab:est} is cross-validated over scenes, with every calibration
fitted on the training scenes of a fold, but its estimators were developed on the same 50 scenes; the depth errors on
the test scenes (\hyperref[S-par:transfer]{Sec.~\ref{S-sec:zedd}}), their \hyperref[S-sec:conftest]{confirmation with a
confidence filter} and the depth errors on the second file of the prototype (Table~\ref{S-tab:coddepth}) are the checks on data
that the development did not see.

\Para{Data and selections in one place}
\emph{Prototype}: the released linear-slide files of the prototype of~\cite{S-luo2025cod}, whole crops of
$400\times400$ pixels and, in one control, 16 windows of $64\times64$ pixels, three iris diameters, pairs of optical powers 0.06\,D apart; intervals from 2\,000
resamples of the slide positions in blocks of three. \emph{DSLR}: the 50 validation scenes of ZEDD~\cite{S-zuo2026} with
LiDAR depth, windows of $64\times64$ pixels, five folds over scenes; the main selection keeps windows with uniform
LiDAR depth in the more textured half of their scene (41.6\% of the samples over the 9.5\,px search), the gate keeps
the windows whose 0.10\,cycles/px energy exceeds a multiple of its first percentile in the scene, chosen on the
training folds (47.6\%), and ``all windows'' are those with valid LiDAR depth, of which a confidence discards a fixed share by a quantity
and a threshold chosen on the training folds (\hyperref[S-sec:confidence]{Depth with a confidence}); intervals from
2\,000 resamples of the scenes. \emph{Phone}: focal sweeps of Learning to Autofocus~\cite{S-herrmann2020}, 47 test captures and 351 training
captures in other rooms; intervals from two-stage resamples of rooms and captures.

\section{Defocus as diffusion}
\label{S-sec:gauss}

\begin{figure}[htbp]
\centering
\includegraphics[width=0.62\linewidth]{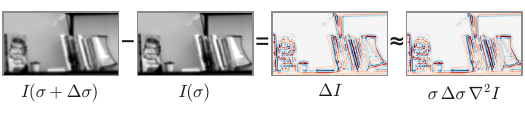}
\caption{\textbf{Defocus as diffusion, on a real texture.} Blurring an image slightly more, from $\sigma$ to
$\sigma+\Delta\sigma$, changes it by a multiple of its own Laplacian, as the heat equation predicts. The ratio of the
change to the Laplacian gives the defocus level $\sigma$, and hence depth. Crop of a ZEDD image with Gaussian blur of
$\sigma=1.6$ and $2.0$\,px; red and blue: positive and negative values on the same scale.}
\label{S-fig:dbc}
\end{figure}

Fig.~\ref{S-fig:dbc} shows the heat equation of main-paper Sec.~\ref{sec:background} on a real texture. The OTF of
Gaussian pupils in wave optics (main-paper Eq.~\eqref{eq:gauss_otf}) is derived in the companion
paper~\cite{S-luo2026theory}, which also shows that only Gaussian pupils make defocus an exact diffusion.

\section{Circular apertures}
\label{S-sec:hard}

\Para[S-par:proofsmall]{\texorpdfstring{Proof of Lemma~\ref{lem:small}}{Proof of the small-defocus lemma}}
For a real, nonnegative, centrally symmetric pupil the overlap $O_{\va}$ is real, nonnegative and even in $\vu$.
Expanding the Fourier transform in~\eqref{eq:otf} in $\psi$, the odd moments vanish and
\begin{equation}
  \log H(\vf;\psi)=\log H(\vf;0)-\tfrac12\Big(\frac{2\pi\psi|\va|}{\lambda}\Big)^2M(\va)+O(\psi^4),
\end{equation}
where $M(\va)$ is the second moment along $\va/|\va|$ of the overlap normalized to unit mass. With $|\va|/\lambda=s|\vf|$,
$\partial_\psi\log H=-4\pi^2s^2|\vf|^2\psi\,M(\va)+O(\psi^3)$, so $c(\vf;0^+)=M(\va)$. \hfill$\square$

\Para[S-par:proofdisc]{\texorpdfstring{Proof of Lemma~\ref{lem:disc}}{Proof of the circular-aperture lemma}}
Two discs of radius $R$ centered at $\pm\va/2$ overlap in a lens-shaped region. Take $R=1$, so that the centers
lie at $\pm q$ with $q=a/2=f/f_c$. A chord of the region perpendicular to $\va$, at distance $x$ from the center,
has the half-length $\sqrt{1-(|x|+q)^2}$, so with $t=|x|+q$ the second moment of the region along $\va$ and its
area are
\begin{equation*}
\begin{aligned}
  4\int_q^1(t-q)^2\sqrt{1-t^2}\,\dd t&=\tfrac12\Big[(1+4q^2)\arccos q-\tfrac q3(13+2q^2)\sqrt{1-q^2}\Big],\\
  4\int_q^1\sqrt{1-t^2}\,\dd t&=2\Big[\arccos q-q\sqrt{1-q^2}\Big].
\end{aligned}
\end{equation*}
Their ratio, divided by its value $1/4$ at $q=0$, is the first line of~\eqref{eq:t4}. For the second line take
polar coordinates about the center of the region,
with $\theta$ measured from $\va$; its boundary is $r_{\max}(\theta)=R-\tfrac a2|\cos\theta|+O(a^2)$. Its second
moment along $\va$ is $\int\cos^2\theta\,r_{\max}^4/4\dd\theta=\pi R^4/4-\tfrac43R^3a+O(a^2)$, and its area is
$\pi R^2-2Ra+O(a^2)$. The ratio is $\tfrac{R^2}{4}\big[1-\tfrac{10}{3\pi}\tfrac aR+O((a/R)^2)\big]$. With
$a/R=2f/f_c$ this gives the second line of~\eqref{eq:t4}. \hfill$\square$

\noindent Stokseth's empirical approximation of the defocused OTF~\cite{S-stokseth1969} puts the defocus in an argument
proportional to $q(1-q)$, $q=f/f_c$, which gives a curvature proportional to $(1-q)^2\approx1-2q$; the exact slope is
$20/(3\pi)\approx2.12$. Table~\ref{S-tab:M} gives exact values. At finite defocus (Table~\ref{S-tab:map}), diffraction
lowers $c$ at high frequencies when the blur is small; as the geometric blur grows, the non-Gaussian shape of the
pillbox raises $c$ again. The two effects cancel along a curve in the $(\psi,f)$ plane.

\begin{table}[htbp]
\centering
\caption{$M(a)/M(0)$ for a disc: exact (second moment of the overlap) against first order.}
\label{S-tab:M}
\begin{tabular}{ccccccccc}
\toprule
$f/f_c$ & 0.010 & 0.025 & 0.050 & 0.075 & 0.100 & 0.150 & 0.200 & 0.250 \\
\midrule
exact & 0.9790 & 0.9478 & 0.8972 & 0.8481 & 0.8006 & 0.7101 & 0.6257 & 0.5470 \\
first order & 0.9788 & 0.9469 & 0.8939 & 0.8408 & 0.7878 & 0.6817 & 0.5756 & 0.4695 \\
\bottomrule
\end{tabular}
\end{table}

\begin{table}[htbp]
\centering
\caption{Validity map for a disc ($R=1$\,mm): $c(\vf;\psi)/(R^2/4)$. Entries are omitted where the geometric blur
times the frequency, $R\psi a/\lambda$, exceeds 0.5.}
\label{S-tab:map}
\begin{tabular}{rcccc}
\toprule
$\psi$ [D] & $a/R=0.05$ & $0.10$ & $0.20$ & $0.40$ \\
\midrule
0.05 & 0.948 & 0.897 & 0.801 & 0.626 \\
0.20 & 0.948 & 0.899 & 0.806 & 0.639 \\
0.50 & 0.951 & 0.908 & 0.837 & 0.729 \\
1.00 & 0.960 & 0.944 & 0.986 & -- \\
2.00 & 1.001 & 1.144 & -- & -- \\
\bottomrule
\end{tabular}
\end{table}

\Para[S-par:proofls]{\texorpdfstring{Proof of Proposition~\ref{prop:ls}}{Proof of the least-squares proposition}}
The least-squares fit of $G*I_\rho=-M_0s^2\psi\,\Lap(G*I)$ over a window gives
$\hat\psi=-\langle G*I_\rho,\Lap G*I\rangle/(M_0s^2\|\Lap G*I\|^2)$. For a large window, Parseval's theorem turns the
inner products into sums over frequencies weighted by the texture spectrum $S_T$ and the pre-filter $|G|^2$, with
$\widehat{I_\rho}=\partial_\rho H\,\hat P$, $\hat I=H\hat P$ and $\widehat{\Lap}=-4\pi^2|\vf|^2$. This gives the ratio
in Proposition~\ref{prop:ls}. For a Gaussian pupil with $M_0=A^2$, $\partial_\rho H=4\pi^2s^2\psi A^2|\vf|^2H$ at
every frequency, so the ratio is 1 for every $S_T$. \hfill$\square$

\noindent Table~\ref{S-tab:ls} evaluates the ratio for textures with $S_T\propto|\vf|^{-\eta}$, a Gaussian
pre-filter of 1\,px and a band limit at the pixel Nyquist frequency, using exact OTFs.

\begin{table}[htbp]
\centering
\caption{$\hat\psi/\psi$ for the conventional least-squares estimate with $M_0=R^2/4$ (disc) or $M_0=A^2$ (Gaussian).}
\label{S-tab:ls}
\begin{tabular}{llcccc}
\toprule
System & Pupil, texture & $\psi=0.1$\,D & $0.3$\,D & $1.0$\,D & $2.0$\,D \\
\midrule
f/6: $R=1$\,mm, $s=12.1$\,mm, & disc, $\eta=1$ & 0.449 & 0.504 & 0.579 & 0.205 \\
$2\,\mu$m px ($f_{\rm Nyq}/f_c=0.83$) & disc, $\eta=2$ & 0.506 & 0.562 & 0.836 & 0.505 \\
& Gaussian, $\eta=2$ & 1.000 & 1.000 & 1.000 & 1.000 \\
\midrule
f/1.8: $R=1.22$\,mm, $s=4.4$\,mm, & disc, $\eta=1$ & 0.704 & 0.769 & 0.966 & 0.408 \\
$1.4\,\mu$m px ($f_{\rm Nyq}/f_c=0.35$) & disc, $\eta=2$ & 0.736 & 0.793 & 1.050 & 0.727 \\
& Gaussian, $\eta=2$ & 0.999 & 1.000 & 1.000 & 1.000 \\
\bottomrule
\end{tabular}
\end{table}

\begin{figure}[htbp]
\centering
\includegraphics[width=0.62\linewidth]{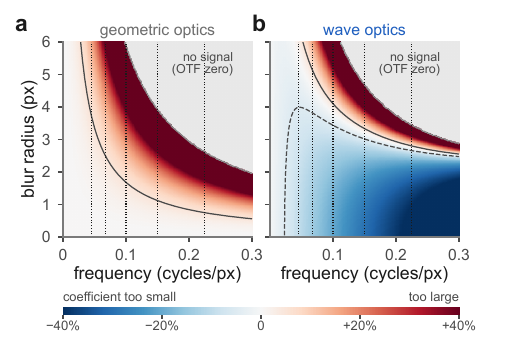}
\caption{\textbf{Where the heat equation fails for a circular aperture.} Error of the diffusion coefficient
$c(f;\psi)$ against the constant $R^2/4$ of the heat equation, for the DSLR at f/16, against spatial frequency and the
geometric blur radius $R s|\psi|$. (a)~Geometric optics: the pillbox makes the coefficient too large as the blur grows.
(b)~Wave optics: diffraction also makes it too small near focus at high frequencies. Lines: $\pm5\%$ (dashed:
$-5\%$); gray: where the OTF has fallen below 0.03, just before its first zero; dotted: the five bands of our DSLR
experiments. Main-paper Fig.~\ref{fig:overlap}c draws (b) on these five bands, gray from the first zero itself.}
\label{S-fig:errmap}
\end{figure}

\Para[S-par:errs]{Where conventional DfDD errs}
Fig.~\ref{S-fig:errmap} maps both errors of the heat equation for the DSLR at f/16 (main-paper
Sec.~\ref{sec:errmap}): the diffusion coefficient of a circular aperture, computed from the exact OTF, relative to the
constant $R^2/4$, as a function of spatial frequency and of the geometric blur radius. In geometric optics
(Fig.~\ref{S-fig:errmap}a) the coefficient depends only on the product of blur and frequency. It equals $R^2/4$ for small
blur and grows, because of the non-Gaussian shape of the pillbox, as the band approaches the first zero of the OTF; this
is the error that limits the working range of conventional DfDD. Wave optics (Fig.~\ref{S-fig:errmap}b) adds the opposite
error near focus. At a blur radius of 0.5\,px the coefficient is 20\% too small at 0.1 cycles/px and 42\% too small at
0.225 cycles/px, and the two errors cancel along a curve in between. Only the low frequencies near focus are free of
error, and that region shrinks as the f-number grows or the pixels shrink.

\Para[S-par:aberr]{Aberrations at low spatial frequency}
For an aberration $W_a$ and a shift $\va=\lambda s\vf$ along $\ve$, the phase across the overlap is
$\Phi=2\pi s|\vf|\,[\psi\,u_e+T(\vu)]+O(a^3)$, with the ray slope $T=\partial_eW_a$. The cumulant expansion
$\log|H|=-\tfrac12\kappa_2(\Phi)+\tfrac1{24}\kappa_4(\Phi)-\dots$ over the normalized overlap gives, for the term
linear in $\psi$,
\begin{equation}
  \frac{\partial_\psi\log|H|}{-(2\pi s|\vf|)^2}=\psi\Big[M(a)-\tfrac12(2\pi s|\vf|)^2\,\kappa_O(u_e,u_e,T,T)\Big]
  +\operatorname{Cov}_O(u_e,T)+\dots
  \label{S-eq:cumulant}
\end{equation}
\begin{itemize}
  \item At second order the slope stays $M(a)$; the focus of the band moves by $-\operatorname{Cov}_O(u_e,T)/M(a)$.
  \item At fourth order the joint cumulant of the pupil coordinate and the aberration slope changes the slope. It
  lowers it when the slope of $W_a$ grows toward the rim along $\ve$, as for coma along $\ve$. On the full pupil,
  coma $W_a=c\lambda\rho^3\cos\phi$ gives $\kappa=\tfrac5{48}(c\lambda)^2$, i.e.\ a relative change $-\tfrac5{24}\beta^2$
  with $\beta=4\pi c\,f/f_c$. The integrand sits at the rim, so the overlap reduces this already at $f\approx0.1f_c$;
  numerically the change is $0.65$ times the full-pupil value at 95 cycles/mm and $0.7$ times at 45 cycles/mm
  ($c\le0.2$).
  \item A focus shift that varies over the orientations inside one band, as for astigmatism at $45^\circ$ to the band
  axis, acts as an incoherent focus spread and lowers the slope by a term of the same order.
\end{itemize}
Both mechanisms grow as $f^2$ and as the square of the aberration, unlike the hard-edge term of
Lemma~\ref{lem:disc}, which is linear in $f$. Primary spherical aberration up to half a wave changes the phone band
ratios by less than 0.01.

\section{Depth estimation in sub-bands}
\label{S-sec:bw}

\begin{figure}[htbp]
\centering
\includegraphics[width=0.62\linewidth]{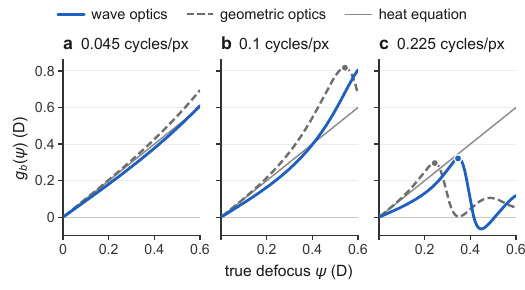}
\caption{\textbf{Each band responds differently to defocus.} Through-focus curves $g_b(\psi)$ of one f/16 image pair
(step 0.06\,D) on the DSLR for three of the five bands, in wave optics (blue) and geometric optics (gray, dashed).
Near focus, geometric optics follows the heat equation (gray) in every band, whereas in wave optics the high band~(c)
responds more weakly, by the factor of Lemma~\ref{lem:disc}. Far from focus, a curve turns over (dots) as the blur
approaches the first zero of the OTF in that band, so the lower bands stay informative longer.}
\label{S-fig:response}
\end{figure}

\Para[S-par:curves]{Through-focus curves}
The band ratio of main-paper Eq.~\eqref{eq:bw_ratio} is the heat equation fitted in one band. With
$X_b=W_b*(I_+-I_-)/\Delta\psi$ and $Y_b=\Lap(W_b*\bar I)$ for the mean image $\bar I$, Parseval's theorem gives
$\sum X_bY_b=(e_b^+-e_b^-)/(2\Delta\psi)$ up to window edges, since the Laplacian is self-adjoint, and $\omega_b$
exceeds $\sum Y_b^2$ by $\tfrac14\sum(\Lap W_b*(I_+-I_-))^2$, the energy of the image change. For a small step, $r_b$
is thus the ratio $\sum X_bY_b/\sum Y_b^2$ of the heat equation, main-paper Eq.~\eqref{eq:dbc}, in one band, and
$r_b=A^2s^2\psi$ in every band for a Gaussian pupil; the curves below contain the finite step.
For a large window, the band energies in~\eqref{eq:bw_ratio} become Parseval sums over frequencies. With
$\hat I_\pm=H(\vf;\psi\pm\tfrac{\Delta\psi}{2})\hat P$, $e_b^\pm=\sum_{\vf}|W_b|^2L\,|H_\pm|^2S_T$ and
$E_b^\pm=\sum_{\vf}|W_b|^2L^2|H_\pm|^2S_T$, which gives~\eqref{eq:bw_response}. Image noise of spectral density $\sigma_n^2$
adds $N_b=\sigma_n^2\sum_{\vf}|W_b|^2L^2$ to $E_b^\pm$ and, on average, nothing to $e_b^+-e_b^-$; the estimator does
not subtract it. Noise therefore scales $r_b$ by $S_b/(S_b+N_b)$ for a signal energy $S_b$ and raises the weight
$\omega_b$, so a band that holds only noise pulls the fit toward focus (Table~\ref{S-tab:bw_summary}). The through-focus curve $g_b(\psi)$ evaluates the same ratio with the exact OTF, the pixel aperture and a nominal
spectrum $S_0\propto|\vf|^{-2}$. For a Gaussian pupil every frequency contributes $A^2s^2\psi\tanh x/x$ with
$x=-A^2s^2\psi\Delta\psi L$, so the finite step lowers the high bands far from focus even there; the curves contain this
factor. Fig.~\ref{S-fig:response} shows the curves of the DSLR at f/16.

\begin{proposition}[Residual texture dependence]
\label{S-prop:bw}
Write~\eqref{eq:bw_response} as a weighted average of the per-frequency response
$\zeta(\vf;\psi)=(|H_+|^2-|H_-|^2)\big/\big(\Delta\psi\,L\,(|H_+|^2+|H_-|^2)\big)$; $g_b$ is the same average with
$S_0$ in place of $S_T$. For any set $\mathcal B$ of frequencies, let $\tau_T$ and $\tau_0$ be the fractions of the two
weights outside $\mathcal B$, and $\zeta_{\max}$ the largest $|\zeta|$ over the sampled frequencies where $W_b\ne0$. Then
\begin{equation*}
  |r_b-g_b(\psi)|\le\max_{\vf\in\mathcal B}\zeta-\min_{\vf\in\mathcal B}\zeta+2\zeta_{\max}(\tau_T+\tau_0).
\end{equation*}
\end{proposition}

\Para[S-par:proofbw]{\texorpdfstring{Proof of Proposition~\ref{S-prop:bw}}{Proof of Proposition S1}}
Write $r_b=\sum_{\vf}\omega_T(\vf)\,\zeta(\vf;\psi)$, an average of the per-frequency response $\zeta$ with
normalized weights
\begin{equation*}
  \omega_T\propto|W_b|^2L^2\big(|H_+|^2+|H_-|^2\big)S_T\ge0 ,
\end{equation*}
and $g_b$ the same average with weights $\omega_0$, which have $S_0$ in place of $S_T$. Split each average into its
part over $\mathcal B$ and the rest: $r_b=(1-\tau_T)a_T+\tau_Tc_T$, where $a_T$ is the mean of $\zeta$ over
$\mathcal B$ and $c_T$ the mean outside it, both under $\omega_T$, and likewise $g_b=(1-\tau_0)a_0+\tau_0c_0$. Then
$r_b-g_b=(a_T-a_0)+\tau_T(c_T-a_T)-\tau_0(c_0-a_0)$. Both $a_T$ and $a_0$ lie between the smallest and the largest
$\zeta$ in $\mathcal B$, and $|c-a|\le2\zeta_{\max}$. \hfill$\square$

\Para{\texorpdfstring{Choice of $\mathcal B$}{Choice of the band set}}
A log-Gabor band has no finite support, so the bound needs a choice of $\mathcal B$. With
$|W_b|^2=\exp(-u^2/\sigma^2)$, $u=\ln(|\vf|/f_b)$ and $\sigma=0.25$, a texture with $S_T\propto|\vf|^{-\eta}$ gives,
apart from the OTF, a weight per unit $u$ proportional to $\exp(-u^2/\sigma^2)\,e^{(6-\eta)u}$, since $L^2$ and the
area element $2\pi|\vf|^2\,\mathrm du$ contribute $|\vf|^6$. This is a Gaussian in $u$ with standard deviation
$\sigma/\sqrt2=0.18$ and center $(6-\eta)\sigma^2/2$, 0.09--0.16 for $\eta=1$--3. For $\mathcal B=\{|u|\le0.7\}$,
an octave on either side of $f_b$, the weight outside $\mathcal B$ is then at most $1.1\times10^{-3}$, and the bound
reduces to the range of $\zeta$ over these two octaves. The factors $|H_+|^2+|H_-|^2$ and the pixel aperture, which
fall with frequency, move weight toward lower frequencies. Near an OTF zero they can move most of the weight out of
$\mathcal B$, and there the bound says nothing; this is where the curves turn over (Fig.~\ref{S-fig:response}).

\Para{First-order texture error}
The bound is loose: on the f/16 geometry the range of $\zeta$ over $\mathcal B$ is 0.7--0.9 of $g_b$ at
0.225\,cycles/px within 0.2\,D for a step of 0.05\,D, and up to 1.4 for a step of 0.2\,D. A sharper statement follows from the same Gaussian. Between two textures its center moves by $(\eta_0-\eta)\sigma^2/2$, so to
first order
\begin{equation*}
  r_b-g_b\approx(\eta_0-\eta)\,\frac{\sigma^2}{2}\,\frac{\partial\zeta}{\partial\ln|\vf|},
\end{equation*}
with the derivative at the center of the nominal weight. For the clear disc at f/16 in wave optics this term gives the
exact error within 12\% of its value, or 0.5 percentage points, for $|\psi|\le0.2$\,D and steps up to 0.2\,D: for
$\eta=1$ and 3 the exact error at 0.225\,cycles/px is $\mp$1.3--3.9\%, at 0.15\,cycles/px $\mp$0.8--2.0\%. Near an
OTF zero the term fails, at 0.225\,cycles/px and 0.35\,D even in sign (\texttt{theory/prop\_s1\_first\_order.py},
\texttt{theory/prop\_s1\_first\_order.txt}). The error shrinks as the band narrows, since it scales with $\sigma^2$, but a
narrow band moves the depth far near a turnover of its curve, where the joint fit relies on the other bands.

\Para[S-par:specratio]{Relation to the spectral ratio}
Frequency-domain depth from defocus compares two images band by band through the spectral ratio
$\varrho_b=(P_b^+-P_b^-)/(P_b^++P_b^-)$ of the plain band energies
$P_b^\pm=\sum(W_b*I_\pm)^2$~\cite{S-pentland1987,S-morganmar2014}. In a band so narrow that $L\approx L_b=-4\pi^2f_b^2$
inside it, $e_b^\pm\approx L_bP_b^\pm$ and $E_b^\pm\approx L_b^2P_b^\pm$, so
\begin{equation}
  r_b\approx\frac{\varrho_b}{\Delta\psi\,L_b},\qquad
  \omega_b\big(r_b-\gamma g_b\big)^2\approx\frac{P_b^++P_b^-}{2\,\Delta\psi^2}\,
  \big(\varrho_b-\gamma\,\Delta\psi L_b\,g_b\big)^2 .
  \label{S-eq:srlimit}
\end{equation}
In this limit the band ratio is the spectral ratio in other units, and the joint fit of main-paper
Eq.~\eqref{eq:bw_estimate} is a fit of spectral ratios weighted by the band energies. Fitting several bands at once
against a model of the OTF is therefore not new. The two statistics part in a band of finite width. Write
$\varrho(\vf)=(|H_+|^2-|H_-|^2)/(|H_+|^2+|H_-|^2)=\Delta\psi\,L\,\zeta$ for the spectral ratio of one frequency and
$w=|W_b|^2(|H_+|^2+|H_-|^2)S_T$ for its energy weight. Then
\begin{equation}
  \varrho_b=\frac{\sum_{\vf}w\,\varrho}{\sum_{\vf}w},\qquad
  r_b=\frac{\sum_{\vf}w\,L\,\varrho}{\Delta\psi\sum_{\vf}w\,L^2},\qquad
  r'_b=\frac{\sum_{\vf}w\,\varrho}{\Delta\psi\sum_{\vf}w\,L}=\frac{P_b^+-P_b^-}{\Delta\psi\,(e_b^++e_b^-)} .
  \label{S-eq:srfinite}
\end{equation}
The spectral ratio is the mean of $\varrho$ over the band. Up to the step $\Delta\psi$, the band ratio $r_b$ is the
slope of the least-squares line through the origin of $\varrho$ against $L$, which is the heat equation fitted inside
the band, and $r'_b$, the energy difference over the mixed energy, is the ratio of the two means. Both are weighted averages of the response $\zeta$ of
Proposition~\ref{S-prop:bw}, with weights $wL^2$ and $w|L|$, so a coefficient that is constant over the band gives the
same value for every texture. The spectral ratio is not such an average. For any band and any pupil
\begin{equation}
  \varrho_b=-\Delta\psi\,r'_b\,m_b,\qquad m_b=\frac{\sum_{\vf}w\,|L|}{\sum_{\vf}w}=-\frac{e_b^++e_b^-}{P_b^++P_b^-},
  \label{S-eq:msf}
\end{equation}
where $m_b$ is the mean of $4\pi^2|\vf|^2$ over the band weighted by the energy of the pair, the mean square
frequency of the texture in the band. The identity holds for any band and any pupil. How strongly $m_b$ follows the
texture depends on the weight $w$. For $S_T\propto|\vf|^{-\eta}$, $\partial\ln m_b/\partial\eta$ is the mean of $u$
under $w$ minus its mean under $w|L|$. For the filter alone, without the OTF and the pixel aperture, $w$ is, per unit
$u$, a Gaussian centered at $(2-\eta)\sigma^2/2$ with variance $\sigma^2/2$, and $\ln m_b$ changes by $-\sigma^2$ per
unit of the exponent, 6.25\% for our filters. With the OTF and the pixel aperture of the f/16 model the change is
5.7--6.2\% within 0.2\,D, and 5.2--6.2\% for a Gaussian pupil of the same second moment, for which $r_b$ is exact for
a small step. The through-focus curve of the
spectral ratio is computed for the nominal exponent and cannot follow $m_b$. For the clear disc at f/16 in wave
optics, per unit of the exponent, within 0.2\,D and for steps of 0.05 and 0.2\,D, the spectral ratio changes by
5.1--6.1\% in the three lower bands, by 4.3--5.2\% at 0.15\,cycles/px and by 2.4--4.6\% at 0.225\,cycles/px, where the
fall of $\zeta$ with frequency offsets part of $m_b$. The band ratio $r_b$ changes by 0.3--0.9\%, 0.8--2.0\% and
1.3--3.9\%, with the opposite sign, and $r'_b$ as $r_b$ within 0.4 points (\texttt{theory/spectral\_ratio\_texture.py},
\texttt{theory/spectral\_ratio\_texture.txt}). \hyperref[S-par:srcontrol]{Section~\ref{S-sec:zedd}} measures $m_b$ on the
DSLR and fits the three statistics with the same curves.

\begin{figure*}[tp]
\centering
\includegraphics[width=\textwidth]{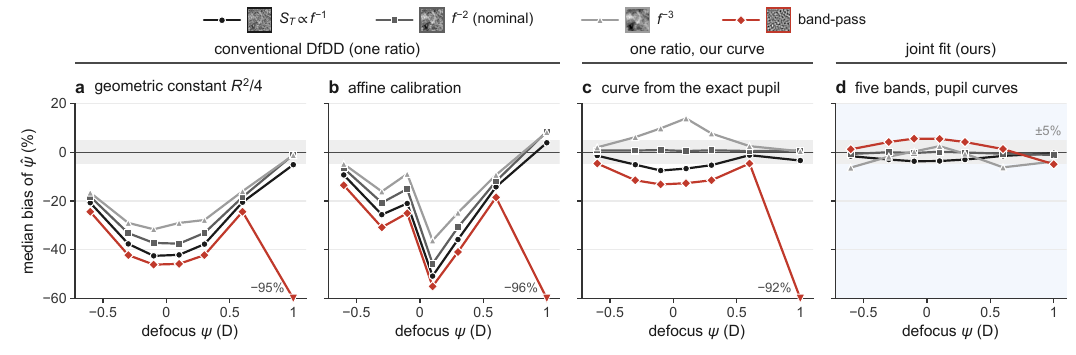}
\caption{\textbf{A diffusion coefficient that falls with frequency makes depth estimates depend on the texture.}
Simulation of an f/6 camera with a circular aperture ($R=1$\,mm, $s=12.1$\,mm, $2\,\mu$m pixels) that images four
textures with different power spectra $S_T$ (patches in the legend) at seven depths (noise 0.002 at mean intensity 1).
(a)~The conventional estimate with the geometric coefficient $R^2/4$ falls short of the true defocus and fails near an OTF
zero (arrowheads: off-scale values). (b)~An affine calibration fitted on the nominal texture shifts the error between
depths but leaves the textures apart. (c)~A through-focus curve computed from the exact pupil is unbiased only for the
nominal texture. (d)~The sub-band estimator of main-paper Sec.~\ref{sec:bw} keeps all textures within $-6.3\%$ and
$+5.6\%$ (gray band: $\pm5\%$). Every estimator uses the band ratio and weight of main-paper Eq.~\eqref{eq:bw_ratio},
as on the DSLR; at noise 0.01, bands that hold only noise pull the sub-band estimate toward focus
(Table~\ref{S-tab:bw_summary}).}
\label{S-fig:sim}
\end{figure*}

\Para[S-par:simsetup]{Simulation setup}
\begin{itemize}
  \item System: clear disc, $R=1$\,mm, $s=12.1$\,mm (f/6), $2\,\mu$m pixels, so $f_c=0.60$ and Nyquist $0.5$
  cycles/px. The OTF is exact (one-dimensional quadrature of the pupil overlap).
  \item Captures: a $\rho$ sweep with $\Delta\psi=0.1$\,D, $256^2$ images, noise standard deviation 0.002 or 0.01 (mean 1,
  texture rms contrast 0.2).
  \item Textures: $S_T\propto|\vf|^{-\eta}$ with $\eta=1,2,3$, and a band-pass texture centered at 0.15 cycles/px.
  Seven depths, $\psi\in\{-0.6,\dots,1.0\}$\,D.
  \item Window: $31\times31$\,px for all methods. The sub-band estimator uses five log-Gabor bands at 0.075 to 0.37 cycles/px
  ($0.12$ to $0.62\,f_c$), log-bandwidth 0.25, with envelope standard deviations of 8.5 to 1.7\,px.
  \item Every estimator forms the band ratio and the weight of main-paper Eq.~\eqref{eq:bw_ratio} from the two
  images, without noise subtraction, as on the DSLR (\texttt{notes/bandwise\_dbc.py}). Baselines (panels a--c of
  Fig.~\ref{S-fig:sim}): one wide band (pre-filter 1.5\,px without its DC term, plus Laplacian) with the geometric
  coefficient $R^2/4$ (conventional DfDD, geometric); the same band with an affine calibration $\psi=\alpha r+\beta$ fitted on
  $\eta=2$ textures at all seven depths (conventional DfDD, affine); the same band with its through-focus curve computed from
  the pupil (one ratio, our curve).
  \item Noise floor (Table~\ref{S-tab:bw_summary}, rows ``noise floor''): the mean window energy $E_b$ of each band on a
  flat field with the same noise, an independent capture, subtracted from $E_b^+$ and $E_b^-$ and clipped at 5\%, as in
  the noise-floor runs of the DSLR (\texttt{bandwise\_dbc.py --floor}).
  \item Noise-aware weights (rows ``noise-aware''): the same floor subtracted, and the weight $S_b^2/(S_b+N_b/2)$ in
  place of $\omega_b$, with $S_b$ the band energy after subtraction and $N_b$ the floor; a band whose energy lies within
  5\% of the floor gets no weight, as in the noise-aware row of Table~\ref{S-tab:baselines} for the DSLR
  (\texttt{bandwise\_dbc.py --nw}). This variant was set before it was run, with the criterion that the largest bias
  and the texture range stay within 10\% for the disc at noise 0.01 and for the Gaussian pupil at both noise levels,
  over four noise seeds (\texttt{--seed 0} to \texttt{3}).
  \item A Gaussian pupil ($A=R/2$) is the control.
\end{itemize}

\begin{table}[htbp]
\centering
\caption{Simulation summary over four textures and seven depths. Bias: median $\hat\psi/\psi-1$ per case. Spread:
robust standard deviation over pixels, relative to $|\psi|$. Texture range: largest spread of the median across
textures at one depth. $|\Delta Z|$: per-case median absolute depth error at $Z_f=0.85$\,m, averaged over cases.
Noise seed 0; the noise-aware rows over four seeds are in the text.}
\label{S-tab:bw_summary}
\small
\begin{tabular}{llrrrr}
\toprule
Pupil, noise & Method & max $|$bias$|$ & mean $|\Delta Z|$ & spread & texture range \\
& & [\%] & [mm] & [\%] & [\%] \\
\midrule
disc, 0.002 & conventional DfDD, geometric & 95.0 & 97.0 & 4.1 & 94.0 \\
& conventional DfDD, affine & 95.9 & 69.2 & 4.5 & 104.4 \\
& one ratio, our curve & 91.9 & 34.5 & 4.8 & 92.6 \\
& joint fit (ours) & 6.3 & 10.3 & 1.0 & 9.3 \\
& our joint fit, noise floor & 5.6 & 6.1 & 1.0 & 9.3 \\
& our joint fit, noise-aware & 5.6 & 6.0 & 1.0 & 9.2 \\
\midrule
disc, 0.01 & conventional DfDD, geometric & 95.1 & 107.0 & 9.2 & 89.3 \\
& conventional DfDD, affine & 95.7 & 76.8 & 10.5 & 102.5 \\
& one ratio, our curve & 92.1 & 50.0 & 9.7 & 87.9 \\
& joint fit (ours) & 67.7 & 67.1 & 3.7 & 65.3 \\
& our joint fit, noise floor & 12.7 & 20.0 & 5.1 & 16.8 \\
& our joint fit, noise-aware & 13.6 & 15.0 & 4.6 & 17.7 \\
\midrule
Gaussian, 0.002 & conventional DfDD, geometric & 8.4 & 22.3 & 4.9 & 8.7 \\
& one ratio, our curve & 7.2 & 20.8 & 5.0 & 8.7 \\
& joint fit (ours) & 91.6 & 69.2 & 0.9 & 39.4 \\
& our joint fit, noise floor & 25.9 & 9.5 & 0.9 & 19.0 \\
& our joint fit, noise-aware & 3.4 & 4.3 & 0.8 & 3.0 \\
\midrule
Gaussian, 0.01 & conventional DfDD, geometric & 27.1 & 38.7 & 8.6 & 22.7 \\
& one ratio, our curve & 25.9 & 36.7 & 8.6 & 22.7 \\
& joint fit (ours) & 99.8 & 213.5 & 2.6 & 81.6 \\
& our joint fit, noise floor & 96.1 & 79.3 & 4.7 & 45.4 \\
& our joint fit, noise-aware & 11.1 & 19.3 & 4.0 & 7.3 \\
\bottomrule
\end{tabular}
\end{table}

\begin{table}[htbp]
\centering
\caption{Clear disc, noise 0.002: median bias of $\hat\psi$ [\%] for one ratio with our curve / the joint fit (ours).}
\label{S-tab:bw_depth}
\begin{tabular}{rcccc}
\toprule
$\psi$ [D] & $\eta=1$ & $\eta=2$ & $\eta=3$ & band-pass \\
\midrule
$-0.60$ & $-1.4$ / $-1.6$ & $0.7$ / $-0.7$ & $2.0$ / $-6.3$ & $-4.6$ / $1.2$ \\
$-0.30$ & $-5.1$ / $-3.0$ & $0.7$ / $0.0$ & $6.2$ / $-1.8$ & $-11.5$ / $4.2$ \\
$-0.10$ & $-7.5$ / $-3.7$ & $0.9$ / $-0.3$ & $9.8$ / $0.3$ & $-13.2$ / $5.6$ \\
$0.10$ & $-6.7$ / $-3.6$ & $0.5$ / $0.2$ & $13.9$ / $2.5$ & $-12.7$ / $5.5$ \\
$0.30$ & $-5.3$ / $-3.0$ & $0.8$ / $-0.1$ & $7.7$ / $-0.6$ & $-11.5$ / $4.2$ \\
$0.60$ & $-1.2$ / $-1.5$ & $0.5$ / $-0.6$ & $2.5$ / $-6.2$ & $-4.6$ / $1.3$ \\
$1.00$ & $-3.4$ / $-0.4$ & $0.7$ / $-1.1$ & $0.4$ / $-3.8$ & $-91.9$ / $-4.9$ \\
\bottomrule
\end{tabular}
\end{table}

\Para{Findings}
\begin{itemize}
  \item With a clear disc, the geometric coefficient underestimates $|\psi|$ by 16 to 46\% at most depths and fails
  near OTF zeros. Calibration cannot fix this: the affine fit trades errors between depths, and its texture range
  stays near 100\%.
  \item The broadband through-focus curve computed from the pupil removes the depth dependence for the nominal spectrum, but keeps a texture
  dependence of $-13$ to $+14\%$. It fails ($-92\%$) for the band-pass texture at $\psi=1$\,D, whose energy lies near
  an OTF zero.
  \item At noise 0.002 the sub-band estimate keeps every texture within $-6.3$ to $+5.6\%$; the extremes are the steep
  spectrum ($\eta=3$) at $\pm0.6$\,D and the band-pass texture near focus. Its spread is the lowest.
  \item Its weights, the band energies of main-paper Eq.~\eqref{eq:bw_ratio}, contain the noise energy, which for
  white noise grows as $f_b^6$ with the band frequency. A band that holds only noise therefore gets a large weight and a ratio near zero
  and pulls the fit toward focus. At noise 0.01 this underestimates the steep spectrum by 26--68\%; with the Gaussian
  pupil, whose OTF removes the high bands far from focus, the bias reaches $-52$ to $-92\%$ at 1\,D already at noise
  0.002. A noise floor measured on a flat field limits the bias to 13\% for the disc at noise 0.01, and for the
  Gaussian pupil at noise 0.002 to $-7$ to $-26\%$ at 1\,D and 4\% elsewhere.
  \item Noise-aware weights keep the largest bias at 5.5--5.8\% for the disc and 3.3--3.7\% for the Gaussian pupil at
  noise 0.002 (four seeds). At noise 0.01 the largest bias is 6.8--13.6\% and the texture range 12.3--17.7\% for the
  disc, and 8.1--19.9\% and 7.3--21.0\% for the Gaussian pupil, so the variant fails its criterion, and the main paper
  keeps Eq.~\eqref{eq:bw_ratio}. With a known floor the floor-corrected ratio of two window sums is nearly unbiased
  down to a signal-to-noise ratio of 1 in the band (mean bias $+0.3\%$ for 400 independent samples per window); below
  0.2 its mean bias exceeds 5\% and the 5\% clip begins to act (\texttt{notes/ratio\_bias\_mc.py}). The remaining
  failures at noise 0.01 thus involve bands far below the noise. The estimator of the main paper, however, subtracts
  no floor, so a band at a signal-to-noise ratio of 1 reads 50\% low; with a floor, its error is the error of the
  floor (a floor 10\% too low leaves $-9\%$ at a ratio of 1), and the first percentile over the windows of a scene lies
  below the mean noise energy of the noise-only windows (for white noise by about 5--12\% at 600--1\,200 independent
  samples per window in the top two bands). On the DSLR, floor subtraction and noise-aware weights change the held-out
  error by $-2$ to $+3\%$ (Table~\ref{S-tab:baselines}), but the least energetic third of the textured windows collapses
  toward focus beyond 3.5\,px with or without them (Table~\ref{S-tab:range_energy}): weak bands carry no signal there.
  On all windows the noise-aware weights lower the error and the share of gross errors, and with a texture gate the
  two weightings err alike (\hyperref[S-par:coverage]{Error against coverage}, Fig.~\ref{S-fig:coverage}).
  \item For the Gaussian control the broadband estimates stay within 8.4\% at noise 0.002.
  \item Cost: five bands instead of one, and a larger spatial support (the lowest band has an envelope of 8.5\,px, plus
  the window). The operation count follows.
\end{itemize}

\Para{\texorpdfstring{The simulation of main-paper Fig.~\ref{fig:teaser}c, d}{The simulation of main-paper Fig. 1c, d}}
\label{S-sec:teasersim}
A simulated camera (circular aperture, f/6, 2\,$\mu$m pixels, wave optics) photographs a flat print 6.1\,cm wide and
67.7\,cm away, with one brick texture at two scales, at two slightly different focus settings ($\psi=0.3\pm0.05$\,D,
noise 0.002). One ratio with a through-focus curve computed from the exact pupil for the nominal spectrum places the
coarse half 0.8\,cm too near and the fine half 0.7\,cm too far. Our joint fit, with the band ratios and weights of
main-paper Eq.~\eqref{eq:bw_ratio}, places both halves 0.3--0.5\,cm too near, so the two textures differ by 0.19\,cm
instead of 1.56\,cm (0.82\,cm too near and 0.74\,cm too far). The numbers are medians over textured windows. The
estimates also scatter within each half, with a robust standard deviation of 0.79 and 0.55\,cm for one ratio against
0.13\,cm for the joint fit, and the surfaces of the figure, drawn at true scale (1\,cm bar), show this scatter smoothed
with a Gaussian of 24\,px for display (\texttt{realdata/teaser\_sim.py}). The simulated bands reach $0.62f_c$, with the
cutoff $f_c$ at 1.2 times the Nyquist frequency; on the DSLR at f/16, whose bands reach $0.23f_c$, the effect is too
small to resolve (main-paper Sec.~\ref{sec:range}).

\Para{Computational cost}
\label{S-sec:flops}
Table~\ref{S-tab:flops} counts floating-point operations per pixel for two images of $2^{20}$ pixels
(\texttt{realdata/flop\_count.py}). A real FFT of $n$ pixels counts as $2.5n\log_2n$ operations, a window sum as four
additions per pixel (separable running sums, independent of the window size), a symmetric separable filter with $t$ taps
as $\lceil t/2\rceil$ multiplications and $t-1$ additions per pass, and a division as one operation. Conventional DfDD
prefilters both images with a 9-tap Gaussian, takes a five-point Laplacian of the mean image, forms two products and two
window sums, and divides. The sub-band estimator filters each image into five log-Gabor bands by FFT, with and without
the Laplacian folded into the filter (four inverse FFTs per band); per band and image it forms the two band energies
of main-paper Eq.~\eqref{eq:bw_ratio} (two products and two window sums), then the ratio and its weight; then it
fits the through-focus curves. The depth maps search all candidates of the grid (about 600); main-paper
Table~\ref{tab:est} and the learning curves search every tenth candidate and then the 21 around the best (82 in all,
\texttt{zedd\_learning\_curve.py}), which costs 2\,900 operations for the joint fit and 1\,600 for one ratio with a curve.
A two-level search over 32 coarse candidates and 16 fine ones around the best gives the same accuracy within the
standard search: on 200\,000 DSLR samples at f/16 with calibrated curves the median error is 0.0241\,D against
0.0242\,D for the full search, although 4.5\% of the estimates differ by more than 0.01\,D
(\texttt{realdata/search\_check.py}). Over the 9.5\,px search of the working range in defocus ($\pm1.02$\,D at f/16, coarse
candidates 0.066\,D apart) it changes 1.8\% of the calibrated estimates by more than 0.05\,D and none of the medians
(\texttt{search\_check.py --search-px 9.5}, \texttt{results/search\_check\_wide.txt}). On our curves, held out on the
samples of main-paper Table~\ref{tab:est} over the 9.5\,px search (\texttt{realdata/zedd\_cascade.py},
\texttt{results/zedd\_cascade.txt}), 32 + 16 candidates keep the near-focus error (21.4\,mD) but change 7.2\% of the
estimates beyond 3.5\,px by more than 0.05\,D, raise the relative error at 1.5--3.5 and 3.5--9.5\,px from 11.3\% and
9.0\% to 11.7\% and 9.5\%, and lower the 20\% limit on the small steps from 6.6 to 3.6\,px. With 64 + 16 candidates
(2\,860 operations) every number of Table~\ref{S-tab:estfull} stays: 21.4\,mD, 11.2\% (11.3\% with the 82 candidates) and 9.0\%, limits
9.5 and 6.6\,px,
and 1.7\% of the estimates beyond 3.5\,px move by more than 0.05\,D. The cost grows linearly with the number of bands
(Table~\ref{S-tab:bandabl}). With wave-optics curves, three bands need about 1\,400 operations, err 3--6\% more than
five on the same samples and keep a working range in defocus of at least 1.2--7.7\,px at f/16. Geometric curves depend more on the
choice of bands, and three bands raise their error at f/16 by 13\%. With five bands their median relative error already reaches
20\% in the bin at 2.1\,px (Table~\ref{S-tab:range}); without the band at 0.15\,cycles/px it exceeds 20\% at
1.2--2.4\,px and again at 3.0--3.6\,px, and the range shrinks to 2.4--3.0\,px. Two bands at 0.10 and 0.225\,cycles/px
cost about 950 operations and err 5\% more within the standard search range, but 33\% and 12\% more at f/16 and
f/5.6 when the search extends to 9.5\,px, and at f/16 they keep only 1.2--4.2\,px. Without a band near 0.1\,cycles/px
the error rises by 30--42\% (\texttt{results/zedd\_band\_ablation\_fold.txt}, \texttt{\_wide\_fold\_f16\_rows.txt},
\texttt{\_wide\_fold\_f56.txt}). For one ratio a table lookup can replace the search wherever its through-focus curve
is monotone; a weighted fit of five ratios has
no such one-dimensional table, and the joint fit keeps the two-level search. With the 82 candidates of main-paper
Table~\ref{tab:est} for both, the 9-tap ratio costs 430 operations against 2\,900 for the
joint fit, 7 times less; with a table lookup for the ratio (106 operations) and a two-level search for the joint fit
(2\,220--2\,860) it costs 21--27 times less (Table~\ref{S-tab:flops}).

The one-ratio rows of main-paper Table~\ref{tab:est} sum the five band statistics (1\,300 operations, 1\,600 with the
curve fit). We also ran the cheap pipeline of conventional DfDD on the same windows: a 9-tap Gaussian prefilter $G$ of
width $\sigma_G$ and a five-point Laplacian of each image, in the energy form of main-paper Eq.~\eqref{eq:bw_ratio},
which needs no pixel-accurate registration. In this form the window mean of $G*I$ must be removed. Since $I$ is the luma over its
local mean, $G*I$ has a mean near 1, and $\sum_W(G*I)\,\Lap(G*I)$ then contains that mean times $\sum_W\Lap(G*I)$, the
flux of the gradient through the window edge. This term is first order in the texture contrast, while the defocus
signal is second order, and in the pair difference it acts as a random gain. The log-Gabor bands have zero mean and
no such term, nor have the baseline of Fig.~\ref{S-fig:sim} and the product form counted in Table~\ref{S-tab:flops}. We
subtract the window mean of $G*I$, at the cost of two more window sums per image (102 operations per pixel before the
curve fit, about 430 with the 82-candidate search of Table~\ref{tab:est}; \texttt{realdata/zedd\_process\_conv.py
--dc window}, curves from \texttt{zedd\_model.ConvModel}; \texttt{run\_r13\_conv.ps1},
\texttt{results/zedd\_r13\_conv.txt}). Without the subtraction the tuned ratio below errs 19\% more than ours near
focus instead of 3\%, and 57\% instead of 50\% beyond 3.5\,px.
With $\sigma_G=1.5$\,px it errs at f/16 within $\pm0.35$\,D by 0.030\,D with a calibrated curve and with its
wave-optics curve (paired ratios to the joint fit 1.41 [1.30, 1.52] and 1.40 [1.31, 1.48]). Its weight then centers
near 0.13\,cycles/px, against 0.24 for the five-band sum, so we also chose $\sigma_G$ on the training scenes of each
fold among 0.35, 0.5, 0.65, 0.8, 1, 1.25, 1.5 and 2\,px (smallest median training error within the test range of the
run; \texttt{zedd\_learning\_curve.py}, methods C72t). The choice falls on 0.65\,px with the wave-optics curve and on
0.8\,px with a calibrated curve in every fold. With its wave-optics curve the tuned ratio errs by 0.0220\,D within
$\pm0.35$\,D (1.03 [1.00, 1.06]; small steps 0.98 [0.96, 1.01]) and by 0.0172\,D within 0.05\,D (0.99 [0.96, 1.02]),
with a calibrated curve by 0.0255\,D (1.19 [1.11, 1.27]); over the 9.5\,px search it errs by 12\% and 50\% at 1.5--3.5
and 3.5--9.5\,px, with a 20\% limit at 3.9\,px (Table~\ref{S-tab:range_energy}). Near focus a cheap ratio with the
wave-optics curve of its filter thus comes within 3\% of the joint fit, as the five-band sum does (0.99,
Table~\ref{S-tab:estfull}); against a cascade of two cheap ratios (below), the 2\,900 operations of the joint fit buy about
10\% at both ends of the range, a third less error at 1.5--3.5\,px and, on the small steps, the range beyond 3\,px.
Chosen instead for range on the
training samples, by the mean binned relative error over 1.5--9.5\,px or over 3.5--9.5\,px among the same widths and 3
and 4\,px (\texttt{--c72-select range}, \texttt{far}), the width rises to 2--3\,px with either curve. With a calibrated
curve the ratio then errs by 19\% at 1.5--3.5\,px and by 15.5--15.9\% at 3.5--9.5\,px (20\% limit 8.9--9.0\,px),
1.72--1.77 times the joint fit there ($+6.9$ [$+3.5$, $+11.0$] and $+6.4$ [$+3.2$, $+10.3$] points, paired over the
scene resamples), and by 0.045--0.047\,D within $\pm0.35$\,D, 2.1--2.2 times the joint fit; with its wave-optics curve
it errs by 16--18\% and 21\% (limit 5.8--5.9\,px). On the small steps beyond 3.5\,px it errs by 28--29\%, against 14\%
for the joint fit. On the small steps, longer prefilters and decimation do not change this. With 17 taps at 2--4\,px or 25 taps at
4--6\,px (150 and 198 operations per pixel; \texttt{--taps}), or with 9 taps after a $2\times2$ block mean at 2--6\,px
(28 operations; \texttt{--decimate 2}, whose wave-optics curve includes the block mean and the aliasing), chosen in
the same way, the calibrated choice over 3.5--9.5\,px falls on 3\,px in every fold. With the window mean removed, a
2--3\,px Gaussian prefilter of 9 or 17 taps, or of 9 taps after $2\times$ decimation, chosen for range keeps the 20\%
limit to 8.0--9.5\,px at 2.1--3.3 times our near-focus error but errs 1.5--1.8 times ours beyond 3.5\,px, and 27--29\%
against 14\% on the small steps; one ratio of the two lowest bands errs about as ours beyond 3.5\,px over all pairs
(Table~\ref{S-tab:lowband}).
On the large steps the 17-tap and decimated ratios err by 9.7\% against our 8.3\% (1.16 [0.95, 1.38] and [0.94,
1.38]), at 2.3--2.5 times our near-focus error. Of all 40 candidates (20 prefilters, two curves; \texttt{--c72-all},
\texttt{results/\allowbreak zedd\_t1\_\allowbreak r13c72all\_f16.txt}) none comes within 20\% of the joint fit both
within $\pm0.35$\,D and beyond 3.5\,px; the closest, 2\,px with 9 or 17 taps and a calibrated curve, errs 1.81--1.85
times ours near focus and 1.84--1.85 times beyond 3.5\,px. A Gaussian prefilter followed by a five-point Laplacian
passes a wide band, so one such ratio depends on the spectrum of the texture.

\Para[S-par:cascade]{A cascade of two cheap ratios (post hoc)}
One ratio is accurate near focus and another far from it, so a designer will switch between them. Near focus we take
the 9-tap ratio above (0.65\,px, window mean removed) with its wave-optics curve and the gain chosen within
$\pm0.35$\,D. Far from it we take the ratio of the two lowest bands, $(A_1+A_2)/(B_1+B_2)$ at 0.045 and
0.0675\,cycles/px, with its wave-optics curve and one gain chosen on the training samples beyond 0.3\,D (0.76--0.82),
which needs no calibration per pair. The cascade keeps the 9-tap estimate where the two-band estimate lies within
$\pm\tau$ and the two-band estimate elsewhere, so a camera computes the switch from the pair itself. We chose $\tau$ per
fold on the training scenes, either by the smaller of the larger ratio to ours within $\pm0.35$\,D and over
3.5--9.5\,px, which gives 0.35--0.45\,D, or with the two components in place of ours, which gives 0.40--0.425\,D.
Gains and $\tau$ use only the training scenes of each fold, and the estimates are held out on the samples of
main-paper Table~\ref{tab:est}; our estimates and the 9-tap estimates equal those of the runs behind Table~\ref{S-tab:estfull}
sample by sample (\texttt{realdata/zedd\_cascade.py}, \texttt{results/zedd\_cascade.txt}; Table~\ref{S-tab:cascade}).
The cascade errs 8--11\% more than ours within $\pm0.35$\,D, all of it beyond 0.1\,D (1.00 times ours within 0.1\,D,
1.05--1.06 at 0.1--0.2\,D, 1.30--1.46 at 0.2--0.35\,D), and 7--10\% more beyond 3.5\,px, but 1.43--1.46 times more at
1.5--3.5\,px, where it switches. On the small steps it errs 4--8\% more near focus and as ours beyond 3.5\,px (14.7\%
and 15.1\% against 14.3\%), but its binned error passes 20\% at 2.6--2.9\,px, against 6.6\,px for ours; on the large
steps it errs 18--20\% more near focus and as ours far from it. Beyond 3.5\,px fewer of its estimates err by more
than half (17.0--18.6\% against 19.7\%). With $\tau$ by the components, its error over all pairs also passes 20\% at
the switch (3.2\,px). A sweep of $\tau$ on the held-out samples shows no better setting: wherever the cascade comes
within 7\% of ours near focus ($\tau\ge0.425$\,D) it errs 1.14--1.88 times ours beyond 3.5\,px, and at every $\tau$ it
errs by at least 13.7\% at 1.5--3.5\,px (ours 11.3\%). With table lookups for both ratios the cascade
costs 683 operations per pixel (102 for the 9-tap ratio, 571 for the two band filters by FFT, 10 for two lookups and
the switch), 0.24--0.31 of the joint fit with a two-level search (2\,220--2\,860); with the 82-candidate search for
every curve, as evaluated, it costs 1\,331, or 0.46 of 2\,900. A lookup inverts the two-band curve to 0.95\,D (8.9\,px)
in every fold and pair and the 9-tap curve over the whole search. Against this cascade the joint fit buys about 10\%
at both ends of the range, a third less error at 1.5--3.5\,px and, on the small steps, the range from 3 to 6.6\,px.

\begin{table*}[tp]
\centering
\caption{\textbf{A cascade of two cheap ratios against the joint fit (post hoc).} DSLR at f/16, one image pair,
held-out scenes, textured windows, 9.5\,px search, the samples of main-paper Table~\ref{tab:est}. Near focus: median
$|\hat\psi-\psi|$ within $\pm0.35$\,D, its ratio to ours [95\% interval over 2\,000 resamples of the scenes], and the
ratio by ring of $|\psi|$ (0--0.1/0.1--0.2/0.2--0.35\,D). Range: mean binned median relative error at 1.5--3.5 and 3.5--9.5\,px (ratio to ours [95\%
interval over 20\,000 resamples]), the 20\% limit over all pairs and on the small steps, and the share of the
estimates beyond 3.5\,px that err by more than half. Operations per pixel with table lookups for the ratios and a
two-level search for the joint fit (32\,+\,16 or 64\,+\,16 candidates) / with the 82-candidate search of
Table~\ref{tab:est} for every curve. Cascade (a), (b): $\tau$ chosen by (a) the larger ratio to ours, (b) the larger ratio to the
two components (\texttt{results/zedd\_cascade.txt}).}
\label{S-tab:cascade}
\footnotesize
\setlength{\tabcolsep}{3pt}
\begin{tabular}{lccccc}
\toprule
& ours & 9-tap, 0.65\,px & two lowest bands & cascade (a) & cascade (b) \\
\midrule
within $\pm0.35$\,D [mD] & 21.4 & 22.0 & 52.5 & 23.9 & 23.1 \\
\quad ratio to ours & -- & 1.03 [1.00, 1.06] & 2.45 [2.23, 2.68] & 1.11 [1.07, 1.16] & 1.08 [1.04, 1.12] \\
\quad ratio by ring & -- & 0.98/1.02/1.11 & 2.72/2.62/1.96 & 1.00/1.06/1.46 & 1.00/1.05/1.30 \\
1.5--3.5\,px [\%] & 11.3 & 13.5 & 20.1 & 16.5 & 16.2 \\
\quad ratio to ours & -- & 1.19 [1.11, 1.29] & 1.77 [1.48, 2.07] & 1.46 [1.34, 1.58] & 1.43 [1.32, 1.56] \\
3.5--9.5\,px [\%] & 9.0 & 57.5 & 9.6 & 9.6 & 9.9 \\
\quad ratio to ours & -- & 6.38 [4.71, 7.32] & 1.07 [0.93, 1.26] & 1.07 [0.94, 1.26] & 1.10 [0.97, 1.29] \\
20\% limit, all / small [px] & 9.5/6.6 & 3.5/3.4 & 9.5/6.8 & 9.5/2.6 & 3.2/2.9 \\
far errors above 50\% [\%] & 19.7 & 43.5 & 12.5 & 17.0 & 18.6 \\
operations & 2\,220--2\,860/2\,900 & 106/430 & 575/899 & 683/1\,331 & 683/1\,331 \\
\bottomrule
\end{tabular}
\end{table*}

\begin{table}[htbp]
\centering
\caption{The joint fit on subsets of the five bands (DSLR, one image pair, held-out scenes, focus calibrated on the
training scenes of each fold; \texttt{zedd\_band\_ablation.py --cals}): held-out median error [D] with the standard search, the 20\% working range in defocus at f/16 with the
search extended to 9.5\,px as in main-paper Table~\ref{tab:est}, and operations per pixel with a two-level search
(\texttt{realdata/zedd\_band\_ablation.py}). Wave: wave-optics curves; geom.: geometric curves.}
\label{S-tab:bandabl}
\small
\begin{tabular}{lcccccr}
\toprule
& \multicolumn{2}{c}{f/16} & f/5.6 & \multicolumn{2}{c}{Range at f/16 [px]} & \\
\cmidrule(lr){2-3}\cmidrule(lr){5-6}
Bands [cycles/px] & wave & geom. & wave & wave & geom. & ops \\
\midrule
0.045, 0.0675, 0.10, 0.15, 0.225 & 0.0214 & 0.0273 & 0.0092 & 1.2--9.5 & 1.2--9.5 & 2\,220 \\
0.045, 0.10, 0.15, 0.225 & 0.0212 & 0.0270 & 0.0092 & 1.2--9.5 & 1.2--9.5 & 1\,796 \\
0.045, 0.0675, 0.10, 0.225 & 0.0229 & 0.0310 & 0.0096 & 1.2--9.5 & 2.4--3.0 & 1\,796 \\
0.045, 0.10, 0.225 & 0.0227 & 0.0308 & 0.0095 & 1.2--7.7 & 2.4--3.0 & 1\,372 \\
0.0675, 0.10, 0.225 & 0.0226 & 0.0309 & 0.0096 & 0.6--9.5 & 2.4--3.0 & 1\,372 \\
0.10, 0.225 & 0.0225 & 0.0308 & 0.0097 & 1.2--4.2 & 2.4--3.0 & 948 \\
0.045, 0.225 & 0.0303 & 0.0449 & 0.0120 & 1.2--2.4 & 2.4--3.0 & 948 \\
\bottomrule
\end{tabular}
\end{table}

\begin{table}[htbp]
\centering
\caption{Floating-point operations per pixel, two images of $2^{20}$ pixels, five bands. Conventional DfDD in the
product form as usually implemented, which we did not evaluate, and in the energy form of main-paper
Eq.~\eqref{eq:bw_ratio} that Tables~\ref{S-tab:estfull} and~\ref{S-tab:range_energy} evaluate (window mean
of $G*I$ removed; 86 without); a table lookup adds 4 to a ratio. The two-level search of 32\,+\,16 candidates keeps
the near-focus accuracy, that of 64\,+\,16 also the working range in defocus (Sec.~\ref{S-sec:flops}).}
\label{S-tab:flops}
\small
\begin{tabular}{lrrr}
\toprule
& \multicolumn{2}{c}{conventional DfDD} & \\
\cmidrule(lr){2-3}
Step & product form & energy form & sub-band estimator \\
\midrule
difference and mean of the two images & 2 & -- & -- \\
prefilter / forward FFTs & 52 & 52 & 100 \\
Laplacian / band filters (inverse FFTs) & 5 & 10 & 1\,040 \\
products, window sums, ratios, weights & 13 & 40 & 120 \\
depth fit, two-level search (32\,+\,16) & -- & -- & 960 \\
\midrule
total & 72 & 102 & 2\,220 \\
with a two-level search of 64\,+\,16 & -- & -- & 2\,860 \\
with 82 candidates (main-paper Table~\ref{tab:est}) & -- & 430 & 2\,900 \\
with all 600 candidates & -- & -- & 13\,260 \\
\bottomrule
\end{tabular}
\end{table}

\section{Phone camera: dual-pixel focal sweeps}
\label{S-sec:pixel}

\Para{Data}
The test split of the Learning to Autofocus dataset~\cite{S-herrmann2020} has 47 focal sweeps from a Pixel~3 main
camera (focal length 4.44\,mm, f/1.8), with 49 focus settings each from 3.91\,m to 0.10\,m (0.26 to 9.8\,D), captured in five
rooms. We use the center phone's raw dual-pixel data. The sum of the left and right views gives the raw green
channel through the full pupil, sampled on a $2016\times756$ grid with a pitch of $2.8\,\mu$m along the rows and $5.6\,\mu$m along
the columns. With $\lambda=0.53\,\mu$m and $s\approx4.54$\,mm the diffraction cutoff is $f_c\approx1025$ cycles/mm.
The bands are log-Gabor filters at 20, 30, 45, 65 and 95 cycles/mm (log-bandwidth 0.25) with an angular window of
$30^\circ$ around the row axis, which avoids the coarser column sampling; they span $f/f_c=0.020$ to $0.093$. The data
need 7.4\,GB of the 108\,GB archive, which we stream once.

\Para{Preprocessing}
\begin{itemize}
  \item \emph{Registration.} Adjacent slices differ by focus breathing and small optical-image-stabilization shifts.
  The breathing scale differs between the center and the edges of the field, so each adjacent pair is registered with
  a cubic displacement field (Gauss--Newton on low-passed images; defocus changes are even and nearly orthogonal to the
  odd gradient terms). Both slices of a pair are warped half-way to a common midpoint frame with quintic splines, so
  that both see the same interpolation transfer function. A one-sided warp would leave up to 1\% amplitude mismatch at
  0.3 cycles/px and bias the high bands near focus.
  \item \emph{Focus-step gauge.} With lens-shift focusing, the magnification change between two slices is proportional
  to $\Delta(1/s)$. The least-squares isotropic scale of each pair's displacement field, accumulated and rescaled to the
  nominal end points, gives the focus position of every slice. The voice-coil steps are irregular: the gauged step is
  0.53 to 1.49 times the nominal one (5th to 95th percentile), and the deviations are uncorrelated between captures
  (median correlation 0.04), so they cannot be calibrated once. On synthetic sweeps the gauge recovers the focus
  positions to within 0.05\,D (rms step error 0.004\,D).
  \item \emph{Frame quality.} In 29\% of the pairs the displacement field departs from a similarity by more than
  0.12\,px rms. These pairs coincide with the irregular steps: 82\% of the pairs whose step deviates from nominal by
  more than a factor $e^{0.3}$ are flagged, against 18\% of the others (rank correlation 0.50). The lens was probably
  still moving during the rolling-shutter readout, so the focus is not uniform across those frames; they are excluded.
  \item \emph{Gain and flat field.} A smooth multiplicative field matches exposure and vignetting within each pair.
  The raw dual-pixel data carry a fixed pattern of 2.3--2.4\% rms. A flat field estimated from all 2303 images removes
  most of it in held-out locations (cross-location correlation from 0.57--0.62 to 0.23--0.24), and it changes the band
  statistics by less than 1\%.
\end{itemize}

\Para[S-par:pixref]{Reference depth}
Depth from focus (DfF) over all 49 slices: the energy in the 45 and 65 cycles/mm bands, normalized by the local mean,
is maximized over the slices, and the peak is refined by a log-parabola in diopters. Only points with a clear peak (at
least twice the median energy) are kept. As an independent check, the dataset's multi-view-stereo depth is aligned to
the raw grid by an affine fit to the RGB images (normalized cross-correlation about 0.8). DfF agrees with it up to one
global scale per capture: median correlation 0.89 over the 47 captures and median absolute difference 0.23\,D after
scaling. The scale between the two inverse depths is 2.14 (interquartile range 2.06 to 2.29); the MVS depth is not
metric, as the dataset documentation warns.

\Para{Statistics}
For a pair $(k,k+1)$ with gauged step $\delta_k$ and band $b$, window sums are taken at fixed scene points (Gaussian
window of $67\,\mu$m): $S_{xy}=\langle X_b\bar Y_b\rangle$ with $X_b=W_b*(I_{k+1}-I_k)$ and $\bar Y_b=\Lap W_b*\bar I$,
and $S_\times=\langle Y_{b,k}Y_{b,k+1}\rangle$ with $Y_{b,k}=\Lap W_b*I_k$. The temporal noise of the two slices is
independent, so $S_\times$ has no noise bias. The pooled ratio of a $\psi$ bin is
\begin{equation}
  \kappa_b=\frac{\sum c_k\,S_{xy}}{\sum S_\times\,\psi_{\mathrm{GT}}},\qquad c_k=\frac{4}{\delta_k(Rs)^2},
  \label{S-eq:kappa_pooled}
\end{equation}
reported as $\kappa_b/\kappa_{20}$ averaged over $\pm\psi$, with 95\% intervals from a bootstrap over captures. Medians
of per-pixel ratios must be avoided: near focus they are biased toward zero by noisy denominators and by
signal-to-noise gating. The error of $\psi_{\mathrm{GT}}$ is one offset $\epsilon_p$ per scene point $p$, while the
differences between the pairs of a point are exact up to the focus-step gauge. The within-point slope
\begin{equation}
  W_b=\frac{\sum_p\sum_kc_kS_{xy,k}\,(g_k-\bar g_p)}{\sum_p\sum_kS_{\times,k}\,(g_k-\bar g_p)^2},\qquad
  \bar g_p=\frac{\sum_kS_{\times,k}\,g_k}{\sum_kS_{\times,k}},
  \label{S-eq:within}
\end{equation}
with $g_k=\psi_{\mathrm{GT}}$ and sums over the pairs $k$ of point $p$ inside a window $\psi_1\le|\psi_{\mathrm{GT}}|<\psi_2$,
removes $\epsilon_p$: since $\mathbb E[c_kS_{xy,k}]=\kappa_b(g_k-\epsilon_p)S_{\times,k}$, the offset drops out by
construction of $\bar g_p$. In the ratio $W_b/W_{20}$ a common scale, such as that of the reference depth, cancels as
well, though the pupil radius still enters through $f/f_c$ and the step.

\Para[S-par:pixcontrols]{Controls}
\begin{itemize}
  \item \emph{Synthetic sweeps} with the same grid, irregular focus steps, breathing, OIS-like shifts and
  Poisson--Gaussian noise, rendered with the exact wave or geometric disc OTF, or with the OTF of a Gaussian pupil
  with the disc's second moment (\texttt{synth\_sweep.py gauss}), and processed by the same code (three sweeps each,
  seeds 1--3). For $|\psi|\le0.55$\,D the pipeline reproduces the model ratios within about 5\%. With the within-point
  slopes it reproduces, for all three pupils, the model with the synthetic reference's own error within 0.02 at
  0.3--0.45\,D and lies up to 0.05 above it at 0.45--0.6\,D (Table~\ref{S-tab:real_xb}).
  \item \emph{Aliasing.} The samples are far coarser than the diffraction cutoff, so near focus texture above Nyquist
  folds into the bands and responds to defocus at its true frequency. A model of the sampled system
  (Table~\ref{S-tab:alias}) shows that the ratios for $|\psi|\le0.25$\,D depend on unknown sampling details, while for
  $0.3\le|\psi|\le0.9$\,D they change by less than 0.06 and the two models stay far apart. The test therefore uses
  $|\psi|=0.3$ to 0.6\,D.
  \item \emph{Reference error.} On the synthetic sweeps the DfF reference has an error of 0.12\,D (robust standard
  deviation against the true depth). A Gaussian reference error of 0.1\,D in the model reproduces the lowering of
  the pooled ratios there; for 95/20 at 0.75\,D: 1.07 (synthetic), 1.09 (model with the error), 1.19 (model without
  it). A Gaussian reference error of 0.3\,D alone would lower the clear-disc 95/20 ratio from 0.91 to 0.64 at 0.38\,D. The within-point slope
  removes this confound: on the synthetic sweeps, 95/20 at 0.38/0.53/0.75\,D is 0.94/1.04/1.22 (wave render),
  1.16/1.30/1.65 (geometric render) and 1.00/1.07/1.04 (Gaussian render), against 0.94/1.00/1.25, 1.18/1.27/1.67 and
  1.02/1.02/1.04 for the model with a 0.12\,D reference error.
\end{itemize}

\begin{table}[htbp]
\centering
\caption{Model ratio $\kappa_{95}/\kappa_{20}$ for the Pixel~3 sampling, without aliasing and with the two limiting
pixel apertures.}
\label{S-tab:alias}
\begin{tabular}{llccccc}
\toprule
Optics & Sampling & 0.15\,D & 0.25\,D & 0.38\,D & 0.53\,D & 0.75\,D \\
\midrule
wave & no aliasing & 0.85 & 0.86 & 0.90 & 0.98 & 1.21 \\
     & $1.4\,\mu$m points & 0.79 & 0.86 & 0.97 & 1.06 & 1.24 \\
     & $2.8\times5.6\,\mu$m box & 0.91 & 0.92 & 0.94 & 1.00 & 1.21 \\
\midrule
geometric & no aliasing & 1.02 & 1.05 & 1.12 & 1.25 & 1.66 \\
     & $1.4\,\mu$m points & 0.80 & 1.00 & 1.13 & 1.29 & 1.69 \\
     & $2.8\times5.6\,\mu$m box & 1.17 & 1.15 & 1.16 & 1.26 & 1.63 \\
\bottomrule
\end{tabular}
\end{table}

\begin{figure}[htbp]
\centering
\includegraphics[width=\linewidth]{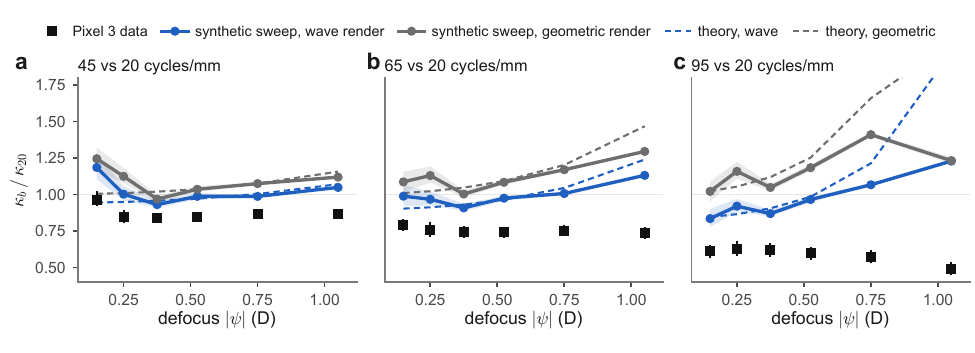}
\caption{\textbf{Phone camera: the pipeline reproduces the model ratios, and the data fall below both models.} Band
ratios against $|\psi|$ (whole field): real data, the pipeline applied to synthetic wave and geometric sweeps (shaded:
95\% intervals), and the clear-disc theory. Below 0.3\,D aliasing affects the ratios (Table~\ref{S-tab:alias}).}
\label{S-fig:crossband}
\end{figure}

\Para[S-par:pixratios]{Band ratios}
Table~\ref{S-tab:real_xb} gives the band ratios for 4.4~million pair samples with the DfF reference, and
Fig.~\ref{S-fig:crossband} the pooled ratios against $|\psi|$ (\texttt{realdata/\allowbreak l2af\_within\_point.py},
\texttt{results/\allowbreak l2af\_within\_point.txt}). The statistic of Fig.~\ref{S-fig:evidence}a is the
within-point slope $W_b$ of Eq.~\eqref{S-eq:within}, which an offset of the reference at a scene point cannot bias; every
model is computed through the same statistic, steps and windows, with its own DfF reference
(\texttt{model\_pupil.\allowbreak group\_stats}). The 47 captures come from five rooms (home1 9, home2 10, microkitchen1 10,
pillowroom1 8, secondfloorgames 10), and captures of one room share scenes and light, so the intervals resample the
rooms with replacement and then the captures within each drawn room (20\,000 resamples).
\begin{itemize}
  \item \emph{Windows and multiplicity.} We chose the bands, the window 0.3--0.6\,D and the three zones with these data
  in view, so the analysis is exploratory; a confirmation on new rooms, registered in advance, follows below
  (Sec.~\ref{S-sec:pixelconfirm}). The conclusion does not hinge on the window: in all four windows from 0.3 to
  1.2\,D and in every zone, the ratios at 65 and 95\,cycles/mm exclude both a constant coefficient and geometric
  optics. The smallest margin to the constant coefficient (model minus the upper end of the interval) is $+0.040$, at
  the center, 65/20, 0.3--0.45\,D ($z=2.7$ over rooms and captures, 3.3 over captures); at 95/20 it is $+0.15$. The
  claim of the main paper rests on twelve cells (65 and 95 against 20\,cycles/mm, three zones, two windows) and holds
  only if every cell excludes the model, an intersection--union test, so each cell is tested at its own one-sided
  2.5\% level without correction.
  \item \emph{Rooms.} The rooms differ most in the middle zone at 95/20: Cochran's $Q$ is 11.0 and 26.1 in the two
  windows (between-room SD $\tau=0.067$ and 0.091), with microkitchen1 at 0.72--0.74 against 0.51--0.61 for the
  others. With any one room left out, every one of the twelve cells still excludes the constant coefficient (smallest
  margin $+0.021$, center, 65/20, 0.3--0.45\,D, secondfloorgames left out; two-stage intervals over the other four
  rooms).
  \item \emph{The reference error.} $W$ removes the offset of the reference but not its effect on which samples fall in a
  window; an error of 0.12\,D raises $W$ at 95/20 for the disc from 0.916 to 0.940 and for geometric optics from 1.138
  to 1.176 in the first window, and leaves the constant coefficient at 1.01--1.02, so the model values at zero error are
  conservative. The pooled ratio $P$ (Eq.~\eqref{S-eq:kappa_pooled}) falls with the error for every pupil. On the data
  $P$ lies 4\% and 8\% below $W$ at 95/20 (whole field), the gap the constant coefficient shows at an error between
  0.06\,D ($-2\%$) and 0.12\,D ($-8\%$ to $-9\%$); the synthetic sweeps measure 0.11--0.12\,D. Through $P$ with that error the
  constant coefficient gives 0.93--0.94 at 95/20, still far above the data.
\end{itemize}

\begin{table}[htbp]
\centering
\caption{Phone camera: band ratios at $|\psi|\approx0.38$ / $0.53$\,D (windows 0.3--0.45 and 0.45--0.6\,D, both sides
of focus), 47 captures in five rooms. Upper part: within-point slopes $W_b/W_{20}$ (Eq.~\eqref{S-eq:within}), the
statistic of Fig.~\ref{S-fig:evidence}a, with 95\% intervals over 20\,000 resamples of the rooms and of the
captures within them (second row of each zone); the models through the same statistic, steps and windows, with their
own DfF reference at zero error; and the same pipeline applied to synthetic sweeps rendered with each pupil (three
each; their reference errs by 0.11--0.12\,D). Lower part: the pooled ratio $P$ (Eq.~\eqref{S-eq:kappa_pooled}), which
an error of the reference lowers for every pupil, and the Gaussian pupil through $P$ without and with a Gaussian
reference error of 0.12\,D (\texttt{realdata/l2af\_within\_point.py}).}
\label{S-tab:real_xb}
\footnotesize
\setlength{\tabcolsep}{3.2pt}
\begin{tabular}{lcccc}
\toprule
& 30/20 & 45/20 & 65/20 & 95/20 \\
\midrule
\multicolumn{5}{l}{\emph{within-point slopes $W$}} \\
data, center ($r<0.3$) & 0.97 / 0.98 & 0.92 / 0.92 & 0.87 / 0.86 & 0.78 / 0.74 \\
\quad 95\% interval & & & [0.77, 0.96] / [0.78, 0.93] & [0.68, 0.86] / [0.67, 0.81] \\
data, middle ($0.3\le r<0.6$) & 0.94 / 0.94 & 0.83 / 0.84 & 0.73 / 0.75 & 0.61 / 0.64 \\
\quad 95\% interval & & & [0.63, 0.81] / [0.65, 0.82] & [0.50, 0.71] / [0.52, 0.73] \\
data, edge ($r\ge0.6$) & 0.90 / 0.95 & 0.75 / 0.83 & 0.63 / 0.67 & 0.50 / 0.53 \\
\quad 95\% interval & & & [0.56, 0.70] / [0.59, 0.75] & [0.40, 0.60] / [0.42, 0.63] \\
data, whole field & 0.94 / 0.95 & 0.84 / 0.86 & 0.75 / 0.77 & 0.64 / 0.65 \\
\quad 95\% interval & & & [0.67, 0.81] / [0.69, 0.82] & [0.56, 0.71] / [0.56, 0.72] \\
clear disc, wave optics & 0.98 / 0.98 & 0.96 / 0.97 & 0.93 / 0.96 & 0.92 / 0.97 \\
clear disc, geometric optics & 1.01 / 1.01 & 1.02 / 1.03 & 1.05 / 1.08 & 1.14 / 1.23 \\
Gaussian pupil (constant coefficient) & 1.00 / 1.00 & 1.00 / 1.00 & 1.00 / 1.01 & 1.01 / 1.02 \\
synthetic sweeps: disc, wave & 0.97 / 1.00 & 0.94 / 1.00 & 0.94 / 1.00 & 0.94 / 1.04 \\
\quad disc, geometric & 0.97 / 1.02 & 1.00 / 1.05 & 1.06 / 1.13 & 1.16 / 1.30 \\
\quad Gaussian pupil & 0.97 / 1.04 & 0.96 / 1.04 & 0.98 / 1.05 & 1.00 / 1.07 \\
\midrule
\multicolumn{5}{l}{\emph{pooled ratios $P$}} \\
data, center & 0.98 / 0.97 & 0.93 / 0.92 & 0.87 / 0.85 & 0.76 / 0.71 \\
data, whole field & 0.94 / 0.94 & 0.84 / 0.85 & 0.74 / 0.74 & 0.62 / 0.60 \\
synthetic sweeps: disc, wave & 0.96 / 1.00 & 0.93 / 0.99 & 0.91 / 0.97 & 0.87 / 0.97 \\
\quad disc, geometric & 0.94 / 1.01 & 0.96 / 1.04 & 1.00 / 1.08 & 1.04 / 1.18 \\
\quad Gaussian pupil & 0.98 / 1.04 & 0.96 / 1.04 & 0.96 / 1.03 & 0.96 / 1.03 \\
Gaussian pupil, no reference error & 1.00 / 1.00 & 1.00 / 1.00 & 1.00 / 1.01 & 1.01 / 1.02 \\
Gaussian pupil, reference error 0.12\,D & 0.99 / 0.99 & 0.98 / 0.98 & 0.96 / 0.96 & 0.93 / 0.94 \\
\bottomrule
\end{tabular}
\end{table}

\begin{figure}[htbp]
\centering
\includegraphics[width=\linewidth]{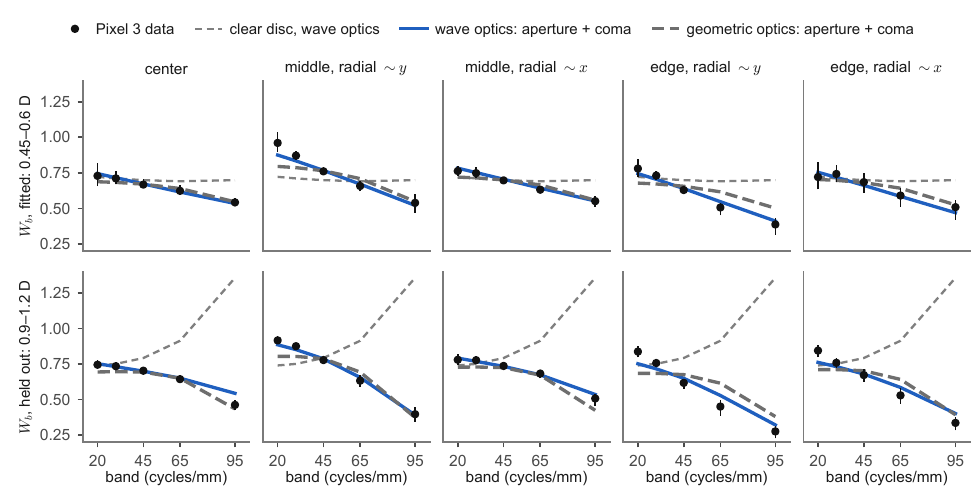}
\caption{\textbf{Phone camera: wave optics with an effective aperture and coma fits every field class.} Absolute
within-point band slopes by field class in a fitted window (top) and the held-out window (bottom): data (95\% intervals
over captures), the clear disc, and the joint fits with an effective aperture and coma in wave and in geometric optics
(one global scale).}
\label{S-fig:aberr}
\end{figure}

\begin{figure}[htbp]
\centering
\includegraphics[width=\linewidth]{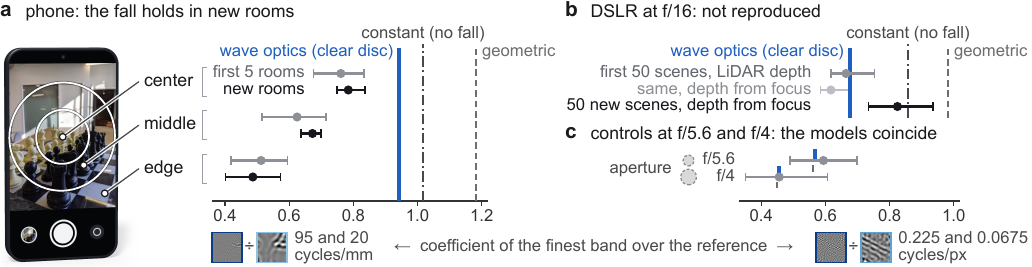}
\caption{\textbf{Slope ratios on the phone and on the DSLR.} Dots: the slope ratio of the finest band to the
reference band (pictures), our measure of the ratio of their coefficients, with 95\% intervals; lines: the same
statistic of three models. Gray: the data in which we found the fall; black: the confirmation fixed in advance.
(a)~Phone, within-point slopes at $|\psi|=0.3$--0.6\,D in three field zones (circles): 47 sweeps in five rooms, then
351 in new rooms; the data fall faster than a clear disc. (b)~DSLR at f/16, small focus steps, $|\psi|<0.2$\,D, models
through a simulated processing chain: 50 validation scenes with LiDAR depth, the same by depth from focus as in the
confirmation (pale), and the 50 test scenes, which refute the fall. (c)~Controls at f/5.6 and f/4 on the validation
scenes, where wave and geometric optics coincide. The axis is a slope ratio, which the finite focus step lowers for
every pupil; this is why a constant coefficient lies at 0.84 and not at 1 in panel b.}
\label{S-fig:evidence}
\end{figure}

\Para[S-par:pixpupil]{Pupil fit}
\begin{itemize}
  \item \emph{Model.} A hard effective aperture of radius $rR$ with Zernike or Seidel aberrations and an optional
  incoherent focus spread, in wave optics (exact overlap integral) or geometric optics (spot-diagram OTF,
  $H=\langle\exp(2\pi i\,s\vf\cdot\nabla W)\rangle$). It uses the same band weights and the same within-point statistic
  as the data, including the DfF window. The OTFs agree with one-dimensional disc quadratures to $10^{-6}$. On the
  synthetic sweeps the model statistic matches the pipeline within 0.03 for $|\psi|\le0.9$\,D.
  \item \emph{Data.} Absolute within-point slopes $W_b$ for five bands and three windows (0.3--0.45, 0.45--0.6 and
  0.6--0.9\,D) in five field classes: the center ($r<0.3$), and the middle ($0.3\le r<0.6$) and edge ($r\ge0.6$) zones
  split by whether the radial direction lies within $45^\circ$ of the band axis. The 0.9--1.2\,D window is held out.
  The covariance is a bootstrap over captures plus a 3\% model floor.
  \item \emph{Scale.} One global factor $g=0.96\,s_q^2$ for all classes, where $s_q$ is the unknown metric scale of the
  focus table and 0.96 the pipeline factor from the synthetic sweeps. Focus breathing bounds it. Over the sweep the
  magnification changes by 5.07\% (interquartile range 4.99--5.17\% over captures). For lens-shift focusing this equals
  $\Delta s/D_{\mathrm{xp}}$, with $\Delta s=0.197$\,mm for the nominal table, so an exit pupil 3.5--5.2\,mm from the
  sensor gives $s_q=0.90$--1.34 and $g\le1.75$.
\end{itemize}
The deficit grows about linearly with frequency: a fit of $1-\alpha f-\beta f^2$ to the band ratios gives
$\beta\approx0$ in every field class, and $\alpha=0$ is worse by $\Delta\chi^2=40$--220. Table~\ref{S-tab:aberr} gives
the joint fits, Table~\ref{S-tab:aberr_par} the parameters of the best wave fit, and Fig.~\ref{S-fig:aberr} the slopes.

\begin{table}[htbp]
\centering
\caption{Joint fit of the absolute within-point slopes (5 field classes $\times$ 5 bands $\times$ 3 windows = 75
values, one global scale $g\le1.75$, per-class pupil parameters). Held out: the 0.9--1.2\,D window (25 values,
$\chi^2$ of the prediction). Aperture $r$ in units of $R$; coma along the band axis.}
\label{S-tab:aberr}
\begin{tabular}{lcccc}
\toprule
& \multicolumn{2}{c}{wave optics} & \multicolumn{2}{c}{geometric optics}\\
Model per class & $\chi^2$/dof & held out & $\chi^2$/dof & held out \\
\midrule
clear disc & 1778/74 & 9874 & 5737/74 & 9832 \\
aperture & 298/69 & 1169 & 986/69 & 2701 \\
coma + astigmatism + focus spread & 97.5/59 & 91.0 & 141.7/59 & 118.1 \\
aperture + coma & \textbf{44.7/64} & \textbf{46.4} & 157.9/64 & 135.9 \\
aperture + coma + focus spread & 41.4/59 & 39.0 & 97.3/59 & 79.0 \\
\bottomrule
\end{tabular}
\end{table}

\begin{table}[htbp]
\centering
\caption{Wave optics, aperture + coma, at the scale bound $g=1.75$: effective aperture, coma at the nominal rim,
rms wavefront inside the effective aperture (after tilt), and $\chi^2$ for the 15 values of each class.}
\label{S-tab:aberr_par}
\begin{tabular}{lcccc}
\toprule
Class & $r$ & coma [waves] & rms [waves] & $\chi^2$/15 \\
\midrule
center & 0.68 & 1.51 & 0.055 & 6.2 \\
middle, radial along $y$ & 0.73 & 2.02 & 0.094 & 10.4 \\
middle, radial along $x$ & 0.69 & 1.56 & 0.061 & 2.4 \\
edge, radial along $y$ & 0.68 & 2.56 & 0.095 & 16.5 \\
edge, radial along $x$ & 0.68 & 2.16 & 0.081 & 9.2 \\
\bottomrule
\end{tabular}
\end{table}

\begin{itemize}
  \item Neither mechanism alone fits. A smaller aperture alone gives the linear term but drives $g$ to its bound and
  fails in the held-out window. Aberrations alone give $f^2$ deficits and miss the low bands where the radial direction
  lies along the band axis.
  \item Aperture and coma fit every class and predict the held-out window ($\chi^2=46.4$ for 25 values). The largest
  misses are 0.06--0.09: at 95 cycles/mm at the center, and at 20 and 65 cycles/mm in the edge zone. On the synthetic
  sweeps the pipeline itself departs from the model by up to this much in that window.
  \item The scale and the aperture trade off. With the nominal focus table ($g=0.96$) the fit gives $r\approx0.9$, coma
  $\approx1$ wave and $\chi^2=76.5$. Across $g=0.96/1.25/1.75/2.2$, wave optics gives $\chi^2=76.5/56.5/44.7/45.3$ and
  geometric optics $225.6/187.5/157.9/146.0$.
  \item The ratio fits per class with a free gain lead to the same ranking.
  \item \emph{Band focus.} The two sides of focus give the focus of each band relative to the DfF reference,
  $\delta_b=\psi(\kappa_--\kappa_+)/(\kappa_-+\kappa_+)$ for a linear response. In the middle zone the low bands focus
  0.06--0.08\,D to one side of the reference where the radial direction lies along the band axis, and 0.05--0.08\,D to
  the other side where it lies across it. These are the tangential and sagittal foci of astigmatism, about 0.16\,D
  apart ($W_{222}\approx0.2$ waves). At the edge the high bands focus 0.10--0.11\,D from the reference. At the center
  all offsets are below 0.03\,D.
  \item \emph{Dual-pixel baseline.} The left and right views see two halves of the pupil, so their disparity per
  diopter measures the distance between the two sub-pupil centroids. The slope is 0.683\,px/D at the center (95\%
  interval [0.665, 0.699]), unchanged through the middle zone, 26\% lower at the top and bottom edges (0.503) and 12\%
  lower at the left and right edges. This indicates a narrowing of the pupil along the radial direction (vignetting), the same
  field pattern as the extra high-band deficit at the top and bottom edges. The baseline is 0.40 times that of a sharp
  split of the nominal pupil into two half discs, so the photodiodes split the pupil softly, and the baseline alone
  cannot fix $r$.
\end{itemize}

\Para{A pre-registered confirmation on new rooms}
\label{S-sec:pixelconfirm}
Because we chose the bands, the window and the zones with the 47 test captures in view, we registered a confirmation
before obtaining any image of the training split of the dataset (\texttt{notes/\allowbreak preregistration\_l2af\_phone.md},
fixed on 2026-11-08, with one amendment to the extractor before any file was kept). The code and its inputs were frozen
by their SHA-256 and checked before and after the run.
\phantomsection\label{S-par:localrecords}The registration files, the hash lists and the file times quoted here and for
the DSLR (Sec.~\ref{S-sec:zedd}) are records on our own machine, and none carries a time stamp from a third party. They
show what was fixed; they cannot prove to a reader when it was fixed. The whole run is one script
(\texttt{realdata/\allowbreak run\_l2af\_confirm.ps1}). Run first on the test captures, laid out as the training split,
it reproduced the processed captures, the sums and the MVS cross-check of this section exactly.
\begin{itemize}
  \item \emph{Data.} Every capture of the seven training archives whose center-phone views are complete for all 49
  slices, streamed once with each archive checked against its MD5. Rooms named like a test room are left out, and the
  decision needs at least 20 rooms. Processing, reference and statistic are those of this section. The flat field is
  applied if the fixed pattern of the new captures correlates with it at 0.12 or more in both views (half the value
  between disjoint halves of the test split), and not applied otherwise.
  \item \emph{Statistic.} $W_b/W_{20}$ (Eq.~\eqref{S-eq:within}) at 65 and 95 cycles/mm in the three zones and the windows
  0.30--0.45 and 0.45--0.60\,D, which gives twelve cells. The 95\% intervals come from 20\,000 two-stage resamples (the
  rooms, then the captures within each drawn room), and $U_c$ is the upper end in cell $c$.
  \item \emph{Predictions.} For a constant coefficient ($G_c$, the Gaussian pupil with the disc's second moment) and for
  geometric optics ($\mathrm{Geo}_c$), each cell takes the lowest of three values. The first is the theory through the
  same statistic at slice steps of 0.10--0.35\,D, which span the gauged steps. The second is the lower one-sided
  97.5\% bound over 30 synthetic sweeps of that pupil through the whole pipeline, whose reference errs by
  0.11--0.13\,D. The third is the theory at 0.20\,D times the aliasing factor of either pixel aperture of
  Table~\ref{S-tab:alias}. A lower prediction makes a confirmation harder.
  \item \emph{Decision.} The result is confirmed if $U_c<G_c$ and $U_c<\mathrm{Geo}_c$ in all twelve cells, an
  intersection--union test at a one-sided level of 2.5\%. It is confirmed at 95 cycles/mm only if all six 95/20 cells
  pass but some 65/20 cell does not, refuted if $U_c\ge G_c$ in some 95/20 cell, and undecided otherwise.
  \item \emph{Effective pupil.} The prediction of the effective pupil of Table~\ref{S-tab:aberr} (wave optics, aperture and
  coma per class, one scale) is reported whatever the outcome. Its 95\% prediction intervals combine its refits, the
  spread between rooms and the resampling of the new data. It counts as inside if the joint $\chi^2$ of the twelve
  cells lies below its 95\% point and no cell lies outside its 99.6\% interval. On the 47 captures it was fitted to,
  the joint $\chi^2$ is 29.1, so it fails this criterion on its own data.
\end{itemize}
These rules were set after the test captures had been analyzed. Applied to the test captures, they give ``confirmed at
95 cycles/mm only''. In the center, 65/20, 0.30--0.45\,D, $U_c$ is 0.961 against $G_c=0.947$, where the synthetic bound
sets the prediction; the comparison above uses the theory, which gives 1.003 in that cell. We simulated the outcome with
the spread between rooms and between captures measured on the test captures. For about 35 new rooms, ``confirmed'' had a
probability of 0.82 and ``confirmed at 95 cycles/mm only'' 0.18 (0.63 and 0.37 with every prediction lowered by 0.03),
and ``refuted'' stayed below 0.01. We first rendered 10 synthetic sweeps per pupil and raised the number to 30 before any
synthetic bound had been computed. We also ran the prediction and decision scripts once on three-sweep predictions to
test their mechanics; on the test captures that run gave the same outcome.

\Para[S-par:pixoutcome]{Outcome on new rooms}
After the registration we streamed the seven training archives once (839\,GB), keeping the center-phone files, and
checked each archive against its MD5. We then checked that none of the 79 frozen files had changed and ran the
confirmation unchanged (\texttt{results/\allowbreak l2af\_confirm\_train.txt}). The list holds 356 captures under 37
room names, none named like a test room; the dataset's readme gives 351 sweeps for this split, and five of the 356
proved to be copies of others (\hyperref[S-par:pixdup]{see below}). All 356 captures were
processed without a failure. The fixed pattern correlates with our flat field at 0.553 and 0.562, so the camera is the
same unit and the flat field was applied.

The outcome is \emph{confirmed at 95 cycles/mm only} (Table~\ref{S-tab:pixelconfirm}). In all six 95/20 cells $U_c$ lies
below both predictions. $W$ is 0.47--0.80, against 0.97--1.01 for $G_c$ and 1.09--1.21 for $\mathrm{Geo}_c$, and the
smallest margin to $G_c$ is 0.083 (center, 0.30--0.45\,D). At 65/20 five of the six cells exclude both predictions. In
the center at 0.30--0.45\,D the ratio is 0.877 [0.827, 0.958], which excludes geometric optics (0.997) but not the
constant coefficient (0.947), as on the test captures. The main paper's answer to Q1 therefore says ``at 95 cycles/mm
only''. The registration asks for the outcome in the abstract, in the section on the phone and in that
answer; the section, which the record calls Sec.~6.2, is main-paper Sec.~\ref{sec:pixel}. The registered
secondary outcomes follow, reported whatever the outcome.
\begin{itemize}
  \item \emph{Clear disc.} The data lie below its value in all twelve cells (0.92--0.97 at 95/20), as on the test
  captures, so the fall is steeper than for a clear disc.
  \item \emph{Effective pupil.} Over the twelve cells its prediction gives a joint $\chi^2$ of 25.7, above the 95\% point
  of 21.0, so by our criterion the new rooms lie outside it. No cell lies outside its 99.6\% interval (largest $|z|$
  2.11). The data lie above the prediction at the center at 95/20 ($z=+2.11$ and $+1.98$) and below it at the edge
  ($-1.01$ to $-1.90$). The clear disc lies outside the 95\% prediction interval in all six 95/20 cells, so the
  criterion discriminates.
  \item \emph{Rooms.} Each cell rests on an effective number of 15.7--19.3 rooms (3.1--3.7 on the test captures), and no
  room carries more than 17\% of the weight of a cell (photobox1). With any one room left out the outcome stays
  ``confirmed at 95 cycles/mm only'' for 35 rooms, and it becomes ``confirmed'' without books or without garage.
  \item \emph{Pooled ratio and reference error.} In the center and middle zones $P/W-1$ at 95/20 is $-0.046$ to $-0.096$
  over the four windows, close to the $-0.072$ to $-0.089$ that the constant coefficient shows at a reference error of
  0.12\,D ($-0.019$ to $-0.025$ at 0.06\,D). At the edge it is $-0.050$ to $+0.053$, which no reference error of the
  model gives.
  \item \emph{MVS cross-check.} The registered step stopped at tableroom\_6, the last capture, and saved nothing. The
  archives hold no scaled RGB image for that capture, and the selection needs only the raw views. We reran the frozen
  script unchanged on the other 355 captures (\texttt{results/\allowbreak mvs\_check\_train\_rgb355.txt}). Four of them
  have fewer than 50 points at which both depths are valid (macrokeg\_6 and \_8, macroobjects\_0 and \_1). On the other
  351 the reference agrees with the multi-view-stereo depth as on the test captures, whose values from the paragraph
  \emph{Reference depth} follow in parentheses. The median correlation is 0.90 (0.89), the median absolute difference
  after scaling 0.22\,D (0.23\,D), and the scale 2.19 (2.14), with interquartile range 2.07 to 2.38 (2.06 to 2.29). Of
  the 351 captures, 27 correlate below 0.5 (3 of 47), 11 of them in the six close-up rooms whose names begin with macro.
  The reference depths of these 27 span 1.8\,D at the median (5th to 95th percentile), against 3.5\,D for the others,
  and differ from the scaled MVS depth by 0.29\,D at the median.
  \item \emph{Other bands and windows.} Table~\ref{S-tab:pixelconfirm_all}. At 30/20 the ratios are 0.90--0.98, against
  0.93--1.00 predicted for a constant coefficient. Beyond 0.6\,D every 95/20 ratio lies between 0.38 and 0.74.
\end{itemize}
Fig.~\ref{S-fig:evidence}a shows the 95\,cycles/mm band only, the new rooms in black and the five rooms in
gray, each as the mean of the two windows with two-stage intervals; the new rooms give 0.78, 0.67 and 0.49 from the
center to the edge, and Tables~\ref{S-tab:real_xb}, \ref{S-tab:pixelconfirm} and~\ref{S-tab:pixelconfirm_all} give each
window and band. The phone drawn there shows capture secondfloorgames\_6 of the five rooms at the focus setting where a
window 96 rows high, centered 62\% down and 45\% across the frame, is sharpest, and the two pictures under the axis
show that window in the 95 and 20\,cycles/mm bands (\texttt{realdata/\allowbreak fig\_evidence\_data.py}). The
decision rests on the twelve cells.

\begin{table*}[tp]
\centering
\caption{\textbf{The pre-registered confirmation on new rooms: confirmed at 95 cycles/mm only.} Within-point slope
ratios $W_b/W_{20}$ of the 356 captures (351 distinct sweeps under 37 room names, see the text), with 95\% intervals
over 20\,000 resamples of the rooms and of the captures within
them, which decide, and over the captures alone; the pooled ratio $P$; the registered predictions of a constant
coefficient ($G_c$) and of geometric optics ($\mathrm{Geo}_c$), each the lowest of three values; the clear disc at a
step of 0.20\,D; and the prediction of the effective pupil of Table~\ref{S-tab:aberr} with its 95\% prediction interval,
and the $z$ of the data against it. Bold: the one cell whose upper end does not lie below $G_c$
(\texttt{results/l2af\_confirm\_train.txt}).}
\label{S-tab:pixelconfirm}
\scriptsize
\setlength{\tabcolsep}{3pt}
\begin{tabular}{lllcccccccc}
\toprule
zone & window [D] & band & $W$ [rooms, captures] & captures only & $P$ & $G_c$ & $\mathrm{Geo}_c$ & disc & effective pupil & $z$ \\
\midrule
center & 0.30--0.45 & 65/20 & \textbf{0.877 [0.827, 0.958]} & [0.834, 0.934] & 0.846 & \textbf{0.947} & 0.997 & 0.935 & 0.822 [0.705, 0.938] & $+0.93$ \\
 & & 95/20 & 0.800 [0.748, 0.886] & [0.754, 0.861] & 0.755 & 0.969 & 1.086 & 0.916 & 0.713 [0.633, 0.793] & $+2.11$ \\
 & 0.45--0.60 & 65/20 & 0.857 [0.825, 0.894] & [0.833, 0.883] & 0.824 & 0.989 & 1.083 & 0.957 & 0.826 [0.764, 0.888] & $+0.98$ \\
 & & 95/20 & 0.769 [0.735, 0.804] & [0.743, 0.795] & 0.711 & 1.005 & 1.210 & 0.969 & 0.718 [0.667, 0.768] & $+1.98$ \\
middle & 0.30--0.45 & 65/20 & 0.753 [0.710, 0.782] & [0.728, 0.775] & 0.737 & 0.994 & 1.043 & 0.935 & 0.792 [0.689, 0.894] & $-0.73$ \\
 & & 95/20 & 0.660 [0.611, 0.693] & [0.628, 0.686] & 0.629 & 1.003 & 1.102 & 0.916 & 0.654 [0.503, 0.805] & $+0.09$ \\
 & 0.45--0.60 & 65/20 & 0.776 [0.733, 0.804] & [0.751, 0.797] & 0.734 & 1.001 & 1.083 & 0.957 & 0.794 [0.682, 0.906] & $-0.31$ \\
 & & 95/20 & 0.685 [0.653, 0.713] & [0.666, 0.703] & 0.620 & 1.005 & 1.210 & 0.969 & 0.652 [0.462, 0.842] & $+0.34$ \\
edge & 0.30--0.45 & 65/20 & 0.578 [0.395, 0.686] & [0.450, 0.670] & 0.615 & 0.989 & 1.026 & 0.935 & 0.743 [0.570, 0.916] & $-1.86$ \\
 & & 95/20 & 0.467 [0.339, 0.578] & [0.374, 0.551] & 0.492 & 1.003 & 1.102 & 0.916 & 0.573 [0.418, 0.729] & $-1.34$ \\
 & 0.45--0.60 & 65/20 & 0.625 [0.521, 0.698] & [0.548, 0.688] & 0.599 & 0.981 & 1.077 & 0.957 & 0.742 [0.621, 0.862] & $-1.90$ \\
 & & 95/20 & 0.507 [0.451, 0.572] & [0.468, 0.551] & 0.482 & 0.997 & 1.210 & 0.969 & 0.564 [0.453, 0.675] & $-1.01$ \\
\bottomrule
\end{tabular}
\end{table*}

\begin{table*}[tp]
\centering
\caption{\textbf{The confirmation's secondary bands and windows.} $W_b/W_{20}$ of the 356 captures with 95\% intervals
over rooms and captures, and in parentheses the registered prediction of a constant coefficient $G_c$. ---: $G_c$ is
negative there, set by synthetic sweeps that carry almost no 95\,cycles/mm signal at that defocus. Geometric optics and
the clear disc for every cell are in \texttt{results/l2af\_confirm\_train.txt}.}
\label{S-tab:pixelconfirm_all}
\footnotesize
\setlength{\tabcolsep}{4pt}
\resizebox{\textwidth}{!}{%
\begin{tabular}{llcccc}
\toprule
zone & window [D] & 30/20 & 45/20 & 65/20 & 95/20 \\
\midrule
center & 0.30--0.45 & 0.954 [0.914, 1.007] (0.932) & 0.924 [0.882, 0.989] (0.934) & 0.877 [0.827, 0.958] (0.947) & 0.800 [0.748, 0.886] (0.969) \\
 & 0.45--0.60 & 0.970 [0.948, 0.990] (0.961) & 0.927 [0.897, 0.962] (0.974) & 0.857 [0.825, 0.894] (0.989) & 0.769 [0.735, 0.804] (1.005) \\
 & 0.60--0.90 & 0.966 [0.950, 0.983] (0.992) & 0.926 [0.906, 0.948] (1.000) & 0.861 [0.835, 0.886] (1.002) & 0.740 [0.706, 0.774] (1.008) \\
 & 0.90--1.20 & 0.979 [0.965, 0.995] (0.990) & 0.950 [0.930, 0.974] (0.992) & 0.885 [0.851, 0.919] (0.993) & 0.667 [0.614, 0.727] (0.993) \\
middle & 0.30--0.45 & 0.932 [0.918, 0.945] (0.972) & 0.821 [0.728, 0.864] (0.974) & 0.753 [0.710, 0.782] (0.994) & 0.660 [0.611, 0.693] (1.003) \\
 & 0.45--0.60 & 0.946 [0.931, 0.958] (1.000) & 0.863 [0.813, 0.891] (1.000) & 0.776 [0.733, 0.804] (1.001) & 0.685 [0.653, 0.713] (1.005) \\
 & 0.60--0.90 & 0.952 [0.940, 0.962] (1.000) & 0.875 [0.851, 0.893] (1.000) & 0.785 [0.757, 0.807] (1.002) & 0.658 [0.623, 0.690] (1.008) \\
 & 0.90--1.20 & 0.960 [0.948, 0.970] (0.994) & 0.894 [0.872, 0.910] (0.993) & 0.793 [0.759, 0.819] (0.987) & 0.562 [0.522, 0.605] (---) \\
edge & 0.30--0.45 & 0.904 [0.850, 0.942] (0.982) & 0.649 [0.384, 0.804] (0.980) & 0.578 [0.395, 0.686] (0.989) & 0.467 [0.339, 0.578] (1.003) \\
 & 0.45--0.60 & 0.923 [0.884, 0.954] (0.981) & 0.725 [0.573, 0.829] (0.971) & 0.625 [0.521, 0.698] (0.981) & 0.507 [0.451, 0.572] (0.997) \\
 & 0.60--0.90 & 0.920 [0.899, 0.937] (0.988) & 0.785 [0.763, 0.809] (0.988) & 0.648 [0.602, 0.688] (0.992) & 0.466 [0.395, 0.533] (1.000) \\
 & 0.90--1.20 & 0.925 [0.903, 0.943] (0.998) & 0.793 [0.764, 0.817] (0.990) & 0.611 [0.557, 0.667] (0.989) & 0.382 [0.322, 0.454] (0.960) \\
\bottomrule
\end{tabular}}
\end{table*}

\Para[S-par:pixdup]{Sweeps that the archives hold twice, and captures without a sample}
We found the following after the confirmation, on 2026-11-13, while checking the test on the two views below; none of
it was registered.
\begin{itemize}
  \item \emph{Five sweeps are held twice.} We compared the raw dual-pixel files of all 356 training and 47 test
  captures (\texttt{realdata/\allowbreak l2af\_dup\_check.py}). Five capture folders of the seventh archive,
  randomsurplus\_1, \_3, \_4, \_5 and \_6, are identical byte for byte, in all 98 raw files each, to apt2\_0 of the
  first archive and to officeroom1\_2, \_3, \_4 and \_5 of the fifth. The training split therefore holds 351 distinct
  sweeps, the number in the dataset's readme, and the confirmation counted five of them twice. No training capture is
  identical to a test capture.
  \item \emph{One folder is not a room.} The other five captures of randomsurplus show five unrelated scenes (a patio,
  a curtain, a model rocket, a desk telephone and a shelf of records). The two-stage resampling treated this folder
  as one of 37 rooms; the split has 36 named rooms and this collection.
  \item \emph{The outcome without them.} We ran the frozen decision script on the list without the five copies (351
  captures) and on the list without the folder (346 captures in 36 rooms); the registered output is untouched
  (\texttt{results/\allowbreak l2af\_confirm\_\allowbreak train\_nodup.txt} and
  \texttt{l2af\_confirm\_\allowbreak train\_\allowbreak nosurplus.txt}). Both give ``confirmed at 95 cycles/mm
  only''. Without the copies $W$ at 95/20 is 0.466--0.803 (registered: 0.467--0.800),
  the smallest margin to $G_c$ is 0.076 (0.083), and the center at 0.30--0.45\,D and 65/20 gives 0.879 [0.828, 0.965]
  (0.877 [0.827, 0.958]) against $G_c=0.947$. Without the folder these are 0.464--0.801, 0.076 and 0.878
  [0.827, 0.965]. The tables and Fig.~\ref{S-fig:evidence}a show the registered run on the 356 captures.
  \item \emph{Captures without a sample.} The within-point slope needs two pairs of a scene point inside a window. In
  20 of the 356 captures the frame-quality rule excludes 40 to 48 of the 48 pairs and no scene point keeps two in any
  window, and four more have no sample in the windows 0.30--0.45 and 0.45--0.60\,D. The twelve cells thus rest on 332
  captures, 328 of them distinct sweeps, from all 37 folders. On the test captures 3 of the 47 have no sample in
  these windows. Of the sweeps that the main paper counts, 44 of 47 and 328 of 351 thus carry samples there.
\end{itemize}

\Para[S-par:pixviews]{A pre-specified test on the two dual-pixel views}
The confirmation above tests the fall of the coefficient through the band energies of one image, the sum of the two
dual-pixel views. The two views themselves allow two further tests of Lemma~\ref{lem:small} on the 356 captures
already on disk. We fixed both in a local record with the SHA-256 of 16 files
(\texttt{notes/\allowbreak preregistration\_l2af\_views.md}, 2026-11-12) before any training capture was processed
for its views. The record was written with the outcome of the confirmation known, and like the others it carries no
time stamp from a third party (\hyperref[S-par:localrecords]{see above}), so we call this test pre-specified and not
registered.
\begin{itemize}
  \item \emph{Coefficient and disparity of a view.} A view sees the pupil through a weight. We take the weight at
  the midpoint of the two pupil points that form the detail at $\vf$, so that view $L$ has the overlap $w_LO_{\va}$
  and view $R$ has $w_RO_{\va}$, with $w_L+w_R=1$ because the two photodiodes share the pixel. For an unaberrated
  pupil these overlaps are real and nonnegative, but they are not centered. For such a measure $\mu$ in place of
  $O_{\va}$ in Eq.~\eqref{eq:otf}, with mean $\bar{\vu}_\mu$ and variance $V_\mu$ along $\hat{\vf}=\vf/|\vf|$,
  \begin{equation}
    \frac{|H(\vf;\psi)|^2}{|H(\vf;0)|^2}=1-(2\pi s\psi)^2|\vf|^2\,V_\mu+O(\psi^4),\qquad
    \arg H(\vf;\psi)=-2\pi s\psi\,\vf\cdot\bar{\vu}_\mu+O(\psi^3).
    \label{S-eq:viewsexp}
  \end{equation}
  The band energies behind the slope $W$ of Eq.~\eqref{S-eq:within} thus measure the variance, which is the coefficient
  of Lemma~\ref{lem:small} for a pupil that need not be centered, and the phase measures the mean, so that the
  detail of a view at $\vf$ moves on the sensor by $s\,\bar{\vu}_\mu$ per diopter.
  \item \emph{The law of total variance.} The overlap of the sum is the mixture of the overlaps of the two views,
  with masses $p_L$ and $p_R=1-p_L$, so at every frequency, for any pupil and any weights,
  \begin{equation}
    c_S=p_L\,c_L+p_R\,c_R+p_L\,p_R\,(\bm\Delta\cdot\hat{\vf})^2,\qquad\bm\Delta=\bar{\vu}_L-\bar{\vu}_R .
    \label{S-eq:totalvar}
  \end{equation}
  Here $c_S$, $c_L$ and $c_R$ are the coefficients of the sum and of the two views, $p_L$ and $p_R$ their shares of
  the band signal, and $s\bm\Delta$ is the disparity between the views per diopter. Eq.~\eqref{S-eq:totalvar} has no
  free parameter and needs no model of the pupil or of the split. It tests that the measured coefficient is a variance
  over the pupil overlap, which is the content of the lemma. It holds in geometric optics too, where the overlap is
  the whole pupil, so by itself it does not separate wave from geometric optics. With aberrations the overlaps become
  complex; the identity then holds for the complex moments, and for the measured quantities up to terms of second
  order in the difference of the wave aberration between the two pupil points, which grows with the frequency. We
  therefore test it in the three lowest bands.
  \item \emph{The disparity per band.} For a split that is linear across the pupil, $w_L=(1+\beta y/R)/2$, the mean
  of $w_LO_{\va}$ along the split is $\beta/R$ times the second moment of $O_{\va}$ along it. For frequencies along
  the split the disparity per diopter $D$ then obeys $D(f)/D(0)=c_S(f)/c_S(0)$, so the disparity of fine detail falls
  with frequency as the coefficient does. For a pupil with Gaussian amplitude the overlap has the same shape at every
  frequency, and in geometric optics it is the whole pupil, so in both the disparity near focus is the same in every
  band, whatever the split. The disparity is a phase between two views with independent noise, so image noise does
  not bias it as it can bias a ratio of energies.
  \item \emph{Statistics.} The streams $S=L+R$, $L$ and $R$ are processed with the registration, the reference, the
  frame quality, the windows and the scene points of the confirmation, read from its files; each view gets its own
  gain field. The five bands of this section lie along $y$, the band axis of Table~\ref{S-tab:aberr_par}, which is also
  the direction of the split (the broadband disparity is 0.68\,px/D along $y$ and 0.001\,px/D along $x$). Three
  further bands lie along $x$, at 20, 30 and 45 cycles/mm, below the Nyquist frequency of 89 cycles/mm of the
  $5.6\,\mu$m columns. For each stream the slope $W$ of Eq.~\eqref{S-eq:within} gives the coefficient in units of
  $R^2/4$. In the bands along $y$ a view moves between the two slices of a pair by half the change of its disparity,
  which the registration of the sum does not remove and which lowers $S_\times$ of a view by the cosine of a phase
  shared by all points of the pair; we take the cross-product of the two slices along that phase, which is zero for
  the sum. The disparity of a slice is
  $d=\operatorname{atan2}(P_{\mathrm{im}},P_{\mathrm{re}})\,P_{\mathrm{re}}/(2Q_1)$, where
  $P_{\mathrm{re}}+iP_{\mathrm{im}}$ is the windowed product of the analytic band signals of the two views and $Q_1$
  the same product weighted by $2\pi|f_y|$, and $D$ is its within-point slope against the slice's own defocus,
  weighted by $P_{\mathrm{re}}$. The shares follow from the pooled denominators $\Sigma_L$ and $\Sigma_R$ of the two
  views, $p_L=\sqrt{\Sigma_L}/(\sqrt{\Sigma_L}+\sqrt{\Sigma_R})$, and the angular width of a band enters through $m$,
  the share of $f_y^2$ in $|\vf|^2$ under the weights of the statistic (0.90--0.92 in the cells of rule A below,
  0.09--0.10 in the bands along $x$).
  In these units Eq.~\eqref{S-eq:totalvar} reads
  \begin{equation}
    \underbrace{W_S-p_LW_L-p_RW_R}_{\mathrm{LHS}}=\underbrace{4\,p_L\,p_R\,m\,D^2/(Rs)^2}_{\mathrm{RHS}},\qquad
    Q=\frac{\mathrm{LHS}}{\mathrm{RHS}},
    \label{S-eq:viewsQ}
  \end{equation}
  with $Q=1$ expected. Both sides scale with the inverse square of the defocus scale, so an error of that scale
  cancels in $Q$. A capture enters a cell only if the denominators of all three streams are positive there. The
  intervals are those of the confirmation, over 20\,000 two-stage resamples.
  \item \emph{Models.} We computed the same statistics on noise-free images of model pupils: a clear disc in wave
  and in geometric optics and a Gaussian pupil, with linear splits of $\beta=0.5$, 0.7, 1 and 8, the last close to a
  split into two half discs, and slice steps of 0.10--0.35\,D (\texttt{realdata/\allowbreak
  l2af\_\allowbreak views\_\allowbreak model.py}). They give $Q=0.98$--1.08 in the bands up to 45 cycles/mm for
  every pupil and split. At 65 and 95 cycles/mm the clear disc
  gives 1.07--1.37, because the finite defocus brings in higher cumulants, and the Gaussian pupil 0.96--1.00.
  $D_b/D_{20}$ at 45 to 95 cycles/mm is 0.986--1.05 for the Gaussian pupil, whose coefficient is constant, and
  1.01--1.14 for geometric optics; at 95 cycles/mm the clear disc in wave optics gives 0.87--0.96. The measured
  disparity corresponds to $\beta=0.7$, and two half discs would give 1.6\,px/D.
  \item \emph{Rule A, the law of total variance along the split.} $Q$ at 20, 30 and 45 cycles/mm in the three zones
  and the windows 0.30--0.45 and 0.45--0.60\,D, 18 cells, with a tolerance of 0.85--1.15. It is confirmed if the 95\%
  interval lies inside the tolerance in all 18 cells, refuted if in some cell the interval lies entirely outside it,
  and undecided otherwise.
  \item \emph{Rule B, the disparity falls with frequency.} $D_b/D_{20}$ for $b=45$, 65 and 95 cycles/mm in the same
  zones and windows, 18 cells, with a threshold of 0.97. It is confirmed if the upper end of the 95\% interval lies
  below 0.97 in all 18 cells, refuted if in some cell the lower end is 0.97 or more, and undecided otherwise.
  \item \emph{Secondary outcomes}, reported whatever the primary ones: $Q$ at 65 and 95 cycles/mm; across the
  split, the views against the sum and the two sides of Eq.~\eqref{S-eq:viewsQ} in the bands along $x$; the quotient of
  $D_b/D_{20}$ and $W_{S,b}/W_{S,20}$, which is 1 at small defocus for a linear split and 0.95--1.09 for the models,
  with a band of 0.85--1.15 for the report; the comparison with the clear disc; the azimuth classes, the whole field
  and intervals over captures only.
  \item \emph{What was known.} We knew the broadband disparity of the test captures (the dual-pixel baseline of the
  pupil fit above) and both splits through the sum, and no statistic of a single view in a band had been computed on
  either split. We wrote the theory and the code first and changed the estimator twice on the model pupils, to the
  disparity per slice and to the cross-product along the common phase. We then explored the 47 test captures, where
  rule A gives ``undecided'' ($Q=0.96$--1.08, with 10 of the 18 intervals inside the tolerance over five rooms) and
  rule B ``confirmed'' (95/20: 0.71 [0.65, 0.75] at the center and 0.49 [0.40, 0.55] at the edge). The condition on
  the denominators was added after this exploration, in which one capture alone turned the slope of a cell to $-12$.
  The cells, the tolerance and the threshold were fixed after the exploration and before any training capture was
  processed. Both rules use models without aberrations, as the confirmation does.
\end{itemize}

\Para[S-par:pixviewsout]{Outcome of the test on the two views}
All 356 captures were processed without a failure. For every capture the sum of the two views through the new code
reproduces the stored statistics of the confirmation to $6\times10^{-6}$ of the band's largest value, and the 16
frozen files were unchanged before and after the run. The processing ran with 16 parallel workers where the record
says 12, which changes the running time only; there is no other deviation. The decision is the one the frozen script
printed (\texttt{results/\allowbreak l2af\_views\_train.txt}).

\emph{Outcome A is undecided} (Table~\ref{S-tab:viewsA}). $Q$ is 1.006--1.094 in the 18 cells, and 17 of the 18
intervals lie inside 0.85--1.15. In the edge zone at 20 cycles/mm and 0.30--0.45\,D the interval is [1.045, 1.169],
whose upper end passes 1.15. No interval lies outside the tolerance.

\emph{Outcome B is confirmed} (Table~\ref{S-tab:viewsB}). In all 18 cells the upper end of the interval lies below
0.97, and the largest is 0.941 (center, 0.45--0.60\,D, 45/20). From 20 to 95 cycles/mm the disparity falls to
0.75--0.77 of its value at the center, to 0.69--0.70 in the middle zone and to 0.54--0.55 at the edge, and the two
windows agree within 0.015 in every zone and band. At 20 cycles/mm the disparity is 0.71\,px/D at the center, 0.70 in
the middle zone and 0.57 at the edge. In the center at 65/20, the cell that the confirmation left open, the disparity
ratio is 0.866 [0.856, 0.877] and 0.874 [0.861, 0.884] in the two windows.

\begin{table*}[tp]
\centering
\caption{\textbf{The law of total variance between the two dual-pixel views and their sum (outcome A: undecided).}
$Q=\mathrm{LHS}/\mathrm{RHS}$ of Eq.~\eqref{S-eq:viewsQ} in the bands along the split, 356 captures, with 95\%
intervals over 20\,000 resamples of the rooms and of the captures within them. The bands at 20, 30 and 45 cycles/mm
are the 18 pre-specified cells, with a tolerance of 0.85--1.15; the model pupils and splits give 0.98--1.08 there.
The bands at 65 and 95 cycles/mm are secondary. Bold: the one cell whose interval does not lie inside the tolerance
(\texttt{results/l2af\_views\_train.txt}).}
\label{S-tab:viewsA}
\footnotesize
\setlength{\tabcolsep}{4pt}
\resizebox{\textwidth}{!}{%
\begin{tabular}{llccccc}
\toprule
 & & \multicolumn{3}{c}{pre-specified cells} & \multicolumn{2}{c}{secondary} \\
\cmidrule(lr){3-5}\cmidrule(l){6-7}
zone & window [D] & 20 cycles/mm & 30 cycles/mm & 45 cycles/mm & 65 cycles/mm & 95 cycles/mm \\
\midrule
center & 0.30--0.45 & 1.043 [1.022, 1.077] & 1.062 [1.024, 1.139] & 1.034 [1.012, 1.064] & 1.036 [1.010, 1.071] & 1.085 [1.041, 1.131] \\
 & 0.45--0.60 & 1.027 [1.015, 1.045] & 1.032 [1.021, 1.047] & 1.041 [1.026, 1.060] & 1.046 [1.024, 1.072] & 1.071 [1.027, 1.120] \\
middle & 0.30--0.45 & 1.040 [1.024, 1.060] & 1.038 [1.020, 1.060] & 1.029 [0.998, 1.059] & 1.039 [1.010, 1.072] & 1.113 [1.031, 1.200] \\
 & 0.45--0.60 & 1.041 [1.030, 1.055] & 1.043 [1.030, 1.060] & 1.050 [1.035, 1.068] & 1.087 [1.061, 1.122] & 1.168 [1.106, 1.234] \\
edge & 0.30--0.45 & \textbf{1.094 [1.045, 1.169]} & 1.051 [1.018, 1.091] & 1.006 [0.942, 1.085] & 1.005 [0.915, 1.086] & 0.898 [0.452, 1.154] \\
 & 0.45--0.60 & 1.061 [1.039, 1.081] & 1.034 [1.008, 1.059] & 1.014 [0.975, 1.041] & 0.982 [0.886, 1.052] & 0.995 [0.534, 1.492] \\
\bottomrule
\end{tabular}}
\end{table*}

\begin{table*}[tp]
\centering
\caption{\textbf{The disparity between the two views falls with frequency (outcome B: confirmed).} Upper part: the
disparity per diopter in a band relative to the 20 cycles/mm band, $D_b/D_{20}$, same captures and intervals as
Table~\ref{S-tab:viewsA}. The columns 45/20, 65/20 and 95/20 are the 18 pre-specified cells; a constant coefficient
and geometric optics give 0.97 or more, and every upper end lies below it. Middle part: the coefficient ratio of
the sum, $W_{S,b}/W_{S,20}$, through the code of this test; it agrees with Table~\ref{S-tab:pixelconfirm_all} within
0.003. Lower part: the quotient of the two, a secondary outcome with a band of 0.85--1.15 for the report; bold: the
two values outside it.}
\label{S-tab:viewsB}
\footnotesize
\setlength{\tabcolsep}{4pt}
\begin{tabular}{llcccc}
\toprule
zone & window [D] & 30/20 & 45/20 & 65/20 & 95/20 \\
\midrule
\multicolumn{6}{l}{\emph{disparity ratio $D_b/D_{20}$}} \\
center & 0.30--0.45 & 0.969 [0.965, 0.974] & 0.927 [0.920, 0.937] & 0.866 [0.856, 0.877] & 0.754 [0.738, 0.769] \\
 & 0.45--0.60 & 0.971 [0.964, 0.976] & 0.932 [0.923, 0.941] & 0.874 [0.861, 0.884] & 0.768 [0.745, 0.787] \\
middle & 0.30--0.45 & 0.955 [0.951, 0.959] & 0.889 [0.876, 0.898] & 0.807 [0.794, 0.819] & 0.689 [0.669, 0.708] \\
 & 0.45--0.60 & 0.957 [0.952, 0.960] & 0.891 [0.874, 0.902] & 0.816 [0.800, 0.828] & 0.704 [0.679, 0.725] \\
edge & 0.30--0.45 & 0.933 [0.923, 0.944] & 0.822 [0.809, 0.836] & 0.694 [0.666, 0.719] & 0.536 [0.479, 0.585] \\
 & 0.45--0.60 & 0.931 [0.917, 0.942] & 0.820 [0.805, 0.837] & 0.707 [0.671, 0.735] & 0.547 [0.481, 0.595] \\
\midrule
\multicolumn{6}{l}{\emph{coefficient ratio of the sum $W_{S,b}/W_{S,20}$}} \\
center & 0.30--0.45 & 0.954 [0.915, 1.007] & 0.924 [0.882, 0.989] & 0.877 [0.827, 0.958] & 0.799 [0.746, 0.885] \\
 & 0.45--0.60 & 0.970 [0.948, 0.990] & 0.927 [0.897, 0.962] & 0.857 [0.825, 0.894] & 0.766 [0.732, 0.802] \\
middle & 0.30--0.45 & 0.932 [0.918, 0.945] & 0.821 [0.728, 0.864] & 0.753 [0.710, 0.782] & 0.657 [0.609, 0.690] \\
 & 0.45--0.60 & 0.946 [0.931, 0.958] & 0.863 [0.813, 0.891] & 0.776 [0.733, 0.804] & 0.684 [0.653, 0.712] \\
edge & 0.30--0.45 & 0.904 [0.850, 0.943] & 0.650 [0.384, 0.804] & 0.578 [0.395, 0.686] & 0.467 [0.338, 0.578] \\
 & 0.45--0.60 & 0.923 [0.884, 0.954] & 0.725 [0.573, 0.829] & 0.625 [0.521, 0.699] & 0.507 [0.450, 0.574] \\
\midrule
\multicolumn{6}{l}{\emph{quotient of the two}} \\
center & 0.30--0.45 & 1.016 [0.963, 1.058] & 1.004 [0.941, 1.047] & 0.988 [0.908, 1.044] & 0.944 [0.857, 1.005] \\
 & 0.45--0.60 & 1.001 [0.982, 1.020] & 1.005 [0.973, 1.035] & 1.020 [0.980, 1.055] & 1.002 [0.957, 1.043] \\
middle & 0.30--0.45 & 1.024 [1.009, 1.040] & 1.082 [1.036, 1.206] & 1.073 [1.039, 1.131] & 1.048 [1.009, 1.125] \\
 & 0.45--0.60 & 1.012 [0.998, 1.027] & 1.034 [1.008, 1.078] & 1.051 [1.023, 1.101] & 1.029 [1.004, 1.059] \\
edge & 0.30--0.45 & 1.032 [0.993, 1.091] & \textbf{1.266 [1.028, 2.129]} & \textbf{1.199 [1.020, 1.785]} & 1.149 [0.953, 1.628] \\
 & 0.45--0.60 & 1.009 [0.981, 1.042] & 1.130 [1.001, 1.414] & 1.131 [1.016, 1.387] & 1.079 [0.981, 1.219] \\
\bottomrule
\end{tabular}
\end{table*}

\begin{table}[tp]
\centering
\caption{\textbf{Across the split (secondary).} Bands along $x$, same captures and intervals as
Table~\ref{S-tab:viewsA}. Upper part: $Q$ of Eq.~\eqref{S-eq:viewsQ}; the disparity term enters these bands only through
the 9--10\% of $f_y^2$ in them. Lower part: $\mathrm{LHS}-\mathrm{RHS}$ in percent of $W_S$; bold: the four intervals
that exclude zero (\texttt{results/l2af\_views\_train\_check.txt}).}
\label{S-tab:viewsC}
\footnotesize
\setlength{\tabcolsep}{4pt}
\begin{tabular}{llccc}
\toprule
zone & window [D] & 20 cycles/mm & 30 cycles/mm & 45 cycles/mm \\
\midrule
\multicolumn{5}{l}{\emph{$Q$}} \\
center & 0.30--0.45 & 0.989 [0.815, 1.124] & 1.044 [0.899, 1.230] & 0.935 [0.790, 1.057] \\
 & 0.45--0.60 & 1.027 [0.919, 1.114] & 0.980 [0.891, 1.035] & 0.937 [0.864, 0.991] \\
middle & 0.30--0.45 & 1.057 [0.940, 1.197] & 0.900 [0.777, 1.020] & 0.734 [0.602, 0.853] \\
 & 0.45--0.60 & 1.054 [0.985, 1.141] & 0.949 [0.869, 1.021] & 0.878 [0.782, 0.981] \\
edge & 0.30--0.45 & 1.060 [0.819, 1.256] & 0.965 [0.788, 1.119] & 0.709 [0.431, 0.952] \\
 & 0.45--0.60 & 1.037 [0.858, 1.211] & 0.910 [0.736, 1.066] & 0.852 [0.633, 1.051] \\
\midrule
\multicolumn{5}{l}{\emph{$\mathrm{LHS}-\mathrm{RHS}$ [\% of $W_S$]}} \\
center & 0.30--0.45 & $-0.02$ [$-0.29$, $+0.19$] & $+0.07$ [$-0.16$, $+0.36$] & $-0.10$ [$-0.32$, $+0.09$] \\
 & 0.45--0.60 & $+0.04$ [$-0.13$, $+0.17$] & $-0.03$ [$-0.17$, $+0.06$] & $\mathbf{-0.10}$ [$\mathbf{-0.21}$, $\mathbf{-0.02}$] \\
middle & 0.30--0.45 & $+0.07$ [$-0.08$, $+0.24$] & $-0.13$ [$-0.29$, $+0.03$] & $\mathbf{-0.35}$ [$\mathbf{-0.52}$, $\mathbf{-0.19}$] \\
 & 0.45--0.60 & $+0.07$ [$-0.02$, $+0.18$] & $-0.07$ [$-0.17$, $+0.03$] & $\mathbf{-0.17}$ [$\mathbf{-0.29}$, $\mathbf{-0.03}$] \\
edge & 0.30--0.45 & $+0.05$ [$-0.16$, $+0.24$] & $-0.03$ [$-0.21$, $+0.11$] & $\mathbf{-0.26}$ [$\mathbf{-0.50}$, $\mathbf{-0.04}$] \\
 & 0.45--0.60 & $+0.04$ [$-0.14$, $+0.19$] & $-0.09$ [$-0.27$, $+0.06$] & $-0.13$ [$-0.33$, $+0.04$] \\
\bottomrule
\end{tabular}
\end{table}

The pre-specified secondary outcomes follow.
\begin{itemize}
  \item \emph{$Q$ at 65 and 95 cycles/mm} (Table~\ref{S-tab:viewsA}). In the center and the middle zone $Q$ is
  1.04--1.09 at 65 and 1.07--1.17 at 95 cycles/mm, between the Gaussian pupil (0.96--1.00) and the clear disc
  (1.07--1.09 and 1.15--1.19 in the first window, 1.13--1.16 and 1.28--1.37 in the second). At the edge the
  intervals at 95 cycles/mm are wide, [0.45, 1.15] and [0.53, 1.49].
  \item \emph{The views against the sum.} Along the split $(p_LW_L+p_RW_R)/W_S$ is 0.834--0.877 at the center,
  0.861--0.896 in the middle zone and 0.905--0.953 at the edge, where the soft splits of the model give 0.79--0.95.
  Across the split it is 0.984--0.994, where the model gives 0.97--0.99. The coefficient of a view thus lies 5--17\%
  below that of the sum along the split and 0.6--1.6\% below it across. Along the split the disparity term is 5--16\%
  of $W_S$ and accounts for the difference up to $-0.5\%$ to $+1.6\%$ of $W_S$.
  \item \emph{Across the split} (Table~\ref{S-tab:viewsC}). At 20 and 30 cycles/mm $Q$ is 0.98--1.04 at the center
  and 0.90--1.06 in the other zones. At 45 cycles/mm it is 0.94 at the center and 0.71--0.88 elsewhere, and four of
  the six intervals lie below 1. As a difference, $\mathrm{LHS}-\mathrm{RHS}$ is $-0.35\%$ to $+0.07\%$ of $W_S$;
  four of the 18 intervals exclude zero, all at 45 cycles/mm, where the left side is smaller by 0.10--0.35\% of
  $W_S$. The model gives equality, and we have not found the cause of this deficit.
  \item \emph{Disparity ratio against coefficient ratio} (Table~\ref{S-tab:viewsB}). The quotient lies inside
  0.85--1.15 in 22 of the 24 cells and is 0.94--1.08 in the center and the middle zone. The two values outside are
  in the edge zone at 0.30--0.45\,D, where the coefficient ratio of the sum is poorly determined (45/20: 0.650
  [0.384, 0.804]).
  \item \emph{Clear disc.} At 95/20 the disparity ratio lies below the 0.87--0.96 of a clear disc in every zone, as
  the coefficient ratio of the sum does.
  \item \emph{Azimuth classes and whole field.} In the classes of Table~\ref{S-tab:aberr_par}, $Q$ at 20--45
  cycles/mm is 1.00--1.05 in the middle zone and 0.98--1.10 at the edge, where two intervals pass 1.15 (1.104
  [1.046, 1.194] at 20 cycles/mm with the radial direction along $y$, and 1.100 [1.016, 1.240] at 45 cycles/mm with
  it along $x$, both at 0.30--0.45\,D). At 95 cycles/mm the classes differ. With the radial direction along $x$,
  $Q$ is 1.21 and 1.27 in the middle zone and 1.35 and 1.19 at the edge; with it along $y$, 0.96 and 0.96 in the
  middle zone and 0.74 and 0.87 at the edge. Over the whole field $Q$ is 1.04--1.06 at 20--45 cycles/mm, and $D_b/D_{20}$ is 0.956, 0.890, 0.813
  and 0.695 at 30, 45, 65 and 95 cycles/mm in the first window and 0.957, 0.888, 0.818 and 0.706 in the second.
  \item \emph{Intervals over captures only.} All 18 intervals of A then lie inside the tolerance, the edge cell
  with [1.057, 1.141], and all 18 of B below 0.97, with a largest upper end of 0.939. The rule is stated on the
  two-stage intervals, so outcome A stays undecided.
  \item \emph{Checks.} Pooled over captures, the slope of the sum from the new files agrees with the stored
  statistics of the confirmation to $7\times10^{-6}$. For a single capture's cell the frozen script prints a largest
  difference of 0.15. It belongs to a cell outside the two windows of the test (lecturehall\_3, center, 95
  cycles/mm, 0.90--1.20\,D) whose denominator is negative and carries $2\times10^{-6}$ of the pooled one; in the two
  windows the largest difference is 0.038, in a cell with $2\times10^{-4}$ of the pooled denominator. The condition
  on the denominators leaves out 566 of some 61\,500 capture-cells with data, most of them at 95 cycles/mm or in
  wider windows. In the 18 cells of A these carried less than 0.005\% of the denominator of the sum, and in those of
  B at most 0.95\% (edge, 0.45--0.60\,D, 95 cycles/mm). As for the confirmation, 24 of the 356 captures have no
  sample in the two windows (\hyperref[S-par:pixdup]{see above}).
\end{itemize}
The following were not in the record and are post hoc (\texttt{realdata/\allowbreak l2af\_views\_check.py}).
\begin{itemize}
  \item \emph{One room left out.} B is confirmed with any one of the 37 room names left out, with a largest upper
  end of 0.938--0.943. A is undecided for 36 of them and confirmed without officetrashshadows (9 captures), where
  the edge cell gives 1.075 [1.039, 1.126].
  \item \emph{Without the copies.} Without the five copied captures, and without the whole folder randomsurplus, A is
  undecided with 17 of 18 intervals inside and B is confirmed. The point values change by at most 0.003, and the edge
  cell gives 1.094 [1.043, 1.168] and 1.095 [1.045, 1.173].
  \item \emph{Wider windows.} At 0.60--0.90 and 0.90--1.20\,D the disparity ratio at 95/20 is 0.770 and 0.698 at the
  center, 0.685 and 0.553 in the middle zone and 0.484 and 0.377 at the edge. Its quotient with the coefficient
  ratio of the sum is 1.00--1.06 there, while the energy of the 95 cycles/mm band relative to the 20 cycles/mm band
  falls to 0.53--0.65 and 0.28--0.42 of its value in the first window. A phase and an energy statistic thus show the
  same fall where the relative band energy differs by a factor of 2.4 to 3.6.
\end{itemize}

\begin{figure}[htbp]
\centering
\includegraphics[width=3.5in]{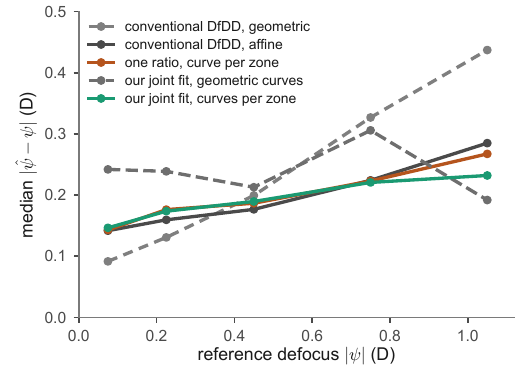}
\caption{\textbf{Phone camera: with calibration, broadband and sub-band estimates are equally accurate, and the
sub-band estimate degrades least with defocus.} Held-out median depth error against the reference defocus.}
\label{S-fig:realerr}
\end{figure}

\Para[S-par:pixdepth]{Depth estimation}
Each estimator uses one slice pair and the same $67\,\mu$m window, and is cross-validated over captures (five folds,
calibration on the training captures only). The broadband estimators use one wide log-Gabor band (45 cycles/mm,
log-bandwidth 0.6); the sub-band estimators use the five narrow bands. Sub-band estimates minimize the errors-in-variables criterion
$\sum_b(A_b-g_b(\psi)B_b)^2/(\operatorname{var}A_b+g_b^2\operatorname{var}B_b)$ with $A_b=c_kS_{xy}$ and $B_b=S_\times$,
which needs no division by the noisy $B_b$ and no gate. The evaluation set has 671\,000 pair samples with
$|\psi_{\mathrm{GT}}|\le1.2$\,D (Table~\ref{S-tab:real_est}, Fig.~\ref{S-fig:realerr}).

\begin{table}[htbp]
\centering
\caption{Phone camera, held-out median $|\hat\psi-\psi_{\mathrm{GT}}|$ [D]. ``Zone'': through-focus curves calibrated
separately in the three field zones.}
\label{S-tab:real_est}
\begin{tabular}{lcccccc}
\toprule
Method & all & 0--0.15 & 0.15--0.3 & 0.3--0.6 & 0.6--0.9 & 0.9--1.2\,D \\
\midrule
conventional DfDD, geometric & 0.223 & \textbf{0.092} & \textbf{0.131} & 0.199 & 0.327 & 0.437 \\
conventional DfDD, affine & 0.194 & 0.142 & 0.160 & \textbf{0.177} & 0.224 & 0.285 \\
one ratio, curve per zone & 0.198 & 0.143 & 0.176 & 0.187 & 0.223 & 0.267 \\
our joint fit, geometric curves & 0.243 & 0.242 & 0.239 & 0.213 & 0.306 & \textbf{0.192} \\
joint fit, clear-disc curves (ours) & 0.243 & 0.246 & 0.249 & 0.228 & 0.240 & 0.259 \\
our joint fit, curves per zone & \textbf{0.193} & 0.146 & 0.174 & 0.190 & \textbf{0.221} & 0.232 \\
\bottomrule
\end{tabular}
\end{table}

\begin{itemize}
  \item With calibration, broadband and sub-band estimates reach the same overall median (0.19\,D). The sub-band
  estimate with curves per zone degrades least with defocus: 0.23\,D at 0.9--1.2\,D against 0.27--0.29\,D for the
  calibrated broadband estimates and 0.44\,D for the geometric coefficient.
  \item Near focus the broadband estimate with the geometric coefficient is best (0.09\,D). The calibrated inversions have a floor of about
  0.14\,D there, where the through-focus curves are flat and the inversion is limited by noise.
  \item Model curves with one global gain fail (fitted gains 0.34--0.44): the real band responses are not those of a
  clear disc, which is the physics result above. On this camera the curves must be calibrated per field zone or
  modeled with aberrations.
  \item The texture dependence of the bias is small for every method (within $\pm4\%$ across texture terciles),
  because the bands lie below $0.1f_c$, where the coefficient varies little within each band.
\end{itemize}

\Para{Caveats}
The reference depth comes from the same sweep (all 49 slices, different statistic); MVS confirms it only up to a
global scale, and the within-point slope removes its per-point error. There is one camera unit (the new rooms of the
confirmation come from the same one), one wavelength band
and one band orientation. The effective aperture $r$ is an equivalent hard edge whose origin cannot be separated here,
and it trades off against the metric scale of the focus table, which breathing bounds only to $\pm20\%$. The coma is
the effective odd aberration along the band axis; trefoil gives nearly the same signature. The model floors (2\% for
ratios, 3\% for absolute slopes) are choices.

\section{DSLR: focus and aperture stacks with LiDAR depth}
\label{S-sec:zedd}

\Para{Data and access}
ZEDD~\cite{S-zuo2026} (CC-BY 4.0) has 100 scenes. We use the 50 scenes of the validation split, which come with LiDAR
depth at $1824\times1216$. The dataset is a single 8.5\,GB zip archive. Its central directory sits at the end, so HTTP
range requests read the directory and then only the 2900 members of the validation split (4.3\,GB), each checked
against its CRC. Each scene was captured with a Sony $\alpha$1\,II (ILCE-1M2), a full-frame mirrorless body, and a
Sony FE 50\,mm F1.4 GM at nine focus runs (lens readings 0.82--8.10\,m) and six apertures (f/1.4, 2, 2.8, 4, 5.6,
16). The dataset paper calls this camera a DSLR~\cite{S-zuo2026}, and we keep that name throughout. The images are JPEG
(quality 95, 4:2:0 chroma, no EXIF), $4104\times2736$ with $8.4\,\mu$m pixels, warped by the dataset authors into one
canonical frame with lens breathing removed. A pixel is therefore a fixed direction in object space: $1/5951$\,rad, from the intrinsics at
image resolution.

\Para{Object-space optics}
The models use the entrance-pupil radius $R=F/(2N)$ with $F=50$\,mm and the defocus $\psi=q-q_{\mathrm{focus}}$.
The OTF of an angular frequency $\nu$ depends only on the pupil shear $\lambda\nu$ and on $\psi R^2$, so lens breathing
does not enter. The OTF is the luma-weighted mean of the clear-disc OTFs over wavelengths of 0.45--0.63\,$\mu$m
(Gaussian weights centered at 0.545\,$\mu$m, standard deviation 0.045\,$\mu$m), in wave optics by exact quadrature or in
geometric optics as $2J_1(x)/x$, multiplied by the pixel aperture.

\Para{Statistics}
The JPEG luma is linearized with the sRGB curve and divided by its local mean (Gaussian of standard deviation 32\,px), so the
statistics are contrast energies and do not depend on the exposure. Bands are isotropic log-Gabor filters at 0.03,
0.045, 0.0675, 0.10, 0.15 and 0.225 cycles/px (log-bandwidth 0.25). The dataset's registration leaves shifts of
0.5--1\,px between focus runs, which would spoil cross products at these frequencies, so the window sums are alignment free
($64\times64$\,px windows on a 32\,px grid). For the band-passed image $B_k$ of run $k$,
\begin{equation}
  E^L_k=\langle B_k\,LB_k\rangle,\quad E^{LL}_k=\langle(LB_k)^2\rangle,\quad
  S_{xy}=\tfrac12(E^L_{\psi+\delta/2}-E^L_{\psi-\delta/2}),\quad S_\times=\tfrac12(E^{LL}_k+E^{LL}_{k+1}),
\end{equation}
with $L$ the Laplacian, and $\psi_{\mathrm{app}}$ follows from~\eqref{eq:psiapp} with $Rs$ replaced by $R$ times the
focal length in pixels. A window must have valid LiDAR depth over at least 98\% of its area with a standard deviation
below 0.01\,D, no clipped pixels in the two images, and an f/16 band energy in the more energetic half of its scene's windows.
Restricting to depth-uniform windows matters at wide apertures: at f/4, windows with a LiDAR standard deviation of
0.01--0.02\,D give slopes up to 50\% lower. With only the top 20\% textures the band ratios change by at most 0.02
(Table~\ref{S-tab:robust}).

\begin{figure}[htbp]
\centering
\includegraphics[width=\linewidth]{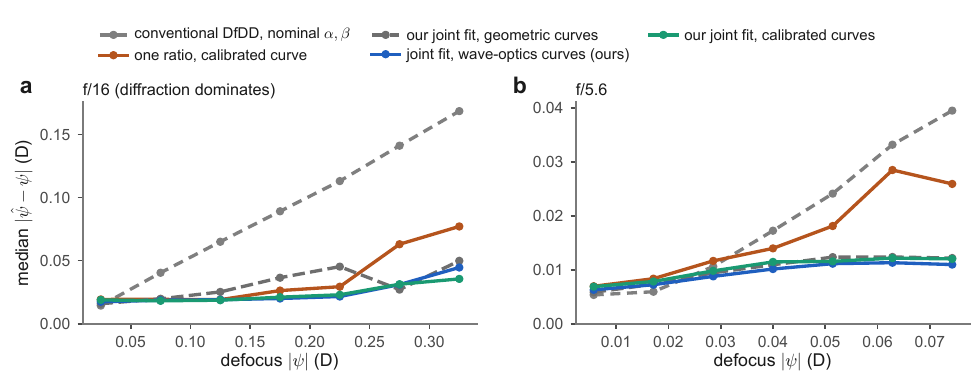}
\caption{\textbf{DSLR: sub-band estimates with wave-optics curves come close to curves calibrated for every pair and
band.} One image pair per estimate: held-out median error against the defocus $|\psi|$ at f/16, where diffraction
dominates, and at f/5.6. Each model-curve estimator has one gain per aperture, the calibrated curves none; focus
calibrated on the training scenes of each fold (\texttt{results/zedd\_estimate\_foldcal.npz}).}
\label{S-fig:zedd_est}
\end{figure}

\Para[S-par:focuscal]{Focus calibration}
The lens readings and the LiDAR disagree (Table~\ref{S-tab:calib}). The energy ratio $E_N/E_{16}$ of a run peaks at the
run's focus; we fit the main lobe of the binned median log ratio against the LiDAR vergence with the disc model of the
ratio, shifted by the peak position, with a free vertical offset and a free width scale. Runs 2--9 are determined to
$\pm0.0005$\,D (formal error of the fit) from f/2, f/2.8 and f/4 with two bands each; a focus shift with aperture is
constant per aperture and drops out of the steps, but not out of the midpoints. Run 1 has too few nearer windows, so it
follows from the zero crossing of the Laplacian ratio of its pair. The zero crossings of all pairs agree with the
midpoints of these foci within 0.011\,D, and across the folds the pair midpoints move by up to 0.017\,D (pair 1:
0.028\,D). In the highest band with crossings the mean crossing lies $+0.009$, $+0.005$, $+0.001$, $-0.003$ and
$-0.009$\,D from the midpoints at f/2, f/2.8, f/4, f/5.6 and f/16, a shift of focus with aperture; at f/16 and
0.225\,cycles/px all eight pairs cross 0.002--0.019\,D below their midpoints. The band ratios of the 2.62\,m pair cross
zero 0.018--0.038\,D below its midpoint in all six bands at f/16 (standard errors 0.0003--0.004\,D), and those of the
0.92\,m pair 0.010--0.028\,D below (\texttt{results/zedd\_calib.json}, \texttt{mids}), so at f/16 a pair midpoint can
be off by up to 0.04\,D, and the midpoint nuisance of Table~\ref{S-tab:fig3b} is $\pm0.04$\,D. A smooth model
$q=1/(z-d)+e$ fits
the midpoints with $d=6.9\pm0.5$\,cm and $e=-0.020\pm0.002$\,D but leaves $\chi^2=1112$ for 63 degrees of freedom, so
we use the per-run foci. The band-ratio tests use the calibration from the LiDAR depth of all 50 scenes. The depth
estimates of the main paper (except the posterior, whose foci are measured at f/16; Sec.~\ref{S-sec:posterior}), of
Fig.~\ref{S-fig:zedd_est} and of Tables~\ref{S-tab:est_bins}, \ref{S-tab:baselines},
\ref{S-tab:basetex}, \ref{S-tab:weights}, \ref{S-tab:texpred}, \ref{S-tab:range}, \ref{S-tab:lowband}, \ref{S-tab:bandabl},
\ref{S-tab:maps}--\ref{S-tab:sameinput} use the focus fitted to the training scenes of their fold
(\texttt{realdata/zedd\_foldcal.py}); the foci of runs 2--9 then change by at most 0.005\,D and that of run 1, which has
few nearer windows, by up to 0.05\,D. Against the calibration from all scenes, the held-out errors change by $-1$ to
$+5\%$ for the joint fit (f/16: 0.0214 against 0.0211\,D with wave-optics curves) and by up to $+8\%$ for one ratio
(0.0099 against 0.0092\,D at f/2.8 with a calibrated curve).

\begin{table}[htbp]
\centering
\caption{Focus calibration: lens reading, nominal vergence, calibrated vergence in the LiDAR frame, and the ratio of
the calibrated to the nominal step to the next run.}
\label{S-tab:calib}
\small
\begin{tabular}{lccccccccc}
\toprule
Run & 1 & 2 & 3 & 4 & 5 & 6 & 7 & 8 & 9 \\
\midrule
lens reading [m] & 0.823 & 1.199 & 1.670 & 2.294 & 3.081 & 3.835 & 4.884 & 5.775 & 8.097 \\
nominal $q$ [D] & 1.2145 & 0.8343 & 0.5988 & 0.4359 & 0.3246 & 0.2607 & 0.2047 & 0.1732 & 0.1235 \\
calibrated $q$ [D] & 1.3313 & 0.8492 & 0.6018 & 0.4295 & 0.3167 & 0.2485 & 0.1856 & 0.1638 & 0.1028 \\
step ratio & 1.27 & 1.05 & 1.06 & 1.01 & 1.07 & 1.12 & 0.69 & 1.23 & -- \\
\bottomrule
\end{tabular}
\end{table}

\Para[S-par:robust]{Band ratios at f/16: robustness}
Table~\ref{S-tab:robust} repeats the f/16 test of Fig.~\ref{S-fig:evidence}b with other reference bands, step subsets,
defocus ranges and texture thresholds. With small steps, the wave prediction lies inside the data interval in every
band and range, and the geometric prediction outside it. In the selections without a step restriction (all pairs,
ref 0.10, top 20\% textures), the wave prediction lies inside or within 0.01 of the interval in all 16 cases, and the
geometric prediction outside it in 9. With large steps both models decline because of the finite difference and
discriminate less. Near focus ($|\psi|<0.2$\,D) these data lie closer to the geometric model in the two high bands:
0.81 [0.77, 0.86] and 0.60 [0.57, 0.63], against 0.75 and 0.53 (wave) and 0.84 and 0.57 (geometric).

\begin{table}[htbp]
\centering
\caption{f/16 band ratios $\kappa_b/\kappa_{0.0675}$ for bands of 0.10, 0.15 and 0.225 cycles/px. Entries: data [95\%
interval over scenes] (wave model, geometric model), both models evaluated on the same windows. ``ref 0.10'': ratios
to the 0.10 cycles/px band.}
\label{S-tab:robust}
\resizebox{\linewidth}{!}{%
\begin{tabular}{llccc}
\toprule
$|\psi|$ & Selection & 0.10 & 0.15 & 0.225 \\
\midrule
$<0.2$\,D & all pairs & 0.96 [0.92, 0.99] (0.92, 1.00) & 0.82 [0.78, 0.86] (0.78, 0.92) & 0.63 [0.60, 0.67] (0.59, 0.74) \\
& ref 0.10 & -- & 0.86 [0.83, 0.88] (0.84, 0.91) & 0.65 [0.63, 0.68] (0.63, 0.71) \\
& steps 0.17--0.48\,D & 0.96 [0.91, 1.00] (0.91, 0.98) & 0.81 [0.77, 0.86] (0.75, 0.84) & 0.60 [0.57, 0.63] (0.53, 0.57) \\
& steps 0.02--0.11\,D & 0.96 [0.92, 1.01] (0.93, 1.02) & 0.85 [0.77, 0.92] (0.83, 1.04) & 0.66 [0.61, 0.74] (0.68, 1.00) \\
& \quad pairs 4--5 (run 5) & 0.97 [0.92, 1.02] (0.93, 1.02) & 0.86 [0.77, 0.94] (0.83, 1.04) & 0.68 [0.61, 0.76] (0.68, 0.98) \\
& \quad pairs 6--8 & 0.92 [0.87, 1.01] (0.93, 1.02) & 0.78 [0.69, 0.87] (0.83, 1.05) & 0.61 [0.53, 0.71] (0.69, 1.10) \\
& top 20\% textures & 0.97 [0.92, 1.02] (0.92, 0.99) & 0.84 [0.78, 0.89] (0.78, 0.92) & 0.64 [0.60, 0.70] (0.59, 0.74) \\
\midrule
$<0.35$\,D & all pairs & 0.93 [0.90, 0.97] (0.91, 0.97) & 0.77 [0.73, 0.82] (0.76, 0.85) & 0.53 [0.50, 0.59] (0.53, 0.57) \\
& ref 0.10 & -- & 0.81 [0.78, 0.85] (0.81, 0.83) & 0.55 [0.52, 0.59] (0.55, 0.54) \\
& steps 0.17--0.48\,D & 0.91 [0.88, 0.95] (0.89, 0.93) & 0.72 [0.67, 0.76] (0.70, 0.73) & 0.46 [0.42, 0.50] (0.44, 0.42) \\
& steps 0.02--0.11\,D & 0.98 [0.94, 1.02] (0.95, 1.05) & 0.87 [0.82, 0.92] (0.87, 1.08) & 0.67 [0.63, 0.73] (0.71, 0.85) \\
& \quad pairs 4--5 (run 5) & 0.98 [0.93, 1.04] (0.95, 1.04) & 0.86 [0.80, 0.94] (0.86, 1.05) & 0.66 [0.60, 0.74] (0.69, 0.84) \\
& \quad pairs 6--8 & 0.97 [0.92, 1.06] (0.96, 1.07) & 0.87 [0.80, 0.99] (0.91, 1.17) & 0.71 [0.60, 0.86] (0.80, 0.88) \\
& top 20\% textures & 0.94 [0.90, 0.98] (0.91, 0.97) & 0.78 [0.73, 0.84] (0.76, 0.85) & 0.54 [0.50, 0.60] (0.53, 0.57) \\
\bottomrule
\end{tabular}}
\end{table}

\Para{The unresampled reference frame}
ZEDD removes lens breathing by warping every focus run to the frame of run~5 (3.08\,m) with OpenCV's bicubic
interpolation, and run~5 itself is not resampled~\cite{S-zuo2026}; the released files are JPEGs with the IJG tables of
quality 95. Pairs 4 and 5 contain run~5, pairs 6--8 do not. Their band ratios agree within the intervals
(Table~\ref{S-tab:robust}), and in both subsets the wave prediction lies inside the data interval at $|\psi|<0.35$\,D and
the geometric prediction outside it in at least two bands.

\Para[S-par:chain]{Control of the measurement chain}
We rendered 96 synthetic scenes, each a fronto-parallel dead-leaves texture (disc radii from 1 to 100\,px with density
$\propto r^{-3}$, random contrast) at one vergence between 0.08 and 1.30\,D, through the nine ZEDD focus settings at f/16
and f/5.6. Each image is rendered at twice the resolution with the polychromatic OTF of the model (wave or geometric),
reduced by $2\times2$ averaging (pixel aperture) and magnified per run by the breathing of a unit-focus 50\,mm lens (up
to 4.7\%, OTF in object space). The chain then adds shot noise (20\,000 electrons at full scale) and read noise, applies
the sRGB curve and 8-bit quantization, a camera JPEG of quality 95, OpenCV's bicubic warp ($a=-0.75$) to the frame of
run~5 for the other runs, and the released JPEG of quality 95. The window statistics, the selection and the analysis
are those of the real data (\texttt{realdata/zedd\_pipeline\_sim.py}, \texttt{zedd\_pipesim\_table.py}).
Table~\ref{S-tab:pipesim} gives the result for three pupils: the clear disc in wave optics and in geometric optics, and a
Gaussian amplitude pupil with the second moment of the disc, whose diffusion coefficient is the same at every
frequency. A second chain adds an unsharp mask at f/16 before the camera JPEG ($e+0.125\,(e-G_1*e)$ on the sRGB
image), sized so that the in-focus energy ratio of f/5.6 and f/16 at 0.225\,cycles/px falls from 0.196 to 0.135,
near the 0.125 of the real data.

From clean renders the analysis recovers the ratios of each model within 0.01 and the absolute scale within 4\%. The
Gaussian pupil does not give a flat line: its finite step lowers the high bands, to 0.99, 0.94 and 0.83 within 0.35\,D,
and the clean renders follow this within 0.01. The chain lowers the ratio of the highest band by 0.08--0.10 for every
pupil, because noise adds energy to the denominator of weak bands, which also explains why weak bands read low on the
real data (Table~\ref{S-tab:texpred}). Through the chain the geometric renders give 1.04, 1.04 and 0.82, outside the data
intervals in all three bands, also with the sharpened chain, and the wave renders 0.95, 0.85 and 0.63, inside all three.
The Gaussian renders give 0.98, 0.92 and 0.74, at the edge of the data intervals in the two highest bands within
0.35\,D, and 0.99, 0.95 and 0.83 within 0.2\,D, outside the data in the two highest bands. Sharpening, a linear filter
of the f/16 images, changes no ratio by more than 0.01 and does not change the absolute scale, so in-camera processing
of this kind could explain the in-focus loss (data at about 60\% of the predicted value) at the highest band without
affecting the test, though not at the lowest (\hyperref[S-par:onepupil]{One effective pupil}); the
forward slope of 0.64 (not the estimation gain of 0.82; see Absolute scale below) stays unexplained by the chain. Noise of the LiDAR reference flattens the slope (Table~\ref{S-tab:eiv}). The
noise of this chain exceeds that of the data: in the 0.225\,cycles/px band its floor is 4.5 times the floor of the
data, so the drop of the highest band through this chain is an upper bound. The comparison below corrects each model
for a chain with less noise, 1.7 times the floor of the data (Table~\ref{S-tab:fig3b}).

\begin{table*}[htbp]
\centering
\caption{Control of the measurement chain at f/16: band ratios $\kappa_b/\kappa_{0.0675}$, data [95\% interval over
1\,000 resamples of the scenes] (wave, geometric and Gaussian-pupil models on the same windows); chain: noise, sRGB, two
JPEGs and the bicubic warps; sharpened: the chain with an unsharp mask at f/16. $g$: absolute scale relative to the
model of the rendered pupil (real data: wave model); in focus: $\log_{10}E_{5.6}/E_{16}$ at 0.225\,cycles/px (wave
model 0.206).}
\label{S-tab:pipesim}
\resizebox{\linewidth}{!}{%
\begin{tabular}{lccccc}
\toprule
& 0.10\,cycles/px & 0.15\,cycles/px & 0.225\,cycles/px & $g$ & in focus \\
\midrule
\multicolumn{6}{l}{\emph{steps of 0.02--0.11\,D, $|\psi|<0.35$\,D}} \\
wave render, clean & 0.95 [0.94, 0.95] (0.95, 1.04, 0.99) & 0.86 [0.85, 0.87] (0.86, 1.07, 0.94) & 0.71 [0.70, 0.73] (0.70, 0.88, 0.82) & 1.04 & 0.204 \\
wave render, chain & 0.95 [0.94, 0.95] (0.95, 1.04, 0.99) & 0.85 [0.83, 0.86] (0.86, 1.08, 0.94) & 0.63 [0.59, 0.65] (0.71, 0.85, 0.82) & 0.99 & 0.196 \\
wave render, sharpened & 0.95 [0.94, 0.95] (0.95, 1.04, 0.99) & 0.85 [0.83, 0.86] (0.86, 1.08, 0.94) & 0.63 [0.59, 0.65] (0.71, 0.85, 0.82) & 0.99 & 0.135 \\
geometric render, clean & 1.04 [1.03, 1.04] (0.95, 1.04, 0.99) & 1.06 [1.05, 1.07] (0.85, 1.06, 0.94) & 0.92 [0.87, 0.94] (0.69, 0.91, 0.84) & 1.04 & 0.00 \\
geometric render, chain & 1.04 [1.03, 1.04] (0.95, 1.04, 0.99) & 1.04 [1.03, 1.06] (0.85, 1.06, 0.94) & 0.82 [0.73, 0.86] (0.70, 0.87, 0.83) & 1.00 & 0.00 \\
geometric render, sharpened & 1.04 [1.04, 1.05] (0.95, 1.04, 0.99) & 1.05 [1.03, 1.06] (0.85, 1.06, 0.94) & 0.82 [0.73, 0.86] (0.70, 0.87, 0.84) & 0.99 & -- \\
Gaussian render, clean & 0.98 [0.98, 0.99] (0.95, 1.04, 0.99) & 0.94 [0.93, 0.95] (0.86, 1.07, 0.94) & 0.84 [0.81, 0.86] (0.69, 0.90, 0.83) & 1.04 & -- \\
Gaussian render, chain & 0.98 [0.98, 0.99] (0.95, 1.04, 0.99) & 0.92 [0.91, 0.93] (0.86, 1.07, 0.94) & 0.74 [0.68, 0.77] (0.70, 0.86, 0.83) & 0.99 & 0.079 \\
Gaussian render, sharpened & 0.98 [0.98, 0.99] (0.95, 1.04, 0.99) & 0.93 [0.91, 0.94] (0.86, 1.07, 0.94) & 0.74 [0.68, 0.77] (0.70, 0.87, 0.83) & 0.99 & 0.018 \\
real data & 0.98 [0.94, 1.03] (0.95, 1.05, 0.98) & 0.87 [0.82, 0.93] (0.87, 1.08, 0.93) & 0.67 [0.63, 0.74] (0.71, 0.85, 0.83) & 0.64 & 0.125 \\
\midrule
\multicolumn{6}{l}{\emph{steps of 0.02--0.11\,D, $|\psi|<0.2$\,D}} \\
wave render, chain & 0.93 [0.92, 0.93] (0.93, 1.02, 0.99) & 0.82 [0.81, 0.83] (0.83, 1.04, 0.96) & 0.65 [0.63, 0.66] (0.68, 0.98, 0.86) & & \\
geometric render, chain & 1.02 [1.02, 1.03] (0.93, 1.02, 0.99) & 1.03 [1.02, 1.04] (0.83, 1.04, 0.96) & 0.95 [0.90, 0.99] (0.68, 0.97, 0.87) & & \\
Gaussian render, chain & 0.99 [0.98, 1.00] (0.93, 1.02, 0.99) & 0.95 [0.94, 0.96] (0.83, 1.04, 0.96) & 0.83 [0.79, 0.86] (0.68, 0.97, 0.86) & & \\
real data & 0.96 [0.92, 1.02] (0.93, 1.02, 0.99) & 0.85 [0.79, 0.93] (0.83, 1.04, 0.96) & 0.66 [0.62, 0.75] (0.68, 1.00, 0.88) & & \\
\midrule
\multicolumn{6}{l}{\emph{steps of 0.17--0.48\,D, $|\psi|<0.2$\,D}} \\
wave render, chain & 0.90 [0.89, 0.91] (0.90, 0.95, 0.92) & 0.70 [0.68, 0.73] (0.71, 0.74, 0.71) & 0.44 [0.40, 0.47] (0.45, 0.42, 0.44) & & \\
real data & 0.96 [0.92, 1.00] (0.91, 0.98, 0.95) & 0.81 [0.76, 0.85] (0.75, 0.84, 0.80) & 0.60 [0.57, 0.63] (0.53, 0.57, 0.56) & & \\
\bottomrule
\end{tabular}}
\end{table*}

\Para[S-par:allpairs]{A test on all image pairs}
Fig.~\ref{S-fig:evidence}b uses the five pairs with small steps. Before running it we fixed a test on all eight pairs
(\texttt{realdata/zedd\_allpairs\_test.py}): the band ratios at 0.10, 0.15 and 0.225\,cycles/px of the data and of the
two clear-disc models, pooled over the pairs as in Fig.~\ref{S-fig:evidence}b; 2\,000 resamples of the 50 scenes, each
with both models recomputed at an f-number and a wavelength scale drawn from $\{0.95,1,1.05\}^2$
(\texttt{zedd\_dbc.py --perturb}); residuals standardized by the bootstrap standard deviation of the data; and the
statistic $T$, the sum over the bands of the squared standardized residuals of geometric optics minus that of wave
optics. Wave optics fits better if the 95\% interval of $T$ lies above zero. Table~\ref{S-tab:allpairs} gives the result.
The primary test, all pairs within 0.35\,D, does not decide, also with nominal models ($T=+19$ [$-12$, $+58$]), so
the indecision comes from the pooling and not from the model perturbation. The pooled ratios weight a pair by the
square of its step and by its spread of defocus, so the three large-step pairs carry 65\% of the weight within
0.35\,D, pair 3 alone 32\%, and there the two models predict nearly the same values. The small-step pairs, the subset
of our earlier analysis, favor wave optics within 0.2 and 0.35\,D. Since we chose this subset after seeing these
scenes, its result is exploratory; a pre-registered confirmation on the 50 test scenes refuted it (end of this section,
Table~\ref{S-tab:confirm}).

\Para{Band ratios against chain-corrected models}
Table~\ref{S-tab:fig3b} and Fig.~\ref{S-fig:evidence}b compare the data with three pupils, each corrected for
the processing chain (\texttt{realdata/zedd\_fig3b\_test.py}): the clear disc in wave and in geometric optics and the
Gaussian pupil with the second moment of the disc, whose coefficient is constant (the line ``constant (no fall)'' of
Fig.~\ref{S-fig:evidence}). All three models are recomputed on
each of 2\,000 resamples of the scenes at an f-number and a wavelength scale from $\{0.95,1,1.05\}^2$, the Gaussian
pupil included; for every pair the step ($\times0.9$, $\times1.1$) and the midpoint ($\pm0.04$\,D, the uncertainty of
the f/16 midpoints above) are nuisance parameters, drawn at random per resample and pair from the nominal and the four
shifted states (\texttt{zedd\_dbc.py} with \texttt{--perturb --gauss --nuisance --nuis-mid 0.03,0.04} on all eight
pairs, \texttt{results/zedd\_dbc\_r11\_m04.npz}; the test keeps the $\pm0.04$\,D states, \texttt{--nuis-use 0.04}). The
models are recomputed on the resamples of all ten seeds (\texttt{--model-seeds 10}). The chain correction of each pupil is the band ratio of the synthetic stacks through the chain minus that
of the clean renders. We set the chain noise from the
0.225\,cycles/px band: the first percentile of the window energy of that band at f/16, where flat windows hold only
noise, is 0.22 times that of the chain of Table~\ref{S-tab:pipesim} (\texttt{zedd\_noise\_match.py}; the clean renders
reach only 4\% of it), so we rerun the chain with the noise standard deviation scaled by 0.47, and by 0.66 for twice
the variance. The 8-bit encoding and the two JPEGs add a floor of their own, so the first of these chains still has
1.7 times the floor of the data in that band (\texttt{results/zedd\_noise\_match\_nonlin.txt}). That floor may
underestimate the noise of textured windows (texture terciles below), so we also give the chain with 4.5 times the floor
of the data. The evidence is the
paired residual, data minus model on the same resamples, whose 95\% interval includes the model uncertainty; the
interval endpoints are averaged over ten seeds of the resampling, and their standard deviation over the seeds
(0.001--0.009) only says how reproducible the endpoints are. Within 0.2\,D on the small steps the residual of the
Gaussian pupil is [$-0.17$, $-0.02$] at 0.15\,cycles/px and [$-0.26$, $-0.09$] at 0.225\,cycles/px ([$-0.25$,
$-0.08$] with twice the variance of the chain, [$-0.23$, $-0.07$] with the chain of Table~\ref{S-tab:pipesim}, 4.5 times
the floor of the data), and that of the wave model contains zero [$-0.07$, $+0.09$], which meets the criterion we
fixed for this test. Within 0.35\,D the small steps reject the Gaussian pupil only at 0.225\,cycles/px and barely
[$-0.17$, $-0.002$]. On all pairs within 0.2\,D the Gaussian pupil lies inside every interval, and the data lie above
wave optics at 0.10\,cycles/px [$+0.002$, $+0.07$]. The large steps reject no model within 0.35\,D and only wave
optics, at 0.225\,cycles/px, within 0.2\,D [$+0.009$, $+0.13$], where pairs 1 and 2 lie above both clear-disc models;
within 0.1\,D on all pairs wave optics fails at 0.10\,cycles/px [$+0.01$, $+0.20$], and the Gaussian pupil and
geometric optics fit every band (\texttt{results/zedd\_fig3b\_test.txt}).

\Para[S-par:nuis]{The nuisances set against the test}
Random draws average over the nuisance parameters, but an error of the step scale, of the midpoint calibration or of
the iris could act on every pair alike. We therefore also choose, on the full sample, the f-number and wavelength scale
(one for all pairs) and the nuisance states of the pairs that bring a model closest to the data, over every combination
of the states ($5^8$ on all pairs), and hold them over the resamples of each of the ten seeds. Within 0.2\,D on the small steps the residual of the Gaussian pupil at
0.225\,cycles/px is then [$-0.20$, $-0.05$] (matched chain), [$-0.17$, $-0.02$] (4.5 times the floor), [$-0.16$,
$-0.008$] (mild denoiser; Monte-Carlo SD of the upper end over the seeds 0.006) and [$-0.11$, $+0.04$] (strong
denoiser), and geometric optics stays below $-0.04$ with every chain. The rejections on other subsets mostly do not
survive: on the large steps within 0.2\,D wave optics then lies inside [$-0.03$, $+0.04$] except with the strong
denoiser [$+0.003$, $+0.07$], and on all pairs within 0.1\,D wave optics stays rejected at 0.10\,cycles/px by a
margin of at most 0.003 (lower ends $+0.001$ to $+0.003$), and not with the mild denoiser. Table~\ref{S-tab:fig3b} marks
the rejections that survive with the matched chain by a double dagger. A model that can reach the data then lies
inside its interval (wave optics on all pairs within 0.35\,D at 0.225\,cycles/px: [$-0.03$, $+0.04$]; moving every
pair as far as it goes would carry it past the data, to [$-0.09$, $-0.02$]). For a model that cannot reach them, as
the Gaussian pupil on the small steps, the closest state is the one that moves it furthest toward the data, and both
definitions give the intervals above (\texttt{results/zedd\_fig3b\_test.txt}, closest model). The closest model takes
its state separately in each band, which is conservative for a rejection but lenient for a model found inside. With
one f-number, one wavelength scale and one set of pair states for all three bands, chosen to minimize the sum over the
bands of the squared distance to the data in units of its bootstrap SD (\texttt{--shared}), no decision about wave
optics or the Gaussian pupil changes, with the linear or the nonlinear chains and with the f-number free or held at
nominal. Wave optics stays inside on all pairs within 0.35\,D ([$-0.026$, $+0.040$], [$-0.049$, $+0.036$] and
[$-0.029$, $+0.041$] at 0.10, 0.15 and 0.225\,cycles/px with the matched chain); the cells against it remain those
within 0.1\,D at 0.10\,cycles/px (lower ends $+0.001$ to $+0.008$) and those of the denoised chains at
0.225\,cycles/px; geometric optics is rejected in 1--3 more of 81 cells (\texttt{results/zedd\_fig3b\_r12shared\_*.txt},
\texttt{\_r12sharedn\_*.txt}). With up to $9\times5^8$ states the closest model can follow the data closely. On all pairs
within 0.35\,D the shared state brings wave optics to 0.929, 0.775 and 0.528 against data of 0.934, 0.767 and 0.530,
and the Gaussian pupil to 0.931, 0.767 and 0.532, and the state chosen per band meets the data to the third decimal
for both pupils (post hoc check, \texttt{realdata/\allowbreak zedd\_supp\_closest.py},
\texttt{results/\allowbreak zedd\_supp\_closest.txt}). ``Inside'' under the closest model thus means not rejected with the
nuisances profiled out, and the evidence for a model found inside lies in the drawn intervals.

\Para[S-par:radius]{The radius of the f/16 pupil}
The effective radius of the f/16 pupil is not known beyond the $\pm5\%$ of the f-number above. Scaling the radius of
every model (\texttt{zedd\_dbc.py --rscan}, 0.8--1.3 times nominal) brings the Gaussian pupil inside the
0.225\,cycles/px interval of the small steps within 0.2\,D at 1.24 times the nominal radius with the matched chain, 1.22
with twice its variance, 1.18 with 4.5 times the floor, and 1.14 and 1.01 with the mild and strong denoisers, while
wave optics stays inside from 0.90 times nominal up (0.95 with 4.5 times the floor, 1.00 with the strong denoiser).
With the nuisance state of every pair set against the test (chosen at the nominal radius and held at every radius) the
Gaussian pupil enters at 1.19, 1.17, 1.12, 1.08 and 0.90 times nominal for the same chains, so the strong denoiser
admits it at the nominal radius (one seed; each entry radius has a Monte-Carlo SD of about 0.02).

Far from focus the band energy of a hard-edged pupil passes through a minimum near the first zero of the defocused
OTF, and the pooled coefficient of a band first rises and then falls through zero; the defocus of this turnover falls
as the radius grows. We measure it on the near side of focus ($\psi>0$), where the small steps dominate; the far side
is populated by the near pairs with large steps, whose crossing lies about half a step further out, beyond 0.6\,D. Our
measure of the radius is the point at which the coefficient falls to half its peak (Table~\ref{S-tab:turnover};
\texttt{realdata/zedd\_turnover.py}, \texttt{results/zedd\_turnover.txt}). It depends less than the zero crossing on how
deep the minimum is, which noise and denoising change, and on the spread of the window depths. The zero crossing of
the data is marginal (the coefficient stays within 0.05 of zero from 0.475 to 0.575\,D, and 5--7\% of the resamples
have none), the denoised stacks do not cross zero on all pairs, and through 4.5 times the floor even the Gaussian
renders cross at 0.52\,D; once noise compresses the coefficient, the zero crossing no longer separates the pupils. The
data pass half their peak later than the nominal pupil, at 0.421 [0.399, 0.442]\,D against 0.393\,D (zero crossing
0.503 against 0.424\,D). The chains move the half peak of the wave renders by $-0.035$ to $-0.003$\,D. Solving for the
radius per resample gives 0.89--0.94 of nominal, with one-sided 95\% upper bounds of 0.92--0.98 for all six chains, the
denoised ones included (0.976 without a chain correction, 0.952 matched, 0.917 with 4.5 times the floor, and 0.950,
0.964 and 0.970 through the color chain without, with the mild and with the strong denoiser), and 0.84--1.01 if every
pair midpoint moves by $\pm0.03$\,D; the zero crossing gives 0.88--0.91 (upper bounds 0.93--0.96). With the f/16 zero
crossings of this band as the pair midpoints (0.002--0.019\,D below $m_k$, Focus calibration above; \texttt{zedd\_dbc.py
--mid-zero 5}) the data pass half their peak at 0.431\,D and the radius falls by 0.01--0.02, to 0.88--0.93 (upper
bounds 0.91--0.96; zero crossing 0.86--0.89 and 0.91--0.94; \texttt{results/zedd\_turnover\_z.txt}). A spread or a
scatter of the focus moves the half peak out and the radius back toward nominal (next paragraphs).

Without noise a Gaussian pupil has no turnover, so the bound above assumes the clear disc. Through a chain, however,
the coefficient of a Gaussian pupil also falls toward the floor of the bands. It has no interior maximum. Through the
matched chain it is 0.95 at 0.025\,D, 0.87 at 0.175\,D and 0.33 at 0.375\,D, whereas the data and the wave model peak
at 0.325\,D (0.66 and 0.94). For a Gaussian pupil we therefore take the defocus at which the coefficient falls to half
its value at 0.175\,D, the first bin of the window in which the peak is sought. The Gaussian renders through the five
chains reach this point at 0.31--0.36\,D on the small steps (wave renders 0.36--0.38\,D, geometric 0.27--0.30\,D;
0.34--0.39\,D for the Gaussian renders on all pairs; \texttt{results/\allowbreak zedd\_turnover\_joint\_r12.txt}). For a
Gaussian pupil this point is set by where the band meets the noise floor, not by a zero of the OTF as for the disc, so
0.31--0.36\,D is a bracket over chains whose floors in this band are 0.74--4.5 times that of the data
(\texttt{results/zedd\_noise\_match.txt}, \texttt{\_nonlin.txt}). A smaller Gaussian pupil reaches the point later,
but by less than the inverse of the radius predicts. Besides the defocus term, its OTF has the in-focus factor
$\exp(-2\pi^2\sigma_d^2|\vf|^2)$ of main-paper Eq.~\eqref{eq:gauss_otf}, and $\sigma_d=\lambda s/(4\pi A)$ depends
on the radius alone, so a smaller pupil starts nearer the floor. We rendered the Gaussian pupil through the five chains
at 0.50--1.08 times the nominal radius (\texttt{realdata/\allowbreak zedd\_supp\_gaussr\_sim.py}, 96 scenes per
radius). Through the matched chain it reaches the half point at 0.418\,D at 0.8 times the nominal radius and at
0.321\,D at 1.08, and through the mild denoiser at 0.414 and 0.335\,D, where the inverse of the radius gives 0.433 and
0.321\,D, and 0.447 and 0.331\,D. It reaches the data's 0.421\,D at 0.79 [0.70, 0.87] of the nominal radius through the
matched chain and at 0.78 [0.68, 0.87] through the mild denoiser, and at 0.69--0.79 over the five chains and both
choices of midpoints, not at the 0.73--0.85 that the inverse of the radius gave (\texttt{realdata/\allowbreak
zedd\_supp\_gaussr.py}, \texttt{results/\allowbreak zedd\_supp\_gaussr.txt}; intervals over 2\,000 resamples of both
the real and the synthetic scenes). At 1.08, the radius at which it enters the test interval with the mild chain, the
five chains put the half point at 0.29--0.33\,D, and through the strong denoiser at its own entry radius of 0.90 it lies
at 0.38\,D, both before the data's lower 2.5\% bound of 0.399\,D. The turnover therefore keeps the radius below the
entry radius of the Gaussian pupil under either hypothesis about the pupil. A radius 0.9 times nominal would account for most of the gain
of 0.82 on the wave-optics curves, which scales as the square of the radius, but curves computed for that radius take a
gain of 1.05 and lose range (The working range in defocus below; with the camera noise subtracted, a disc of $0.92R$
needs a gain of 1.01, Sec.~\ref{S-sec:signs}), and the loss in focus favors a larger pupil, so no one clear disc
fits the gain, the turnover and the loss in focus together. We claim only that the pupil is no wider than nominal,
which moves the Gaussian pupil away from the data. Geometric optics would need a radius below 0.8 of nominal to turn
over this late, and it predicts no loss in focus.

\Para{A spread of the focus within a window}
\label{S-sec:spread}
A window whose parts lie at slightly different depths mixes their power spectra, and a lens with longitudinal chromatic
aberration focuses the wavelengths of the luma at slightly different distances, which averages their OTFs before the
power is taken. Either spread of the focus, of standard deviation $\sigma_\psi$, flattens the through-focus behavior of
the high bands and moves the Gaussian pupil toward the data. We gave every model of the test a Gaussian spread
(\texttt{zedd\_dbc.py --spread-depth}, applied to $|H|^2$, and \texttt{--spread-chroma}, applied to $H$;
\texttt{zedd\_model.BandModel}) for $\sigma_\psi=0.01$--0.04\,D and reran the test with ten model seeds
(Table~\ref{S-tab:spread}; \texttt{run\_r11\_scan.ps1}). With the midpoint nuisance of $\pm0.04$\,D the worst case of the mild
chain reaches zero at $\sigma_\psi=0.012$\,D of depth and 0.017\,D of chromatic spread ($\pm0.03$\,D: 0.017 and
0.025\,D), and that of the matched chain at 0.031\,D of depth and beyond 0.04\,D of chromatic spread; a chromatic spread
acts about half as strongly, as the OTFs are averaged before squaring. Both parts are bounded. The SD of the LiDAR inverse
depth within the selected windows, which also contains the LiDAR noise, has an RMS of 0.0023--0.0027\,D (weighted as in
the slope sums; 90th percentile 0.004\,D) on the small steps within 0.2\,D (\texttt{realdata/zedd\_spread\_bound.py},
\texttt{results/zedd\_spread\_bound.txt}). For the chromatic part we located the focus of the R, G and B channels at
f/1.4--f/2.8 (\texttt{realdata/zedd\_lca.py}, \texttt{results/zedd\_lca.txt}): per window and channel the vertex of a
parabola through the log energy of a band at the three settings around its maximum. At f/1.4 and 0.03\,cycles/px R and
B focus 0.0040 [0.0035, 0.0045] and 0.0147 [0.0137, 0.0157]\,D nearer than G (mean over 50 scenes of the per-scene
median). The differences fall with frequency, to 0.0006 and 0.0024\,D at 0.10\,cycles/px, because demosaicing and chroma
filtering in the camera take part of the color difference from G; a Gaussian attenuation fitted to the two lowest bands
extrapolates them to 0.0047 and 0.0172\,D at zero frequency. With the luma weights 0.299, 0.587 and 0.114, and the spread
within each channel from a quadratic focus through 600, 535 and 465\,nm over channel sensitivities of SD 40\,nm, this is
a spread of 0.010\,D over the luma, 0.011\,D at the ends of the intervals; f/2 and f/2.8 give less (0.009 and
0.008\,D), the luma spectrum of the models less again (0.005\,D). The chroma filtering also lowers the share of R and B
in the high bands of the luma, so the bound errs on the high side. The bound rests on two assumptions. First, the f/16
pupil, one eleventh the radius of the f/1.4 pupil, has no more chromatic spread over the luma than the f/1.4 pupil,
which we did not measure at f/16; the B$-$G difference falls as the aperture closes (0.0147, 0.0131 and 0.0086\,D at
f/1.4, f/2 and f/2.8 and 0.03\,cycles/px), as an aberration that varies over the pupil makes it, though R$-$G does not
(0.0040, 0.0034 and 0.0055\,D). If the chromatic difference of focus of a zone grows as the square of its height in
the pupil, the focus seen through f/$N$ is linear in $1/N^2$. Extrapolated in this way from f/1.4--f/2.8 to f/16, the
zero-frequency foci give R$-$G of 0.0063 [0.0054, 0.0075]\,D, more than at f/1.4, and B$-$G of 0.0089
[0.0069, 0.0116]\,D, a spread over the luma of 0.0072 [0.0062, 0.0088]\,D (0.0077 [0.0063, 0.0096]\,D from f/2 and
f/2.8 alone), against 0.0103\,D at f/1.4 (\texttt{realdata/\allowbreak zedd\_supp\_lca\_zone.py},
\texttt{results/\allowbreak zedd\_supp\_lca\_zone.txt}). Second, most of the bound comes from the assumed spread within each
channel, not from the measured channel foci, which alone give 0.0054\,D. With the f-number held at nominal the mild
chain needs 0.034\,D of chromatic spread (Table~\ref{S-tab:spread}), three times the bound. With both spreads at their bounds (0.003\,D of depth
and 0.012\,D of chromatic spread) the worst case of the mild chain stays at $-0.004$ ($-0.012$ with $\pm0.03$\,D;
Monte-Carlo error of the ten-seed mean 0.002), and that of the matched chain at $-0.043$. With a spread of 0.04\,D of either kind
the turnover of Table~\ref{S-tab:turnover} gives a radius of 0.89--0.97 on the small steps (zero crossing; upper bounds
0.94--0.995) and 0.90--0.97 (half peak; upper bounds 0.93--1.007), so its point estimates put the pupil no wider than
nominal also then (\texttt{results/zedd\_turnover\_sd0.04.txt}, \texttt{\_sc0.04.txt}). The f/2.8--f/5.6 controls of
Table~\ref{S-tab:controls} reject a depth spread of 0.03\,D, which takes the wave model out of 12 of their 18 intervals
(17 at 0.04\,D, 2 at 0.02\,D), but no chromatic spread up to 0.04\,D (at most 1 of 18;
\texttt{results/zedd\_controls\_ci\_r11\_*.txt}), so the LiDAR bound is the tighter one.

\Para{Radius and smoothing fitted together, and a scatter of the focus between scenes}
A smoothing of the focus delays the zero crossing of the turnover more than its half peak, so the two turnovers of
Table~\ref{S-tab:turnover} together could determine the radius and a smoothing $s$
(\texttt{realdata/\allowbreak zedd\_turnover\_joint.py}, \texttt{results/\allowbreak zedd\_turnover\_joint\_r12.txt}).
We took the f/16 zero crossings as midpoints and three kinds of smoothing,
the two spreads above and a scatter of the reference vergence between windows (\texttt{zedd\_dbc.py --scatter}: every
model at the LiDAR vergence plus a normal error), over radii of 0.8--1.3 and $s$ of 0--0.06\,D, and fitted both
turnovers on every resample of the scenes. The fit is poorly posed. As $s$ grows the models lose their zero crossing
(with a depth spread above $r=1.10$ at 0.03\,D, above 1.00 at 0.04\,D and at every radius at 0.06\,D), and the bilinear
fit cannot enter those cells. We call a fit interior if it lies neither at the edge of the region where the models
still cross zero nor at an end of the grid in $s$ or $r$. Of the 93\% of the resamples that have a zero crossing (95\%
with the midpoints $m_k$), the share of interior fits with a depth spread, a scatter and a depth spread with the
midpoints $m_k$ is 8\%, 10\% and 12\% without a chain correction, 62\%, 67\% and 55\% with the matched chain, 42\%, 36\%
and 26\% with 4.5 times the floor, and 60\%, 65\% and 52\% through the color chain without a denoiser. Only with the
chromatic spread, whose models keep their zero crossing, do most fits lie inside (86\%, 90\%, 45\% and 90\% for the same
four chains). Through the mild and the strong denoiser the wave renders do not cross zero, so these chains have no
joint fit (\texttt{realdata/\allowbreak zedd\_supp\_turnover\_interior.py}, \texttt{results/\allowbreak
zedd\_supp\_turnover\_interior.txt}). We therefore give no radius from the joint fit and take the radius from the half
peak alone, which the smoothing moves little. With a spread or a
scatter at its bound it gives one-sided 95\% upper bounds of 0.91--0.97 for all six chains, the denoised ones
included; at a depth spread of 0.04\,D, more than ten times the LiDAR bound, 0.93--0.99 (0.94--1.01 with the
midpoints $m_k$, where the point estimates are at most 0.97), and at 0.06\,D 0.97--1.05, where its
point estimates are at most 1.00 (scatter up to 0.06\,D: at most 0.96). The pupil stays no wider than nominal. The
smoothing of the joint fit with the matched chain, whose full-sample fits are interior, is 0.026--0.038\,D. It exceeds
the bounds of the spread within a window, so most of it, if it is real, is something else. A scatter between windows fits the turnovers as well, but it moves the band ratios of the test little:
at 0.02 and 0.04\,D it lowers the Gaussian pupil at 0.225\,cycles/px by 0.004 and 0.015, against 0.085 for a depth
spread of 0.04\,D, because it moves every band alike to first order. The zero crossings of single scene-pairs at 0.10
and 0.225\,cycles/px scatter by 0.015--0.018\,D, shared by both bands (0.010--0.013\,D by DerSimonian--Laird; 86
scene-pairs of the small steps; \texttt{realdata/zedd\_scene\_focus.py}, \texttt{results/zedd\_scene\_focus.txt}), an
upper bound on a scatter of the reference; pairs of one scene share about 0.023\,D, as an offset of the LiDAR reference
or of the focus of a scene would. We therefore add a scatter of up to 0.02\,D to every model. In the worst case
above the f-number takes the state N$\times$0.95, a pupil 5\% wider than the turnover allows, so we hold it at nominal
there (\texttt{zedd\_fig3b\_test.py --n-fixed}; Table~\ref{S-tab:spread}, right). With both spreads at their bounds and a
scatter of 0.02\,D the worst case is then $-0.021$ with the mild chain, $-0.060$ with the matched chain and $+0.025$
with the strong denoiser; the mild chain reaches zero only at 0.024\,D of depth or 0.034\,D of chromatic spread, 8 and
3 times the bounds. With the f-number free, as in the protocol, the same stack gives $-0.000$ (mild) and $-0.039$
(matched), and the drawn nuisances $-0.042$ and $-0.081$. In this cell the bootstrap distribution is off-center (data
0.664, bootstrap median 0.675; one scene, val\_0001, holds 23\% of the summed Laplacian energy of the
0.225\,cycles/px band, which weights the windows in the slopes, these shares $p_s$ give $1/\sum_s p_s^2=10$ scenes,
and leaving one scene out moves the data between 0.650 and 0.685; \texttt{results/\allowbreak zedd\_supp\_twosets.txt}), so
BCa ends, and bootstrap-t ends whose standard error comes from a jackknife over the scenes within each resample, both
lie about 0.04 lower ($-0.059$ and $-0.058$ mild, $-0.098$ and $-0.097$ matched, tied; \texttt{--bca}, \texttt{--boott}).
Post hoc, we calibrated the interval ends by a second level of resampling, in which each resample of the scenes serves
as a sample that is resampled again 1\,000 times, with the full-sample value as the truth and the closest state held
fixed (\texttt{realdata/\allowbreak zedd\_supp\_dboot.py}). Over 1\,000 such samples a nominal 97.5\% upper end covers
the truth in 98.7\% of them for the percentile end, 98.4\% for bootstrap-t, 96.9\% for BCa and 96.6\% for the basic end
(Monte-Carlo SD 0.4--0.6 points), so the percentile end of the protocol is mildly conservative. Over 8\,000 samples its
calibrated level is 0.939 (Monte-Carlo SD 0.006), which moves the upper end 0.022 lower, to $-0.043$ (mild) and
$-0.082$ (matched) (Table~\ref{S-tab:spread}; \texttt{results/\allowbreak zedd\_supp\_dboot\_sbx.txt},
\texttt{\_crn1.txt}); second-order ends thus lie 0.02--0.04 below the percentile end. Leaving one scene out, with the
closest model chosen again on the other 49 (\texttt{--loo}), shows the influence of single scenes and is not a further
uncertainty, since the resampling of the scenes already contains it. It moves data minus the Gaussian pupil by
$-0.012$ to $+0.019$, against a jackknife standard error of 0.038, and the upper end from $-0.062$ to $+0.003$ with the
mild chain and from $-0.101$ to $-0.035$ with the matched chain (ten seeds, post hoc;
\texttt{results/\allowbreak zedd\_supp\_loo.txt}).
The highest mild ends come without val\_0036 ($+0.003$) and without val\_0035 ($+0.001$), each with a Monte-Carlo SD
of 0.001, so the latter lies at zero; without val\_0036 and with no smoothing the end is $-0.004$
(Table~\ref{S-tab:spread}, last columns). The mild margin of 0.021 is thus about half a standard error; the matched
chain, the one of the refutation rule, keeps a margin of at least 0.035.
With the f/16 zero crossings as midpoints and no smoothing, the tied worst case of the mild chain
is $-0.034$ ($-0.014$ untied) and the Gaussian pupil enters at 1.21 and 1.10 times the nominal radius with the matched
and mild chains (\texttt{results/zedd\_fig3b\_r12n\_z\_m0.04.txt}, \texttt{\_r11z\_m0.04.txt}).

\begin{table}[htbp]
\centering
\caption{The test of Fig.~\ref{S-fig:evidence}b with a spread of the focus in every model, or a scatter of
the reference vergence between windows: small steps within 0.2\,D, 0.225\,cycles/px, midpoint nuisance $\pm0.04$\,D,
upper end of the 95\% interval of data minus the Gaussian pupil (mean over ten seeds; Monte-Carlo SD 0.005--0.006),
by the SD of the smoothing in D; both bounds: 0.003\,D of depth and 0.012\,D of chromatic spread; all three: both
bounds and a scatter of 0.02\,D. Left, the f-number varies by $\pm5\%$ as in the protocol: drawn nuisances and worst
case, whose state is N$\times$0.95, $\lambda\times$0.95 and step $\times$1.1 with midpoint $+0.04$\,D on every
small-step pair. Right, the f-number held at nominal, as the turnover allows (\texttt{--n-fixed}): worst case, and for
the mild chain its BCa and bootstrap-t ends (\texttt{--bca}, \texttt{--boott}), its percentile end at the level
calibrated by a double bootstrap (cal.; post hoc, 8\,000 resamples of the scenes resampled 1\,000 times each, the same
in every row; Monte-Carlo SD 0.002; \texttt{realdata/\allowbreak zedd\_supp\_dboot.py},
\texttt{results/\allowbreak zedd\_supp\_dboot\_crn*.txt}), and the lowest and highest upper end with one scene left out
and the closest model chosen again (\texttt{--loo}, extended post hoc to ten seeds; Monte-Carlo SD of an end 0.001;
\texttt{realdata/\allowbreak zedd\_supp\_loo.py}, \texttt{results/\allowbreak zedd\_supp\_loo.txt}). With $\pm0.03$\,D every worst
case is 0.007--0.009 lower. Negative: the Gaussian pupil is rejected (\texttt{results/zedd\_fig3b\_r11\_*.txt},
\texttt{\_r12n\_*.txt}, \texttt{\_r12x\_*.txt}; \texttt{zedd\_spread\_table.py}).}
\label{S-tab:spread}
\small
\setlength{\tabcolsep}{2.4pt}
\begin{tabular}{lccccccccccc}
\toprule
& \multicolumn{4}{c}{f-number $\pm5\%$} & \multicolumn{7}{c}{f-number at nominal} \\
\cmidrule(lr){2-5}\cmidrule(lr){6-12}
& \multicolumn{2}{c}{matched chain} & \multicolumn{2}{c}{mild denoiser} & matched & \multicolumn{6}{c}{mild denoiser} \\
\cmidrule(lr){2-3}\cmidrule(lr){4-5}\cmidrule(lr){6-6}\cmidrule(lr){7-12}
Smoothing & drawn & worst & drawn & worst & worst & worst & BCa & cal. & boot-t & \multicolumn{2}{c}{one scene out} \\
\midrule
none & $-0.088$ & $-0.046$ & $-0.049$ & $-0.008$ & $-0.067$ & $-0.028$ & $-0.066$ & $-0.049$ & $-0.065$ & $-0.070$ & $-0.004$ \\
depth 0.01 & $-0.083$ & $-0.041$ & $-0.044$ & $-0.003$ & $-0.062$ & $-0.023$ & $-0.061$ & $-0.044$ & $-0.060$ & $-0.065$ & $+0.001$ \\
depth 0.02 & $-0.067$ & $-0.027$ & $-0.028$ & $+0.012$ & $-0.047$ & $-0.009$ & $-0.047$ & $-0.030$ & $-0.046$ & $-0.051$ & $+0.015$ \\
depth 0.03 & $-0.042$ & $-0.003$ & $-0.003$ & $+0.035$ & $-0.025$ & $+0.014$ & $-0.025$ & $-0.006$ & $-0.024$ & $-0.029$ & $+0.037$ \\
depth 0.04 & $-0.010$ & $+0.027$ & $+0.029$ & $+0.066$ & $+0.005$ & $+0.043$ & $+0.005$ & $+0.023$ & $+0.006$ & $+0.001$ & $+0.066$ \\
chromatic 0.01 & $-0.086$ & $-0.044$ & $-0.047$ & $-0.005$ & $-0.064$ & $-0.025$ & $-0.064$ & $-0.047$ & $-0.063$ & $-0.068$ & $-0.001$ \\
chromatic 0.02 & $-0.078$ & $-0.037$ & $-0.039$ & $+0.002$ & $-0.057$ & $-0.018$ & $-0.057$ & $-0.040$ & $-0.056$ & $-0.061$ & $+0.005$ \\
chromatic 0.03 & $-0.065$ & $-0.025$ & $-0.026$ & $+0.014$ & $-0.046$ & $-0.007$ & $-0.046$ & $-0.029$ & $-0.044$ & $-0.050$ & $+0.017$ \\
chromatic 0.04 & $-0.048$ & $-0.009$ & $-0.009$ & $+0.030$ & $-0.030$ & $+0.009$ & $-0.030$ & $-0.012$ & $-0.029$ & $-0.034$ & $+0.032$ \\
both bounds & $-0.084$ & $-0.043$ & $-0.045$ & $-0.004$ & $-0.063$ & $-0.024$ & $-0.062$ & $-0.045$ & $-0.061$ & $-0.066$ & $+0.000$ \\
scatter 0.01 & $-0.087$ & $-0.046$ & $-0.049$ & $-0.007$ & $-0.066$ & $-0.027$ & $-0.065$ & $-0.049$ & $-0.064$ & $-0.069$ & $-0.003$ \\
scatter 0.02 & $-0.085$ & $-0.043$ & $-0.046$ & $-0.004$ & $-0.063$ & $-0.024$ & $-0.063$ & $-0.047$ & $-0.062$ & $-0.066$ & $-0.000$ \\
all three & $-0.081$ & $-0.039$ & $-0.042$ & $-0.000$ & $-0.060$ & $-0.021$ & $-0.059$ & $-0.043$ & $-0.058$ & $-0.062$ & $+0.003$ \\
\bottomrule
\end{tabular}
\end{table}

\begin{table*}[htbp]
\centering
\caption{f/16 band ratios $\kappa_b/\kappa_{0.0675}$: data [95\% interval over 2\,000 resamples of the scenes, mean over
ten seeds] and the chain-corrected predictions of wave optics / geometric optics / a Gaussian pupil at the lowest chain
noise (1.7 times the floor of the data at 0.225\,cycles/px), all three with $\pm5\%$ uncertainty of f-number and
wavelength and the step ($\pm10\%$) and midpoint ($\pm0.04$\,D) of every pair as nuisance parameters drawn at random;
models on the resamples of all ten seeds (\texttt{results/zedd\_dbc\_r11\_m04.npz}, \texttt{zedd\_fig3b\_test.py
--model-seeds 10 --nuis-use 0.04}).
$^\dagger$: the paired residual, data minus model, excludes zero; $^\ddagger$: it excludes zero also with the
nuisances set against the test.}
\label{S-tab:fig3b}
\small
\setlength{\tabcolsep}{3pt}
\resizebox{\linewidth}{!}{%
\begin{tabular}{llcccccc}
\toprule
& & \multicolumn{2}{c}{0.10\,cycles/px} & \multicolumn{2}{c}{0.15\,cycles/px} & \multicolumn{2}{c}{0.225\,cycles/px} \\
\cmidrule(lr){3-4}\cmidrule(lr){5-6}\cmidrule(lr){7-8}
Pairs & $|\psi|$ & data & w / g / G & data & w / g / G & data & w / g / G \\
\midrule
small steps (4--8) & $<0.1$\,D & 1.00 [0.91, 1.13] & 0.92 / 1.01 / 1.00 & 0.91 [0.79, 1.09] & 0.81 / 1.02 / 0.98 & 0.73 [0.61, 0.93] & 0.65 / 1.00$^\ddagger$ / 0.93 \\
small steps (4--8) & $<0.2$\,D & 0.96 [0.92, 1.02] & 0.94 / 1.03$^\ddagger$ / 1.00 & 0.85 [0.79, 0.94] & 0.83 / 1.05$^\ddagger$ / 0.96$^\ddagger$ & 0.66 [0.62, 0.75] & 0.68 / 0.98$^\ddagger$ / 0.86$^\ddagger$ \\
small steps (4--8) & $<0.35$\,D & 0.98 [0.94, 1.04] & 0.96 / 1.05$^\dagger$ / 0.98 & 0.86 [0.82, 0.94] & 0.87 / 1.07$^\ddagger$ / 0.92 & 0.67 [0.63, 0.74] & 0.67 / 0.77 / 0.77$^\dagger$ \\
all pairs & $<0.1$\,D & 1.02 [0.94, 1.13] & 0.93$^\ddagger$ / 1.01 / 0.99 & 0.91 [0.80, 1.05] & 0.81 / 1.00 / 0.95 & 0.73 [0.63, 0.88] & 0.64 / 0.89 / 0.83 \\
all pairs & $<0.2$\,D & 0.96 [0.93, 0.99] & 0.92$^\dagger$ / 1.00$^\dagger$ / 0.97 & 0.82 [0.78, 0.87] & 0.79 / 0.94$^\dagger$ / 0.88 & 0.63 [0.60, 0.67] & 0.59 / 0.75$^\dagger$ / 0.69 \\
all pairs & $<0.35$\,D & 0.93 [0.91, 0.97] & 0.92 / 0.99$^\dagger$ / 0.94 & 0.77 [0.72, 0.82] & 0.77 / 0.86$^\dagger$ / 0.79 & 0.53 [0.49, 0.59] & 0.53 / 0.55 / 0.58 \\
large steps (1--3) & $<0.1$\,D & 1.04 [0.94, 1.16] & 0.93$^\ddagger$ / 1.01 / 0.99 & 0.91 [0.79, 1.09] & 0.80 / 0.98 / 0.93 & 0.73 [0.62, 0.91] & 0.62 / 0.81 / 0.76 \\
large steps (1--3) & $<0.2$\,D & 0.96 [0.91, 1.00] & 0.92 / 0.98 / 0.95 & 0.81 [0.76, 0.85] & 0.76 / 0.86 / 0.81 & 0.60 [0.57, 0.63] & 0.53$^\dagger$ / 0.58 / 0.57 \\
large steps (1--3) & $<0.35$\,D & 0.91 [0.88, 0.95] & 0.90 / 0.95 / 0.91 & 0.71 [0.67, 0.77] & 0.71 / 0.74 / 0.72 & 0.46 [0.42, 0.50] & 0.44 / 0.42 / 0.46 \\
\bottomrule
\end{tabular}}
\end{table*}

\begin{table}[htbp]
\centering
\caption{Turnover of the 0.225\,cycles/px band at f/16 on the near side of focus (small steps): defocus [D] at which
the pooled coefficient falls to half its peak, our measure of the radius, and at which it crosses zero, for the data
[95\% interval over 2\,000 resamples of the scenes], for the models on the same windows at $r$ times the nominal pupil
radius, and for the renders through the five simulated chains (for the Gaussian renders, whose coefficient has no
interior maximum, the defocus at which it falls to half its value at 0.175\,D); below, the shift of the turnover of
the wave renders through the chains and the radius that matches the data with its one-sided 95\% upper bound over the
resamples, over all chains, also with 0.04\,D of depth spread (\texttt{zedd\_turnover.py},
\texttt{zedd\_turnover\_sd0.04.txt}); last block, the same with the f/16 zero crossings as pair midpoints, the half
peak alone with a smoothing of the focus at its bound and at 0.04\,D of depth spread (Sec.~\ref{S-sec:spread};
\texttt{zedd\_turnover\_joint.py}); in both blocks, the radius at which Gaussian renders through the five chains reach
their half point at the data's (renders at 0.50--1.08 times the nominal radius, \texttt{zedd\_supp\_gaussr.py}).}
\label{S-tab:turnover}
\small
\setlength{\tabcolsep}{4pt}
\begin{tabular}{lcc}
\toprule
& half of peak & zero crossing \\
\midrule
data & 0.421 [0.399, 0.442] & 0.503 [0.472, 0.525] \\
wave, $r=0.85$ & 0.486 & 0.521 \\
wave, $r=0.90$ & 0.449 & 0.489 \\
wave, $r=0.95$ & 0.415 & 0.461 \\
wave, $r=1$ & 0.393 & 0.424 \\
wave, $r=1.10$ & 0.347 & 0.395 \\
wave, $r=1.25$ & 0.297 & 0.349 \\
geometric, $r=1$ & 0.294 & none \\
Gaussian pupil, $r=1$ & none & none \\
renders through the chains: wave & 0.36--0.38 & 0.42--0.47 or none \\
\phantom{renders through the chains:} geometric & 0.27--0.30 & none \\
\phantom{renders through the chains:} Gaussian & 0.31--0.36 & none (4.5$\times$: 0.52) \\
\midrule
chain shift & $-0.035$ to $-0.003$ & $+0.012$ to $+0.018$ \\
radius $r$ & 0.89--0.94 & 0.88--0.91 \\
95\% upper bound & 0.92--0.98 & 0.93--0.96 \\
upper bound, depth spread 0.04\,D & 0.94--1.01 & 0.99--0.995 \\
Gaussian renders at the data's half point & $r=0.71$--0.79 & \\
\midrule
\multicolumn{3}{l}{midpoints at the f/16 zero crossings of the band} \\
data & 0.431 [0.408, 0.446] & 0.514 [0.477, 0.565] \\
radius $r$ & 0.88--0.93 & 0.86--0.89 \\
95\% upper bound & 0.91--0.96 & 0.91--0.94 \\
upper bound, smoothing at its bound & 0.91--0.97 & \\
upper bound, depth spread 0.04\,D & 0.93--0.99 & \\
Gaussian renders at the data's half point & $r=0.69$--0.77 & \\
\bottomrule
\end{tabular}
\end{table}

\Para[S-par:nonlin]{A nonlinear camera chain}
A linear filter shared by the two images of a pair scales the numerator and the denominator of $r_b$ alike, so the
sharpened chain of Table~\ref{S-tab:pipesim} could not have changed the ratios. A camera also samples the image through a
color filter array, interpolates the missing colors with rules that depend on the image, and denoises with filters
that depend on the local contrast. We rendered 96 scenes of colored dead leaves (each leaf with a random tint) per
color channel with the OTF of that channel (Gaussian sensitivities at 600, 535 and 465\,nm), sampled them through an
RGGB Bayer mosaic, added the noise of the lowest chain above to the raw values, interpolated the colors along the
direction of the smaller gradient (Hamilton--Adams green, bilinear color differences), formed the JPEG luma, and applied
a bilateral denoiser of $5\times5$\,px with spatial $\sigma$ of 1\,px and a range $\sigma$ of 0, 2/255 or 5/255 before
the camera JPEG, the warps and the released JPEG (\texttt{zedd\_pipeline\_sim.py --color}). The denoiser smooths
texture whose contrast lies below its range $\sigma$ and keeps edges, so it acts more far from focus. The floor of the
chain at 0.15 and 0.225\,cycles/px is 1.9 and 1.7 times that of the data without the denoiser, 1.16 and 0.88 times
with the mild one, the closest to the data, and 0.96 and 0.74 times with the strong one, which thus smooths the highest
band more than the camera does.
Table~\ref{S-tab:nonlin} gives the band ratios within 0.2\,D on the small steps. Wave renders stay inside the data
intervals in every band. Geometric renders stay outside in the two highest bands with every denoiser. Gaussian renders
stay outside in the two highest bands without the denoiser and with the mild one; with the strong one the 0.15 band
reaches the edge of its interval (0.926 against 0.927) and the 0.225 band stays above it (0.771 against 0.752). Corrected
for these chains as in Table~\ref{S-tab:fig3b}, the residual of the Gaussian pupil at 0.225\,cycles/px is [$-0.26$,
$-0.09$] without the denoiser, [$-0.22$, $-0.05$] with the mild one and [$-0.17$, $-0.004$] with the strong one, that of
geometric optics lies below $-0.08$, and that of the wave model contains zero, so the criterion of the previous
paragraph holds for each chain (\texttt{results/zedd\_fig3b\_nonlin.txt}). On all pairs within 0.2\,D the chains
decide less: with the mild denoiser, whose floor is closest to the data, the Gaussian pupil lies inside every interval
([$-0.13$, $+0.04$] at 0.225\,cycles/px); the strong denoiser rejects wave optics in every band ($+0.005$ to $+0.12$)
and geometric optics in none; on all pairs within 0.1\,D the mild denoiser rejects wave optics in every band ($+0.004$
to $+0.28$) and fits the Gaussian pupil and geometric optics; and the large steps reject wave optics at
0.225\,cycles/px with every chain. With the nuisances set against the test these rejections of wave optics disappear,
except with the strong denoiser on the large steps and, at 0.10 and 0.225\,cycles/px, on all pairs within 0.1\,D.

\Para[S-par:dslrconfirm]{A pre-registered confirmation}
Because we chose the small-step subset after seeing the validation scenes, its result is exploratory. We fixed a
confirmation before obtaining any further data (\texttt{notes/preregistration\_zedd\_smallstep.md}, 2026-11-07): on ZEDD
scenes outside the 50 validation scenes, with the same pipeline, focus calibration, nuisance parameters and chains, the
small-step result is confirmed if at 0.225\,cycles/px the paired interval of data minus the Gaussian pupil excludes zero
with both the matched linear chain and the mild nonlinear chain while that of wave optics contains zero, refuted if the
Gaussian interval contains zero with the matched chain for drawn nuisances, and undecided otherwise. The test on all pairs is repeated and
reported whatever its outcome. The released file list shows, however, that the 50 test scenes come with camera
intrinsics only and no depth (\texttt{data/zedd/\_members.tsv}: \texttt{gt/K.txt}), so the confirmation cannot run as
written. Before any download we amended it (same file, Amendment~1). The focus positions stay those calibrated on the
validation scenes, since the test scenes use the same nine settings. The reference defocus is depth from focus at f/2.8,
the vertex of a parabola through the log energy of the 0.10\,cycles/px band at the three settings around its maximum.
A window is kept if the same vertex on its four quadrants has an SD below 0.01\,D, in place of the LiDAR mask. The
models carry the midpoint nuisance of $\pm0.04$\,D, and also the chromatic spread of focus at its 95\% upper bound
(Sec.~\ref{S-sec:spread}) and, by a second amendment, the scatter of the reference between scenes of 0.02\,D.

A third amendment, written after we had implemented and checked the reference on the validation scenes and before any
download, fixes what the first left open (\texttt{realdata/zedd\_dff\_ref.py}, \texttt{zedd\_dbc.py --ref};
\texttt{run\_r12\_ref.ps1}). The vertex is corrected by its median error against LiDAR in 0.02\,D bins of the vertex,
a table frozen from the validation scenes (median $|$error$|$ 0.011\,D raw, 0.007\,D corrected; 90th percentile 0.054
and 0.031\,D; \texttt{results/zedd\_dff\_ref.txt}). The error follows the spacing of the focus settings, within
0.006\,D up to 0.38\,D, up to $+0.07$\,D at 0.6\,D and $+0.15$\,D at 0.9\,D, where the settings lie 0.17 and 0.48\,D
apart. The table holds the 34 bins with at least 100 windows, 0.13--0.99\,D, is constant beyond them and interpolates
linearly across 0.70--0.90\,D, where no bin has 100 windows ($+0.07$\,D at 0.80\,D). No window of the deciding cell,
whose references reach 0.58\,D, has its vertex there; in the test on all pairs within 0.35\,D such windows make up
0.09\% of the windows on the validation scenes and 0.43\% on the test scenes (0.35\% and 1.17\% of the weight of the
0.225\,cycles/px band; post hoc, \texttt{realdata/\allowbreak zedd\_supp\_biastable.py},
\texttt{results/\allowbreak zedd\_supp\_biastable.txt}). The quadrant mask keeps 53\% of all windows and 58\% of those with
uniform LiDAR depth. On the windows of both masks the deciding cell gives the same decision with LiDAR and with the
reference (confirmed), but the reference lowers the upper end of data minus the Gaussian pupil by 0.023 (with the raw
vertex 0.032 for drawn nuisances and 0.025 for the closest model), and
the ends of data minus wave optics by 0.021 and 0.011, as the models are evaluated at the reference; the data ratio
itself moves by $-0.011$ ($-0.002$ with the raw vertex; \texttt{realdata/zedd\_ref\_check.py},
\texttt{results/zedd\_ref\_check.txt}). The
confirmation therefore adds these shifts to every interval end before the rules apply. On the image-based mask alone,
as on the test scenes, the validation scenes give data minus the Gaussian pupil of [$-0.29$, $-0.14$] (matched) and
[$-0.25$, $-0.11$] (mild) for drawn nuisances and [$-0.22$, $-0.10$] in the worst case, before the shift. Fixed as
well: the chromatic spread at 0.012\,D, the worst case as the closest model, percentile intervals over ten seeds, and
the f-number set of the protocol. The result is confirmed if, at 0.225\,cycles/px, data minus the Gaussian pupil
excludes zero with both chains for drawn nuisances and in the worst case with the matched chain, with and without that
spread and that scatter, while data minus wave optics contains zero; it is refuted if the Gaussian interval contains
zero with the matched chain for drawn nuisances, and undecided otherwise. We downloaded the test scenes only after the
fourth amendment.

A fourth amendment, also before any download, adds the code for scenes without LiDAR and runs it dry
(\texttt{run\_r13\_confirm.ps1}, with \texttt{zedd\_process.py --no-depth}, \texttt{zedd\_dff\_ref.py apply} for the
frozen table, \texttt{zedd\_dbc.py --split} and \texttt{zedd\_confirm\_decide.py}). On the 50 validation scenes laid
out as the test scenes will arrive (f/2.8 and f/16 JPEGs only, no depth) it reproduces the band statistics, the
reference, the model sums and every test output of the validation pipeline exactly, and it gives the same outcome,
confirmed in all four runs (\texttt{results/\allowbreak zedd\_dryrun\_check.txt}). The amendment also states what a confirmed outcome tests. The image mask
keeps other windows than the LiDAR mask. With the LiDAR vergence the deciding data ratio is 0.664 on the LiDAR mask,
0.655 once the vertex and its four quadrant vertices are required, and 0.613 once the SD of these vertices must lie
below 0.01\,D; that last step removes 22\% of the LiDAR-masked windows and carries $-0.042$ [$-0.115$, $-0.005$] of the
change of $-0.051$ [$-0.100$, $-0.008$]. With the corrected reference on the image mask alone, as on the test scenes,
the ratio is 0.617, $-0.047$ [$-0.101$, $-0.009$] from the LiDAR mask (paired intervals over the same resamples of the
scenes, computed after the outcome; \texttt{realdata/\allowbreak zedd\_supp\_maskeffect.py},
\texttt{results/\allowbreak zedd\_supp\_maskeffect.txt}). Between these window sets the medians of the drawn models move
by at most 0.010 for wave optics and the Gaussian pupil and by 0.018 for geometric optics. On the image mask wave
optics reaches the data only with the nuisances at their bounds;
its closest model with the matched chain, 0.618 against 0.617, takes N$\times$1.05, $\lambda\times$1.05 and step
$\times$1.1 with midpoint $-0.04$\,D on all five pairs, and no other f-number or wavelength scale brings it within 0.01
of the data. The Gaussian pupil lies further from the data than on the LiDAR mask, so the rule that binds is the one
for wave optics. A confirmed outcome thus says that on windows uniform at the 0.01\,D level
of the depth-from-focus vertex the 0.225\,cycles/px band falls below a constant coefficient and agrees with wave
optics; it is not the LiDAR-masked test of Fig.~\ref{S-fig:evidence}b repeated on new scenes. With the bias
table cross-fitted over five folds of the validation scenes, the reference shifts the Gaussian ends by $-0.025$ to
$-0.027$ instead of $-0.023$ (\texttt{zedd\_ref\_crossfit.py}, \texttt{results/zedd\_ref\_xfit\_shift.txt}), and the
rules now add $+0.027$ to every Gaussian upper end and $+0.023$ and $+0.011$ to the wave-optics ends. The shift was
measured without the chromatic spread and the scatter and is added in all four runs; measured with both, after the
outcome, it rounds to the same values ($-0.027$ for the Gaussian ends, $-0.023$ and $-0.011$ for wave optics;
\texttt{realdata/\allowbreak zedd\_supp\_refshift.py}, \texttt{results/\allowbreak zedd\_supp\_refshift.txt}). The pass rule
of the validation check was written after the check, so the check could not fail; its use lies in the measured shift.
The SHA-256 hashes of the 18 scripts, 5 inputs and 24 chain renders that the confirmation runs or reads, and of its
command lines, are recorded with the amendment (\texttt{notes/amendment4\_sha256.txt}, 05:54 on 8 November; this time
and the file times below are \hyperref[S-par:localrecords]{local records without a third-party time stamp}). The list
omits \texttt{l2af\_common.py}, which \texttt{zedd\_model.py} imports for the two clear-disc OTFs. That file was last
written on 4 November and not changed before the run; it was edited after the outcome for the phone confirmation, and
with the edited file \texttt{zedd\_dbc.py} reproduces all 37 arrays of the model sums of the validation scenes through
the confirmation path (\texttt{results/\allowbreak zedd\_supp\_freeze.txt}).

\Para[S-par:dslroutcome]{Outcome}
After the fourth amendment and its hashes we downloaded the f/2.8 and f/16 stacks of the 50 test scenes from the
archive the validation scenes came from (950 files, 1.56\,GB, each checked against its CRC; no depth), checked that
none of the 47 frozen files had changed (\texttt{hash\_verify.py}, written at 06:05 during the download, 06:04--06:06,
so the check fell between 06:05 and the start of the run at 06:07; it was repeated after the run; file times in
\texttt{results/\allowbreak zedd\_supp\_freeze.txt}), and ran the confirmation unchanged
(\texttt{results/\allowbreak zedd\_confirm\_r13test.txt}). The reference keeps 53.0\% of the windows (validation
53.1\%). The outcome is \emph{refuted}. In all four runs and with either shift, data minus the Gaussian pupil
contains zero for drawn nuisances with the matched chain. In the deciding cell (Table~\ref{S-tab:confirm}) the data give 0.824
[0.734, 0.935] at 0.225\,cycles/px, against 0.617 [0.584, 0.678] on the validation scenes through the same path. Data
minus the Gaussian pupil is [$-0.123$, $+0.112$] with the matched chain, [$-0.084$, $+0.151$] with the mild chain and
[$-0.094$, $+0.114$] for the closest model. Data minus wave optics is [$+0.056$, $+0.271$] ([$+0.013$, $+0.215$] for the
closest model), so wave optics is rejected, and data minus geometric optics is [$-0.239$, $+0.011$]; all ends are
before the shifts. Holding the f-number at nominal changes no decision. At 0.10 and 0.15\,cycles/px the data give 1.054
and 1.007 (validation 0.931 and 0.797). In every cell of the full test the ratio at 0.225\,cycles/px lies above its
validation value, for example 0.725 against 0.608 on all pairs within 0.2\,D, and so does the ratio at
0.15\,cycles/px; at 0.10\,cycles/px it does in 3 of the 9 cells. Counted after the outcome over its 27 cells (three
windows of $|\psi|$, three sets of pairs, three bands; matched chain, nuisances drawn), the full test rejects wave
optics in 6 cells of the test scenes, always with the data above the model, geometric optics in 4 and the Gaussian
pupil in 1, and with the closest model in 4, 2 and 1; through the same path the validation scenes reject them in 1, 18
and 14 cells (0, 15 and 9). With the mild chain the counts are 11, 4 and 0 on the test scenes and 1, 16 and 13 on the
validation scenes (\texttt{results/\allowbreak zedd\_supp\_fulltest.txt}). The test on all pairs is undecided in every cell
(primary cell $T=-0.8$ [$-24.7$, $+34.1$]). The model lines of Fig.~\ref{S-fig:evidence}b are computed on the
validation windows; on the windows of the test scenes the models lie within 0.02 of them, and geometric optics within
0.04 (\texttt{results/\allowbreak zedd\_fig3b\_r13test\_base\_m0.04.txt}).

\Para[S-par:twosets]{How far the two sets differ (post hoc)}
Everything in this paragraph, too, was computed after the outcome was known, on both sets through the same path
(\texttt{realdata/\allowbreak zedd\_supp\_twosets.py}, \texttt{results/\allowbreak zedd\_supp\_twosets.txt}). In the
deciding cell the test scenes exceed the validation scenes by $+0.207$ [$+0.091$, $+0.315$] over independent resamples
of the two sets. Over 200\,000 random splits of the 100 scenes into two sets of 50, a difference this large in either
direction occurs with $p=0.0013$ ($z=2.8$). The choice of the cell explains a small part of it. If one half of every
split picks, among the nine cells at 0.225\,cycles/px, the one where its data lie furthest below the nominal Gaussian
pupil, as a model of how the small-step subset was chosen after seeing the validation scenes, the other half's ratio
exceeds the picking half's in that cell by $+0.035$ on average and by $+0.207$ or more in 3\% of 20\,000 splits. This
rule is not the path we took (through this path it picks the small steps within 0.1\,D on the validation scenes), and a
choice among the three sets of pairs within 0.2\,D alone falls on the small steps in every split. Over 2\,000 random
splits, with 1\,000 resamples of each half, the other half lies inside a half's 95\% interval in 72\% of cases, where a
normal statistic with equal standard errors would give 83\%. The bootstrap SD of a half ranges over 0.025--0.065 (5th
to 95th percentile); that of the validation scenes, 0.024, is smaller than that of 96\% of the halves, and only 2 of the
2\,000 random splits reach the $z$ of the observed one (3.6, with the bootstrap SDs of the two sets). Treated as a
random effect of the set, the difference $d$ gives a between-set SD $\tau=0.14$, from $\tau^2=(d^2-s_T^2-s_V^2)/2$ with
the bootstrap SDs $s_T=0.052$ and $s_V=0.024$. It rests on one degree of freedom, so its 95\% interval runs from 0.05
to beyond any value of interest; its point value is close to the separation of the Gaussian pupil and wave optics
through the matched chain in this cell (0.18 on the validation windows, 0.17 on the test windows). An interval over 50
of these scenes therefore understates how far 50 new scenes can lie, and this bears on every such interval in this
section.

\Para[S-par:posthoc]{Post hoc diagnostics}
Everything in this paragraph was computed after the outcome was known and does not change it
(\texttt{realdata/zedd\_posthoc\_*.py}, \texttt{results/zedd\_posthoc\_*.txt}; both sets through the same path;
intervals from 2\,000 resamples of the scenes of each set). (i)~Relative to wave optics on the same windows, the slopes
of the two sets agree at 0.15\,cycles/px (test minus validation $+0.01$ [$-0.05$, $+0.06$]), agree within 0.09 at
0.225\,cycles/px ($+0.05$ [$-0.02$, $+0.09$]), and differ increasingly toward the coarse bands ($-0.08$ [$-0.12$, $-0.01$] at 0.10,
$-0.16$ [$-0.23$, $-0.06$] at 0.0675, $-0.25$ [$-0.39$, $-0.07$] at 0.045 and $-0.47$ [$-0.76$, $-0.11$] at
0.03\,cycles/px; Table~\ref{S-tab:confirm}, lower part; \texttt{zedd\_posthoc\_slopes.py}). The ratio of the two finest
bands, which no coarse band enters, is 0.774 [0.760, 0.816] on the validation scenes and 0.817 [0.800, 0.835] on the
test scenes, a difference of $+0.04$ [$-0.00$, $+0.06$]; through the matched chain on these windows wave optics gives
0.81 and 0.80 and the Gaussian pupil 0.89 and 0.88, and with drawn nuisances the Gaussian pupil is excluded on the
validation scenes and not on the test scenes (Table~\ref{S-tab:confirm}). A change of local
contrast between the two exposures of a pair, by a factor $1+\epsilon$ in every band, adds
$4\epsilon\,\langle e_b/E_b\rangle/(\delta(AF_x)^2)$ to the apparent defocus of band $b$, with
$e_b/E_b\approx-0.83/(2\pi f_b)^2$ for our bands; for pair~4 a change of 1\% moves it by $-0.019$\,D at 0.0675 and by
$-0.002$\,D at 0.225\,cycles/px. A shortfall of this origin is the same in every band once multiplied by $f_b^2$. On
the test scenes the shortfall of the coarse bands below the level of the two finest, times $f_b^2$, is 0.50, 0.54, 0.51
and 0.42 $\times10^{-3}$ at 0.03, 0.045, 0.0675 and 0.10\,cycles/px ([0.24, 0.71], [0.26, 0.76], [0.22, 0.75] and
[0.05, 0.65]); on the validation scenes it is 0.09, 0.03, $-0.21$ and $-0.42$. A power law $f_b^{-p}$ fitted to the
shortfall of the four coarse bands by generalized least squares, with the covariance over the scene resamples, gives
$p=1.99$ [1.65, 2.38] on the test scenes ($\chi^2=0.98$ on 2 degrees of freedom; 14.5 at $p=1$ and 18.3 at $p=0$,
both rejected), and $p=1.35$ [0.67, 1.78] for the difference between the sets, for which $p=2$ is rejected
($\chi^2=9.4$ against 1.6); the validation scenes follow no power law ($\chi^2\ge14.4$ on 2 degrees of freedom). The
shortfall of the test scenes thus has the frequency dependence of a contrast change, the difference between the sets
only roughly. Within 0.1\,D the two sets give 0.99 and 0.90 against 0.99 and 0.87 at 0.10 and
0.15\,cycles/px, and 0.74 against 0.66 at 0.225\,cycles/px. (ii)~Leaving one test scene out gives 0.78--0.84. The
largest move from one scene is 0.040 (test\_0034, 19\% of the difference between the sets); the three scenes that
lower the ratio most account for 38\% of it and five for 52\%, while removing the validation scenes that raise their
ratio most closes 24\% and 30\% of it (\texttt{results/\allowbreak zedd\_supp\_twosets.txt}). In pair~4, which carries 75\% of the weight on the test scenes, the median
within-scene coarse slope is 0.73 of wave optics against 0.83 on the validation scenes. In two test scenes it is negative, $-0.11$ in a scene with 14\% of the
pair's weight (a bicycle rack before a hedge) and $-0.58$ in another (a vending machine); one validation scene has 0.38
(a portable toilet before a brick wall). The test scenes are of the same kinds as the validation scenes. (iii)~Other
checks find the sets alike in most respects. The nine focus motor positions in the file names and the JPEG
quantization tables are identical. The pooled f/16 zero crossings against the reference agree within $\pm0.016$\,D in
pairs 4 and 5, which carry 88--93\% of the deciding weight (test minus validation at 0.10\,cycles/px $+0.003$
[$-0.007$, $+0.011$] and $-0.001$ [$-0.010$, $+0.010$]\,D), but differ in pair~6 at 0.10\,cycles/px ($+0.017$
[$+0.002$, $+0.036$]\,D; \texttt{zedd\_posthoc\_r13.txt}), and the focus scatter between scenes shared by two bands is
0.013\,D on both. The reference keeps the same share of
windows with the same median quadrant SD (0.0034 and 0.0035\,D); its error against depth cannot be measured on the test
scenes and is assumed to transfer. An intercept per scene gives 0.817 (validation 0.653). Keeping only windows whose
f/2.8 vertices in three bands agree within 0.01\,D, which on the validation scenes selects windows with smaller
reference error, gives 0.949 (validation 0.593). The change of the coarsest band between consecutive f/16 exposures, a
measure of motion and light, answers three questions differently. Its medians agree (robust SD over the windows of
pair~4, 0.041 and 0.045 in log energy; test minus validation $-0.004$ [$-0.019$, $+0.011$]). Its tails do not: three
test scenes lie above every validation scene (robust SD of the change at 0.03\,cycles/px in pair~4: test\_0016 0.26,
test\_0047 0.19, test\_0004 0.17, against at most 0.16); test\_0016 and test\_0047 are the two least textured test
scenes, test\_0004 the sixth. And the part that varies with $\psi$, which alone moves a slope, differs: a $1/f_b^2$
term fitted with a free gain to the six pooled slopes is $0.52$ [0.25, 0.73] $\times10^{-3}$ on the test scenes and
$0.12$ [$-0.02$, 0.25] $\times10^{-3}$ on the validation scenes, a difference of $+0.40$ [$+0.09$, $+0.64$]
$\times10^{-3}$ (in the units of the shortfall times $f_b^2$ above; \texttt{zedd\_posthoc\_slopes.txt},
\texttt{zedd\_posthoc\_stability.txt}). (iv)~With the 0.10 or the 0.15\,cycles/px band as
reference instead (\texttt{zedd\_posthoc\_bref.py}), two further looks at the same data that cannot confirm anything,
the Gaussian pupil is excluded on the validation scenes and not on the test scenes; with 0.10\,cycles/px the test scenes
still lie above wave optics for drawn nuisances, though not for its closest state, and with 0.15\,cycles/px the
Gaussian pupil and wave optics are both inside. Both references reject geometric optics for drawn nuisances on both
sets (data minus geometric optics on the test scenes [$-0.229$, $-0.019$] and [$-0.144$, $-0.012$]), on the test
scenes not for its closest state ([$-0.125$, $+0.023$] and [$-0.051$, $+0.022$]; Table~\ref{S-tab:confirm};
\texttt{results/\allowbreak zedd\_supp\_geocol.txt}). On the test scenes the reference band carries most of the
scene-to-scene variance of the pre-registered statistic. With each resample of the scenes applied to one band at a
time, its share of the bootstrap variance of the deciding ratio is 94\% (the covariance of the ratio resampled in this
band alone with the ratio resampled in both, over the variance of the latter; 95\% interval over resamples of the
scenes 82\% to more than 100\%; 92\% by the jackknife), against 55\% on the validation scenes (29\% to more than
100\%; 30\% by the jackknife). With the 0.15\,cycles/px band as reference the bootstrap SD of the ratio is 0.010
instead of 0.052 on the test scenes and 0.015 instead of 0.024 on the validation scenes
(\texttt{results/\allowbreak zedd\_supp\_twosets.txt}).
We do not know why the coarse bands fall short on the test scenes. The small-step result of
Fig.~\ref{S-fig:evidence}b thus holds on the validation scenes and not on the test scenes: by the slopes of the band
ratios the DSLR does not answer Q1. Sec.~\ref{S-sec:signs} gives two later tests on the test scenes that rest on the
positions of the turnovers and of the sign changes and not on the slopes.

\begin{table*}[tp]
\centering
\caption{\textbf{The pre-registered confirmation on the 50 test scenes, and post hoc diagnostics.} Deciding cell (f/16,
small steps, $|\psi|<0.2$\,D, 0.225\,cycles/px); both sets through the path of the confirmation (depth-from-focus
reference on the image mask, no LiDAR); matched chain, 2\,000 resamples $\times$ 10 seeds. Upper part: the data ratio
[95\% interval], and the 95\% intervals of data minus each pupil with the f-number, wavelength scale and pair states
drawn on every resample (drawn) or set to the values that bring the model nearest the data on the full sample
(closest). The rows with the 0.0675\,cycles/px reference are the protocol; the others are post hoc. Ends before the
shift for the reference error, which the protocol adds ($+0.027$ to the Gaussian upper end, $+0.023$ and $+0.011$ to
the wave-optics ends) and which was measured for the protocol's reference band only. Lower part: slopes relative to
wave optics at nominal on the same windows, pooled with the weights of the ratio, with 95\% intervals over 2\,000
resamples of the scenes of each set (\texttt{results/zedd\_confirm\_r13test.txt},
\texttt{zedd\_fig3b\_r13\{val,test\}\_base\_m0.04.txt}, \texttt{zedd\_posthoc\_bref\{3,4\}\_r13\{val,test\}.txt},
\texttt{zedd\_supp\_geocol.txt}, \texttt{zedd\_posthoc\_slopes.txt}).}
\label{S-tab:confirm}
\footnotesize
\setlength{\tabcolsep}{4pt}
\begin{tabular}{llcccc}
\toprule
& & \multicolumn{2}{c}{validation scenes} & \multicolumn{2}{c}{test scenes} \\
\cmidrule(lr){3-4}\cmidrule(lr){5-6}
reference & & drawn & closest & drawn & closest \\
\midrule
0.0675 & data & \multicolumn{2}{c}{0.617 [0.584, 0.678]} & \multicolumn{2}{c}{0.824 [0.734, 0.935]} \\
 & data $-$ Gaussian & [$-0.290$, $-0.144$] & [$-0.223$, $-0.101$] & [$-0.123$, $+0.112$] & [$-0.094$, $+0.114$] \\
 & data $-$ wave & [$-0.107$, $+0.021$] & [$-0.033$, $+0.057$] & [$+0.056$, $+0.271$] & [$+0.013$, $+0.215$] \\
 & data $-$ geometric & [$-0.419$, $-0.248$] & [$-0.331$, $-0.199$] & [$-0.239$, $+0.011$] & [$-0.138$, $+0.073$] \\
\midrule
0.10 & data & \multicolumn{2}{c}{0.661 [0.639, 0.709]} & \multicolumn{2}{c}{0.786 [0.730, 0.830]} \\
 & data $-$ Gaussian & [$-0.240$, $-0.110$] & [$-0.177$, $-0.070$] & [$-0.138$, $+0.035$] & [$-0.069$, $+0.058$] \\
 & data $-$ wave & [$-0.096$, $+0.009$] & [$-0.029$, $+0.037$] & [$+0.003$, $+0.136$] & [$-0.028$, $+0.080$] \\
 & data $-$ geometric & [$-0.338$, $-0.181$] & [$-0.256$, $-0.136$] & [$-0.229$, $-0.019$] & [$-0.125$, $+0.023$] \\
\midrule
0.15 & data & \multicolumn{2}{c}{0.773 [0.759, 0.816]} & \multicolumn{2}{c}{0.818 [0.800, 0.839]} \\
 & data $-$ Gaussian & [$-0.149$, $-0.043$] & [$-0.094$, $-0.015$] & [$-0.102$, $+0.003$] & [$-0.027$, $+0.032$] \\
 & data $-$ wave & [$-0.055$, $+0.018$] & [$-0.013$, $+0.041$] & [$-0.015$, $+0.049$] & [$-0.024$, $+0.026$] \\
 & data $-$ geometric & [$-0.186$, $-0.064$] & [$-0.130$, $-0.031$] & [$-0.144$, $-0.012$] & [$-0.051$, $+0.022$] \\
\bottomrule
\end{tabular}

\vspace{6pt}
\begin{tabular}{lccc}
\toprule
slope / wave optics & validation & test & test $-$ validation \\
\midrule
0.03\,cycles/px & 0.66 [0.51, 0.83] & 0.20 [$-0.04$, 0.51] & $-0.47$ [$-0.76$, $-0.11$] \\
0.045 & 0.75 [0.65, 0.85] & 0.50 [0.39, 0.66] & $-0.25$ [$-0.39$, $-0.07$] \\
0.0675 & 0.82 [0.77, 0.87] & 0.67 [0.61, 0.75] & $-0.16$ [$-0.23$, $-0.06$] \\
0.10 & 0.82 [0.78, 0.85] & 0.74 [0.71, 0.80] & $-0.08$ [$-0.12$, $-0.01$] \\
0.15 & 0.79 [0.74, 0.83] & 0.80 [0.77, 0.83] & $+0.01$ [$-0.05$, $+0.06$] \\
0.225 & 0.74 [0.72, 0.80] & 0.80 [0.77, 0.83] & $+0.05$ [$-0.02$, $+0.09$] \\
\bottomrule
\end{tabular}
\end{table*}

\Para[S-par:transfer]{Depth on the test scenes (post hoc)}
The depth estimators of main-paper Table~\ref{tab:est} scored here, every estimator of
Table~\ref{S-tab:transfer} but the Gaussian-pupil curves, were fixed before the download. After the outcome was known we
fitted them on the 50 validation scenes with LiDAR (focus from the calibration of those scenes, one gain per curve
family chosen as in Table~\ref{tab:est}, curves calibrated per pair and band, the perceptron trained for 12 epochs)
and scored them on the test scenes against the confirmation's reference, with the validation scenes through the same
reference as the control (\texttt{realdata/\allowbreak zedd\_\allowbreak posthoc\_\allowbreak transfer\_r15.py},
\texttt{results/\allowbreak zedd\_\allowbreak posthoc\_\allowbreak transfer\_r15.txt}; Tables~\ref{S-tab:transfer} and~\ref{S-tab:transfer_range}). The test
scenes are estimated by the models fitted on all 50 validation scenes, the validation scenes by the models of the five
folds of Table~\ref{tab:est} refitted with this focus, so no row is in sample; in sample the perceptron would err 15\%
less on the validation scenes, ours and calibrated curves at most 4\% less. Three things limit this check. The reference
errs against LiDAR by a median of 0.007\,D, which adds to every error. It covers vergences up to about 1\,D, so its last
populated bin of blur lies at 8.6\,px over all pairs and at 7.4\,px on the small steps, and the range beyond cannot be
tested. And the image mask keeps easier windows than the LiDAR mask, so the comparison that counts is the one between
the two sets through the reference.

Through it, ours errs by 20.5\,mD within 0.35\,D on the test scenes and by 20.2\,mD on the validation scenes (ratio 1.02
[0.88, 1.18]; 1.01 with the cross-fitted bias table), and by 9.6\% and 8.0\% against 9.9\% and 8.5\% below and beyond
3.5\,px over all pairs. Calibrated curves transfer as ours (0.99 [0.86, 1.13] times their validation error near focus;
9.8\% and 5.8\% against 9.8\% and 6.8\%). The perceptron errs 8\% less on the test scenes (0.92 [0.78, 1.10]), and its
lead over ours grows from 0.84 to 0.75 [0.72, 0.80] near focus; beyond 3.5\,px it errs by 3.9\% against 4.5\%. By ring
of $|\psi|$ (0--0.1, 0.1--0.2 and 0.2--0.35\,D) calibrated curves err 0.91, 0.98 and 1.03 times ours on the test scenes
(validation 1.02, 1.03 and 1.01) and the perceptron 0.72, 0.77 and 0.73 times (validation 0.85, 0.89 and 0.76). The
other estimators keep their order. By step group and by thirds of the in-focus energy, every estimator errs on the test
scenes as on the validation scenes or less (Table~\ref{S-tab:transfer_range}); on the least energetic third ours errs by
21\% and 23\% at 1.5--3.5 and 3.5--9.5\,px against 21\% and 37\%.

Beyond 3.5\,px ours collapses onto the pair midpoint (scene median of $\hat\psi/\psi$ below 0.2) in three test scenes,
test\_0047, test\_0016 and test\_0032, with relative errors of 95--99\%, and calibrated curves collapse in the same three
(91--97\%); the perceptron collapses in none, and its worst test scene errs by 18\%. By the median in-focus 0.10\,cycles/px
energy of the windows of the reference's mask, test\_0016 and test\_0047 are the two least textured test scenes and
test\_0032 the 13th (on all windows 1st, 4th and 24th; on their own samples beyond 3.5\,px 1st, 2nd and 7th). Their far
windows are weakly textured. Take the far samples (beyond 3.5\,px) of the large-step pairs whose 0.045\,cycles/px energy
lies within ten times its first percentile in the scene: 7.4\% of those samples on the test scenes, 5.2\% on the
validation scenes through the reference and 6.8\% against LiDAR, half of them in test\_0047, test\_0004 and test\_0032,
and in val\_0028 and val\_0009 (val\_0009, val\_0028 and val\_0040 against LiDAR). There ours gives $\hat\psi/\psi$ of
$+0.02$, $+0.12$ and $+0.01$ on the three truths, with 72\%, 59\% and 78\% of its estimates in error by more than half,
calibrated curves $+0.07$, $+0.30$ and $+0.06$, the perceptron $+1.01$, $+0.97$ and $+0.94$ (1--7\% above half). The
threshold is sharp: at 10--18 times the floor ours gives $+0.93$, $+0.89$ and $+0.77$, and outside this regime 3.8\%,
6.4\% and 11.4\% of its far large-step estimates err by more than half. The collapses are thus the failure on weak texture
(Table~\ref{S-tab:range_energy}), which on the test scenes gathers into whole scenes. In this regime
the 0.15 and 0.225\,cycles/px bands carry 97--98\% of the weight of main-paper Eq.~\eqref{eq:bw_estimate} (medians
over the samples) but respond at 0.11--0.12 of their through-focus curves on the test scenes and against LiDAR (0.33--0.40
through the reference on the validation scenes), while the two lowest bands respond at 0.26--0.69 and carry almost no
weight; outside it the two finest bands carry 79--81\% and respond at 0.53--0.99
(\texttt{realdata/zedd\_posthoc\_collapse\_mech.py}). A band that holds no defocus signal pulls the fit toward focus
(main-paper Sec.~\ref{sec:bw}).

The absolute criterion of \hyperref[S-par:coverage]{Error against coverage} removes them as a gate: no estimate unless the pair's
0.10\,cycles/px energy is at least $10^{1.5}$ times the first percentile of its scene. On the textured windows it keeps
82\% of the test samples within 0.35\,D and 76\% beyond 3.5\,px (validation through the reference 88\% and 73\%), ours
then errs by 8.6\% and 7.3\% below and beyond 3.5\,px (validation 9.1\% and 7.7\%), 3.8\% of its far estimates err by
more than half instead of 10.1\% (validation 5.5\% instead of 11.5\%), and no scene collapses. Of the far samples of
test\_0016, test\_0032 and test\_0047 it keeps 21\%, 48\% and 9\%, on which ours errs by 4\%, 25\% and 6\%. On all
windows it keeps 60\% on both sets through the reference (48\% against LiDAR, as in Table~\ref{S-tab:range_energy});
without it four test scenes collapse there (test\_0012 joins the three), with it none. The gate does not cure
val\_0006, a separate failure. Against LiDAR its far samples lie on the small steps (10\,349 of 10\,450), the gate keeps
90\% of them and ours errs by 96\% on those (perceptron 63\%); through the reference, on the windows of the image mask,
val\_0006 does not fail (7\%). On all windows against LiDAR val\_0045 fails in the same way (all far samples kept, 97\%).

A gate on whole scenes, set post hoc on two events, uses the median
over a scene of the fit
residual $J(\hat\psi)/\sum_b\omega_br_b^2$, with $J$ the cost of main-paper Eq.~\eqref{eq:bw_estimate} at the
estimate, which the pair alone gives. With the most lenient threshold that flags both validation collapses against
LiDAR (0.112), it flags 6 of 50 validation scenes against LiDAR, 3 through the reference, and 4 of 50 test scenes:
test\_0032 (0.30), test\_0016 (0.22), test\_0047 (0.18) and test\_0028 (0.14), the three collapses first, against 0.020
for the median scene. Without the flagged scenes, 94\% of the test samples, ours errs by 9.2\% and 7.6\%, and 7.7\% of
its far estimates err by more than half.

Curves of a Gaussian pupil with the second moment of the disc, whose coefficient is constant, err 1.49 times ours near
focus on the test scenes (validation 1.45). With their radius chosen on the validation scenes against LiDAR (0.70 times
nominal, gain 1.15) they err 1.24 [1.18, 1.31] times ours (validation 1.24), all of it beyond 0.1\,D. Within 0.1\,D
they err 0.96--0.97 times ours on both sets, at 0.1--0.2\,D 1.20 and 1.25 times and at 0.2--0.35\,D 1.85 and 1.79 times
(test, validation). This compares the shapes of the through-focus curves away from focus (Q2); it says nothing about the
coefficient near focus (Q1).

\Para[S-par:tone]{The tone curve (post hoc)}
ZEDD gives JPEGs, and we linearize the luma $Y'$ with the sRGB curve, not with the camera's own curve, which we do not
know. With a residual pointwise curve $C$ between the radiance $L$ and the analyzed image, defocus as diffusion in $L$,
$\partial_tL=\tfrac12\nabla^2L$ with $t$ the variance of the blur, gives
$\partial_t C(L)=\tfrac12\nabla^2C(L)-\tfrac12C''(L)|\nabla L|^2$. The loss of fine contrast under defocus thus moves
the local mean, a coarse-scale change whose share in band $b$ grows roughly as $1/f_b^2$ and whose sign depends on
$C''$ and on where the texture sits in the coarse structure. We recomputed the statistics of pair~4, which carries 65--75\% of the
deciding weight, for all 100 scenes with the luma taken as $Y'$, $Y'^{1.8}$, sRGB and $Y'^{2.8}$, on the windows of the
confirmation path (\texttt{realdata/zedd\_posthoc\_luma.py}, \texttt{results/zedd\_posthoc\_luma.txt}; the sRGB
statistics equal those of the paper exactly). On the validation scenes the slope ratio of pair~4 is 0.60, 0.62, 0.60
and 0.62 for the four curves. On the test scenes it is 0.72 [0.65, 0.79], 0.90, 0.88 [0.75, 1.03] and 1.25
[0.82, 2.56], against 0.67 for nominal wave optics. The coarse bands carry the change; on the test scenes the
0.03\,cycles/px slope relative to wave optics goes from 0.54 without linearization to 0.02 with sRGB and $-0.56$ with
$Y'^{2.8}$, while that at 0.225\,cycles/px moves from 0.87 to 0.80 and 0.73. Within power laws of 1.8--2.8, sRGB
included, the pair-4 ratio of the test scenes stays at 0.88--1.25, above nominal wave optics (0.67); without
linearization it is 0.72 [0.65, 0.79]. Under no curve tried do the two sets agree (test minus validation $+0.121$
[$+0.015$, $+0.202$] for $Y'$ to $+0.629$ [$+0.158$, $+1.840$] for $Y'^{2.8}$, over independent resamples of the two
sets; Table~\ref{S-tab:tonecurve}, last column). This recomputes pair~4 against nominal
models, not the registered cell with its chains and nuisances. On the validation scenes the curve scales all bands
alike (slope relative to wave optics at 0.03\,|\,0.225\,cycles/px: 0.81\,|\,0.80 for $Y'$, 0.69\,|\,0.75 for sRGB,
0.61\,|\,0.69 for $Y'^{2.8}$), and going from $Y'$ to $Y'^{2.8}$ changes the ratio by $+0.02$ [$-0.08$, $+0.14$]; on the
test scenes it moves the coarse band (0.54\,|\,0.87, 0.02\,|\,0.80, $-0.56$\,|\,0.73) and the ratio by $+0.53$
[$+0.15$, $+1.77$] (paired over the same resamples), and on these scenes the coarse reference band measures the
nonlinearity of the encoding as much as the optics. Power laws do not act oppositely in shadows and highlights, as a
camera's S-curve does; the next paragraph tests such curves on the registered cell.

\Para[S-par:tonebracket]{A registered bracket of tone curves (post hoc)}
After the analysis above, and before any new curve was applied to an image, we registered a bracket of global curves
(Amendment~5 of \texttt{notes/\allowbreak preregistration\_zedd\_smallstep.md}; its code, check and inputs hashed at
11:46 on 10 November in \texttt{notes/\allowbreak amendment5\_sha256.txt}, again a
\hyperref[S-par:localrecords]{local record without a third-party time stamp}; \texttt{realdata/\allowbreak
zedd\_supp\_tonecurve.py}, \texttt{results/\allowbreak zedd\_supp\_tonecurve.txt}). ZEDD does not say how its images
were encoded, and the files carry no EXIF. The camera's defaults are the sRGB color space, the Creative Look ST, for
which Sony publishes no curve, and the D-Range Optimizer on Auto, which sets brightness and gradation area by area. The
alignment tool that ZEDD cites, focus-stack, by default matches contrast between images by a smooth gain and white
balance by a gain and an offset per channel, on the 8-bit values. Either would change the local contrast between the
two exposures of a pair, and no global curve represents such a change, so the rules speak about global curves only.
Each curve replaces sRGB$^{-1}$ in the f/16 statistics of both sets; the reference, its bias table, both shifts and the
chain corrections are held. The eight curves are sRGB, the protocol; four camera-like curves, namely the inverse BT.709
transfer function, sRGB after a contrast S-curve with slope 1.15 or 1.30 at mid-gray (0.18), a power law about
mid-gray on either side that holds black and white, and sRGB after a filmic shoulder from $Y'=200$ that puts white at
twice the sRGB white; $Y'^{1.8}$ and $Y'^{2.8}$, for continuity with the analysis of pair~4; and $Y'$, for reference and
outside the rules. The two slopes come from an internal review; the form of the S-curve and the shoulder
are ours. The refutation
would depend on the encoding if under some camera-like curve the test scenes rejected wave optics in none of the four
runs (base, spread, scatter, both) and the 95\% interval of the difference between the sets in the deciding ratio
contained zero. It would be robust to the encoding, for global curves, if under each of the seven curves other than
$Y'$ the test scenes rejected wave optics in all four runs. Otherwise the bracket does not decide, and under every
reading the registered outcome stays refuted.

Neither rule holds, so the bracket does not decide (Table~\ref{S-tab:tonecurve}). Under each camera-like curve the test
scenes reject wave optics in all four runs, and the sets still differ, by $+0.126$ [$+0.044$, $+0.186$] to $+0.199$
[$+0.091$, $+0.289$] against $+0.207$ [$+0.089$, $+0.312$] under sRGB. These curves bring the test ratio to
0.738--0.801 and narrow the intervals to 0.64--0.91 of their sRGB width. The corrected lower end of data minus wave
optics, the lowest over the four runs, falls from $+0.066$ under sRGB to $+0.034$ under BT.709, $+0.035$ and $+0.011$
under the two S-curves and $+0.048$ under the shoulder. The second rule fails on $Y'^{2.8}$ alone. Under it the test
ratio rises to 1.106 [0.609, 1.592], further above wave optics than under sRGB, while the interval of data minus wave
optics widens to 4.6 times its sRGB width and so takes in zero. On the validation scenes every curve leaves the outcome
confirmed, with ratios of 0.602--0.631. The $1/f_b^2$ term of the test scenes drops from $0.52\times10^{-3}$ under sRGB
to 0.30--0.45$\times10^{-3}$ under the camera-like curves; its difference between the sets excludes zero under each of
them except the S-curve with slope 1.30 ($+0.23$ [$-0.03$, $+0.43$]$\times10^{-3}$). Pair~4 on its fixed windows
differs between the sets under every curve. Without linearization, outside the rules, the test scenes do not reject
wave optics, the sets agree ($+0.080$ [$-0.007$, $+0.142$]) and the registered outcome on the test scenes is undecided;
$Y'$ takes the 8-bit code values as proportional to radiance. The bracket does not reach a local operator. If the
D-Range Optimizer or the matching of the alignment tool acted on these images, no global curve settles the question,
and the released files cannot tell whether either did.

\begin{table*}[tp]
\centering
\caption{\textbf{A registered bracket of tone curves (post hoc).} Each curve replaces sRGB$^{-1}$ in the f/16
statistics of both sets; the reference, its bias table, both shifts and the chain corrections are held. Deciding cell
(f/16, small steps, $|\psi|<0.2$\,D, 0.225\,cycles/px; matched chain, nuisances and f-number drawn). Data ratio of each
set [95\% interval]. Registered outcome on the test scenes; on the validation scenes every curve gives
\emph{confirmed} with wave optics inside in all four runs, and no outcome changes with the shift (cross-fitted or in
sample) or with the f-number held at nominal. Runs: the number of the four runs (base, spread, scatter, both) in
which the test scenes reject wave optics after the cross-fitted shift. Test minus validation of the data ratio over
independent resamples of the two sets (2\,000 per seed, ten seeds per set, ends averaged over the seeds; Monte-Carlo
SD of the ends at most 0.012). Width: of the test scenes' interval of data minus wave optics (base run, upper line) and
of the difference interval (lower line), each relative to sRGB. Post hoc descriptors: test minus validation of the
$1/f_b^2$ term fitted with a free gain to the six pooled slopes of pairs 4--8, in the units of the
\hyperref[S-par:posthoc]{post hoc diagnostics}; and test minus validation of the pair-4 ratio on the fixed windows of the
\hyperref[S-par:tone]{analysis of pair~4} against nominal models, over independent resamples. For sRGB the difference of
the data ratio uses the registered seeds; the paragraph \hyperref[S-par:twosets]{on how far the two sets differ} gives
$+0.207$ [$+0.091$, $+0.315$] from other resamples. Every run, both shifts, both f-number
modes, the $1/f_b^2$ term of each set, the ratio of each pair alone and the full tests are in
\texttt{results/\allowbreak zedd\_supp\_tonecurve.txt} and \texttt{results/\allowbreak zedd\_tonecurve/}.}
\label{S-tab:tonecurve}
\scriptsize
\setlength{\tabcolsep}{3pt}
\begin{tabular}{lcccccccc}
\toprule
& \multicolumn{2}{c}{data ratio} & \multicolumn{2}{c}{test scenes} & & & $1/f_b^2$ term [$10^{-3}$] & pair~4 \\
\cmidrule(lr){2-3}\cmidrule(lr){4-5}
curve & validation & test & outcome & runs & test $-$ validation & width & test $-$ validation & test $-$ validation \\
\midrule
sRGB & 0.617 & 0.824 & refuted & 4 & $+0.207$ & 1.00 & $+0.40$ & $+0.282$ \\
\textit{protocol} & [0.584, 0.678] & [0.734, 0.935] & & & [$+0.089$, $+0.312$] & 1.00 & [$+0.09$, $+0.64$]
& [$+0.127$, $+0.434$] \\[2pt]
BT.709 & 0.617 & 0.768 & refuted & 4 & $+0.151$ & 0.73 & $+0.26$ & $+0.213$ \\
\textit{camera-like} & [0.587, 0.676] & [0.703, 0.838] & & & [$+0.057$, $+0.220$] & 0.73 &
[$+0.01$, $+0.47$] & [$+0.095$, $+0.304$] \\[2pt]
S-curve, 1.15 & 0.614 & 0.772 & refuted & 4 & $+0.158$ & 0.76 & $+0.30$ & $+0.221$ \\
\textit{camera-like} & [0.585, 0.672] & [0.704, 0.847] & & & [$+0.063$, $+0.231$] & 0.75 &
[$+0.03$, $+0.52$] & [$+0.099$, $+0.322$] \\[2pt]
S-curve, 1.30 & 0.612 & 0.738 & refuted & 4 & $+0.126$ & 0.65 & $+0.23$ & $+0.181$ \\
\textit{camera-like} & [0.585, 0.666] & [0.682, 0.797] & & & [$+0.044$, $+0.186$] & 0.64 &
[$-0.03$, $+0.43$] & [$+0.077$, $+0.260$] \\[2pt]
shoulder & 0.602 & 0.801 & refuted & 4 & $+0.199$ & 0.91 & $+0.38$ & $+0.265$ \\
\textit{camera-like} & [0.572, 0.654] & [0.716, 0.894] & & & [$+0.091$, $+0.289$] & 0.89 &
[$+0.08$, $+0.63$] & [$+0.126$, $+0.397$] \\[2pt]
$Y'^{1.8}$ & 0.631 & 0.848 & refuted & 4 & $+0.217$ & 1.27 & $+0.43$ & $+0.288$ \\
\textit{continuity} & [0.596, 0.693] & [0.731, 0.992] & & & [$+0.075$, $+0.357$] & 1.26 &
[$+0.05$, $+0.77$] & [$+0.103$, $+0.485$] \\[2pt]
$Y'^{2.8}$ & 0.608 & 1.106 & refuted & 0 & $+0.498$ & 4.59 & $+0.79$ & $+0.629$ \\
\textit{continuity} & [0.532, 0.712] & [0.609, 1.592] & & & [$-0.011$, $+0.980$] & 4.45 &
[$+0.26$, $+1.25$] & [$+0.158$, $+1.840$] \\[2pt]
$Y'$ & 0.614 & 0.694 & undecided & 0 & $+0.080$ & 0.66 & $+0.12$ & $+0.121$ \\
\textit{reference} & [0.582, 0.671] & [0.635, 0.753] & & & [$-0.007$, $+0.142$] & 0.67 &
[$-0.26$, $+0.47$] & [$+0.015$, $+0.202$] \\
\bottomrule
\end{tabular}
\end{table*}

\Para[S-par:pair4]{Where the difference sits (post hoc)}
As a property of the registered design rather than a result: the deciding weights put 65\% (validation) and 75\%
(test) of the cell on pair~4 (runs 4 and 5, step 0.11\,D), so the confirmation tested, in effect, one exposure pair
across scenes. On the pooled data slopes of the confirmation path (\texttt{zedd\_posthoc\_slopes.py}; pairs 4--8
together 0.62 and 0.81), test minus validation is $+0.28$ [$+0.12$, $+0.43$] in pair~4 and $+0.00$ [$-0.10$, $+0.08$]
in pairs 5--8 together. Pairs 5--8 are a subset chosen after the outcome, the kind of choice that produced the
small-step finding; we compare no model with them and draw no conclusion from them. A cause of the refutation must
explain why pair~4 moved; a global tone curve acts on every pair.

\begin{table*}[tp]
\centering
\caption{\textbf{Depth estimators fitted on the validation scenes, scored on the test scenes (post hoc).} f/16, the
texture, clip and range rules of main-paper Table~\ref{tab:est}; with a reference, its image mask. The test scenes are
estimated by the models fitted on all 50 validation scenes, the validation scenes by the models of the five folds of
Table~\ref{tab:est} refitted with the same focus, so no row is in sample. Near focus: median $|\hat\psi-\psi|$ within
0.35\,D (standard search) [95\% interval over 2\,000 resamples of the scenes] and the paired ratios to ours of curves
calibrated per pair and band, the perceptron, geometric curves, Gaussian-pupil curves at the disc's second moment and
with their radius fitted, and one ratio of the five bands with our curve. Range (search to 9.5\,px), ours:
mean of the binned median relative errors at 1.5--3.5 and 3.5--9.5\,px over all pairs, in 16 bins of $|\psi|$ up to
the search end (with the bins of Table~\ref{S-tab:estfull}, which end at the largest $|\psi|$, 0.955\,D through the
reference: 10.2 / 8.4\% and 9.6 / 7.7\% for the validation and test scenes); the 20\% limit (${>}8.6$: below 20\% up to
the last populated bin, not testable beyond); and the share of the estimates beyond 3.5\,px that err by more than half.
Over 20\,000 resamples the 3.5--9.5\,px mean of the test scenes has the 95\% interval [7.2, 17.7]\% and exceeds 10\% in
3.3\% of the resamples, those that draw the collapsed scenes often (validation through the reference [7.5, 10.0]\%,
2.5\%). The far error is mostly a scale bias set by the gain, which is chosen near focus: with gains of 0.80--0.85
instead of 0.83 it is 6.3--9.3\% on the test scenes and 7.1--9.9\% on the validation scenes, and the difference between
the sets is $-0.6$ to $-0.8$ points at every gain (\texttt{results/zedd\_posthoc\_transfer\_r15.txt},
\texttt{zedd\_posthoc\_transfer\_bins.txt}).}
\label{S-tab:transfer}
\footnotesize
\setlength{\tabcolsep}{2pt}
\begin{tabular}{lcccccccccc}
\toprule
scenes & truth & ours [mD] & calib. & perc. & geom. & Gauss. & Gauss., $r$ fit & one ratio & ours [\%] & limit / $>50\%$ \\
\midrule
validation & LiDAR & 21.1 [19.1, 23.7] & 1.02 & 0.85 & 1.26 & 1.37 & 1.21 & 0.98 & 11.4 / 9.6 & 9.5\,px / 20.1\% \\
validation & reference & 20.2 [18.4, 22.4] & 1.03 & 0.84 & 1.30 & 1.45 & 1.24 & 0.98 & 9.9 / 8.5 & ${>}8.6$\,px / 11.5\% \\
validation & reference$^\ast$ & 20.3 [18.5, 22.6] & 1.03 & 0.84 & 1.30 & 1.45 & 1.25 & 0.98 & 10.1 / 8.5 & ${>}8.6$\,px / 11.5\% \\
test & reference & 20.5 [18.4, 23.1] & 0.99 & 0.75 & 1.32 & 1.49 & 1.24 & 0.97 & 9.6 / 8.0 & ${>}8.6$\,px / 10.1\% \\
\bottomrule
\end{tabular}

\vspace{2pt}
{\footnotesize $^\ast$With the bias table cross-fitted over five folds of the validation scenes.}
\end{table*}

\begin{table*}[tp]
\centering
\caption{\textbf{The range on the test scenes by estimator (post hoc).} Through the reference, 9.5\,px search, textured
windows; validation scenes held out as in Table~\ref{S-tab:transfer}. Calibrated: our joint fit with curves calibrated
per pair and band; perceptron: our perceptron on the band statistics. Mean binned median relative error [\%] at 1.5--3.5 /
3.5--9.5\,px over all pairs, by step group (in brackets the 20\% limit on the small steps; ${>}7.4$: not testable
beyond), and on the least and the most energetic third of each set's samples by in-focus band energy. Per scene: the
median relative error beyond 3.5\,px of the scenes with at least 30 such samples, summarized as the median, 90th
percentile and maximum over scenes, and the number of scenes above 50\%. Weak regime: the far samples of the large-step
pairs whose 0.045\,cycles/px energy lies within ten times its first percentile in the scene, median $\hat\psi/\psi$
(share in error by more than half). Gated: all pairs with the absolute criterion of \hyperref[S-par:coverage]{Error against coverage}
(\texttt{results/zedd\_posthoc\_transfer\_r15.txt}).}
\label{S-tab:transfer_range}
\scriptsize
\setlength{\tabcolsep}{2.7pt}
\begin{tabular}{llcccccccc}
\toprule
estimator & scenes & all pairs & small steps & large steps & least third & most third & per scene & weak & gated \\
\midrule
ours & validation & 9.9/8.5 & 11.7/10.7 (7.2) & 7.4/8.6 & 21.2/37.2 & 7.1/6.9 & 9.1/20.7/40.0; 0 & $+0.12$ (59\%) & 9.1/7.7 \\
 & test & 9.6/8.0 & 11.0/7.3 (${>}7.4$) & 7.0/8.0 & 21.3/22.9 & 6.3/6.7 & 8.5/23.4/99.2; 3 & $+0.02$ (72\%) & 8.6/7.3 \\
calibrated & validation & 9.8/6.8 & 11.5/11.2 (7.3) & 7.7/6.0 & 20.6/19.6 & 7.5/4.6 & 7.3/18.0/28.6; 0 & $+0.30$ (52\%) & 9.0/5.3 \\
 & test & 9.8/5.8 & 11.5/9.5 (${>}7.4$) & 7.3/4.9 & 20.2/15.5 & 7.1/3.7 & 6.6/18.9/96.6; 3 & $+0.07$ (66\%) & 8.6/4.5 \\
perceptron & validation & 7.5/4.5 & 8.0/15.4 (6.7) & 6.4/3.7 & 10.0/6.1 & 6.7/4.6 & 4.8/10.8/24.7; 0 & $+0.97$ (1\%) & 7.1/4.1 \\
 & test & 6.8/3.9 & 7.0/9.5 (${>}7.4$) & 6.3/3.5 & 10.3/5.3 & 5.2/3.4 & 4.1/7.8/18.1; 0 & $+1.01$ (1\%) & 6.3/3.6 \\
\bottomrule
\end{tabular}
\end{table*}

\begin{table}[htbp]
\centering
\caption{The nonlinear camera chain at f/16: band ratios $\kappa_b/\kappa_{0.0675}$ of rendered stacks within 0.2\,D on
the small steps, through color sampling and edge-directed demosaicing, without and with a bilateral denoiser (range
$\sigma$ 2/255, mild; 5/255, strong), and the real data [95\% interval over 1\,000 resamples of the scenes]
(\texttt{zedd\_pipesim\_table.py pipesim4\_color ... --chains clean nlnone nlmild nlstrong}).}
\label{S-tab:nonlin}
\small
\setlength{\tabcolsep}{4pt}
\begin{tabular}{llccc}
\toprule
Pupil & Chain & 0.10\,c/px & 0.15\,c/px & 0.225\,c/px \\
\midrule
wave & clean & 0.928 & 0.833 & 0.692 \\
     & no denoiser & 0.935 & 0.837 & 0.685 \\
     & mild & 0.934 & 0.829 & 0.664 \\
     & strong & 0.934 & 0.825 & 0.641 \\
geometric & clean & 1.016 & 1.040 & 0.989 \\
     & no denoiser & 1.021 & 1.042 & 0.969 \\
     & mild & 1.015 & 1.023 & 0.929 \\
     & strong & 1.008 & 0.994 & 0.863 \\
Gaussian & clean & 0.987 & 0.957 & 0.867 \\
     & no denoiser & 0.996 & 0.964 & 0.855 \\
     & mild & 0.989 & 0.943 & 0.816 \\
     & strong & 0.981 & 0.926 & 0.771 \\
\midrule
real data & & 0.957 [0.919, 1.017] & 0.846 [0.790, 0.927] & 0.664 [0.616, 0.752] \\
\bottomrule
\end{tabular}
\end{table}

\begin{table}[htbp]
\centering
\caption{Test on the f/16 band ratios with model uncertainty: median of $T$ [95\% interval] and the fraction of
resamples with $T>0$; $T>0$ favors wave optics.}
\label{S-tab:allpairs}
\small
\setlength{\tabcolsep}{3pt}
\begin{tabular}{lccc}
\toprule
Pairs & $|\psi|<0.1$\,D & $|\psi|<0.2$\,D & $|\psi|<0.35$\,D \\
\midrule
all eight & $+3$ [$-37$, $+32$], 0.558 & $+55$ [$-35$, $+211$], 0.857 & $+18$ [$-13$, $+75$], 0.852 \\
small steps (4--8) & $+12$ [$-29$, $+36$], 0.748 & $+120$ [$+40$, $+192$], 0.995 & $+82$ [$+15$, $+157$], 0.989 \\
large steps (1--3) & $-2$ [$-34$, $+19$], 0.424 & $-13$ [$-49$, $+64$], 0.259 & $+7$ [$-7$, $+28$], 0.865 \\
\bottomrule
\end{tabular}
\end{table}

\Para[S-par:controls]{\texorpdfstring{Controls at f/2.8--f/5.6}{Controls at f/2.8–f/5.6}}
At these apertures the finite step dominates the band ratios, and both models give nearly the same values
(Table~\ref{S-tab:controls}). Both lie inside the 95\% intervals of the data in every band and window. At f/4 and f/5.6
the data lie at or above the wave model in all twelve entries, by up to 6\%; at f/2.8 in the narrowest window they lie
13--25\% above both models, with intervals that are wider still (\texttt{realdata/zedd\_controls\_ci.py},
\texttt{results/zedd\_controls\_ci.txt}).

\begin{table}[htbp]
\centering
\caption{Band ratios $\kappa_b/\kappa_{0.0675}$ at f/2.8--f/5.6, all image pairs: data [95\% interval over 2\,000
resamples of the scenes] / wave / geometric.}
\label{S-tab:controls}
\small
\setlength{\tabcolsep}{3pt}
\resizebox{\linewidth}{!}{%
\begin{tabular}{llccc}
\toprule
Aperture & $|\psi|$ & 0.10 c/px & 0.15 c/px & 0.225 c/px \\
\midrule
f/5.6 & $<0.04$\,D & 0.99 [0.93, 1.07] / 0.98 / 1.00 & 0.85 [0.75, 0.96] / 0.85 / 0.88 & 0.59 [0.49, 0.69] / 0.57 / 0.56 \\
     & $<0.08$\,D & 0.93 [0.88, 0.96] / 0.89 / 0.90 & 0.68 [0.61, 0.75] / 0.64 / 0.64 & 0.39 [0.34, 0.46] / 0.37 / 0.36 \\
f/4   & $<0.03$\,D & 1.01 [0.91, 1.16] / 0.97 / 0.98 & 0.79 [0.65, 0.97] / 0.75 / 0.75 & 0.46 [0.35, 0.60] / 0.46 / 0.45 \\
     & $<0.06$\,D & 0.85 [0.80, 0.88] / 0.82 / 0.82 & 0.54 [0.48, 0.60] / 0.52 / 0.51 & 0.28 [0.23, 0.33] / 0.27 / 0.27 \\
f/2.8 & $<0.02$\,D & 1.12 [0.83, 1.35] / 0.92 / 0.92 & 0.79 [0.45, 1.05] / 0.63 / 0.63 & 0.42 [0.23, 0.57] / 0.38 / 0.37 \\
     & $<0.04$\,D & 0.76 [0.69, 0.81] / 0.74 / 0.74 & 0.43 [0.36, 0.51] / 0.42 / 0.42 & 0.21 [0.16, 0.27] / 0.21 / 0.20 \\
\bottomrule
\end{tabular}}
\end{table}

\Para[S-par:infocus]{In focus}
Table~\ref{S-tab:infocus} lists the in-focus energy ratios for all bands and three apertures.
Windows lie within 0.008, 0.006 and 0.004\,D of the best focus of f/5.6, f/4 and f/2.8.

\begin{table}[htbp]
\centering
\caption{In-focus $\log_{10}$ band-energy ratio $E_N/E_{16}$ of f/$N$ and f/16 at the same focus run: median over
windows [95\% interval over scenes], and the clear-disc wave prediction with the nominal f/16 aperture. Geometric
optics predicts 0. With an f/16 aperture of 0.8 times nominal, the f/5.6 prediction would be larger still: $+0.032$,
$+0.049$, $+0.076$, $+0.116$, $+0.184$ and $+0.297$ from 0.03 to 0.225 cycles/px.}
\label{S-tab:infocus}
\small
\setlength{\tabcolsep}{4pt}
\begin{tabular}{lcccccc}
\toprule
& \multicolumn{2}{c}{f/5.6} & \multicolumn{2}{c}{f/4} & \multicolumn{2}{c}{f/2.8} \\
\cmidrule(lr){2-3}\cmidrule(lr){4-5}\cmidrule(lr){6-7}
c/px & data & wave & data & wave & data & wave \\
\midrule
0.03   & $+0.010$ [0.006, 0.014] & $+0.023$ & $+0.011$ [0.006, 0.016] & $+0.027$ & $+0.016$ [0.012, 0.020] & $+0.029$ \\
0.045  & $+0.021$ [0.016, 0.025] & $+0.035$ & $+0.023$ [0.018, 0.028] & $+0.041$ & $+0.030$ [0.024, 0.034] & $+0.045$ \\
0.0675 & $+0.035$ [0.030, 0.040] & $+0.054$ & $+0.040$ [0.034, 0.045] & $+0.062$ & $+0.048$ [0.043, 0.053] & $+0.068$ \\
0.10   & $+0.054$ [0.048, 0.060] & $+0.083$ & $+0.060$ [0.054, 0.067] & $+0.095$ & $+0.071$ [0.064, 0.078] & $+0.104$ \\
0.15   & $+0.080$ [0.073, 0.089] & $+0.129$ & $+0.089$ [0.081, 0.096] & $+0.148$ & $+0.101$ [0.091, 0.111] & $+0.161$ \\
0.225  & $+0.125$ [0.113, 0.138] & $+0.206$ & $+0.136$ [0.124, 0.147] & $+0.234$ & $+0.145$ [0.132, 0.164] & $+0.255$ \\
\bottomrule
\end{tabular}
\end{table}

\Para[S-par:scale]{Absolute scale}
On the 50 validation scenes the forward slopes of $\psi_{\mathrm{app}}$ on the LiDAR defocus scale the wave model by one
factor per aperture, nearly constant across bands: $g=0.64$ at f/16 (data$/(g\cdot$model$)$ 0.94--1.04 across the bands
of 0.0675--0.225\,cycles/px, all pairs within 0.1\,D) and 0.82, 0.79 and 0.66 at f/5.6, f/4 and f/2.8. With geometric
optics at f/16, $g=0.47$, and data$/(g\cdot$model$)$ falls from 1.09 to 0.84 across bands. These forward slopes are
not the gains of the depth estimates of the main paper, 0.82 and 0.52, which are chosen for the smallest depth
error. The test scenes have no
LiDAR. Post hoc, through the depth-from-focus reference of the confirmation, the same statistic gives $g=0.710$
[0.655, 0.790] on the test scenes and 0.731 [0.685, 0.790] on the validation scenes. From the lowest to the highest of
these bands, data$/(g\cdot$model$)$ changes by a factor of 1.144 [0.971, 1.276] for wave optics and 0.810
[0.679, 0.911] for geometric optics on the test scenes, and by 0.950 [0.849, 1.106] and 0.654 [0.584, 0.767] on the
validation scenes (1.101 and 0.770 against LiDAR; \texttt{realdata/\allowbreak zedd\_supp\_scope.py},
\texttt{results/\allowbreak zedd\_supp\_scope.txt}). These slopes pool all pairs, of which the large steps carry
57--73\% of the weight on the validation scenes and 75--84\% on the test scenes; the band ratios of the small steps did
not replicate on the test scenes (Table~\ref{S-tab:confirm}).
Both variables are noisy, however. The forward slope is attenuated by LiDAR noise; the inverse of the reverse slope
($\psi_{\mathrm{GT}}$ on $\psi_{\mathrm{app}}$) is inflated by the noise of $\psi_{\mathrm{app}}$. Table~\ref{S-tab:eiv}
gives the resulting bracket. On the validation scenes an aperture of 0.8 times the nominal f/16 pupil does not explain
the scale: with it the wave model needs $g=1.09$, but it predicts small-step band ratios of 0.92, 0.79 and 0.61 within
0.35\,D, below the data intervals of these scenes in the two highest bands (Table~\ref{S-tab:robust}), and a larger
in-focus loss (Table~\ref{S-tab:infocus}). These band ratios did not replicate. Through the depth-from-focus reference of
the confirmation the small steps within 0.35\,D give 0.962 [0.925, 1.037], 0.927 [0.833, 1.026] and 0.746
[0.649, 0.845] on the test scenes, against 0.950 [0.919, 0.982], 0.835 [0.796, 0.881] and 0.647 [0.616, 0.693] on the
validation scenes (\texttt{results/\allowbreak zedd\_fig3b\_r13test\_base\_full\_m0.04.txt},
\texttt{\_r13val\_base\_full\_m0.04.txt}), and the in-focus loss was measured on the validation scenes only. The
simulated measurement chain leaves the scale at 0.99 (Table~\ref{S-tab:pipesim}).

\begin{table}[htbp]
\centering
\caption{Bracket of the absolute scale of the diffusion coefficient relative to the wave model: forward slope -- inverse reverse slope, focus
pairs pooled, weights $S_\times$.}
\label{S-tab:eiv}
\begin{tabular}{llcccc}
\toprule
Aperture & $|\psi|$ & 0.0675 c/px & 0.10 c/px & 0.15 c/px & 0.225 c/px \\
\midrule
f/16  & $<0.1$\,D  & 0.68--3.64 & 0.68--2.58 & 0.67--1.71 & 0.66--1.05 \\
     & $<0.2$\,D  & 0.72--2.08 & 0.74--1.49 & 0.74--1.14 & 0.72--0.88 \\
     & $<0.35$\,D & 0.76--1.59 & 0.78--1.23 & 0.76--0.99 & 0.72--0.88 \\
f/5.6 & $<0.04$\,D & 0.76--1.43 & 0.78--1.19 & 0.76--1.01 & 0.76--0.92 \\
     & $<0.08$\,D & 0.80--1.14 & 0.81--1.00 & 0.81--0.94 & 0.80--1.03 \\
f/4   & $<0.03$\,D & 0.77--1.38 & 0.80--1.20 & 0.77--1.03 & 0.76--0.98 \\
     & $<0.06$\,D & 0.84--1.11 & 0.85--0.98 & 0.83--0.98 & 0.80--1.09 \\
f/2.8 & $<0.02$\,D & 0.75--1.51 & 0.81--1.22 & 0.78--1.10 & 0.69--1.08 \\
     & $<0.04$\,D & 0.82--1.03 & 0.82--0.97 & 0.82--1.02 & 0.81--1.15 \\
\bottomrule
\end{tabular}
\end{table}

\Para[S-par:onepupil]{One effective pupil}
Can a single clear disc explain the in-focus loss, the band ratios and the scale of the validation scenes together? We
fitted a clear disc in wave optics whose f/16 radius is $r$ times the nominal one, with a Gaussian spread of the focus
of standard deviation $\sigma_\psi$ shared by both apertures (for example a chromatic focal shift), to the six in-focus
ratios of f/5.6 in Table~\ref{S-tab:infocus}, the three band ratios of the small steps within 0.35\,D and the estimation
gain 0.82, all from the 50 validation scenes with LiDAR, by a grid search over $r=0.80$--1.40 and
$\sigma_\psi=0$--0.10\,D (\texttt{realdata/zedd\_pupil\_fit.py}). The family holds clear discs only. The band ratios
of the validation scenes reject the Gaussian pupil, whose coefficient is constant (Table~\ref{S-tab:fig3b}), and those of
the test scenes, which did not replicate, do not (Table~\ref{S-tab:confirm}). The two kinds of data
pull the aperture in opposite directions. The band ratios and the gain are fitted by $r=0.925$ without spread (scale
0.84; $\chi^2=4.5$ for four values), but then the in-focus ratios miss by $\chi^2=977$ for six values. The in-focus
ratios and the band ratios are fitted by an aperture 25\% larger, $r=1.25$ with $\sigma_\psi=0.01$\,D ($\chi^2=15$ for
nine values), which would make the scale 1.66 instead of 0.82. The best compromise for all ten values, $r=0.95$ with
$\sigma_\psi=0.04$\,D, leaves $\chi^2=130$. No clear disc of this family explains the in-focus loss and the scale together.
Processing that acts on the two apertures differently could explain the in-focus loss, but not the sharpening we
simulated (an unsharp mask of 1\,px and amount 0.125 at f/16 only, Table~\ref{S-tab:pipesim}): it raises the f/16
energy by 0.002, 0.004, 0.009, 0.019, 0.038 and 0.066 in $\log_{10}$ from 0.03 to 0.225\,cycles/px, against
shortfalls of 0.013, 0.014, 0.019, 0.029, 0.049 and 0.081 at f/5.6, so it accounts for 15\% of the loss at the lowest
band and 80\% at the highest. Above 0.03\,cycles/px the loss is a nearly constant fraction of the predicted one
(data/wave 0.55--0.71 at f/2.8--f/5.6; 0.41--0.55 at 0.03\,cycles/px), which a scale fits better than a filter. Two readings thus stay open: an f/16 iris 7.5\% smaller than
nominal (about a quarter stop; $r=0.925$ above), which fits the band ratios and the gain but leaves the in-focus loss
to processing of an unknown kind, or processing that sets both. The turnover of the 0.225\,cycles/px band far from
focus favors the first: on its own it gives $r=0.88$--0.94 (Table~\ref{S-tab:turnover}), independently of the gain and
of the in-focus ratios. The forward factor is not an optical scale. Within 0.35\,D on the small steps it is $0.77\pm0.17$ in the reference band, against 0.64 within
0.1\,D over all pairs and bands; errors in the reference vergence attenuate a slope more over a narrower range
(Table~\ref{S-tab:eiv}), and the joint fit is consistent with it ($\chi^2\le0.5$).

\Para[S-par:depth]{Depth estimation with one image pair}
Samples are (window, image pair) with $\psi_{\mathrm{GT}}$ inside the search range of the pair ($\pm0.6$\,D at f/16,
$\pm0.15$\,D at f/5.6, $\pm0.11$\,D at f/4, $\pm0.07$\,D at f/2.8). Per band, $A_b=\psi_{\mathrm{app},b}S_\times$ and
$B_b=S_\times$; sub-band estimates minimize $\sum_bB_b(A_b/B_b-G_b(\psi))^2$ on a grid of 600 points. The weight
$B_b$, the Laplacian energy of the band in the window, is the inverse variance of the band ratio for image noise of
equal spectral density when the signal dominates, and needs no ground truth (Table~\ref{S-tab:weights} compares other weights). The six methods are:
\begin{itemize}
  \item broadband (the sum of the bands) with nominal optics, $\hat\psi=A/B$;
  \item broadband with an affine calibration per pair;
  \item broadband with a calibrated curve per pair, inverted;
  \item sub-band with geometric disc curves and one gain per aperture;
  \item sub-band with clear-disc curves and one gain per aperture;
  \item sub-band with calibrated curves per pair and band.
\end{itemize}
A calibrated curve is fitted on the training scenes of the fold, for one image pair and, in the sub-band estimate,
one band. It is the ratio $\sum A_b/\sum B_b$ of the training samples in bins of the LiDAR defocus, kept where a bin
holds at least 40 samples and interpolated linearly onto the grid; a defocus outside the bins that are kept is not a
candidate of the search. The bins are 0.025\,D wide at f/16, 0.043\,D with the search to 9.5\,px, and
0.004--0.008\,D at f/2.8--f/5.6.
The gain of a model-curve method is chosen on the training scenes for the smallest median depth error. With the focus
of each fold it is 0.81--0.83 at f/16 and 0.92--1.02 at f/2.8--f/5.6 for wave curves, and 0.50--0.54 and 0.85--0.99
for geometric curves. The pooled
least-squares gain is dominated by high-energy windows near focus and gives larger errors. Table~\ref{S-tab:est_bins}
resolves the f/16 errors by defocus, and Fig.~\ref{S-fig:zedd_est} plots them for f/16 and f/5.6.

\begin{table}[htbp]
\centering
\caption{One image pair at f/16: held-out median $|\hat\psi-\psi_{\mathrm{GT}}|$ [D] by $|\psi_{\mathrm{GT}}|$, 953\,939
pair samples, focus calibrated on the training scenes of each fold (\texttt{zedd\_foldcal.py},
\texttt{zedd\_est\_report.py}). Last row: median relative bias $\hat\psi/\psi-1$ for $|\psi|\ge0.1$\,D in the texture
terciles with the least / the most high-frequency energy.}
\label{S-tab:est_bins}
\small
\begin{tabular}{lcccccc}
\toprule
$|\psi_{\mathrm{GT}}|$ [D] & \multicolumn{2}{c}{conventional DfDD} & one ratio & \multicolumn{3}{c}{our joint fit} \\
& nominal & affine & calib.\ curve & geometric & wave (ours) & calib.\ curves \\
\midrule
0--0.07 & 0.019 & 0.022 & 0.019 & 0.017 & 0.018 & 0.018 \\
0.07--0.14 & 0.055 & 0.031 & 0.019 & 0.023 & 0.019 & 0.019 \\
0.14--0.21 & 0.088 & 0.046 & 0.025 & 0.037 & 0.020 & 0.021 \\
0.21--0.28 & 0.123 & 0.072 & 0.037 & 0.037 & 0.024 & 0.026 \\
0.28--0.35 & 0.165 & 0.106 & 0.082 & 0.046 & 0.043 & 0.036 \\
\midrule
bias by texture & $-52/-53\%$ & $+18/+18\%$ & $+4/+2\%$ & $-2/-4\%$ & $-9/-4\%$ & $-2/+3\%$ \\
\bottomrule
\end{tabular}
\end{table}

\Para[S-par:baselines]{Baselines and ablations}
Main-paper Table~\ref{tab:est} and its form in diopters, Table~\ref{S-tab:estfull}, add, on the same samples and folds (the
methods of \texttt{zedd\_baselines.py}, run by \texttt{zedd\_learning\_curve.py} with the focus of each fold): one ratio
inverted with a model curve, the ratio of the summed model numerators and denominators of the five bands
for the nominal spectrum, with one gain; $\alpha$ and $\beta$ fitted only in the linear range (blur radius up to
1.5\,px) by iteratively reweighted least squares with a Huber loss, instead of over the whole search range; the joint
fit with one gain per band, the median of $r_b/g_b(\psi_{\mathrm{GT}})$ over training samples where the model curve
exceeds half its largest magnitude, times one common factor chosen like the gain; per-band estimates from affine
calibrations in the linear range, fused with weights proportional to the Laplacian energy of each band, in the manner
of the multi-filter DfDD of Guo et~al.~\cite{S-guo2019}; spectral ratios~\cite{S-pentland1987,S-morganmar2014} $(E_k-E_{k+1})/(E_k+E_{k+1})$ of the band
energies of the two images with calibrated curves per pair and band, fitted jointly with energy weights (and with our
curves and gain in \hyperref[S-par:srcontrol]{the control below}); and a perceptron (four layers of 256 units, L1 loss) trained on the band ratios, the logarithms of the band energies and the
pair index of the training scenes (\texttt{zedd\_mlp.py}). Two more perceptrons choose their width (32, 128 or 256
units) and stopping step on held-out training scenes, a fifth of the training scenes, without the scenes they are scored on; the second of
them takes our estimate as an extra input and learns a correction to it, with its last layer initialized to zero. The
calibrated curves, finally, also get one gain per aperture chosen by the criterion of the model curves, the smallest
median error on the training samples; their gain is 1.05--1.11. Table~\ref{S-tab:baselines} gives the errors with
intervals and paired ratios, and Table~\ref{S-tab:basetex} the texture dependence (\texttt{zedd\_learning\_curve.py
--mlp-variants cr}, tag \texttt{extra} of \texttt{zedd\_table1.py}; the other methods of that run repeat the fold run
exactly). With the gain, the calibrated curves err 7\% less than ours in the median defocus error at f/16 (0.93
[0.88, 0.97]) and 22\% less at f/2.8, so in this error wave-optics curves come within 8\% of a calibration of every
pair and band at f/16 but do not replace it; in the mean relative depth error of main-paper Table~\ref{tab:est}
the gap is 3\% (0.97 [0.91, 1.02]).

\begin{table*}[htbp]
\centering
\caption{One image pair on the DSLR, the estimators of Table~\ref{S-tab:estfull} with intervals at every aperture: held-out median
$|\hat\psi-\psi|$ [D] (paired ratio of median errors to the joint fit with wave-optics curves on the same windows, 95\%
interval over 2\,000 resamples of the scenes), with the focus calibrated on the training scenes of each fold
(\texttt{zedd\_learning\_curve.py --focus fold}, \texttt{zedd\_table1.py}). Evaluation within $\pm0.35$, 0.08, 0.06 and
0.04\,D. Bottom: the same estimators on all windows with valid LiDAR depth and no clipped pixels (2.2 times as many),
and our estimate with a noise floor subtracted and with noise-aware weights (Sec.~\ref{S-sec:bw}; ratio to ours on the
same windows).}
\label{S-tab:baselines}
\setlength{\tabcolsep}{3pt}
\resizebox{\linewidth}{!}{%
\begin{tabular}{llcccc}
\toprule
& & f/16 & f/5.6 & f/4 & f/2.8 \\
\midrule
one ratio & conventional DfDD, nominal $\alpha,\beta$ & 0.0709 (3.32 [3.02, 3.63]) & 0.0170 (1.85 [1.68, 2.07]) & 0.0159 (1.81 [1.62, 2.03]) & 0.0129 (1.58 [1.36, 1.82]) \\
& conventional DfDD, $\alpha,\beta$, linear range & 0.0327 (1.53 [1.38, 1.69]) & 0.0099 (1.08 [0.98, 1.19]) & 0.0095 (1.08 [0.97, 1.21]) & 0.0078 (0.96 [0.81, 1.17]) \\
& geometric curve & 0.0293 (1.37 [1.26, 1.50]) & 0.0113 (1.23 [1.14, 1.35]) & 0.0119 (1.36 [1.25, 1.51]) & 0.0119 (1.46 [1.30, 1.66]) \\
& our curve & 0.0212 (0.99 [0.97, 1.01]) & 0.0108 (1.17 [1.09, 1.28]) & 0.0115 (1.31 [1.21, 1.45]) & 0.0117 (1.43 [1.28, 1.61]) \\
& calibrated curve & 0.0257 (1.20 [1.12, 1.28]) & 0.0133 (1.45 [1.34, 1.59]) & 0.0126 (1.43 [1.27, 1.68]) & 0.0099 (1.22 [1.07, 1.41]) \\
\midrule
other estimators & band fusion~\cite{S-guo2019} & 0.0294 (1.38 [1.24, 1.52]) & 0.0098 (1.07 [0.97, 1.17]) & 0.0091 (1.04 [0.93, 1.17]) & 0.0070 (0.86 [0.74, 1.02]) \\
& spectral ratio~\cite{S-pentland1987,S-morganmar2014}, calibrated curves & 0.0241 (1.13 [1.06, 1.20]) & 0.0105 (1.14 [1.04, 1.23]) & 0.0091 (1.04 [0.94, 1.15]) & 0.0065 (0.80 [0.68, 0.95]) \\
& spectral ratio, our curves and gain & 0.0247 (1.16 [1.12, 1.20]) & 0.0097 (1.06 [1.02, 1.10]) & 0.0088 (1.00 [0.93, 1.06]) & 0.0080 (0.98 [0.93, 1.03]) \\
& our perceptron & 0.0180 (0.84 [0.78, 0.92]) & 0.0084 (0.91 [0.84, 0.99]) & 0.0075 (0.86 [0.79, 0.94]) & 0.0057 (0.70 [0.61, 0.83]) \\
& our perceptron, held-out selection & 0.0181 (0.85 [0.79, 0.92]) & 0.0083 (0.90 [0.82, 0.98]) & 0.0073 (0.84 [0.76, 0.93]) & 0.0061 (0.75 [0.64, 0.88]) \\
& our perceptron, correcting ours & 0.0175 (0.82 [0.76, 0.90]) & 0.0079 (0.87 [0.79, 0.95]) & 0.0070 (0.80 [0.72, 0.89]) & 0.0060 (0.73 [0.62, 0.88]) \\
\midrule
our joint fit & geometric & 0.0272 (1.27 [1.18, 1.38]) & 0.0095 (1.03 [1.01, 1.06]) & 0.0089 (1.01 [1.00, 1.03]) & 0.0082 (1.01 [0.99, 1.02]) \\
& geometric, 5 gains & 0.0246 (1.15 [1.07, 1.23]) & 0.0096 (1.05 [1.02, 1.07]) & 0.0087 (1.00 [0.98, 1.02]) & 0.0082 (1.01 [0.98, 1.02]) \\
& wave (ours) & 0.0214 & 0.0092 & 0.0088 & 0.0082 \\
& calibrated & 0.0215 (1.01 [0.94, 1.07]) & 0.0099 (1.08 [1.00, 1.15]) & 0.0086 (0.99 [0.89, 1.09]) & 0.0067 (0.82 [0.72, 0.95]) \\
& calibrated, one gain & 0.0198 (0.93 [0.88, 0.97]) & 0.0086 (0.94 [0.87, 1.01]) & 0.0081 (0.92 [0.84, 1.01]) & 0.0064 (0.78 [0.68, 0.91]) \\
& wave, focus fitted per pair & 0.0182 (0.85 [0.81, 0.90]) & 0.0072 (0.79 [0.73, 0.86]) & 0.0067 (0.77 [0.70, 0.84]) & 0.0056 (0.69 [0.59, 0.82]) \\
& wave, noise floor subtracted & 0.0219 (1.02 [0.99, 1.08]) & 0.0092 (1.00 [0.99, 1.01]) & 0.0087 (0.99 [0.97, 1.01]) & 0.0080 (0.98 [0.96, 1.00]) \\
& wave, noise-aware weights & 0.0219 (1.03 [0.99, 1.08]) & 0.0092 (1.00 [0.99, 1.01]) & 0.0087 (0.99 [0.97, 1.01]) & 0.0080 (0.98 [0.96, 1.00]) \\
\midrule
all windows & one ratio, our curve & 0.0319 (0.99 [0.98, 1.01]) & 0.0140 (1.18 [1.11, 1.27]) & 0.0145 (1.29 [1.22, 1.40]) & 0.0140 (1.38 [1.27, 1.51]) \\
& one ratio, calibrated curve & 0.0401 (1.25 [1.17, 1.33]) & 0.0178 (1.51 [1.39, 1.64]) & 0.0158 (1.41 [1.27, 1.60]) & 0.0119 (1.18 [1.06, 1.32]) \\
& our perceptron & 0.0239 (0.74 [0.70, 0.80]) & 0.0110 (0.93 [0.83, 1.04]) & 0.0097 (0.87 [0.79, 0.97]) & 0.0078 (0.77 [0.69, 0.87]) \\
& our joint fit, geometric & 0.0376 (1.17 [1.12, 1.23]) & 0.0122 (1.04 [1.02, 1.05]) & 0.0114 (1.02 [1.01, 1.03]) & 0.0101 (1.00 [0.99, 1.00]) \\
& our joint fit, geometric, 5 gains & 0.0336 (1.05 [1.00, 1.10]) & 0.0121 (1.03 [1.01, 1.05]) & 0.0110 (0.98 [0.97, 1.00]) & 0.0098 (0.97 [0.94, 0.99]) \\
& joint fit, wave (ours) & 0.0321 & 0.0118 & 0.0112 & 0.0101 \\
& our joint fit, calibrated & 0.0320 (1.00 [0.96, 1.04]) & 0.0138 (1.17 [1.10, 1.25]) & 0.0117 (1.04 [0.98, 1.13]) & 0.0090 (0.89 [0.80, 0.99]) \\
\bottomrule
\end{tabular}}
\end{table*}

\begin{table*}[htbp]
\centering
\caption{The estimators of main-paper Table~\ref{tab:est} in diopters (the geometric discs with a fitted or a
frequency-scaled radius are in Table~\ref{S-tab:attrctl}, the posterior in Table~\ref{S-tab:posterior}), at all four
apertures, with the rows that the main paper leaves out,
among them conventional DfDD (conv.) with the nominal
constants; the 9-tap ratio with our curve (window mean removed, prefilter width chosen on the training
scenes); our perceptron that corrects ours; our joint fit with an offset of the focus fitted per pair; and ours on
all windows with valid LiDAR depth. Columns: held-out median error $|\hat\psi-\psi|$ [mD] within $\pm0.35$,
0.08, 0.06 and 0.04\,D of focus at f/16, f/5.6, f/4 and f/2.8, with the paired ratio to ours at f/16 [95\% interval];
median relative defocus error [\%] at f/16 over all pairs, at 1.5--3.5 and 3.5--9.5\,px of blur; operations per pixel
(Table~\ref{S-tab:flops}). One gain per aperture unless stated.}
\label{S-tab:estfull}
\small
\setlength{\tabcolsep}{3.7pt}
\begin{tabular}{llcccccccr}
\toprule
& & \multicolumn{2}{c}{f/16} & & & & \multicolumn{2}{c}{rel.\ defocus error} & \\
\cmidrule(lr){3-4}\cmidrule(lr){8-9}
& Curve & error & ratio & f/5.6 & f/4 & f/2.8 & 1.5--3.5 & 3.5--9.5 & Ops \\
\midrule
one ratio & conv., nominal $\alpha,\beta$ & 70.9 & 3.32 [3.02, 3.63] & 17 & 16 & 13 & 52 & 88 & 1.3k \\
& conv., calib.\ $\alpha,\beta$ & 32.7 & 1.53 [1.38, 1.69] & 9.9 & 9.5 & 7.8 & 33 & 69 & 1.3k \\
& geometric & 29.3 & 1.37 [1.26, 1.50] & 11 & 12 & 12 & 38 & 56 & 1.6k \\
& our curve & 21.2 & 0.99 [0.97, 1.01] & 11 & 12 & 12 & 19 & 66 & 1.6k \\
& calibrated & 25.7 & 1.20 [1.12, 1.28] & 13 & 13 & 9.9 & 27 & 55 & 1.6k \\
& 9-tap, our curve & 22.0 & 1.03 [1.00, 1.06] & -- & -- & -- & 12 & 50 & 0.4k \\
\midrule
other & band fusion~\cite{S-guo2019} & 29.4 & 1.38 [1.24, 1.52] & 9.8 & 9.1 & 7.0 & 28 & 73 & 1.3k \\
estimators & spectral ratio~\cite{S-pentland1987,S-morganmar2014}, calib. & 24.1 & 1.13 [1.06, 1.20] & 11 & 9.1 & 6.5 & 14 & 9.3 & 2.3k \\
& spectral ratio, our curves & 24.7 & 1.16 [1.12, 1.20] & 9.7 & 8.8 & 8.0 & 14 & 13 & 2.3k \\
& our perceptron & 18.0 & 0.84 [0.78, 0.92] & 8.4 & 7.5 & 5.7 & 7.8 & 5.6 & 400k \\
& our perc., correcting ours & 17.5 & 0.82 [0.76, 0.90] & 7.9 & 7.0 & 6.0 & 7.8 & 5.7 & 400k \\
\midrule
our joint fit & geometric & 27.2 & 1.27 [1.18, 1.38] & 9.5 & 8.9 & 8.2 & 16 & 8.6 & 2.9k \\
& geom., 5 gains & 24.6 & 1.15 [1.07, 1.23] & 9.6 & 8.7 & 8.2 & 15 & 11 & 2.9k \\
& wave (ours) & 21.4 & -- & 9.2 & 8.8 & 8.2 & 11 & 9.0 & 2.9k \\
& calibrated & 21.5 & 1.01 [0.94, 1.07] & 9.9 & 8.6 & 6.7 & 11 & 8.3 & 2.9k \\
& calibrated, gain & 19.8 & 0.93 [0.88, 0.97] & 8.6 & 8.1 & 6.4 & 11 & 9.6 & 2.9k \\
& wave, focus offset & 17.9 & 0.84 [0.80, 0.88] & 7.2 & 6.8 & 5.8 & 10 & 8.0 & 2.9k \\
& wave (ours), all windows & 32.1 & -- & 12 & 11 & 10 & 24 & 23 & 2.9k \\
\bottomrule
\end{tabular}
\end{table*}

\begin{table}[htbp]
\centering
\caption{Texture dependence of one-pair estimates: relative bias in the tercile with the least minus the tercile with
the most high-frequency energy [percentage points], 95\% interval over 500 resamples of the scenes; held-out samples of
main-paper Table~\ref{tab:est}, focus calibrated per fold (\texttt{zedd\_t1\_texture.py}; the rows of the two
perceptrons with held-out selection and of the calibrated curves with one gain from \texttt{--tag extra}).}
\label{S-tab:basetex}
\small
\setlength{\tabcolsep}{2.5pt}
\resizebox{\linewidth}{!}{%
\begin{tabular}{lcccc}
\toprule
& f/16 & f/5.6 & f/4 & f/2.8 \\
\midrule
conventional DfDD: one ratio, nominal $\alpha,\beta$ & $+0.3$ [$-3.5$, $+3.6$] & $+7.5$ [$+0.6$, $+15.5$] & $+10.1$ [$+4.4$, $+18.4$] & $+15.6$ [$+10.0$, $+20.1$] \\
conventional DfDD: one ratio, $\alpha,\beta$ & $+0.2$ [$-8.3$, $+8.8$] & $+11.0$ [$+5.2$, $+19.9$] & $+18.8$ [$+11.8$, $+27.1$] & $+35.9$ [$+24.3$, $+48.3$] \\
one ratio, our curve & $+0.5$ [$-3.8$, $+5.6$] & $+17.9$ [$+8.9$, $+34.2$] & $+37.0$ [$+18.8$, $+54.1$] & $+55.6$ [$+37.0$, $+69.3$] \\
one ratio, calibrated curve & $+1.9$ [$-2.6$, $+7.0$] & $+22.0$ [$+11.3$, $+38.3$] & $+39.8$ [$+22.1$, $+58.2$] & $+61.5$ [$+34.6$, $+86.3$] \\
band fusion~\cite{S-guo2019} & $-7.8$ [$-15.2$, $-0.1$] & $+4.1$ [$-1.7$, $+12.0$] & $+5.6$ [$-0.1$, $+12.2$] & $+14.5$ [$+5.3$, $+24.9$] \\
spectral ratio~\cite{S-pentland1987,S-morganmar2014}, calibrated curves & $-16.6$ [$-20.6$, $-12.9$] & $-25.2$ [$-30.8$, $-17.2$] & $-21.9$ [$-30.4$, $-11.2$] & $-4.8$ [$-18.0$, $+9.7$] \\
spectral ratio, our curves and gain & $-17.8$ [$-21.7$, $-13.6$] & $-19.9$ [$-25.8$, $-12.6$] & $-12.4$ [$-21.0$, $-2.5$] & $+4.2$ [$-17.0$, $+18.9$] \\
our perceptron & $+4.3$ [$+1.6$, $+8.0$] & $+1.0$ [$-2.5$, $+4.8$] & $-3.4$ [$-9.1$, $+3.1$] & $-2.8$ [$-10.9$, $+3.9$] \\
our perceptron, held-out selection & $+4.1$ [$+1.5$, $+7.9$] & $+1.6$ [$-2.3$, $+6.1$] & $-2.1$ [$-8.4$, $+4.3$] & $-5.9$ [$-15.3$, $+2.6$] \\
our perceptron, correcting ours & $+4.4$ [$+1.2$, $+8.6$] & $-0.5$ [$-5.5$, $+5.4$] & $-1.3$ [$-8.2$, $+4.3$] & $-6.9$ [$-17.4$, $+1.4$] \\
our joint fit, geometric & $+2.0$ [$-0.9$, $+5.9$] & $-1.9$ [$-8.2$, $+6.4$] & $+3.3$ [$-5.8$, $+14.4$] & $+17.3$ [$+1.0$, $+35.4$] \\
joint fit, wave (ours) & $-5.1$ [$-8.2$, $-0.7$] & $-1.7$ [$-7.4$, $+6.2$] & $+6.1$ [$-3.1$, $+17.4$] & $+20.4$ [$+2.9$, $+39.6$] \\
our joint fit, calibrated & $-4.5$ [$-7.8$, $-0.8$] & $-3.7$ [$-9.5$, $+5.6$] & $+3.8$ [$-6.3$, $+14.4$] & $+21.2$ [$+7.6$, $+36.4$] \\
our joint fit, calibrated, one gain & $-4.5$ [$-7.5$, $-0.8$] & $+2.3$ [$-3.0$, $+9.8$] & $+7.9$ [$-1.7$, $+19.7$] & $+22.5$ [$+9.9$, $+37.1$] \\
\bottomrule
\end{tabular}}
\end{table}

\Para[S-par:srcontrol]{The spectral ratio with the same curves}
The spectral-ratio row of the tables above uses curves calibrated for every pair and band, so it does not show what
the statistic itself contributes; in a narrow band the two statistics coincide
(\hyperref[S-par:specratio]{Sec.~\ref{S-sec:bw}}). We therefore fitted the spectral ratio $\varrho_b$ with the forward
model of our estimator, that is, with the same five bands, curves computed from the same pupil model for the nominal
spectrum, one gain per aperture chosen by the criterion of ours on the same training subsample, the weights
$(P_b^++P_b^-)/2$ of Eq.~\eqref{S-eq:srlimit}, and the folds, focus calibrations, samples and two-level search of
main-paper Table~\ref{tab:est}. The ratio of means $r'_b$ of Eq.~\eqref{S-eq:srfinite} is fitted the same way, with the weights
$(e_b^++e_b^-)^2/(2(P_b^++P_b^-))$, which equal $\omega_b$ in a narrow band. The script computes ours and the
calibrated spectral ratio again and reproduces their saved estimates sample by sample
(\texttt{realdata/zedd\_sr\_control.py}, \texttt{zedd\_sr\_control\_table.py};
\texttt{results/zedd\_sr\_control\_table.txt}). Table~\ref{S-tab:srcontrol} gives the result.

At f/16, within 0.35\,D, the spectral ratio with our curves and gain errs by 24.7\,mD, 16\% more than ours (1.16
[1.12, 1.20]; 1.16 on the small steps and 1.14 on the large), as much as with calibrated curves (1.02 [0.97, 1.08]
times those), and 12\% more on all windows (1.12 [1.09, 1.14]). The ratio of means errs 2\% more than ours (1.02
[1.01, 1.04]). The factor $m_b$ of Eq.~\eqref{S-eq:msf} is all that separates the spectral ratio from the ratio of
means (1.13 [1.10, 1.17] between the two), so it accounts for 13 of the 16 points. The texture dependence splits the
same way. Between the terciles with the least and the most high-frequency energy the relative bias differs by $-17.8$
[$-21.7$, $-13.6$] points for the spectral ratio, against $-5.1$ [$-8.2$, $-0.7$] for ours and $-4.2$ [$-7.5$, $+0.3$]
for the ratio of means ($-16.6$ with calibrated curves, Table~\ref{S-tab:basetex}). Windows with steep spectra read
low, as $m_b$ predicts.

Each statistic has its own weights in Table~\ref{S-tab:srcontrol}, so the table compares complete estimators. The
weights are not what separates them. With the weights of ours carried over to the other two by the band-center
constant $L_b$ ($\omega_b$ for $r'_b$ and $\omega_b/L_b^2$ for $\varrho_b$), the spectral ratio errs 1.16 [1.12, 1.20]
times ours and the ratio of means 1.02 [1.01, 1.04] times. With the weights of the spectral ratio carried over to the
other two ($(P_b^++P_b^-)L_b^2/2$), the spectral ratio errs 1.16 [1.12, 1.20] times ours and the ratio of means 1.02
[1.01, 1.03] times. Over the 9.5\,px search the errors of Table~\ref{S-tab:srcontrol} move by at most 0.6 points (14.3\%
and 13.9\% for the spectral ratio with the weights of ours, 11.7\% and 9.0\% for ours with those of the spectral
ratio; \texttt{realdata/\allowbreak zedd\_sr\_weights.py}, \texttt{results/\allowbreak zedd\_sr\_weights\_f16.txt},
\texttt{zedd\_sr\_weights\_wide\_f16.txt}).

We also measured $m_b$ on the same samples. Relative to the model value at the same defocus and pair, its logarithm
has a robust standard deviation over the windows of 11.0, 10.6, 9.3, 7.9 and 6.8\% in the five bands, from the lowest
to the highest (2.3--4.5\% between the medians of the scenes), and its median differs by 8--16 points between the two
extreme terciles. Against the spectral exponent estimated from the two band energies that define the terciles (median
1.5, robust standard deviation 1.1 over the windows) it falls by 3.7--7.2\% per unit (Theil--Sen; 3.3--6.5\% by least
squares), where a power law through our filters gives 5.9--6.2\% (\texttt{realdata/\allowbreak
zedd\_sr\_meanfreq.py}, \texttt{results/\allowbreak zedd\_sr\_meanfreq.txt}).

Over the 9.5\,px search the spectral ratio with our curves errs by 14.0\% and 13.3\% at 1.5--3.5 and 3.5--9.5\,px,
$+2.7$ [$+1.7$, $+3.6$] and $+4.3$ [$+2.8$, $+5.1$] points above ours. The excess beyond 3.5\,px lies on the large
steps ($+4.5$ [$+2.5$, $+5.5$]; small steps $+0.0$ [$-4.2$, $+2.5$], with a 20\% limit at 5.6\,px against 6.6\,px), and
calibrated curves remove it (9.3\%). The comparison is not one-sided. Beyond 3.5\,px fewer estimates of the spectral
ratio err by more than half, 17.3\% against 19.7\% ($-2.4$ [$-3.3$, $-1.5$] points); on the least energetic third of
the textured windows it errs by 64\% against 72\% beyond 3.5\,px ($-8.0$ [$-16.2$, $-1.8$]); and on all windows it errs
by 24.0\% and 23.6\%, as ours does (24.1\% and 22.9\%). At f/5.6 it errs 6\% more than ours within 0.08\,D (1.06
[1.02, 1.10]) and 2.4 and 2.2 points more over the range. At f/4 and f/2.8 the two err alike (1.00 [0.93, 1.06] and
0.98 [0.93, 1.03]), although the texture dependence of the spectral ratio remains at f/5.6 and f/4 ($-19.9$ and
$-12.4$ points, against $-1.7$ and $+6.1$ for ours).

With geometric curves the order changes. The spectral ratio then errs by 25.4\,mD at f/16, 7\% less than ours with
the same curves (0.93 [0.89, 0.97]), and wave-optics curves do not improve it significantly (0.97 [0.92, 1.03]),
while they improve ours by 21\% and the ratio of means by 18\%. Its texture dependence is $-6.7$ [$-9.5$, $-3.9$]
points with geometric curves against $-17.8$ with wave-optics curves, and that of ours moves the same way, from $-5.1$
to $+2.0$. We read this as two errors that offset. Geometric curves overrate the high bands relative to the low
ones, so with one gain they read windows that rely on the low bands higher than windows rich in fine detail, and
$m_b$ reads windows with steep spectra low. The benefit of wave-optics curves therefore shows only in a statistic
that is free of $m_b$.

The small need for training data belongs to the curves, not to the statistic. Fitted on two scenes that also
calibrate the focus, the protocol of Fig.~\ref{S-fig:depthold}c, the spectral ratio with our curves errs by
25.2\,mD, 2\% more than from all 40 scenes, and ours by 22.6\,mD, 5\% more. Over four draws of the two scenes these
errors span 24.7--26.4 and 21.5--23.5\,mD, which are ranges of the draws and not intervals
(\texttt{zedd\_sr\_control.py --focus kscene}, \texttt{results/\allowbreak zedd\_sr\_control\_kscene\_f16.txt}; the
run reproduces the learning curve of ours exactly).

\begin{table*}[htbp]
\centering
\caption{Three statistics of the same five bands, fitted jointly with the same curves and one gain per aperture
(spectral ratio with calibrated curves: no gain): held-out median $|\hat\psi-\psi|$ [mD] with the paired ratio to ours
[95\% interval over 2\,000 resamples of the scenes], within $\pm0.35$, 0.08, 0.06 and 0.04\,D, on the samples of
main-paper Table~\ref{tab:est}. Texture: relative bias in the tercile with the least minus the tercile with the most
high-frequency energy at f/16 [points], as Table~\ref{S-tab:basetex}. Range: mean of the binned median relative defocus
error [\%] over 1.5--3.5 / 3.5--9.5\,px at f/16. The heat-equation ratio $r_b$ is main-paper
Eq.~\eqref{eq:bw_ratio}; $r'_b$ and $\varrho_b$ are defined in Eq.~\eqref{S-eq:srfinite}
(\texttt{results/zedd\_sr\_control\_table.txt}).}
\label{S-tab:srcontrol}
\setlength{\tabcolsep}{3pt}
\resizebox{\linewidth}{!}{%
\begin{tabular}{llcccccc}
\toprule
Curves & Statistic & f/16 & f/5.6 & f/4 & f/2.8 & texture & range \\
\midrule
wave & heat-equation ratio $r_b$ (ours) & 21.4 & 9.2 & 8.8 & 8.2 & $-5.1$ [$-8.2$, $-0.7$] & 11.3 / 9.0 \\
& ratio of means $r'_b$ & 21.9 (1.02 [1.01, 1.04]) & 9.0 (0.98 [0.97, 0.99]) & 8.8 (1.00 [0.99, 1.01]) & 8.2 (1.01 [1.00, 1.02]) & $-4.2$ [$-7.5$, $+0.3$] & 11.2 / 9.6 \\
& spectral ratio $\varrho_b$ & 24.7 (1.16 [1.12, 1.20]) & 9.7 (1.06 [1.02, 1.10]) & 8.8 (1.00 [0.93, 1.06]) & 8.0 (0.98 [0.93, 1.03]) & $-17.8$ [$-21.7$, $-13.6$] & 14.0 / 13.3 \\
\midrule
calibrated & spectral ratio $\varrho_b$ & 24.1 (1.13 [1.06, 1.20]) & 10.5 (1.14 [1.04, 1.23]) & 9.1 (1.04 [0.94, 1.15]) & 6.5 (0.80 [0.68, 0.95]) & $-16.6$ [$-20.6$, $-12.9$] & 13.9 / 9.3 \\
\midrule
geometric & heat-equation ratio $r_b$ & 27.2 (1.27 [1.18, 1.38]) & 9.5 (1.03 [1.01, 1.06]) & 8.9 (1.01 [1.00, 1.03]) & 8.2 (1.01 [0.99, 1.02]) & $+2.0$ [$-0.9$, $+5.9$] & 16.4 / 8.6 \\
& ratio of means $r'_b$ & 26.6 (1.25 [1.17, 1.33]) & 9.5 (1.03 [1.01, 1.05]) & 8.9 (1.01 [0.99, 1.02]) & 8.1 (1.00 [0.98, 1.01]) & $+1.5$ [$-1.8$, $+5.3$] & 15.6 / 8.3 \\
& spectral ratio $\varrho_b$ & 25.4 (1.19 [1.13, 1.25]) & 10.2 (1.11 [1.07, 1.16]) & 8.9 (1.01 [0.93, 1.09]) & 8.0 (0.98 [0.92, 1.04]) & $-6.7$ [$-9.5$, $-3.9$] & 13.7 / 11.1 \\
\bottomrule
\end{tabular}}
\end{table*}

\Para[S-par:attribution]{Where each improvement comes from}
Table~\ref{S-tab:attribution} collects, from Tables~\ref{S-tab:estfull} and~\ref{S-tab:srcontrol}, what each
ingredient changes in the median defocus error when the rest is held fixed (f/16, textured windows). The curve gives the accuracy near focus.
Wave-optics curves lower the error of the joint fit by 21\%, and one ratio with that curve is already as accurate as
the joint fit. They do not lower the error of the spectral ratio significantly (3\%, 0.97 [0.92, 1.03]), which we
read above as two errors that offset with geometric curves. Dividing the spectral ratio by the mean square frequency
lowers its error by 11\%, and the ratio of means does this as well as $r_b$ does, within 2\%. Fitting the five bands
jointly gives the range and nothing near focus. What is specific to $r_b$ is therefore small. What this paper adds to
a fit of spectral ratios is the curve computed from the pupil, with its small need for labeled data, and the
division by the mean square frequency, without which that curve brings little. More labeled data lower the error of
ours further. An offset of the focus per pair lowers it by 16\%, curves calibrated for every pair and band with a
gain by 7\%, and a perceptron by 16\%.

\begin{table}[htbp]
\centering
\caption{What each ingredient changes at f/16 on the textured windows: held-out median $|\hat\psi-\psi|$ within
$\pm0.35$\,D [mD] and mean of the binned median relative defocus error over 1.5--3.5 / 3.5--9.5\,px [\%], before $\to$
after. Above the rule one ingredient changes and the rest is held fixed; below it, what labeled data add to ours.
Numbers of Tables~\ref{S-tab:estfull} and~\ref{S-tab:srcontrol}, where the intervals are given
(Table~\ref{S-tab:baselines}; for the offset, \hyperref[S-par:perpair]{Errors per pair}).}
\label{S-tab:attribution}
\small
\setlength{\tabcolsep}{4pt}
\begin{tabular}{@{}llcc@{}}
\toprule
Ingredient & Comparison & near focus & range \\
\midrule
wave-optics curve & our joint fit, geometric $\to$ wave & $27.2\to21.4$ & $16$ / $8.6\to11$ / $9.0$ \\
& spectral ratio, geometric $\to$ wave & $25.4\to24.7$ & $14$ / $11\to14$ / $13$ \\
division by $m_b$ & spectral ratio $\to$ ratio of means & $24.7\to21.9$ & $14$ / $13\to11$ / $9.6$ \\
least-squares slope & ratio of means $\to$ $r_b$ & $21.9\to21.4$ & $11$ / $9.6\to11$ / $9.0$ \\
our joint fit & one ratio of the summed bands $\to$ joint fit & $21.2\to21.4$ & $19$ / $66\to11$ / $9.0$ \\
\midrule
focus offset per pair & ours $\to$ ours with the offset & $21.4\to17.9$ & $11$ / $9.0\to10$ / $8.0$ \\
calibrated curves & ours $\to$ curves per pair and band, one gain & $21.4\to19.8$ & $11$ / $9.0\to11$ / $9.6$ \\
learning & ours $\to$ our perceptron & $21.4\to18.0$ & $11$ / $9.0\to7.8$ / $5.6$ \\
\bottomrule
\end{tabular}
\end{table}

\Para[S-sec:attrctl]{Attribution controls: wave optics or an effective aperture?}
In the median defocus error, geometric curves err 27\% more than ours at f/16 with one gain and 15\% more with a
gain per band (Table~\ref{S-tab:estfull}; in the mean relative depth error of the main paper, 22\% and 8\%). A geometric disc smaller than the nominal aperture would lower the gain and move
the turnover of the finest band outward at once. The gain of 0.52 is $0.72^2$, and the turnover of the 0.225\,cycles/px band lies
at 0.42\,D where the nominal geometric disc puts it at 0.29\,D (Table~\ref{S-tab:turnover}). If such a disc matched our
curves, the depth result would not go beyond geometric optics. We fixed the control and its rule in a local record
before computing any of its estimates (\texttt{notes/\allowbreak prespec\_\allowbreak attribution\_\allowbreak
20261114.md}, with the SHA-256 of 72 files, checked after the runs; \texttt{realdata/\allowbreak
zedd\_\allowbreak attribution.py}, \texttt{zedd\_\allowbreak attribution\_\allowbreak table.py},
\texttt{results/\allowbreak zedd\_\allowbreak attribution\_\allowbreak table.txt}). At that time the
only depth estimates with geometric curves were those at the nominal radius.
\begin{itemize}
  \item \emph{Protocol.} That of main-paper Table~\ref{tab:est} at f/16: the same folds, focus calibrations,
  windows, search and training subsample. The script recomputes ours and the nominal geometric curves and stops unless
  they equal the saved estimates sample by sample; they did, to the last digit. Every free quantity of a curve family
  is chosen per fold on the 20\,000 training samples that fix our gain, by our criterion, the smallest median
  $|\hat\psi-\psi|$. A radius is searched from 0.50 to 1.20 of the nominal one in steps of 0.05, then of 0.01
  around the best.
  \item \emph{Families.} A geometric disc with a fitted radius, with and without a gain; a wave-optics disc with a
  fitted radius, with and without a gain; a geometric disc whose radius in band $b$ is $1-f_b/f_c$ times the nominal
  one, the factor of Stokseth's approximation of the defocused OTF~\cite{S-stokseth1969}, with one gain and with none;
  a Gaussian pupil, whose coefficient is constant, with a fitted radius and gain; and ours, the nominal geometric
  curves and the fitted geometric disc with one focus offset per image pair, the median residual of the training
  samples of the pair.
  \item \emph{Rule.} On the held-out median error within 0.35\,D of focus, the statement that the advantage of our
  curves is specific to wave optics and not an effective aperture is confirmed if the 95\% interval of the error of
  the geometric disc with a fitted radius and gain, divided by ours, lies above 1.05, refuted if it lies below
  1.05, and undecided otherwise. The same rule is reported for the disc with a fitted radius alone.
  \item \emph{Predictions written with the rule.} A fitted radius of 0.70--0.76 with a gain near 1 and a ratio of
  1.03--1.15, so that the rule might not decide; the Stokseth-scaled disc within 5\% of ours; a wave-optics radius of
  0.90--1.00.
\end{itemize}
\emph{Outcome} (Table~\ref{S-tab:attrctl}). With a fitted radius alone, 0.70--$0.71R$ in the five folds, the geometric disc errs
1.14 [1.09, 1.18] times ours, which by the rule confirms the statement for this family. With a radius and a gain,
$0.63R$ and 1.25, it errs 1.06 [1.03, 1.09] times ours, and the rule does not decide. Our prediction of the radius was
wrong, since the fit takes a smaller disc and a gain above 1. Within 0.1\,D the fitted geometric disc is 3\% more accurate
than ours, at 0.1--0.35\,D 14\% less. Away from focus it fails. Over the 9.5\,px search its relative defocus error is
19\% and 28\% at 1.5--3.5 and 3.5--9.5\,px against 11\% and 9\% for ours ($+7.9$ [$+6.9$, $+9.5$] and $+19.2$
[$+15.5$, $+30.6$] points), 36\% of its estimates beyond 3.5\,px are off by more than half, and its 20\% limit lies at
2.8\,px [2.5, 3.0]; the nominal geometric disc with one gain, which errs more near focus, keeps the range (16\% and
9\%). One radius moves the turnovers of all bands by the same factor, and the factor that suits the finest band near
focus misplaces the coarse bands that carry the range.

The geometric disc whose radius shrinks with frequency as $1-f_b/f_c$ errs 1.02 [1.01, 1.04] times ours near focus
with one gain of 0.79, and its range equals ours (11\% and 9\%). This is Stokseth's approximation of wave optics, and
its radii, 0.95 to 0.77 for our five bands, are those that the overlap of two displaced pupils prescribes. Without a
gain it errs 1.71 times ours, as our own curves do without theirs (1.52). A Gaussian pupil with a fitted radius and
gain, a constant coefficient, errs 1.20 [1.14, 1.27] times ours.

A clear $0.92R$ disc with no gain matches ours near focus (1.00 [0.98, 1.02]); a radius of $0.96R$ with a
gain of 0.89 is 1\% more accurate (0.99 [0.97, 1.00]). A disc 8\% smaller than the nominal aperture thus accounts for
the gain of 0.82 near focus, and is our only account of it; beyond 3.5\,px that disc errs by 15\% against 9\%
($+6.4$ [$+4.4$, $+9.5$] points), so it is not the pupil either (\hyperref[S-par:onepupil]{One effective pupil}).

With one focus offset per image pair, ours errs by 17.9\,mD, the fitted geometric disc by 18.6\,mD (1.04 [0.99, 1.09]
times) and the nominal geometric curves by 23.9\,mD (1.33 [1.24, 1.46] times); beyond 3.5\,px the offsets change
nothing (31\% for the fitted geometric disc).

In sum, near focus most of the advantage of our curves over nominal geometric curves can be had from a geometric
disc with two fitted numbers, and by our rule the remaining 6\% is undecided. What the fitted disc cannot give is
curves that hold from focus to 9.5\,px; those need an aperture that shrinks with frequency, which is what wave optics
computes from the nominal radius.

\begin{table}[htbp]
\centering
\caption{\textbf{Attribution controls at f/16} (pre-specified; textured windows, held out, the protocol of main-paper
Table~\ref{tab:est}). Fitted radius in units of the nominal one and gain, means over the five folds; held-out
median $|\hat\psi-\psi|$ within 0.35\,D [mD]; its paired ratio to ours within 0.35\,D, within 0.1\,D and at
0.1--0.35\,D [95\% interval over 2\,000 scene resamples]; from the 9.5\,px search, the mean binned median relative
defocus error over 1.5--3.5 / 3.5--9.5\,px [\%] and the share of estimates beyond 3.5\,px in error by more than half
[\%]. The rule applies to the ratio within 0.35\,D of the two rows with a fitted radius of a geometric disc. Over
the 9.5\,px search with a confidence filter, each curve filtered by the confidence of its own fit, the MRE of the
geometric disc with a fitted radius and gain is 1.87 [1.57, 2.23] times ours and that of the disc whose radius
shrinks as $1-f_b/f_c$ 1.08 [1.04, 1.12] times (Table~\ref{S-tab:conft1b}).}
\label{S-tab:attrctl}
\footnotesize
\setlength{\tabcolsep}{3pt}
\resizebox{\linewidth}{!}{%
\begin{tabular}{@{}lcccccccc@{}}
\toprule
& & & & \multicolumn{3}{c}{ratio to ours} & & \\
\cmidrule(lr){5-7}
curves of our joint fit & radius & gain & error & within 0.35\,D & within 0.1\,D & 0.1--0.35\,D & range & $>$ half \\
\midrule
clear disc, nominal radius, one gain (ours) & 1 & 0.82 & 21.4 & -- & -- & -- & 11.3 / 9.0 & 19.7 \\
geometric disc, nominal radius, one gain & 1 & 0.52 & 27.2 & 1.27 [1.18, 1.38] & 0.98 [0.96, 1.01] & 1.51 [1.37, 1.68] & 16.4 / 8.6 & 13.8 \\
clear disc, nominal radius, gain 1 & 1 & 1 & 32.5 & 1.52 [1.43, 1.60] & 0.99 [0.94, 1.05] & 1.86 [1.73, 1.99] & 23.4 / 25.0 & 26.9 \\
geometric disc, nominal radius, gain 1 & 1 & 1 & 72.3 & 3.38 [3.06, 3.71] & 1.55 [1.36, 1.77] & 4.15 [3.70, 4.59] & 46.0 / 34.3 & 31.8 \\
\midrule
geometric disc, fitted radius, gain 1 & 0.71 & 1 & 24.3 & 1.14 [1.09, 1.18] & 0.96 [0.95, 0.98] & 1.29 [1.23, 1.36] & 18.3 / 22.6 & 27.4 \\
geometric disc, fitted radius and gain & 0.63 & 1.25 & 22.7 & 1.06 [1.03, 1.09] & 0.97 [0.96, 0.98] & 1.14 [1.09, 1.19] & 19.3 / 28.2 & 36.5 \\
clear disc, fitted radius, gain 1 & 0.92 & 1 & 21.4 & 1.00 [0.98, 1.02] & 1.01 [1.01, 1.02] & 0.99 [0.95, 1.02] & 12.3 / 15.5 & 24.3 \\
clear disc, fitted radius and gain & 0.96 & 0.89 & 21.1 & 0.99 [0.97, 1.00] & 1.02 [1.01, 1.02] & 0.97 [0.95, 0.99] & 11.5 / 10.2 & 21.5 \\
geometric disc, radius $\times(1-f_b/f_c)$ per band, one gain & -- & 0.79 & 21.9 & 1.02 [1.01, 1.04] & 0.99 [0.98, 1.00] & 1.05 [1.03, 1.07] & 10.9 / 8.7 & 18.6 \\
the same, gain 1 & -- & 1 & 36.6 & 1.71 [1.60, 1.83] & 1.03 [0.96, 1.11] & 2.11 [1.95, 2.27] & 25.4 / 25.3 & 27.1 \\
Gaussian pupil, fitted radius and gain & 0.68 & 1.21 & 25.7 & 1.20 [1.14, 1.27] & 0.97 [0.94, 1.00] & 1.41 [1.31, 1.53] & 20.6 / 46.5 & 37.8 \\
\midrule
clear disc, nominal radius, one gain, offset per pair & 1 & 0.82 & 17.9 & 0.84 [0.80, 0.88] & 0.80 [0.73, 0.87] & 0.86 [0.81, 0.90] & 10.3 / 8.0 & 19.6 \\
geometric, nominal radius, one gain, offset per pair & 1 & 0.52 & 23.9 & 1.12 [1.05, 1.21] & 0.90 [0.83, 0.99] & 1.27 [1.16, 1.37] & 14.6 / 15.8 & 13.7 \\
geometric, fitted radius and gain, offset per pair & 0.63 & 1.25 & 18.6 & 0.87 [0.83, 0.92] & 0.75 [0.68, 0.82] & 0.96 [0.90, 1.03] & 18.1 / 30.7 & 36.9 \\
\bottomrule
\end{tabular}}
\end{table}

\Para{Gains fitted to geometric curves}
With one gain per band, the joint fit learns from the training scenes how much each band departs from geometric
optics. Relative to the lowest band (0.045\,cycles/px), the fitted gains of the four higher bands are 0.97, 0.89, 0.76
and 0.69 at f/16 (means over the five folds, each within 0.04 of the mean), against 0.94, 0.85, 0.78 and 0.76 for the
same statistic of the wave-optics curve on the same samples (\texttt{results/zedd\_t1\_r9\_f16.txt}; the run with the
focus calibrated on all 50 scenes gives the same values). At f/5.6, f/4 and f/2.8 they stay between 0.95 and 1.02,
where the model statistic lies between 0.95 and 1.02 as well. Wave-optics curves need no such correction: their fitted gains at f/16 are 1.04, 1.02, 0.96 and 0.89, and a gain
per band does not lower their error significantly (paired ratio 0.987 [0.967, 1.004], in the run with the focus
calibrated on all scenes, \texttt{results/zedd\_baselines.txt}); the lower gain of the
highest band matches the effect of noise in weak bands (Table~\ref{S-tab:pipesim}).

\Para[S-par:perpair]{Errors per pair and the focus of each pair}
Table~\ref{S-tab:perpair} splits the f/16 errors of Table~\ref{S-tab:estfull}, on the samples of main-paper
Table~\ref{tab:est}, by image pair, named by the distance of its calibrated midpoint. Ours is more accurate than curves calibrated for every pair and band on five of
the eight pairs and less accurate on the pairs at 0.92, 1.37 and 2.62\,m (0.060 against 0.031\,D, 0.020 against
0.018\,D and 0.029 against 0.020\,D). Fitting on the training scenes only an offset of the focus of each pair, the
median residual of our training estimates, brings all pairs to 0.0179\,D, a paired ratio to ours of 0.84 [0.80, 0.88],
as much as an offset and a scale (0.0182\,D, 0.85 [0.81, 0.90]); the offset of the 2.62\,m pair alone gives 0.87, most
of the improvement. That
offset, $-0.026$\,D ($-0.019$ to $-0.030$\,D over the folds), matches the zero crossings of the f/16 band ratios of the
pair, 0.018--0.038\,D below its midpoint from the wider apertures (Focus calibration above), and that of our own
estimates of the pair ($-0.026$\,D, from the median estimates within $\pm0.1$\,D), so this deficit lies in the f/16
focus of the pair. On seven of the eight pairs the offset agrees with the zero crossing of our estimates within
0.014\,D (\texttt{realdata/zedd\_r11b\_supp.py}, \texttt{results/zedd\_r11b\_supp.txt}), though on the 7.76\,m pair the
intervals separate the two. There the offset, $+0.0075$\,D ($+0.004$ to $+0.012$\,D over the folds), and the crossing,
$-0.0064$ [$-0.0088$, $-0.0039$]\,D over 1\,000 resamples of the pair's scenes, differ in sign, and the offset raises
the error of the pair from 0.020 to 0.021\,D (Table~\ref{S-tab:perpair}; \texttt{results/\allowbreak zedd\_r12b\_supp.txt}). The
0.92\,m pair (1.3\% of the samples, a step of 0.48\,D, 86\% of its samples beyond
focus) is different: its offset changes sign with the aperture ($-0.054$, $+0.017$, $+0.012$ and $-0.006$\,D at f/16,
f/5.6, f/4 and f/2.8), ranges from $-0.013$ to $-0.078$\,D over the folds and in one fold raises the error from 0.033
to 0.057\,D; its signed errors per scene span $-0.03$ to $+0.20$\,D, and with its large step its 0.225\,cycles/px
ratios lie above wave optics (Table~\ref{S-tab:fig3b}). Its offset is mostly not a shift of its focus, though the
evidence nearer than focus is thin: 1\,814 of its samples lie there, from three scenes, 1\,612 from one. Its f/16 band ratios
cross zero 0.010--0.028\,D below its midpoint, but in line fits over $\pm0.2$\,D in which three quarters of the windows
lie beyond focus. Our own estimates of the pair cross zero at $+0.002$\,D, with an interval of [$-0.031$, $+0.012$]\,D
over 1\,000 resamples of its 18 scenes (a crossing in 86\% of them) that contains the band crossings, so a shift of
its focus by up to 0.03\,D is not excluded (\texttt{realdata/zedd\_r12b\_supp.py}, \texttt{results/zedd\_r12b\_supp.txt}).
The estimates are pulled toward focus on both sides, by about 0.025\,D within 0.1\,D and by 0.07--0.08\,D at
0.15--0.35\,D beyond focus, and by 0.07\,D at 7.5--9.5\,px; beyond focus this shortfall is neither constant nor
proportional to the defocus. The median over its samples, most of them beyond focus, gives the offset of $-0.054$\,D (the mean over the folds;
pooled over the held-out estimates $-0.057$ [$-0.083$, $-0.018$]\,D over the same 1\,000 resamples of the scenes),
which thus corrects the shortfall beyond focus more than a shift of the focus would; this holds for the point
estimates, since a shift at the lower end of the crossing interval would be 55\% of the offset. A gain on our estimates
corrects the pair as well as the offset. Fitted like the offset on the training estimates of each fold (the gain with
the smallest summed absolute error, the counterpart of the median), it ranges from 1.06 to 1.42 over the folds and errs
by 0.035\,D against 0.034\,D (ratio 1.03 [0.35, 1.80] over 2\,000 resamples of the pair's scenes); chosen instead on the
other folds' held-out estimates, the gain (1.19--1.49) errs by 0.053\,D (0.046\,D with the criterion above) and the
offset by 0.049\,D. Fitted within a fold, either correction also follows that fold's focus calibration of the pair,
whose midpoint moves by 0.041\,D between the folds. In the fold with the lowest midpoint (1.062\,D against
1.086--1.103\,D) the offset is $-0.013$\,D and the gain 1.06, in the others $-0.058$ to $-0.078$\,D and 1.31--1.42
(post hoc; \texttt{realdata/\allowbreak zedd\_supp\_pair092.py},
\texttt{results/\allowbreak zedd\_supp\_pair092.txt}). In the fold that holds 1\,670 of the pair's 1\,814 samples
nearer than focus, where the offset raises the error from 0.033 to 0.057\,D, the gain lowers it to 0.030\,D. These data
do not tell an offset of this pair from a gain. At f/5.6, f/4 and f/2.8 the
offset alone gives 0.0072, 0.0068 and 0.0058\,D (ratios 0.79, 0.77 and 0.71), within 0.0002\,D of the affine map and
below every calibrated method. With the 9.5\,px search it gives 0.86 [0.82, 0.89] at f/16, against 0.95 [0.90, 1.00]
for the affine map, whose scale is fitted within $\pm0.35$\,D and extrapolated; at f/2.8 it gives 0.0129\,D, still
16\% and 22\% above calibrated curves and spectral ratios (\texttt{realdata/zedd\_r9\_supp.py},
\texttt{results/zedd\_r9\_supp.txt}, \texttt{zedd\_r10\_supp.txt}).

\begin{table}[htbp]
\centering
\caption{f/16, one image pair, held-out scenes: median error [D] per pair (focus distance, share of the samples): ours,
ours with an offset or an offset and a scale of the focus of each pair fitted on the training scenes, curves calibrated
for every pair and band, and one ratio with a calibrated curve.}
\label{S-tab:perpair}
\small
\setlength{\tabcolsep}{3pt}
\begin{tabular}{lcccccc}
\toprule
Pair & share & ours & offset & offset, scale & calibrated & one ratio \\
\midrule
0.92\,m & 0.013 & 0.060 & 0.034 & 0.035 & 0.031 & 0.072 \\
1.37\,m & 0.097 & 0.020 & 0.018 & 0.016 & 0.018 & 0.027 \\
1.94\,m & 0.152 & 0.016 & 0.015 & 0.015 & 0.018 & 0.020 \\
2.62\,m & 0.173 & 0.029 & 0.014 & 0.014 & 0.020 & 0.021 \\
3.54\,m & 0.159 & 0.016 & 0.015 & 0.015 & 0.020 & 0.025 \\
4.85\,m & 0.143 & 0.018 & 0.017 & 0.017 & 0.020 & 0.024 \\
5.80\,m & 0.136 & 0.036 & 0.035 & 0.040 & 0.042 & 0.060 \\
7.76\,m & 0.126 & 0.020 & 0.021 & 0.022 & 0.022 & 0.025 \\
\midrule
all & 1 & 0.0214 & 0.0179 & 0.0182 & 0.0215 & 0.0257 \\
\bottomrule
\end{tabular}
\end{table}

\Para[S-par:budget]{What a calibration needs}
That two scenes suffice (Fig.~\ref{S-fig:depthold}c) is a statement about labeled focal stacks, and
Table~\ref{S-tab:budget} spells it out. Nothing about the pupil is fitted. The curves are those of a clear disc of the
nominal f-number, and the labeled scenes fix only the state of the focus and one gain per aperture. The focus is
calibrated against LiDAR depth (\hyperref[S-par:focuscal]{Focus calibration}), the steps between the nine settings from
the peaks of the energy ratios of f/2, f/2.8 and f/4 to f/16, and the midpoint of each pair, which sets the absolute
level, from the zero crossing of its band ratios at f/2.8, f/4 and f/5.6. A labeled scene therefore means the nine
focus settings at f/2, f/2.8, f/4, f/5.6 and f/16 with LiDAR depth, 45 images, and two scenes are 90 images. With two
such scenes the median draw determines four of the eight pair midpoints (the worst draw two), 0.011\,D from those of 40
scenes in the median over pairs (worst 0.09\,D); the others fall back on the lens readings, shifted to the measured
midpoints. The gain at f/16 is then 0.88 on average over the folds and draws (0.79--0.97), against 0.81--0.83 from 40
scenes (\texttt{results/\allowbreak zedd\_learning\_curve\_kscene\_f16.txt}). An estimate needs two images of one
focus pair, the calibrated focus of that pair and the gain of its aperture, and no depth. Focus and gain belong to the
lens setting and the aperture, not to the scene, but we have not tested them on another lens or another copy of this
one. The gain is a scale and does not identify the pupil. A disc of 0.92 times the nominal radius fits the gain and
the turnover, the in-focus loss one of 1.25 (\hyperref[S-par:onepupil]{One effective pupil}).

\begin{table}[htbp]
\centering
\caption{What the one-pair estimator of the main paper sets without labeled data, what it fits on labeled scenes,
and what an estimate needs. A labeled scene is a set of focal stacks with LiDAR depth, 45 images, not a pair of
photos.}
\label{S-tab:budget}
\small
\begin{tabular}{@{}>{\raggedright\arraybackslash}p{1.5cm}>{\raggedright\arraybackslash}p{4.6cm}%
>{\raggedright\arraybackslash}p{4.9cm}>{\raggedright\arraybackslash}p{3.9cm}@{}}
\toprule
& set without labeled data & fitted on labeled scenes & needed for an estimate \\
\midrule
pupil & clear disc of the nominal f-number, 50\,mm focal length, luma-weighted wavelengths of 0.45--0.63\,$\mu$m &
nothing & -- \\
sampling & $8.4\,\mu$m per pixel, pixel aperture, five log-Gabor bands, $64\times64$\,px windows, nominal spectrum
$|\vf|^{-2}$ & nothing & -- \\
focus & lens readings, only where the scenes leave a setting undetermined & vergences of the nine focus settings,
midpoints of the eight pairs & the step and the midpoint of the pair \\
gain & -- & one per aperture & the gain of the aperture \\
images & -- & per scene, with LiDAR depth, the nine focus settings at f/2, f/2.8, f/4, f/5.6 and f/16 & two images of
one focus pair, no depth \\
\bottomrule
\end{tabular}
\end{table}

\Para[S-par:training]{Training data}
Fig.~\ref{S-fig:depthold}c asks how much ground truth each one-pair estimator needs
(\texttt{realdata/\allowbreak zedd\_\allowbreak learning\_\allowbreak curve.py}). Each method is fitted on $K$ scenes drawn at random from the 40 training
scenes of a fold, $K=1$, 2, 5, 10, 20 and 40 with five, four, three, three, two and one draws, and scored on up to
30\,000 samples per test fold within the evaluation range. The errors are those of the vergence $1/Z$, so an error of
the assumed focus counts in full. The focus of the nine settings is calibrated from the same $K$ scenes
(\texttt{--focus kscene}): the zero crossing of a pair where these scenes populate both of its sides, per-run positions
from the aperture-ratio peaks where at least two runs of a set are determined, and elsewhere the lens readings, shifted
as a whole to the measured midpoints. With one scene the median draw measures two of the eight pair midpoints and misses
the 40-scene calibration by 0.014\,D (median over pairs; worst 0.11\,D); with five scenes, seven pairs and 0.008\,D;
with 40 it reproduces the calibration of the fold. Two further runs bracket this protocol: the focus calibrated once on
the 40 training scenes and shared by every method at every $K$ (\texttt{--focus fold}), and the lens readings without
any calibration (\texttt{--focus lens}, f/16). The inversions search every tenth grid point and then the 21 points
around the best, which gives the median errors of the exhaustive search to within 0.0002\,D. A test sample whose pair
has no calibration gets the pair midpoint: with one scene, 25\% of the samples for curves calibrated per pair at f/16 and
62--74\% at f/2.8--f/5.6, where the search range is narrow; with five scenes at most 12\%.

Besides the methods of Table~\ref{S-tab:baselines} we fit our estimate of each pair with an offset and a scale
(Huber, on the training samples of that pair: ``focus fitted per pair''), and train three perceptrons on the band
features: the network of \texttt{zedd\_mlp.py} for 12 epochs and for at least 300 steps, and a network whose width (32,
128 or 256) and stopping step are chosen on held-out training data (20\% of the $K$ scenes, or 20\% of the windows of a
single scene). For the perceptron we report, per draw, the best of the three on the held-out scenes, which favors it. A
fourth network of the same kind takes our estimate as an additional input and learns a correction to it,
with its last layer initialized to zero (``perceptron correcting ours'').

\begin{figure}[htbp]
\centering
\includegraphics[width=\linewidth]{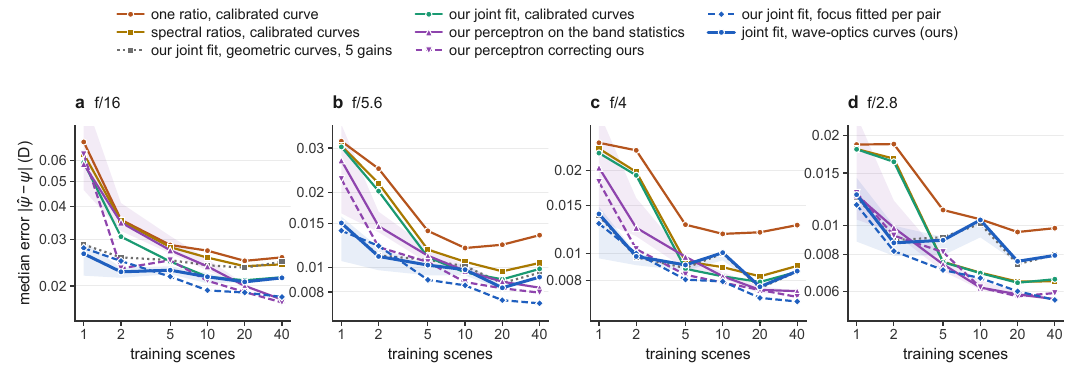}
\caption{Held-out median error of one-pair estimates against the number of training scenes, at the four apertures,
with the focus calibrated from the same scenes. Median over random draws of the training scenes; shaded: range over
draws for the wave-optics curves and the perceptron (per draw the best of three training variants). The spectral ratios are computed as
in~\cite{S-pentland1987,S-morganmar2014}.}
\label{S-fig:learning}
\end{figure}

At f/16 the wave-optics curves err by 0.0265\,D from one scene, 0.0226\,D from two and 0.0215\,D from 40. Over the 20
two-scene calibrations of the five folds the error lies 4\% above the fold's 40-scene error in the median, at most 17\%
above it in 90\% of them, and 33\% above it at worst; with one scene these figures are 14\%, 47\% and 229\%, because a
single scene fixes the focus of few pairs. With the focus shared from 40 scenes, one scene already gives 0.0233\,D; with the lens readings and no calibration the error stays at
0.024--0.027\,D, and fitting the focus of each pair brings it to 0.022\,D from five scenes and 0.018\,D from 40. Curves
calibrated per pair and band need 5--10 scenes to reach 0.023\,D at f/16, also with a gain chosen like ours (0.0239\,D
from five, 0.0214\,D from ten and 0.0199\,D from 40 scenes, 7\% below ours;
\texttt{results/zedd\_learning\_curve\_kscene\_calg\_f16.txt}), the perceptron 10--20; geometric curves stay
17--27\% worse than wave-optics curves at every $K$ with one gain and 8--15\% worse with one gain per band. With 40
scenes the best of the three perceptrons is 17\% more accurate than the wave-optics curves at f/16 (the perceptron of
main-paper Table~\ref{tab:est}, 16\% in this median error and 25\% in its mean relative depth error) and 10--28\%
at f/2.8--f/5.6, about as
accurate as the wave-optics curves with the focus fitted per pair (0.018\,D at f/16), and the correcting perceptron is
3\% better still (0.017\,D). From one scene the correcting perceptron overfits (0.063\,D at f/16), because its held-out
windows come from the same scene; from two scenes on it stays within 9\% of the wave-optics curves, and from ten scenes
on it improves on them at every aperture.
At f/2.8--f/5.6, where diffraction is small, the best data-driven method overtakes the model curves after 5--10 scenes
(Fig.~\ref{S-fig:learning}, Table~\ref{S-tab:learning}). With all 40 training scenes, this experiment with the focus
calibrated per fold on its full held-out folds gives the estimates of main-paper Table~\ref{tab:est} and
Table~\ref{S-tab:estfull} (\texttt{realdata/zedd\_table1.py}).

\begin{table}[htbp]
\centering
\caption{Held-out median error [D] against the number of training scenes (median over draws), f/16 and f/2.8, with the
focus calibrated from the same scenes. Perceptron: per draw the best of three training variants; held-out selection:
width and stopping step chosen on held-out training scenes. Calibrated curves with one gain: f/16 only
(\texttt{results/zedd\_learning\_curve\_kscene\_calg\_f16.txt}).}
\label{S-tab:learning}
\small
\setlength{\tabcolsep}{3.5pt}
\begin{tabular}{llcccccc}
\toprule
& & \multicolumn{6}{c}{training scenes} \\
\cmidrule(lr){3-8}
& Method & 1 & 2 & 5 & 10 & 20 & 40 \\
\midrule
f/16 & joint fit, wave-optics curves (ours) & 0.0265 & 0.0226 & 0.0230 & 0.0217 & 0.0207 & 0.0215 \\
& our joint fit, wave, focus fitted per pair & 0.0279 & 0.0249 & 0.0217 & 0.0192 & 0.0189 & 0.0182 \\
& our joint fit, geometric curves & 0.0311 & 0.0270 & 0.0277 & 0.0263 & 0.0261 & 0.0273 \\
& our joint fit, geometric curves, 5 gains & 0.0287 & 0.0257 & 0.0251 & 0.0240 & 0.0235 & 0.0247 \\
& our joint fit, calibrated curves & 0.0590 & 0.0308 & 0.0248 & 0.0217 & 0.0210 & 0.0216 \\
& our joint fit, calibrated curves, one gain & 0.0687 & 0.0308 & 0.0239 & 0.0214 & 0.0201 & 0.0199 \\
& one ratio, our curve & 0.0265 & 0.0228 & 0.0230 & 0.0214 & 0.0205 & 0.0213 \\
& one ratio, calibrated curve & 0.0706 & 0.0357 & 0.0286 & 0.0272 & 0.0249 & 0.0257 \\
& conventional DfDD: one ratio, $\alpha,\beta$ & 0.0532 & 0.0312 & 0.0280 & 0.0294 & 0.0313 & 0.0326 \\
& band fusion~\cite{S-guo2019} & 0.0536 & 0.0307 & 0.0277 & 0.0275 & 0.0283 & 0.0294 \\
& spectral ratios~\cite{S-pentland1987,S-morganmar2014}, calibrated curves & 0.0636 & 0.0354 & 0.0281 & 0.0256 & 0.0238 & 0.0242 \\
& our perceptron & 0.0582 & 0.0349 & 0.0273 & 0.0237 & 0.0200 & 0.0178 \\
& our perceptron, held-out selection & 0.0630 & 0.0560 & 0.0299 & 0.0239 & 0.0205 & 0.0178 \\
& our perceptron, correcting ours & 0.0633 & 0.0233 & 0.0250 & 0.0209 & 0.0190 & 0.0173 \\
\midrule
f/2.8 & joint fit, wave-optics curves (ours) & 0.0127 & 0.0087 & 0.0089 & 0.0104 & 0.0076 & 0.0079 \\
& our joint fit, wave, focus fitted per pair & 0.0117 & 0.0082 & 0.0071 & 0.0067 & 0.0060 & 0.0056 \\
& our joint fit, geometric curves & 0.0127 & 0.0088 & 0.0088 & 0.0103 & 0.0075 & 0.0080 \\
& our joint fit, geometric curves, 5 gains & 0.0125 & 0.0089 & 0.0091 & 0.0101 & 0.0074 & 0.0080 \\
& our joint fit, calibrated curves & 0.0180 & 0.0164 & 0.0075 & 0.0069 & 0.0064 & 0.0066 \\
& one ratio, our curve & 0.0153 & 0.0125 & 0.0125 & 0.0135 & 0.0112 & 0.0115 \\
& one ratio, calibrated curve & 0.0187 & 0.0188 & 0.0113 & 0.0105 & 0.0095 & 0.0098 \\
& conventional DfDD: one ratio, $\alpha,\beta$ & 0.0156 & 0.0131 & 0.0093 & 0.0092 & 0.0083 & 0.0079 \\
& band fusion~\cite{S-guo2019} & 0.0166 & 0.0126 & 0.0094 & 0.0086 & 0.0074 & 0.0070 \\
& spectral ratios~\cite{S-pentland1987,S-morganmar2014}, calibrated curves & 0.0181 & 0.0168 & 0.0076 & 0.0070 & 0.0065 & 0.0065 \\
& our perceptron & 0.0126 & 0.0098 & 0.0074 & 0.0062 & 0.0059 & 0.0057 \\
& our perceptron, held-out selection & 0.0138 & 0.0123 & 0.0081 & 0.0066 & 0.0063 & 0.0060 \\
& our perceptron, correcting ours & 0.0129 & 0.0093 & 0.0081 & 0.0062 & 0.0058 & 0.0059 \\
\bottomrule
\end{tabular}
\end{table}

\Para[S-par:weights]{Band weights}
The sub-band estimates in Table~\ref{S-tab:est_bins} and main-paper Table~\ref{tab:est} weight each band by its
Laplacian energy $S_{\times,b}$, the inverse variance of the band ratio for image noise of equal spectral density when
the signal dominates, which needs no ground truth. Table~\ref{S-tab:weights} compares other weights, each with its own gain chosen as before
(\texttt{realdata/zedd\_weights\_test.py}; \texttt{zedd\_estimate.py --weights train|selfcons}). The inverse squared
spread of the band ratios about the calibrated curves on the training scenes, which uses their ground truth, gives
larger errors at f/16 and f/5.6 (0.024 against 0.021\,D at f/16), slightly smaller ones at f/4 and f/2.8, and the same
working range in defocus. Equal weights, the band energy over the noise floor of each band, and weights from the residuals about
the model curves at a first estimate are worse at f/16, where the bands differ most in signal. These estimates search
all 600 candidates of the grid and read the range at bin edges; the runs of main-paper Table~\ref{tab:est} search 82
and give the Laplacian-energy weights 1.2--9.5\,px, with an upper end of 9.5 [7.3, 9.5] at f/16 and 9.5 [2.6, 9.5] at
f/5.6 interpolated between bin centers (Tables~\ref{S-tab:range} and~\ref{S-tab:lowband}).

\begin{table}[htbp]
\centering
\caption{Band weights of the sub-band estimate with model curves: held-out median $|\hat\psi-\psi|$ [D] with
wave-optics and geometric curves, and for wave-optics curves the 20\% working range in defocus with a search to 9.5\,px
(95\% interval of its upper limit). Laplacian energy: the weights of the main paper. Training residuals: inverse
spread of the band ratios about the calibrated curves, the only weights that use the ground truth of the training
scenes. Focus calibrated on the training scenes of each fold (\texttt{zedd\_weights\_test.py --cals};
\texttt{zedd\_foldcal.py --weights selfcons}).}
\label{S-tab:weights}
\small
\begin{tabular}{llccc}
\toprule
Aperture & Weights & wave & geometric & range, wave [px] \\
\midrule
f/16  & training residuals & 0.024 & 0.029 & 1.2--9.5 [7.1, 9.5] \\
      & Laplacian energy & \textbf{0.021} & \textbf{0.027} & 1.2--9.5 [7.1, 9.5] \\
      & residuals at a first estimate & 0.032 & 0.033 & \\
      & equal & 0.040 & 0.041 & \\
      & energy over noise floor & 0.043 & 0.044 & \\
f/5.6 & training residuals & 0.0097 & 0.0098 & 1.8--9.5 [9.5, 9.5] \\
      & Laplacian energy & \textbf{0.0092} & \textbf{0.0096} & 1.8--9.5 [2.4, 9.5] \\
      & residuals at a first estimate & 0.0107 & 0.0107 & \\
f/4   & training residuals & \textbf{0.0086} & \textbf{0.0087} & \\
      & Laplacian energy & 0.0088 & 0.0089 & \\
      & residuals at a first estimate & 0.0088 & 0.0089 & \\
f/2.8 & training residuals & 0.0080 & 0.0079 & \\
      & Laplacian energy & 0.0082 & 0.0082 & \\
      & residuals at a first estimate & \textbf{0.0076} & \textbf{0.0076} & \\
\bottomrule
\end{tabular}
\end{table}

\Para[S-par:texpred]{Texture dependence predicted from the band weights}
The broadband apparent defocus of a window is exactly the mean of its band values weighted by $S_{\times,b}$, so
Proposition~\ref{prop:ls} predicts it from the weights of the window and the response of each band
(\texttt{realdata/zedd\_texture\_test.py}). As band responses we take the calibrated band curves of the training folds,
which isolates the weighting, or the wave-optics curves times one gain, which uses no calibration of the bands; the
prediction then goes through the same broadband methods as the measurement. Table~\ref{S-tab:texpred} compares the change
of the relative bias between the texture terciles, which split the windows by spectral slope (the energy at 0.225
against that at 0.0675 cycles/px, in focus at f/16). At f/16 the highest band carries 58\%, 66\% and 76\% of the weight in
the three terciles, and the prediction is 2--7 points. The measured change is smaller because the ratio of a weak band
reads low. At f/16 the 0.225 cycles/px band reads 41\% below its curve in the decile of windows with the lowest
signal-to-noise ratio (band energy over a noise floor, the first percentile of the band energy over the windows of the
scene) and within 5\% of it from the fifth decile on. In the tercile with the least high-frequency energy this band
reads 7.2\% low, against 0.4\% high in the tercile with the most, which offsets the weighting. Subtracting the noise
floor from $S_\times$ more than halves the deficit of the lowest decile but leaves the difference between the terciles
(6.7 instead of 7.6 points), and the measured changes stay at 0--2 points. For comparison, the sub-band estimates
change by $+2$, $-2$, $+3$ and $+17$ points with geometric curves, $-5$, $-2$, $+6$ and $+20$ with wave-optics curves
and $-5$, $-4$, $+4$ and $+21$ with calibrated curves at f/16, f/5.6, f/4 and f/2.8 (Table~\ref{S-tab:basetex}); at
f/2.8 the weights favor the high bands, which lie near the first OTF zero. A noise floor set by the flattest windows
underestimates the noise of textured JPEG windows, and in-camera noise reduction and sharpening may depend on the
contrast; these data cannot separate the two. The optical effect within a band has no fixed sign: near focus,
diffraction makes a band read high for a texture whose energy sits at its low end, whereas near an OTF zero the same
texture reads low.

\begin{table}[htbp]
\centering
\caption{Texture dependence of conventional DfDD on the DSLR: change of the relative bias $\hat\psi/\psi-1$ between
the texture terciles with the least and the most high-frequency energy [percentage points], measured and predicted from
the band weights of each window (Proposition~\ref{prop:ls}) with the calibrated band curves or with wave-optics
curves and one gain. Last column: measured after subtracting a noise floor from $S_\times$. Last row: our joint fit with
calibrated curves. Calibrated $\alpha,\beta$: an affine map per pair fitted on all its training
samples, not in the linear range as in main-paper Table~\ref{tab:est}. These estimates search all 600 candidates of
the grid; the runs of main-paper Table~\ref{tab:est} search 82, on the samples that every method estimates, and give
the calibrated curve $+39.8$ and $+61.5$ at f/4 and f/2.8 (Table~\ref{S-tab:basetex}). Focus calibrated per fold
(\texttt{results/zedd\_texture\_test\_fold.txt}).}
\label{S-tab:texpred}
\small
\begin{tabular}{llrrrr}
\toprule
& & & \multicolumn{2}{c}{predicted from the weights} & measured, noise \\
\cmidrule(lr){4-5}
Aperture & Method & measured & calibrated bands & wave curves & floor subtracted \\
\midrule
f/16  & nominal $\alpha,\beta$ & $+0.3$ & $+1.8$ & $+2.2$ & $+0.2$ \\
      & calibrated $\alpha,\beta$ & $-0.1$ & $+7.1$ & $+6.2$ & $-0.4$ \\
      & calibrated curve & $+1.9$ & $+5.4$ & $+4.3$ & $+2.1$ \\
f/5.6 & nominal $\alpha,\beta$ & $+7.5$ & $+5.7$ & $+6.1$ & $+7.5$ \\
      & calibrated $\alpha,\beta$ & $+12.5$ & $+5.6$ & $+6.9$ & $+12.8$ \\
      & calibrated curve & $+22.0$ & $+8.2$ & $+17.0$ & $+20.6$ \\
f/4   & nominal $\alpha,\beta$ & $+10.1$ & $+7.9$ & $+6.9$ & $+10.1$ \\
      & calibrated $\alpha,\beta$ & $+14.1$ & $+13.0$ & $+8.9$ & $+14.9$ \\
      & calibrated curve & $+42.2$ & $+16.7$ & $+31.8$ & $+40.0$ \\
f/2.8 & nominal $\alpha,\beta$ & $+15.6$ & $+9.0$ & $+10.6$ & $+15.3$ \\
      & calibrated $\alpha,\beta$ & $+48.5$ & $+28.1$ & $+34.8$ & $+49.8$ \\
      & calibrated curve & $+63.1$ & $+35.0$ & $+43.7$ & $+64.3$ \\
\midrule
f/16  & our joint fit, calibrated & $-4.6$ & & & $-4.4$ \\
\bottomrule
\end{tabular}
\end{table}

\begin{figure*}[tp]
\centering
\includegraphics[width=\textwidth]{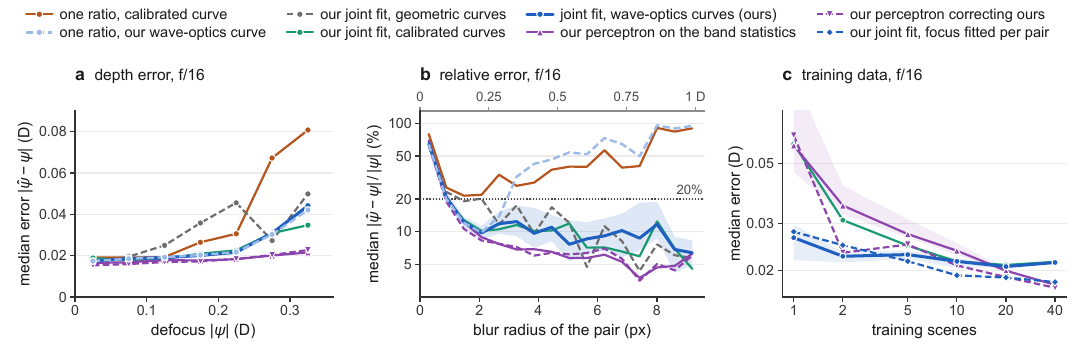}
\caption{\textbf{The curves behind Fig.~\ref{S-fig:depthold}.} DSLR at f/16, one image pair, held-out scenes,
textured windows. (a)~Median error against defocus; one ratio with our curve (light blue) is as accurate
as the joint fits with wave-optics or calibrated curves (blue, green), geometric curves (gray) are worse at
0.10--0.25\,D, and the perceptrons (purple) are best. (b)~Median relative error against the blur radius of the pair
(top: defocus), all pairs, 9.5\,px search, 16 bins (shaded: 95\% interval over scenes for ours); one ratio exceeds
20\% (dotted) once its high bands turn over. (c)~Median error against the number of training scenes, which also
calibrate the focus (median over draws, shaded: their range for ours and the perceptron; perceptron: best of three
variants, chosen on the held-out scenes). Fig.~\ref{S-fig:zedd_est} adds f/5.6 to (a), Fig.~\ref{S-fig:learning} the
other apertures to (c), and Tables~\ref{S-tab:range} and~\ref{S-tab:range_pairs} give (b) with intervals and by pair.
Fig.~\ref{S-fig:depthold}b keeps only the median errors within 0.35\,D, its panel a the 20\% limits as
distances (\hyperref[S-par:range]{working range in defocus}), and its panel c the number of scenes read off (c).}
\label{S-fig:depth_curves}
\end{figure*}

\Para[S-par:range]{The working range in defocus}
For the working range in defocus the single-pair estimates are recomputed with a search range of 9.5\,px geometric blur radius
at every aperture (1.02\,D at f/16, 0.36\,D at f/5.6), with the focus of each fold, and binned by the blur radius
$R|\psi|$ in 16 bins (\texttt{zedd\_table1.py}). Table~\ref{S-tab:range} gives the mean of the binned median relative
error $|\hat\psi-\psi|/|\psi|$ over intervals of the blur radius, with 95\% intervals over 2\,000 resamples of the
scenes, and the 20\% limit: the blur radius at which the binned error, after falling below 20\%, rises above it again,
interpolated between bin centers. Fig.~\ref{S-fig:depth_curves} draws these binned errors at f/16 over all pairs,
together with the error against defocus and against the number of training scenes, the curves behind
Fig.~\ref{S-fig:depthold}. Fig.~\ref{S-fig:depthold}a gives three of the limits as distances for a camera
focused at 1\,m. At f/16 the
geometric blur radius grows by 9.30\,px per diopter of defocus (\texttt{zedd\_range.blur\_px\_per\_D}), so the limits
of one ratio over all pairs (2.87\,px), of the joint fit on the small steps (6.59\,px, pairs 4--8) and of the joint fit
on the large steps (9.50\,px, the end of the search, pairs 1--3) are 0.309, 0.709 and 1.022\,D, or 0.76--1.45\,m,
0.59--3.44\,m and 0.49\,m to infinity. The statistic pools both sides of focus by $|\psi|$, and the lens readings of
the nine focus runs span 0.82--8.10\,m, so the focus at 1\,m is a way of drawing the limits, not one of the captures
(\texttt{realdata/\allowbreak make\_main\_figures.py}). The limits are those of the relative defocus error. To first
order a relative defocus error $\epsilon$ is a relative depth error $\epsilon|\psi|Z$, or $\epsilon|Z-Z_f|/Z_f$ for a
focus at $Z_f$, which is below $\epsilon$ nearer than $2Z_f$ and above it beyond. For $\epsilon=20\%$ it is 4.7\% and
8.9\% at the near and far limits of one ratio and 8.3\% and 48.8\% at those of the joint fit on the small steps
(exactly $-7.7\%$ to $+9.0\%$ and $-32.8\%$ to $+95.3\%$). On the large steps it is 10.1\% at 0.49\,m and has no
bound toward infinity, which a focus at 1\,m blurs by 9.30\,px, inside the search (\texttt{paper\_cvpr/\allowbreak
tools/\allowbreak defocus\_to\_depth.py}). The limit hinges on single bins near 20\%. The geometric curves reach 20.0\% in the
bin at 2.1\,px and then fall to 12\%, so their limit at f/16 is 2.0\,px although they stay below 20\% beyond (interval
[1.6, 9.5]\,px); the means over intervals have no such threshold. At f/4 and f/2.8 the same search shows where the pupil
model falls short of calibration. Over the 9.5\,px search the joint fit with wave-optics curves errs by 0.0147 and
0.0157\,D, jointly fitted spectral ratios with calibrated curves by 0.0135 and 0.0106\,D (at f/2.8 a paired ratio of
0.68 [0.61, 0.75]), the joint fit with calibrated curves by 0.0128 and 0.0111\,D (0.71 [0.64, 0.78]), and the perceptron
by 0.0107 and 0.0091\,D (\texttt{results/zedd\_table1.txt}, tag \texttt{fold\_wide}). With an offset of the focus of
each pair fitted on the training scenes (next paragraph) the joint fit improves to 0.0128 and 0.0129\,D, as good as
calibrated curves at f/4 but still 16\% and 22\% worse than calibrated curves and spectral ratios at f/2.8, so there
the clear-disc curves, not only the focus, limit the range over large blur. The main paper therefore states the
working range for f/16 only.

The range needs texture. Table~\ref{S-tab:range_energy} repeats the f/16 search on all windows with valid LiDAR depth,
2.4 times as many samples, and splits the windows by their in-focus band energy (the sum over the five bands of the
largest energy over the nine focus settings). On all windows the joint fit errs by 24\% and 23\% over 1.5--3.5 and
3.5--9.5\,px, with a 20\% limit of 2.4\,px; noise-aware weights lower this to 20\% and 17\%. Beyond 3.5\,px the other
joint fits err less there: geometric curves by 16\%, calibrated curves by 18\% and spectral ratios by 18\% (other minus
ours, paired over the resamples: $-6.9$ [$-21.6$, $-3.9$], $-4.5$ [$-18.6$, $-2.2$] and $-5.1$ [$-20.7$, $-2.5$]
percentage points). In the third of the textured windows with the least energy ours errs by 29\% and 72\%, more than
geometric curves (24\%, 41\%; $-32$ [$-38$, $-9$] points), calibrated curves (27\%, 50\%; $-22$ [$-30$, $-5$]) and
spectral ratios (27\%, 34\%; $-38$ [$-43$, $-13$]), and every method but geometric curves exceeds 20\% beyond
2.2--2.3\,px; in the least energetic third of all windows every method errs by 89--98\% beyond 3.5\,px. Over all pairs
much of the error of ours is a bias toward focus, $-5\%$ and $-8\%$ at 1.5--3.5 and 7.5--9.5\,px (calibrated curves
within 2\%), and at 7.5--9.5\,px 7--10\% of the estimates of every fitted curve fall on the wrong side of focus, against
0.1--0.2\% for the perceptrons (0.4--0.6\% without the pair index). At 7.5--9.5\,px
on the textured windows the bias of ours toward focus splits by in-focus energy into $-7\%$, $-5\%$ and $-99\%$ from
the most to the least energetic third, with 0.6\%, 6.2\% and 24\% of the estimates on the wrong side of focus;
calibrated curves give $-1\%$, $+1\%$ and $-94\%$ (0.6\%, 6.5\% and 16\%). The wrong-side estimates thus come from weak
bands. The $-7\%$ of the most energetic third comes from one pair: 91\% of the samples at 7.5--9.5\,px come from the
0.92\,m pair, whose estimates fall short beyond focus (Errors per pair above), and with an offset of each pair fitted
within $\pm0.35$\,D the bias falls to $-0.4\%$
(Table~\ref{S-tab:range_pairs}).

\Para[S-par:rangepair]{The working range in defocus by pair}
The pairs differ in their step and in their share of the far samples (Table~\ref{S-tab:range_pairs},
\texttt{realdata/zedd\_r11\_supp.py}, \texttt{results/zedd\_r11\_supp.txt}). The three pairs with large steps
(0.17--0.48\,D, 0.92--1.94\,m) supply 63\% of the samples beyond 3.5\,px and 83\% of those beyond 0.6\,D, and the
0.92\,m pair alone 91\% at 7.5--9.5\,px; on the small steps only 17 scenes have samples beyond 0.6\,D. On the large
steps every joint fit stays below 20\% to 9.5\,px. On the five pairs with small steps (0.02--0.11\,D) ours errs by
12.8\% and 14.3\% at 1.5--3.5 and 3.5--9.5\,px and reaches 20\% at 6.6\,px [2.9, 9.2], geometric curves keep the same
range, calibrated curves and spectral ratios err 7--8 points more beyond 3.5\,px, all of it at 6.5--9.5\,px and not
significantly ($+8.3$ [$-3.3$, $+10.0$] and $+7.1$ [$-4.8$, $+9.2$]; Table~\ref{S-tab:range_pairs} gives every difference
with its interval), and the perceptrons, the most accurate near focus (8.0--8.3\%),
err 13--14 points more beyond 3.5\,px than ours (perceptron with held-out selection $+13.6$ [$+2.4$, $+15.9$], perceptron correcting ours
$+12.8$ [$+1.4$, $+15.7$]); their range lead over all pairs (5.7\% against 9.0\%) comes from the large steps. That
deficit of the perceptrons lies entirely beyond 6.5\,px: at 3.5--6.5\,px they err by 13.0\% and 12.7\% against 12.2\%
($+0.8$ [$-15.2$, $+3.4$] and $+0.5$ [$-15.5$, $+4.2$] points), at 6.5--9.5\,px by 42.8\% and 41.5\% against 16.3\%
($+26.5$ [$+10.4$, $+31.2$] and $+25.2$ [$+10.4$, $+29.3$]), and beyond 6.5\,px lie 1.3\% of the small-step samples,
from 13 scenes, two of which hold 81\%. Ours there errs by 12.9, 24.5 and 9.7\% in the bins at 6.2, 6.8 and 7.4\,px, so
its 20\% limit rests on one bin; ours and the two perceptrons all reach 20\% at 6.2--6.6\,px, but beyond 3.5\,px 26.9\% [15.9, 39.3] of our small-step
estimates err by more than half, against 11.9\% and 12.5\% for the two perceptrons ($-15.0$ [$-23.4$, $-8.0$] and
$-14.4$ [$-22.3$, $-7.8$] points), 19.2\% for geometric curves ($-7.6$ [$-10.6$, $-4.7$]) and 22.5\% for calibrated
curves ($-4.4$ [$-6.3$, $-2.7$]). At 3.5--6.5\,px, which holds 92\% of these samples, the shares are 26.5\% against
9.8\% and 10.6\%; beyond 6.5\,px, where the median lead of ours lies, they are 31.0\%
against 36.7\% and 34.6\% ($+5.7$ [$-8.9$, $+15.1$] and $+3.6$ [$-8.2$, $+11.0$] points; second block of
Table~\ref{S-tab:range_pairs}). On the small steps the bias toward focus at 1.5--3.5 and 7.5--9.5\,px is $-6\%$ and
$-14\%$ (calibrated curves $+2\%$ and $-37\%$, perceptron $-2\%$ and $-49\%$). On the least energetic third of the
textured windows the perceptrons err by 12.0--12.7\% and 10.5--10.8\%, and a constant per pair (the median defocus of
its samples in the other folds) by 48\% and 34\%; beyond 3.5\,px 19.7\% of our estimates err by more than half, against
5.2--5.7\% for the perceptrons and 17.4\% for calibrated curves (\texttt{realdata/zedd\_r12\_supp.py},
\texttt{results/zedd\_r12\_supp.txt}). The perceptrons also see the pair index and the band energies. Without the pair
index (\texttt{zedd\_learning\_curve.py --mlp-no-pair}; \texttt{results/zedd\_t1\_r12nopair\_f16.txt},
\texttt{\_r12nopair\_wide\_f16.txt}) they lose between half and three quarters of their lead near focus (0.54 [0.28, 1.08] and 0.63 [0.46, 0.95] of it,
0.52--0.74 over the three training seeds below), 0.0199\,D within $\pm0.35$\,D against
0.0181 and 0.0175\,D with it (0.93 [0.86, 1.01] and 0.93 [0.88, 0.99] of ours), but keep their range over all pairs
as point estimates (6.3\% and 5.8\% at 3.5--9.5\,px, $-2.7$ [$-5.0$, $+0.9$] and $-3.2$ [$-5.4$, $+0.2$] points against
ours) and on the least energetic third (14\% and 15\%, against 72\% for ours), and on the
small steps they err by 24\% and 23\% beyond 3.5\,px ($+10.1$ [$-1.3$, $+14.4$] and $+8.7$ [$+0.4$, $+12.6$] points
against ours). Within $\pm0.35$\,D their estimates on the wrong side of focus fall from 15.5\% to 8.7\% on the side of
focus where a pair has fewer samples (ours 6.4\%) and rise from 2.6\% to 4.9\% on the other (ours 5.5\%), so a preference for one side remains ($+3.8$ [$+1.2$, $+6.3$]
and $+3.4$ [$+0.8$, $+5.8$] points between the two sides, ours $+0.9$ [$-1.3$, $+3.0$];
\texttt{results/\allowbreak zedd\_supp\_nopair.txt}). On the 2\,911 samples of the 5.80\,m pair farther than focus
beyond 0.1\,D, from four scenes, 30\% and 29\% of their estimates fall on the wrong side (35\% with the index), against
16\% of ours, and more than ours in each of the four scenes; on the 448
samples of the 0.92\,m pair nearer than focus beyond 0.1\,D, from three scenes, they fall from 92\% to 14\% (ours
0\%), and at 7.5--9.5\,px they stay at 0.4--0.6\% (ours 10\%; \texttt{realdata/zedd\_r12b\_supp.py},
\texttt{results/zedd\_r12b\_supp.txt}). With the index their lead near focus lies mostly on the 2.62\,m pair, whose focus error the index absorbs as our
offset per pair does. On that pair they err by 0.0148 and 0.0146\,D with the index and by 0.0194 and 0.0228\,D without
it (1.31 [1.14, 1.47] and 1.56 [1.38, 1.74] times), ours by 0.0288\,D and with the offset by 0.0143\,D; on the other
seven pairs they err 0.96 [0.89, 1.05] and 0.93 [0.85, 1.01] times ours with the index and 1.02 [0.94, 1.10] and 0.98
[0.92, 1.05] times without it, and over all pairs ours with the offset errs by 0.0179\,D, as they do with the index
(post hoc; \texttt{realdata/\allowbreak zedd\_supp\_nopair.py}, \texttt{results/\allowbreak zedd\_supp\_nopair.txt}). The pair index thus gave
the perceptrons a prior on the side of focus and part of their accuracy near focus. That their range and their
tolerance of weak texture come from the band energies is a hypothesis; the retraining removes the index, not the
energies, and a network on the band ratios alone would test it. The energies may also carry the side prior that
remains, because they tell a large step from a small one; the two images of a large-step pair are never both sharp.
Each network was trained once per fold, with one seed. Over that seed and two more per fold, the networks err within
$\pm0.35$\,D by 0.0174--0.0182\,D with the index and 0.0196--0.0205\,D without it, and at 3.5--9.5\,px by 5.4--6.7\%
over all pairs and 4.5--5.7\% on the large steps; their shares on the wrong side of focus within $\pm0.35$\,D move by
at most 0.8 points. On the small steps they err at 3.5--9.5\,px by 24--34\% with the index and 22--30\% without it,
against 14.3\% for ours, and 11--16\% of their estimates beyond 3.5\,px err by more than half, against 26.9\% of ours
(\texttt{realdata/\allowbreak zedd\_supp\_seed\_run.py}, \texttt{zedd\_supp\_seed.py};
\texttt{results/\allowbreak zedd\_supp\_seed.txt}). The intervals over scenes leave this variation out. On the small steps one ratio and the 9-tap ratio exceed 20\% beyond 2.8--4.6\,px, on the large steps beyond 4.5\,px
(one ratio, interval [2.6, 4.8]\,px) and 4.7--5.9\,px (the 9-tap ratio; Table~\ref{S-tab:range_pairs}, intervals in
\texttt{results/\allowbreak zedd\_supp\_tails.txt}). Per fold of ten scenes ours
errs at 3.5--9.5\,px by 6.7--16.8\% over all pairs, 6.2--13.4\% on the large steps and 12--19\% on the small steps,
except in the fold of val\_0006, whose small steps beyond 3.5\,px reach 54\%, and 5.0\% without that scene. The
small-step samples of val\_0006 beyond 3.5\,px, all nearer than focus, are a third of the fold's and 6\,946 of its
7\,527 at 4.7--6.5\,px. There our estimates collapse onto the pair midpoint (median $\hat\psi/\psi$ of $+0.08$, median
relative error 93\%, 43\% on the wrong side of focus), as do those of calibrated curves ($+0.15$, 86\%), while the
perceptron with held-out selection errs by 66\% ($+0.34$). The absolute criterion keeps 90\% of these samples, on which
ours errs by 97\%; this is the separate failure of val\_0006 in \hyperref[S-par:transfer]{Depth on the test scenes}
above. Without val\_0006, 23\% of our small-step estimates beyond 3.5\,px err by more than half, against 8\% for the
two perceptrons (post hoc; \texttt{realdata/\allowbreak zedd\_supp\_tails.py}, \texttt{results/\allowbreak zedd\_supp\_tails.txt}). The last block of Table~\ref{S-tab:range_pairs} gives
ours on the criterion windows of the next paragraph and at f/5.6 with its own texture mask, by step: on the criterion
windows of the small steps it errs by 12.5\% and 26.0\% at 1.5--3.5 and 3.5--9.5\,px, with its 20\% limit at 6.3\,px.

\begin{table}[htbp]
\centering
\caption{f/16, textured windows, 9.5\,px search, by pair group: median relative error [\%] over intervals of blur [px],
with the 95\% interval over 2\,000 resamples of the scenes at 3.5--9.5\,px, the 20\% limit [px], and the difference to
ours [percentage points, 95\% interval paired over the resamples] at 3.5--6.5, 6.5--9.5 and 3.5--9.5\,px. The
perceptrons are the two networks with held-out selection of Table~\ref{S-tab:baselines}, not the perceptron of main-paper
Table~\ref{tab:est}. Second block: on the samples of the first, the share of estimates in error by more than half
[\%], pooled over the samples whose blur lies in each interval (\texttt{realdata/\allowbreak zedd\_supp\_tails.py},
\texttt{results/\allowbreak zedd\_supp\_tails.txt}). Last block:
ours and the joint fit with calibrated curves on the criterion windows (all windows with valid LiDAR depth and
$e_{10}\ge1.5$) and at f/5.6 with its own texture mask (\texttt{realdata/zedd\_r12b\_supp.py}, \texttt{zedd\_r12c\_supp.py};
\texttt{results/zedd\_r12b\_supp.txt}, \texttt{zedd\_r12c\_supp.txt}; 9-tap rows with the window mean removed,
\texttt{zedd\_r13\_conv.py}, \texttt{results/zedd\_r13\_conv.txt}).}
\label{S-tab:range_pairs}
\resizebox{\linewidth}{!}{%
\setlength{\tabcolsep}{3pt}
\begin{tabular}{lcccccccc}
\toprule
& \multicolumn{5}{c}{relative error [\%]} & \multicolumn{3}{c}{minus ours [points]} \\
\cmidrule(lr){2-6}\cmidrule(lr){7-9}
Method & 1.5--3.5 & 3.5--6.5 & 6.5--9.5 & 3.5--9.5 & limit & 3.5--6.5 & 6.5--9.5 & 3.5--9.5 \\
\midrule
\multicolumn{9}{l}{\textit{small steps, pairs 4--8 (0.02--0.11\,D)}} \\
ours & 12.8 & 12.2 & 16.3 & 14.3 [10.9, 31.0] & 6.6 & & & \\
our joint fit, wave, focus offset per pair & 11.2 & 12.6 & 16.5 & 14.5 [11.2, 31.1] & 6.6 & $+0.3$ [$-0.2$, $+0.5$] & $+0.2$ [$-1.0$, $+1.0$] & $+0.3$ [$-0.4$, $+0.7$] \\
our joint fit, geometric curves & 15.8 & 13.4 & 14.0 & 13.7 [12.1, 26.4] & 6.7 & $+1.2$ [$-9.7$, $+4.1$] & $-2.3$ [$-16.4$, $+1.9$] & $-0.6$ [$-8.2$, $+2.5$] \\
our joint fit, calibrated curves & 11.8 & 14.5 & 30.7 & 22.6 [14.1, 34.7] & 7.0 & $+2.3$ [$-10.1$, $+3.1$] & $+14.3$ [$-6.9$, $+19.2$] & $+8.3$ [$-3.3$, $+10.0$] \\
spectral ratios~\cite{S-pentland1987,S-morganmar2014}, calibrated & 14.7 & 15.2 & 27.6 & 21.4 [15.7, 33.5] & 7.1 & $+3.0$ [$-12.1$, $+4.0$] & $+11.3$ [$-9.9$, $+16.3$] & $+7.1$ [$-4.8$, $+9.2$] \\
one ratio, our curve & 23.4 & 67.9 & 93.0 & 80.4 [68.1, 84.9] & 2.8 & $+55.7$ [$+37.7$, $+57.5$] & $+76.6$ [$+43.5$, $+79.8$] & $+66.2$ [$+46.3$, $+68.0$] \\
9-tap, tuned, our curve & 12.0 & 49.8 & 85.9 & 67.8 [49.1, 75.7] & 3.7 & $+37.6$ [$+24.5$, $+39.6$] & $+69.5$ [$+30.4$, $+72.2$] & $+53.6$ [$+31.4$, $+55.5$] \\
9-tap, 1.5\,px, our curve & 13.2 & 25.2 & 55.5 & 40.4 [27.4, 51.8] & 4.6 & $+13.0$ [$+5.5$, $+16.9$] & $+39.2$ [$+10.7$, $+41.9$] & $+26.1$ [$+10.6$, $+28.6$] \\
our perceptron, held-out selection & 8.0 & 13.0 & 42.8 & 27.9 [21.2, 38.3] & 6.3 & $+0.8$ [$-15.2$, $+3.4$] & $+26.5$ [$+10.4$, $+31.2$] & $+13.6$ [$+2.4$, $+15.9$] \\
our perceptron, correcting ours & 8.3 & 12.7 & 41.5 & 27.1 [20.6, 37.1] & 6.2 & $+0.5$ [$-15.5$, $+4.2$] & $+25.2$ [$+10.4$, $+29.3$] & $+12.8$ [$+1.4$, $+15.7$] \\
our joint fit, wave, far gain & 12.4 & 13.2 & 12.3 & 12.8 [9.7, 27.5] & 4.9 & $+1.0$ [$-6.5$, $+2.0$] & $-4.0$ [$-9.1$, $-0.6$] & $-1.5$ [$-5.2$, $-0.0$] \\
our joint fit, wave, curves at $0.9R$ & 12.8 & 20.3 & 9.7 & 15.0 [11.0, 39.4] & 3.4 & $+8.1$ [$+3.1$, $+15.0$] & $-6.6$ [$-6.9$, $+15.4$] & $+0.8$ [$-1.4$, $+12.2$] \\
\midrule
\multicolumn{9}{l}{\textit{small steps, share of estimates in error by more than half [\%]}} \\
ours & 10.3 & 26.5 & 31.0 & 26.9 [15.9, 39.3] & & & & \\
our joint fit, geometric curves & 11.8 & 18.6 & 27.3 & 19.2 [10.6, 29.4] & & $-8.0$ [$-11.1$, $-4.9$] & $-3.7$ [$-6.1$, $-3.2$] & $-7.6$ [$-10.6$, $-4.7$] \\
our joint fit, calibrated curves & 10.6 & 21.6 & 32.7 & 22.5 [12.5, 33.6] & & $-4.9$ [$-6.8$, $-3.1$] & $+1.7$ [$-7.6$, $+7.7$] & $-4.4$ [$-6.3$, $-2.7$] \\
spectral ratios~\cite{S-pentland1987,S-morganmar2014}, calibrated & 13.1 & 20.4 & 27.4 & 21.0 [12.1, 31.4] & & $-6.1$ [$-9.0$, $-3.5$] & $-3.6$ [$-11.9$, $+1.4$] & $-5.9$ [$-8.6$, $-3.5$] \\
our perceptron, held-out selection & 4.6 & 9.8 & 36.7 & 11.9 [5.2, 21.3] & & $-16.7$ [$-25.6$, $-9.0$] & $+5.7$ [$-8.9$, $+15.1$] & $-15.0$ [$-23.4$, $-8.0$] \\
our perceptron, correcting ours & 5.1 & 10.6 & 34.6 & 12.5 [5.8, 22.0] & & $-15.9$ [$-24.1$, $-8.5$] & $+3.6$ [$-8.2$, $+11.0$] & $-14.4$ [$-22.3$, $-7.8$] \\
\midrule
\multicolumn{9}{l}{\textit{large steps, pairs 1--3 (0.17--0.48\,D)}} \\
ours & 8.9 & 8.3 & 8.4 & 8.3 [7.0, 13.3] & 9.5 & & & \\
our joint fit, wave, focus offset per pair & 9.0 & 7.7 & 6.4 & 7.1 [5.4, 11.7] & 9.5 & $-0.6$ [$-1.7$, $+0.6$] & $-1.9$ [$-4.1$, $+0.6$] & $-1.3$ [$-2.5$, $+0.2$] \\
our joint fit, geometric curves & 19.0 & 9.1 & 5.5 & 7.3 [6.5, 9.7] & 9.5 & $+0.8$ [$-1.1$, $+2.8$] & $-2.9$ [$-8.2$, $-1.2$] & $-1.1$ [$-3.8$, $+0.3$] \\
our joint fit, calibrated curves & 9.2 & 7.6 & 6.2 & 6.9 [5.5, 9.7] & 9.5 & $-0.7$ [$-1.9$, $+0.4$] & $-2.2$ [$-7.2$, $-0.2$] & $-1.4$ [$-4.3$, $-0.2$] \\
spectral ratios~\cite{S-pentland1987,S-morganmar2014}, calibrated & 12.5 & 8.5 & 7.1 & 7.8 [6.3, 9.7] & 9.5 & $+0.2$ [$-1.0$, $+1.4$] & $-1.3$ [$-8.2$, $+0.4$] & $-0.5$ [$-4.3$, $+0.6$] \\
one ratio, our curve & 13.2 & 34.4 & 72.4 & 53.4 [40.2, 64.7] & 4.5 & $+26.1$ [$+17.6$, $+35.3$] & $+64.0$ [$+43.5$, $+78.6$] & $+45.1$ [$+32.5$, $+55.0$] \\
9-tap, tuned, our curve & 12.6 & 25.8 & 55.5 & 40.6 [29.4, 54.5] & 4.7 & $+17.4$ [$+12.0$, $+25.7$] & $+47.2$ [$+29.8$, $+66.3$] & $+32.3$ [$+21.8$, $+44.3$] \\
9-tap, 1.5\,px, our curve & 10.3 & 17.6 & 29.3 & 23.4 [17.4, 35.7] & 5.9 & $+9.3$ [$+6.5$, $+13.0$] & $+20.9$ [$+11.9$, $+37.9$] & $+15.1$ [$+9.7$, $+24.7$] \\
our perceptron, held-out selection & 7.4 & 4.9 & 4.2 & 4.6 [4.0, 5.2] & 9.5 & $-3.5$ [$-5.4$, $-2.0$] & $-4.1$ [$-11.8$, $-2.5$] & $-3.8$ [$-8.3$, $-2.6$] \\
our perceptron, correcting ours & 6.8 & 5.1 & 4.0 & 4.6 [4.0, 5.3] & 9.5 & $-3.2$ [$-5.2$, $-1.8$] & $-4.4$ [$-12.1$, $-2.8$] & $-3.8$ [$-8.2$, $-2.7$] \\
our joint fit, wave, far gain & 17.7 & 8.5 & 4.5 & 6.5 [5.3, 8.5] & 1.9 & $+0.1$ [$-1.4$, $+1.2$] & $-3.8$ [$-8.8$, $-2.3$] & $-1.8$ [$-4.8$, $-0.8$] \\
our joint fit, wave, curves at $0.9R$ & 12.6 & 19.7 & 13.7 & 16.7 [13.6, 26.8] & 3.8 & $+11.4$ [$+7.6$, $+14.1$] & $+5.3$ [$+3.6$, $+19.5$] & $+8.4$ [$+6.3$, $+15.7$] \\
\midrule
\multicolumn{9}{l}{\textit{criterion windows, f/16}} \\
ours, small steps & 12.5 & 10.9 & 41.0 & 26.0 [14.5, 38.5] & 6.3 & & & \\
our joint fit, calibrated, small steps & 11.6 & 9.6 & 40.8 & 25.2 [12.6, 36.7] & 6.5 & $-1.4$ [$-8.2$, $+0.3$] & $-0.2$ [$-14.6$, $+8.1$] & $-0.8$ [$-7.7$, $+3.1$] \\
ours, large steps & 10.6 & 8.1 & 9.2 & 8.6 [7.3, 17.4] & 9.5 & & & \\
our joint fit, calibrated, large steps & 10.6 & 6.4 & 4.8 & 5.6 [4.5, 10.5] & 9.5 & $-1.7$ [$-3.6$, $-0.8$] & $-4.4$ [$-13.4$, $+0.1$] & $-3.0$ [$-8.0$, $-0.9$] \\
\multicolumn{9}{l}{\textit{f/5.6, its own texture mask}} \\
ours, small steps & 16.7 & 10.8 & 11.5 & 11.1 [9.6, 15.3] & 9.5 & & & \\
our joint fit, calibrated, small steps & 13.5 & 10.2 & 9.7 & 9.9 [8.3, 12.8] & 9.5 & $-0.6$ [$-1.8$, $+0.5$] & $-1.8$ [$-4.3$, $-1.0$] & $-1.2$ [$-2.8$, $-0.5$] \\
ours, large steps & 14.9 & 15.5 & 13.8 & 14.6 [11.2, 22.3] & 9.5 & & & \\
our joint fit, calibrated, large steps & 15.9 & 13.7 & 9.3 & 11.5 [9.6, 15.1] & 9.5 & $-1.7$ [$-4.4$, $+1.1$] & $-4.4$ [$-14.3$, $-2.2$] & $-3.1$ [$-8.4$, $-1.0$] \\
\bottomrule
\end{tabular}}
\end{table}

\Para[S-par:gainfar]{A gain chosen far from focus}
Is the $-7\%$ a matter of the gain? We chose the gain of ours on the training samples at least 0.6\,D from the pair's
focus instead of within $\pm0.35$\,D (option \texttt{--far-gain 0.6} of \texttt{zedd\_learning\_curve.py}; summary
in \texttt{results/zedd\_r10b\_supp.txt}). The far gain is 0.68--0.74 over the folds against 0.82 near focus. On the textured windows at 7.5--9.5\,px it brings the bias of the most energetic third
from $-7\%$ to $-2.5\%$ and of the middle third from $-4.5\%$ to $-1.5\%$, leaves the least energetic third at $-99\%$
with 24\% on the wrong side, and lowers the error at 3.5--9.5\,px from 9.0\% to 7.1\%; within $\pm0.35$\,D, however,
it errs 30\% more (0.0279 against 0.0214\,D, 1.30 [1.22, 1.39]). On all windows the far gain (0.60--0.67) brings the
error at 3.5--9.5\,px from 23\% to 17\%, as low as the other joint fits, and the 20\% limit from 2.4 to 6.5\,px, at
60\% more error within $\pm0.35$\,D. One gain cannot serve both ranges, but the far gain mostly absorbs the shortfall of
the 0.92\,m pair beyond focus, which holds most far samples: on that pair it lowers the error at 3.5--9.5\,px from
10.1\% to 8.5\%, as its offset does (8.3\%), and on the small steps only from 14.3\% to 12.8\%
(Table~\ref{S-tab:range_pairs}). The gap between the far and the near gain is thus at most weak evidence that the data
grow more slowly than the curves on all pairs: on the small steps the most energetic third at 7.5--9.5\,px is biased by
$-12.5\%$, but from 831 samples in 7 scenes.
Curves computed for 0.9 times the nominal radius, the radius of the turnover (Table~\ref{S-tab:turnover}), take a gain
of 1.05 instead of 0.82, err 3\% more within $\pm0.35$\,D (1.03 [1.00, 1.05]) and 17\% instead of 9\% at 3.5--9.5\,px,
and their far gain is again lower (0.85--0.98). Most of that loss of range lies on the 0.92\,m pair (19.9\% against
10.1\% at 3.5--9.5\,px); on the other pairs they err by 12.1\% against 12.6\%, though on the small steps their 20\%
limit falls from 6.6 to 3.4\,px. An offset per pair removes most of the far bias; we keep one gain.

The texture of the main paper is relative: the more textured half of each scene's windows by the pair's
0.10\,cycles/px Laplacian energy at f/16. The same energy over its first percentile in the scene, the noise floor of
\hyperref[S-par:texpred]{Texture dependence} above, gives a criterion that a camera can apply to any window (Table~\ref{S-tab:texcrit}):
above 31.6 times the floor ($10^{1.5}$), which 48\% of all windows reach (53\% of the samples at 1.5--3.5\,px, 37\%, 33\%
and 24\% at 3.5--5.6, 5.6--7.5 and 7.5--9.5\,px), ours errs by 11.8\% [9.8, 14.7] and 9.4\% [7.9, 16.4] at 1.5--3.5 and
3.5--9.5\,px, calibrated curves by 11.1\% and 6.5\% and spectral ratios by 14.0\% and 7.0\% (paired, $-2.9$ [$-7.9$,
$-0.4$] and $-2.3$ [$-7.7$, $-0.6$] points beyond 3.5\,px), so on these windows ours is the least accurate of the joint
fits far from focus. That threshold is the smallest that gives 12\% on the held-out samples themselves. We
therefore report the threshold chosen on the other four folds by the same rule. It lies at 25--50 times the floor and
gives 11.7\% and 9.4\% [8.0, 18.5] on the held-out folds, but 8.0--28.4\% at 3.5--9.5\,px per fold, above 20\% in two of
five (\texttt{results/zedd\_r11\_supp.txt}; \hyperref[S-par:coverage]{Error against coverage} below adds the share of
gross errors, the noise-aware weights and the scenes). In the textured half of the main paper the energy
lies above 5 times the floor in 95\% of the samples (median 140 times) (\texttt{realdata/zedd\_r10\_supp.py},
\texttt{results/zedd\_r10\_supp.txt}).

\begin{figure}[htbp]
\centering
\includegraphics[width=\linewidth]{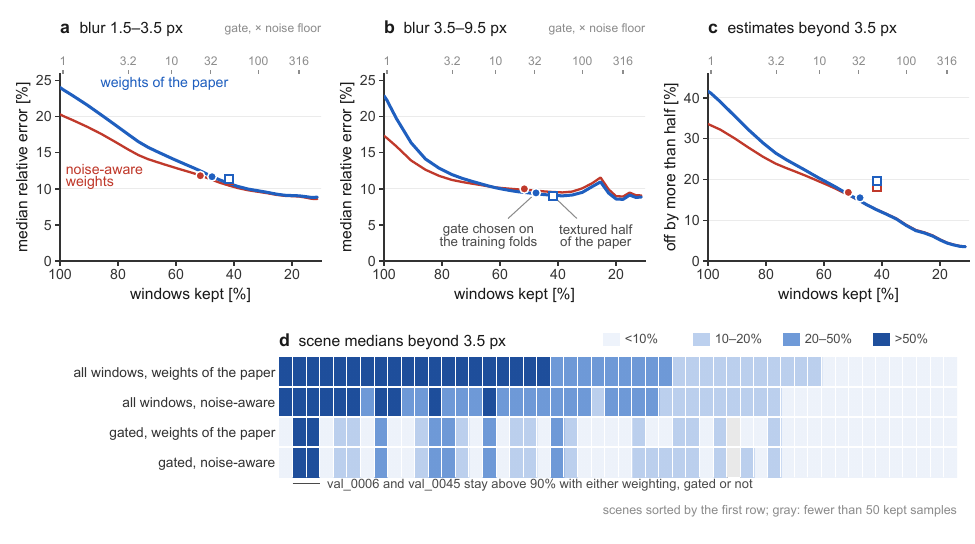}
\caption{\textbf{Error against coverage of one image pair at f/16.} All windows with valid LiDAR depth, held out,
9.5\,px search; a gate keeps the windows whose 0.10\,cycles/px energy exceeds a multiple of the noise floor (top
scale). (a,~b)~Median relative defocus error on the kept windows at 1.5--3.5 and 3.5--9.5\,px of blur, for the band
weights of the main paper (blue) and for noise-aware weights (red). (c)~Share of the kept estimates beyond 3.5\,px in
error by more than half. Dots: the gate chosen on the training folds; squares: the textured half of the main paper, a
relative rule. The weightings differ only where the gate keeps windows within about ten times the floor. (d)~One
column per scene: its median relative error beyond 3.5\,px without and with the gate of its fold; two scenes fail with
either weighting, gated or not (Table~\ref{S-tab:coverage}).}
\label{S-fig:coverage}
\end{figure}

\Para[S-par:coverage]{Error against coverage}
A gate trades windows for error, and the band weights decide how the estimate fails on the windows that a gate would
remove. Fig.~\ref{S-fig:coverage} and Table~\ref{S-tab:coverage} show the trade on all windows with valid LiDAR depth over
the 9.5\,px search, for the weights of main-paper Eq.~\eqref{eq:bw_ratio} and for the noise-aware weights of
Sec.~\ref{S-sec:bw}. They give the share of the windows a gate keeps, the error on them, the share of the kept estimates
in error by more than half, and the scenes whose median error beyond 3.5\,px exceeds 20\% or one half
(\texttt{realdata/zedd\_err\_coverage.py}, \texttt{results/zedd\_err\_coverage.txt}, from the saved held-out
estimates; nothing is estimated again). The gate is the absolute criterion above, a threshold on the pair's
0.10\,cycles/px energy over the noise floor; Table~\ref{S-tab:texcrit} gives the same sweep for calibrated curves and
spectral ratios.

Without a gate the noise-aware weights err less, by 20.3\% and 17.4\% against 24.1\% and 22.9\% at 1.5--3.5 and
3.5--9.5\,px ($-3.7$ [$-5.7$, $-1.4$] and $-5.6$ [$-18.8$, $-2.9$] points), with 33.6\% instead of 41.6\% of the
estimates beyond 3.5\,px in error by more than half ($-8.0$ [$-9.9$, $-6.2$]) and 10 instead of 20 scenes above one
half. Their lead shrinks as the gate rises. With the windows below 3.2 times the floor removed (77\% kept) it is
$-2.1$ [$-3.2$, $-1.0$] and $-0.7$ [$-1.4$, $-0.3$] points, at 10 times (62\% kept) $-1.2$ [$-1.7$, $-0.7$] and
$+0.0$ [$-0.1$, $+0.3$], and at 32 times (48\% kept) $-0.5$ [$-0.8$, $-0.1$] and $+0.4$ [$+0.2$, $+0.6$]. The two
weightings thus differ on the windows within about ten times the noise floor, which a gate removes; choosing between
them is choosing how to fail without a gate.

With the gate chosen on the other four folds (25--50 times the floor, 47.6\% of the windows kept) the weights of the
main paper err by 11.7\% [9.7, 14.5] and 9.4\% [8.0, 18.5], with 9.4\% [6.1, 13.4] and 15.5\% [8.8, 22.6] of the
estimates in error by more than half, and by 8.0--28.4\% at 3.5--9.5\,px per fold. With the same gate the
noise-aware weights err by 11.2\% [9.4, 14.0] and 9.8\% [8.3, 18.7] (9.4\% and 15.4\% above one half). With a gate
chosen for them by the same rule (16--40 times the floor, 51.6\% kept) they err by 11.8\% [9.9, 14.8] and 10.0\% [8.5,
15.8] (10.1\% and 16.8\%) and by 8.8--27.8\% per fold, with 20.0\% and 27.8\% in the two folds where the weights of
the main paper give 26.3\% and 28.4\%.

Scene by scene (Fig.~\ref{S-fig:coverage}d), without a gate the median error beyond 3.5\,px exceeds 20\% in 29 of the
50 scenes and one half in 20 (noise-aware weights: 27 and 10). With the gate, 7 of 49 scenes exceed 20\% with either
weighting (val\_0006, val\_0007, val\_0014, val\_0023, val\_0029, val\_0041 and val\_0045), and two exceed 90\%,
val\_0006 and val\_0045, the failures of \hyperref[S-par:transfer]{Depth on the test scenes}. The gate keeps 32\% and
100\% of their far samples, and neither weighting helps. The scene with the largest error without a gate, val\_0040
(99\%), keeps 2\% of its far samples and errs by 4\% on them.

The textured half of the main paper is a relative rule and holds 41.6\% of these samples. On it the two weightings
err alike (11.3\% and 9.0\%), but 19.7\% [14.0, 25.6] of the estimates beyond 3.5\,px err by more than half, against
12.5\% with the absolute gate at the same share of windows (50 times the floor), and 10 of 50 scenes exceed 20\%,
4 of them one half. A relative rule also keeps the better half of a weakly textured scene.

\begin{table}[htbp]
\centering
\caption{f/16, all windows with valid LiDAR depth, 9.5\,px search: error against coverage for the band weights of
main-paper Eq.~\eqref{eq:bw_ratio} and for noise-aware weights. Gate: the pair's 0.10\,cycles/px energy at least $k$
times its first percentile in the scene. Kept: share of the samples. Error: mean of the binned median relative defocus
error [\%] over 1.5--3.5 / 3.5--9.5\,px. Above half: share of the kept estimates beyond 3.5\,px in error by more than
half [\%]. Scenes: those whose median relative error beyond 3.5\,px exceeds 20\% / one half, among the scenes with at
least 50 kept samples there (50; 49 at $k=32$ and 100, 46 at $k=316$). Last rows: the gate chosen on the other four
folds for the weights of the main paper, the noise-aware weights with a gate chosen for them by the same rule, and
the textured half of the main paper (\texttt{realdata/zedd\_err\_coverage.py}).}
\label{S-tab:coverage}
\small
\setlength{\tabcolsep}{3.5pt}
\begin{tabular}{lrcccccc}
\toprule
& & \multicolumn{3}{c}{weights of the main paper} & \multicolumn{3}{c}{noise-aware weights} \\
\cmidrule(lr){3-5}\cmidrule(lr){6-8}
$k$ & kept & error & above half & scenes & error & above half & scenes \\
\midrule
no gate & 100\% & 24.1 / 22.9 & 41.6 & 29 / 20 & 20.3 / 17.4 & 33.6 & 27 / 10 \\
3.2 & 77\% & 17.6 / 12.0 & 26.5 & 14 / 6 & 15.5 / 11.3 & 23.9 & 11 / 5 \\
10 & 62\% & 14.2 / 10.2 & 20.3 & 9 / 4 & 13.1 / 10.2 & 19.5 & 8 / 4 \\
32 & 48\% & 11.8 / 9.4 & 15.0 & 7 / 2 & 11.3 / 9.8 & 14.9 & 7 / 2 \\
100 & 32\% & 9.7 / 9.5 & 8.7 & 3 / 2 & 9.6 / 10.1 & 8.9 & 3 / 2 \\
316 & 18\% & 9.0 / 8.5 & 4.4 & 1 / 0 & 8.9 / 8.9 & 4.5 & 1 / 0 \\
\midrule
from training folds & 48\% & 11.7 / 9.4 & 15.5 & 7 / 2 & 11.2 / 9.8 & 15.4 & 7 / 2 \\
\quad its own gate & 52\% & & & & 11.8 / 10.0 & 16.8 & 6 / 3 \\
textured half & 42\% & 11.3 / 9.0 & 19.7 & 10 / 4 & 11.3 / 9.0 & 18.2 & 9 / 4 \\
\bottomrule
\end{tabular}
\end{table}

\Para[S-sec:depthunits]{The working range in distance}
A defocus cue measures vergence, and the tables above give errors in diopters or relative to the defocus. A user
wants distance. The DfDD prototypes define their working range by it, as the distances at which the mean depth error
stays below 10\%~\cite{S-luo2025cod} or 5\%~\cite{S-luo2025fs} of the depth. We therefore turn every saved held-out
estimate into a distance, $\hat Z=1/(m+\hat\psi)$ with $m$ the calibrated focus of the image pair, and compare it with
the LiDAR distance (\texttt{realdata/\allowbreak zedd\_depth\_units.py}, \texttt{results/\allowbreak
zedd\_depth\_units.txt}; post hoc, nothing is estimated again). The estimated vergence is clipped to 0.04--2.0\,D, the
range of the LiDAR distances, as depth benchmarks clip predictions to the range of their data.

\emph{Criterion.} Without a confidence filter a fifth of our estimates beyond 3.5\,px are off by more than half, so
a mean over all windows is set by that tail; the prototypes discard their least confident pixels before they take
the mean. Here we use the median relative depth error; \hyperref[S-sec:confidence]{Depth with a confidence} gives the
fit a confidence and takes the mean. For each image pair it is taken in 31 bins of the true defocus,
0.066\,D wide, the middle one centered on the focus, and the interval of an estimator is the run of bins around the
focus in which it stays below 10\%, its two limits interpolated between bin centers. Bins with fewer than 200
samples or five scenes are left out; where the bins end before the error reaches 10\%, the limit is marked as the end
of the data. Intervals come from 2\,000 resamples of the scenes, and differences between two estimators are paired
over the same resamples. The extent of an interval is its width in diopters.

\begin{figure}[htbp]
\centering
\includegraphics[width=\textwidth]{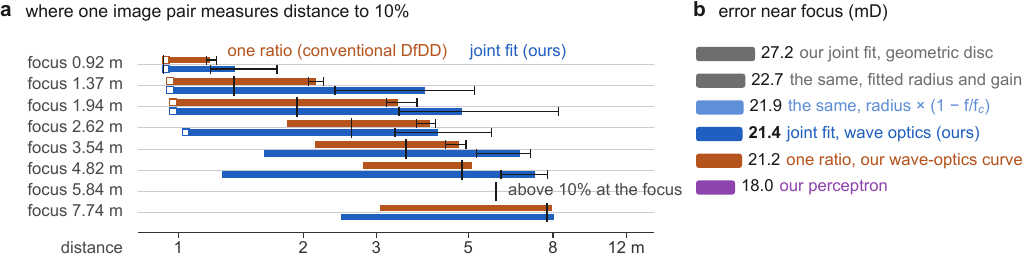}
\caption{\textbf{Where one f/16 image pair measures distance to 10\% by the median error} (post hoc; the criterion
of Table~\ref{S-tab:depthrange}, textured windows). (a)~For each image pair (tick: its focus), the distances over which
the median relative depth error stays below 10\%, for conventional DfDD and the joint fit, with the 95\% interval of
the far limit where every resample has one; open ends: the windows of the data end first. (b)~Median error within
0.35\,D of focus [mD] for the curves of Table~\ref{S-tab:attrctl}.}
\label{S-fig:rangemedian}
\end{figure}

\begin{figure}[htbp]
\centering
\includegraphics[width=\linewidth]{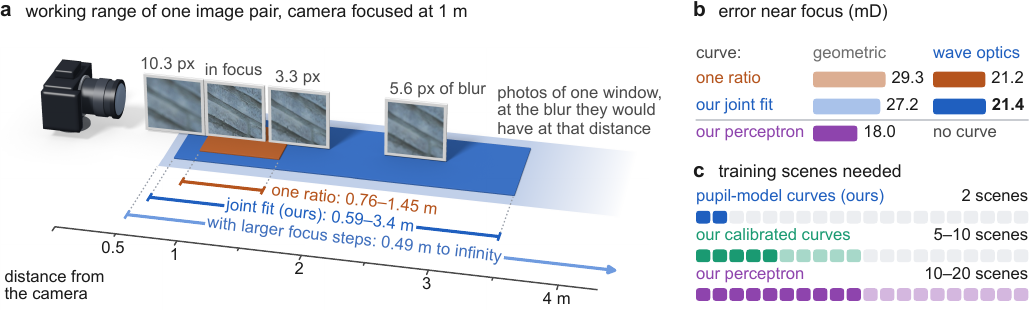}
\caption{\textbf{The working range drawn from limits in defocus, which the main paper does not use.} DSLR at
f/16, one image pair, held-out scenes, textured windows. (a)~Where the median relative defocus error stays below
20\%, for a focus at 1\,m: to 2.9\,px of blur for one ratio (all pairs) and to 6.6 and 9.5\,px for the joint fit with
focus steps of 0.02--0.11 and 0.17--0.48\,D (9.3\,px per diopter); 9.5\,px is the end of the search, not a measured
limit, and infinity lies at 9.3\,px; the 95\% interval of the 6.6\,px is [2.9, 9.2]\,px. Cards: the window of
main-paper Fig.~\ref{fig:method}. (b)~Median error within 0.35\,D of focus. (c)~The scenes that each calibration,
which also sets the focus, needs to match our curves fitted on two (pale: the range; perceptron: best of three
variants, chosen on held-out scenes).}
\label{S-fig:depthold}
\end{figure}

\emph{Outcome} (Table~\ref{S-tab:depthrange}, Fig.~\ref{S-fig:rangemedian}). On the textured windows of
main-paper Table~\ref{tab:est}, seven of the eight pairs have such an interval. In each of them it is wider for the
joint fit than for conventional DfDD, one ratio with calibrated $\alpha$ and $\beta$, by 0.08--0.49\,D, and every
paired interval lies above zero. Against one ratio with our curve it is wider in five pairs and undecided in two;
it does not differ from that of the perceptron where both exist, and the perceptron has none at 0.92 and 7.74\,m.
The far limits are uncertain, 3.9\,m [2.4, 5.2] for the pair focused at 1.37\,m, and the near limits of the four
nearest pairs are the nearest windows of the data. For the pair focused at 5.84\,m, whose focus step of 0.02\,D is
the smallest of the eight, no estimator reaches 10\% even at the focus, where the joint fit errs by 20\% and
conventional DfDD by 24\%. With the 3\,mm aperture of this lens at f/16, the farthest limit of the joint fit lies at
8.0\,m.

The selection matters. With the gate of \hyperref[S-par:coverage]{Error against coverage}, which uses the images
alone and keeps 47.6\% of all windows with valid LiDAR depth, the interval of the joint fit is wider than that of
conventional DfDD in six of the seven pairs. On all windows it is wider at 1.37\,m only ($+0.06$\,D [$+0.03$,
$+0.12$]), and the three farthest pairs have no interval for any estimator. With a mean in place of the median, on
the textured windows, the joint fit and conventional DfDD have intervals for the four nearest pairs only, and neither
leads (to 2.3\,m against 2.1\,m for the pair at 1.37\,m, to 2.8\,m against 3.2\,m at 1.94\,m).

\emph{Standard metrics} (Table~\ref{S-tab:depthunits}). Without any selection by texture, one image pair at f/16 is not
a depth map: over the 9.5\,px search on all windows, 62\% of the joint fit's distances lie within a factor 1.25 of the
truth and 22\% are off by more than half, and the mean relative error, 74\%, is set by that tail. With the
image-only gate these numbers are 80\%, 8\% and 27\%, with a median relative depth error of
7.3\%, against 14.1\% for conventional DfDD. The perceptron, where we ran it, is ahead on every metric.

\begin{table}[htbp]
\centering
\caption{\textbf{Where one f/16 image pair measures distance to 10\%} (post hoc). For each image pair, named by the
distance of its focus, the distances over which the median relative depth error stays below 10\%, for conventional
DfDD (one ratio with calibrated $\alpha$, $\beta$) and for the joint fit with wave-optics curves, with the 95\%
interval of the far limit in brackets ($<$: the windows of the data end before the error reaches 10\%; --: no
interval in some resamples); and the extent of the interval of ours minus that of three other estimators, in
diopters, paired over 2\,000 scene resamples. 9.5\,px search, held out.}
\label{S-tab:depthrange}
\footnotesize
\setlength{\tabcolsep}{4pt}
\resizebox{\linewidth}{!}{%
\begin{tabular}{@{}lccccc@{}}
\toprule
& \multicolumn{2}{c}{interval [m]} & \multicolumn{3}{c}{extent of ours minus [D]} \\
\cmidrule(lr){2-3}\cmidrule(lr){4-6}
focus [m] & conventional DfDD & joint fit (ours) & conventional DfDD & one ratio, our curve & our perceptron \\
\midrule
\multicolumn{6}{@{}l}{\emph{Textured windows with uniform LiDAR depth (the selection of main-paper Table~\ref{tab:est})}} \\
0.92 & $<$0.95--1.20 [1.2, 1.2] & $<$0.95--1.37 [1.2, 1.7] & $+0.11$ [$+0.01$, $+0.26$] & $+0.15$ [$+0.03$, $+0.29$] & -- \\
1.37 & $<$0.97--2.15 [2.1, 2.2] & $<$0.97--3.94 [2.4, 5.2] & $+0.21$ [$+0.05$, $+0.28$] & $+0.18$ [$+0.04$, $+0.27$] & $-0.02$ [$-0.19$, $+0.06$] \\
1.94 & $<$0.99--3.39 [3.2, 3.8] & $<$0.99--4.82 [3.4, 8.2] & $+0.09$ [$+0.01$, $+0.34$] & $+0.11$ [$+0.01$, $+0.29$] & $-0.02$ [$-0.30$, $+0.09$] \\
2.62 & 1.83--4.05 [3.8, 4.2] & $<$1.06--4.23 [3.3, 5.7] & $+0.41$ [$+0.10$, $+0.47$] & $+0.22$ [$-0.04$, $+0.26$] & $-0.08$ [$-0.39$, $+0.05$] \\
3.54 & 2.14--4.75 [4.4, 4.9] & 1.61--6.67 [5.2, 7.1] & $+0.21$ [$+0.15$, $+0.57$] & $+0.05$ [$+0.01$, $+0.40$] & $-0.33$ [$-0.38$, $+0.05$] \\
4.82 & 2.79--5.10 [--] & 1.27--7.22 [6.0, 7.8] & $+0.49$ [$+0.19$, $+0.67$] & $+0.29$ [$+0.01$, $+0.47$] & $+0.11$ [$-0.34$, $+0.30$] \\
5.84 & none & none & -- & -- & -- \\
7.74 & 3.06--7.93 [--] & 2.46--8.03 [--] & $+0.08$ [$+0.02$, $+0.41$] & $+0.00$ [$-0.02$, $+0.30$] & -- \\
\midrule
\multicolumn{6}{@{}l}{\emph{Image-only gate chosen on the training folds, 47.6\% of all windows with valid LiDAR depth}} \\
0.92 & $<$0.95--1.22 [1.2, 1.2] & $<$0.95--1.21 [1.1, 1.5] & $-0.00$ [$-0.05$, $+0.17$] & $+0.08$ [$+0.01$, $+0.21$] & -- \\
1.37 & $<$0.97--2.17 [2.1, 2.3] & $<$0.97--4.50 [2.2, 5.3] & $+0.24$ [$+0.02$, $+0.27$] & $+0.23$ [$+0.01$, $+0.26$] & -- \\
1.94 & $<$0.99--3.44 [3.1, 3.9] & $<$0.99--4.76 [3.8, 8.6] & $+0.08$ [$+0.04$, $+0.30$] & $+0.09$ [$-0.00$, $+0.23$] & -- \\
2.62 & 1.79--4.09 [3.8, 4.2] & $<$0.99--4.30 [3.3, 5.6] & $+0.46$ [$+0.10$, $+0.52$] & $+0.27$ [$-0.04$, $+0.30$] & -- \\
3.54 & 2.06--4.76 [4.4, 4.9] & 1.17--6.07 [4.8, 7.0] & $+0.41$ [$+0.13$, $+0.49$] & $+0.28$ [$+0.02$, $+0.33$] & -- \\
4.82 & 2.76--5.29 [--] & 1.26--7.09 [5.3, 8.0] & $+0.48$ [$+0.19$, $+0.59$] & $+0.29$ [$+0.01$, $+0.41$] & -- \\
5.84 & none & none & -- & -- & -- \\
7.74 & 3.24--7.81 [--] & 2.62--7.94 [--] & $+0.08$ [$+0.02$, $+0.47$] & $-0.01$ [$-0.02$, $+0.34$] & -- \\
\midrule
\multicolumn{6}{@{}l}{\emph{All windows with valid LiDAR depth}} \\
0.92 & $<$0.84--1.18 [--] & $<$0.84--1.15 [1.1, 1.3] & $-0.02$ [$-0.06$, $+0.09$] & $+0.06$ [$+0.02$, $+0.17$] & -- \\
1.37 & 0.99--2.04 [1.9, 2.1] & 0.98--2.27 [2.1, 2.5] & $+0.06$ [$+0.03$, $+0.12$] & $+0.05$ [$+0.02$, $+0.12$] & -- \\
1.94 & 1.32--3.12 [2.8, 3.4] & 1.29--3.08 [2.8, 3.7] & $+0.01$ [$-0.01$, $+0.17$] & $+0.03$ [$+0.00$, $+0.19$] & -- \\
2.62 & 1.79--3.62 [3.2, 4.1] & 1.53--3.06 [2.8, 4.0] & $+0.04$ [$-0.02$, $+0.12$] & $-0.00$ [$-0.02$, $+0.01$] & -- \\
3.54 & 2.23--4.16 [3.8, 4.6] & 1.81--4.29 [3.8, 6.1] & $+0.11$ [$-0.02$, $+0.17$] & $-0.01$ [$-0.07$, $+0.03$] & -- \\
4.82, 5.84, 7.74 & none & none & -- & -- & -- \\
\bottomrule
\end{tabular}}
\end{table}

\begin{table}[htbp]
\centering
\caption{\textbf{One f/16 image pair in the metrics of depth estimation} (post hoc; held out, pooled over the eight
pairs, estimates clipped to 0.5--25\,m). Median relative depth error $|\hat Z-Z|/Z$ [\%], share with
$\max(\hat Z/Z,Z/\hat Z)<1.25$ [\%] and share in error by more than half [\%], each with its 95\% interval over 2\,000
scene resamples; mean relative depth error (AbsRel) [\%]; median $|\hat\psi-\psi|$ [mD]. The perceptron was not run on
all windows over the 9.5\,px search.}
\label{S-tab:depthunits}
\footnotesize
\setlength{\tabcolsep}{4pt}
\begin{tabular}{@{}lccccc@{}}
\toprule
estimator & median rel. error & $\delta_{1.25}$ & $>$ half & AbsRel & $|\hat\psi-\psi|$ \\
\midrule
\multicolumn{6}{@{}l}{\emph{Standard search, within 0.35\,D of focus, textured windows (main-paper Table~\ref{tab:est})}} \\
conventional DfDD: one ratio, $\alpha,\beta$ & 10.2 [8.9, 11.5] & 75 [70, 79] & 8 [5, 12] & 22 & 32.7 \\
one ratio, our curve & 6.1 [5.2, 7.2] & 85 [80, 89] & 5 [3, 7] & 15 & 21.2 \\
spectral ratio~\cite{S-pentland1987,S-morganmar2014}, calibrated & 7.3 [6.2, 8.6] & 81 [76, 85] & 7 [5, 9] & 17 & 24.1 \\
our joint fit, geometric & 8.2 [7.4, 9.1] & 83 [79, 87] & 6 [4, 8] & 19 & 27.2 \\
joint fit, wave (ours) & 6.1 [5.3, 7.2] & 85 [80, 89] & 5 [3, 7] & 16 & 21.4 \\
our joint fit, calibrated & 6.3 [5.4, 7.6] & 83 [79, 88] & 6 [4, 8] & 16 & 21.5 \\
our perceptron & 5.2 [4.3, 6.2] & 87 [83, 92] & 3 [2, 5] & 12 & 18.0 \\
\midrule
\multicolumn{6}{@{}l}{\emph{Standard search, within 0.35\,D of focus, all windows with valid LiDAR depth}} \\
conventional DfDD: one ratio, $\alpha,\beta$ & 12.3 [10.9, 14.0] & 68 [63, 72] & 13 [10, 16] & 35 & 43.7 \\
one ratio, our curve & 9.0 [7.7, 10.8] & 73 [68, 78] & 11 [8, 14] & 29 & 31.9 \\
spectral ratio~\cite{S-pentland1987,S-morganmar2014}, calibrated & 10.3 [8.9, 12.2] & 71 [66, 75] & 13 [10, 16] & 30 & 35.3 \\
our joint fit, geometric & 10.7 [9.5, 12.3] & 72 [67, 77] & 12 [9, 15] & 36 & 37.6 \\
joint fit, wave (ours) & 9.0 [7.7, 10.7] & 73 [68, 78] & 11 [9, 14] & 31 & 32.1 \\
our joint fit, calibrated & 9.2 [8.0, 11.0] & 72 [68, 77] & 12 [9, 15] & 28 & 32.0 \\
our perceptron & 6.7 [5.6, 8.1] & 80 [75, 84] & 7 [5, 10] & 15 & 23.9 \\
\midrule
\multicolumn{6}{@{}l}{\emph{9.5\,px search, textured windows}} \\
conventional DfDD: one ratio, $\alpha,\beta$ & 16.4 [14.6, 19.0] & 57 [53, 61] & 20 [16, 25] & 45 & 58.2 \\
one ratio, our curve & 11.0 [9.2, 13.7] & 64 [59, 69] & 20 [17, 24] & 41 & 36.1 \\
spectral ratio~\cite{S-pentland1987,S-morganmar2014}, calibrated & 9.0 [7.6, 10.8] & 75 [70, 80] & 11 [8, 14] & 29 & 30.2 \\
our joint fit, geometric & 9.8 [8.8, 11.0] & 77 [72, 82] & 10 [7, 14] & 38 & 34.2 \\
joint fit, wave (ours) & 7.5 [6.3, 9.1] & 78 [72, 83] & 10 [7, 13] & 32 & 26.7 \\
our joint fit, calibrated & 7.7 [6.5, 9.5] & 77 [72, 82] & 11 [8, 14] & 28 & 26.4 \\
our perceptron & 6.4 [5.3, 7.8] & 83 [78, 87] & 6 [4, 8] & 15 & 22.7 \\
\midrule
\multicolumn{6}{@{}l}{\emph{9.5\,px search, image-only gate (47.6\% of all windows)}} \\
conventional DfDD: one ratio, $\alpha,\beta$ & 14.1 [12.5, 16.3] & 62 [58, 67] & 15 [10, 19] & 35 & 49.7 \\
one ratio, our curve & 9.6 [8.2, 11.7] & 69 [64, 74] & 14 [11, 19] & 32 & 33.0 \\
spectral ratio~\cite{S-pentland1987,S-morganmar2014}, calibrated & 8.2 [7.0, 10.0] & 78 [72, 83] & 9 [6, 13] & 24 & 28.7 \\
our joint fit, geometric & 9.4 [8.5, 10.8] & 80 [74, 85] & 8 [5, 13] & 32 & 33.2 \\
joint fit, wave (ours) & 7.3 [6.3, 8.7] & 80 [74, 85] & 8 [5, 12] & 27 & 26.5 \\
our joint fit, calibrated & 7.3 [6.2, 8.9] & 80 [74, 85] & 8 [5, 12] & 22 & 25.8 \\
\midrule
\multicolumn{6}{@{}l}{\emph{9.5\,px search, all windows with valid LiDAR depth}} \\
conventional DfDD: one ratio, $\alpha,\beta$ & 23.2 [20.2, 26.8] & 48 [44, 51] & 30 [26, 34] & 81 & 92.7 \\
one ratio, our curve & 20.8 [17.1, 25.4] & 51 [47, 55] & 31 [27, 34] & 76 & 76.2 \\
spectral ratio~\cite{S-pentland1987,S-morganmar2014}, calibrated & 14.0 [11.8, 16.8] & 62 [57, 66] & 21 [18, 25] & 53 & 51.2 \\
our joint fit, geometric & 14.0 [12.2, 16.3] & 63 [58, 68] & 22 [18, 26] & 86 & 53.0 \\
joint fit, wave (ours) & 13.0 [10.8, 15.9] & 62 [57, 67] & 22 [19, 25] & 74 & 48.7 \\
our joint fit, calibrated & 12.7 [10.6, 15.6] & 63 [58, 68] & 21 [18, 25] & 53 & 46.2 \\
\bottomrule
\end{tabular}
\end{table}

\Para[S-sec:confidence]{Depth with a confidence, in the convention of the DfDD prototypes}
The DfDD prototypes report the mean absolute depth error after their least confident pixels are discarded, and
they define the working range as the depths at which the mean error stays below 10\%~\cite{S-luo2025cod} or
5\%~\cite{S-luo2025fs} of the depth; the share discarded is the sparsity. The errors above are medians of the
vergence without a confidence filter. Here we give the fit a confidence and repeat the evaluation in that
convention (\texttt{realdata/\allowbreak zedd\_confidence.py}, \texttt{zedd\_mae\_mre.py},
\texttt{zedd\_mae\_mre\_table1.py}; post hoc, 14 November). No estimate changes. The scripts compute the estimates
again with the functions of the pipeline and stop unless they equal the saved ones sample by sample; they did, with
a largest difference of zero. Distances are $\hat Z=1/(m+\hat\psi)$ with the estimated vergence clipped to
0.04--2.0\,D, as in \hyperref[S-sec:depthunits]{the working range in distance}; MAE is the mean of $|\hat Z-Z|$ and
MRE the mean of $|\hat Z-Z|/Z$.

\emph{Confidence.} The joint fit minimizes $c(\psi)=\sum_b\omega_b\,(r_b-\gamma g_b(\psi))^2$ over a grid. Four
quantities of this cost are candidates for its confidence. The \emph{residual fraction} is
$c(\hat\psi)/\sum_b\omega_b r_b^2$, the cost left at the minimum over the cost of a model that is zero. The
\emph{second minimum} is the lowest other local minimum of the cost on the coarse grid of the search, over
$c(\hat\psi)$. The \emph{margin} is the difference of those two costs over $\sum_b\omega_b r_b^2$, and the
\emph{curvature} is the second difference of the cost at the minimum over $c(\hat\psi)$. Four more candidates use
the image energies alone and are open to every estimator. They are the 0.10\,cycles/px Laplacian energy of the pair
over its first percentile in the scene, which is the gate of \hyperref[S-par:coverage]{Error against coverage}; the
Laplacian energy summed over the five bands, the denominator of one ratio; the magnitude of the summed cross term of
image change and Laplacian, its numerator; and the square of the numerator over the denominator, the part of the
image change that the Laplacian explains. The prototype of~\cite{S-luo2025fs} thresholds the squared image change.
For a held-out fold, an estimator and a sparsity $X$, the confidence is the candidate whose $X$-quantile threshold
leaves the smallest MRE on the other four folds, and the threshold is that quantile; nothing of the held-out fold
enters the choice. A joint fit with model curves chooses among all eight candidates, every other estimator among
the four that use the images alone. On all windows the joint fit with wave-optics curves takes the second minimum
at $X=50\%$ and the residual fraction at $X=75\%$, in all five folds; on the textured windows it takes the residual
fraction at both. Conventional DfDD takes the summed Laplacian energy at 75\%.

\emph{All windows} (Table~\ref{S-tab:confpooled}, main-paper Fig.~\ref{fig:depth}). The samples are all windows
with valid LiDAR depth, without a selection by texture or by the uniformity of the LiDAR depth, with the 9.5\,px
search. With nothing discarded no estimator is usable. The MRE lies between 74\% and 86\% (52.5\% for the joint fit
with calibrated curves, which searches only within its calibrated bins), since a vergence error far from a distant focus is a large error in distance and 21--31\% of the distances are off by more than half.
With the least confident 75\% discarded the joint fit with wave-optics curves has an MRE of 8.3\% [7.3, 9.5] and an
MAE of 0.33\,m, conventional DfDD 20.1\% [16.0, 24.7] and 0.98\,m, and the joint fit with geometric curves, filtered by
the confidence of its own fit, 14.6\% [12.7, 17.0]. With an image-energy confidence the joint fit has 17.3\%
[11.7, 24.3], so three quarters of its gap to conventional DfDD come with the confidence of the fit. The
comparison is not one of easy
windows (Table~\ref{S-tab:confsame}). On the windows kept for ours conventional DfDD errs by 20.6\%, on those kept
for conventional DfDD ours errs by 13.7\% against 20.1\%, and on those kept for both by 6.7\% against 16.8\%. The
joint fit is ahead in each of the three ranges of blur and for each of the eight pairs
(Tables~\ref{S-tab:confblur} and~\ref{S-tab:confpair}); beyond 3.5\,px it keeps 17\% of the windows with an MRE of
13.5\%, conventional DfDD 6\% with 47\%.

\emph{The working range by the mean error} (Table~\ref{S-tab:confpair}). Per pair and bin of distance, with 75\%
discarded, the mean relative error of the joint fit stays below 10\% over a wider total span of bins in all eight
pairs, in seven with a paired interval above zero; summed over the pairs the span is
4.61\,D against 2.24\,D. Three results go the other way. With nothing discarded the joint fit is wider in no pair.
With 50\% discarded and a criterion of 5\% conventional DfDD keeps the larger span, 0.99\,D against 0.66\,D in
total ($-0.33$ [$-0.53$, $-0.07$]). And by the definition of the prototypes, which takes the region contiguous from
the focus, the joint fit has none for the pairs focused at 2.62 and 3.54\,m, because the bin at the focus has a
mean error of 11\% and 19\%, although bins beside it lie below 10\%.

\emph{Main-paper Fig.~\ref{fig:depth}} (\texttt{make\_fig\_range\_scene.py}; a drawing of 16 November, nothing is
estimated again). Panels a and b draw the bins of Table~\ref{S-tab:confpair} as distances. A bin, 0.066\,D wide and
with at least 200 kept windows from five or more scenes, is drawn where the mean relative error of the windows that the
estimator keeps stays below 10\%. In meters these bins add up to 0.37, 1.32, 2.50, 2.76, 1.77, 0.89, 0 and 0\,m for
conventional DfDD and to 0.86, 3.27, 5.48, 3.14, 1.62, 2.98, 1.31 and 4.37\,m for the joint fit, in the order of the
table, 9.6 and 23.0\,m in all; the medians are 1.11 and 3.06\,m. For the pair focused at 3.54\,m conventional DfDD
has the longer span in meters (1.77 against 1.62\,m) and the joint fit the wider one in diopters (0.33 against
0.20\,D), because its bins lie nearer the camera; panel b has a logarithmic axis of distance. Panel a draws one
pair, the one whose two spans lie closest to these medians by the sum of the two absolute differences. This is the
pair focused at 4.82\,m, with 0.89 and 2.98\,m and a sum of 0.30\,m, before the pair focused at 1.37\,m with
0.42\,m (0.97--2.29\,m against 0.86--0.91 and 0.97--4.19\,m); relative differences choose the same pair, and the
spans in diopters would choose the pair focused at 0.92\,m. For the pair drawn, conventional DfDD stays below 10\%
from 3.27 to 4.16\,m and the joint fit from 1.04 to 1.43\,m and from 1.57 to 4.16\,m. Neither does at the focus: in
the bin from 4.16 to 5.73\,m conventional DfDD has 13.8\% and the joint fit 25.2\%, and in the next one, to
9.20\,m, 17.2\% and 11.5\%. The cards show the window of main-paper Fig.~\ref{fig:method} in the photographs of
its scene with a blur of 5.6, 3.3, 1.3 and 0\,px; they stand nearer than the focus where a window has that blur for
this pair, at 1.23, 1.78 and 2.84\,m, and at the focus, by 9.3\,px per diopter. Panel c gives the pooled errors of
Table~\ref{S-tab:confpooled}.

\emph{Main-paper Table~\ref{tab:est} in full} (Tables~\ref{S-tab:conft1} and~\ref{S-tab:conft1b}). The table of the
main paper gives mean relative depth errors. Table~\ref{S-tab:estfull} gives the same estimates as median errors of
the vergence in diopters, the quantity in which the rule of the attribution controls was fixed, and
Tables~\ref{S-tab:conft1} and~\ref{S-tab:conft1b} give them in distance with every column. The order of the estimators near focus is the same in both. Two
numbers lie outside these tables (\texttt{realdata/\allowbreak zedd\_mae\_mre\_extra.py}). With
one gain for every band the geometric disc has an MRE of 16.8\% within 0.35\,D, 1.08 [1.04, 1.12] times ours, the
8\% of main-paper Sec.~\ref{sec:depth}, and
the two perceptrons trained without the pair index (\hyperref[S-par:rangepair]{The working range in defocus by pair}) have
16.6\% and 18.3\%, 1.06 [0.81, 1.53] and 1.17 [0.90, 1.61] times ours. With a
confidence filter over the 9.5\,px search, a geometric disc with a fitted radius and gain errs 1.87 [1.57, 2.23] times
ours and the nominal geometric disc 1.71 [1.50, 1.97] times, each filtered by the confidence of its own fit, while the
disc whose radius shrinks as $1-f_b/f_c$ stays within 8\% (1.08 [1.04, 1.12]). Curves calibrated per pair and band
have no fit confidence here, since they are undefined outside the bins that are kept; with an image-energy
confidence they err 2.05 times ours, and on the windows kept for ours 1.12 [1.04, 1.22] times, or 1.03
[0.99, 1.07] with a gain. The perceptron has no confidence of its own. With nothing discarded it errs half as much
as ours over the search (15.2\% against 32.1\%); with an image-energy confidence it errs by 11.6\%, and on the
windows kept for ours by 6.1\% against 6.0\% (1.02 [0.91, 1.16]). The confidence thus removes the estimates on
which the perceptron was ahead, and the perceptron still gives usable estimates on windows that the joint fit
discards.

\emph{Limits.} The confidence, its candidates and this evaluation were designed after all results above were
known. A \hyperref[S-sec:conftest]{confirmation on the test scenes}, fixed in advance, follows the tables. The thresholds are quantiles over many scenes, where a camera would fix one threshold in advance. The clip
of the vergence at 0.04\,D enters the means, since an estimate beyond infinity counts as 25\,m.

\begin{table}[htbp]
\centering
\caption{\textbf{Depth with a confidence at f/16, all windows with valid LiDAR depth} (post hoc; 9.5\,px search, held out by fold, 3\,329\,706 samples of 50 scenes). For each share of discarded windows: the share kept [\%], the mean absolute distance error (MAE) [m], the mean relative depth error (MRE) [\%] with its 95\% interval over 2\,000 scene resamples, the median relative depth error [\%], the share of distances within a factor 1.25 of the truth [\%], the share off by more than half [\%], and the MRE over that of conventional DfDD [95\% interval]. The confidence of each estimator and its threshold are chosen on the other four folds.}
\label{S-tab:confpooled}
\footnotesize
\setlength{\tabcolsep}{4pt}
\resizebox{\textwidth}{!}{%
\begin{tabular}{@{}lccccccc@{}}
\toprule
estimator and confidence & kept & MAE & MRE & median & $\delta_{1.25}$ & $>$ half & MRE over conventional \\
\midrule
\multicolumn{8}{@{}l}{\emph{nothing discarded}} \\
conventional DfDD & 100 & 2.07 & 81.4 [65.0, 100.4] & 23.2 & 48 & 30 & -- \\
one ratio, our curve & 100 & 1.88 & 75.8 [60.4, 93.6] & 20.8 & 51 & 31 & 0.93 [0.92, 0.94] \\
our joint fit, geometric & 100 & 2.24 & 85.7 [69.0, 104.4] & 14.0 & 63 & 22 & 1.05 [0.98, 1.13] \\
our joint fit, calibrated & 100 & 1.48 & 52.5 [43.2, 63.4] & 12.7 & 63 & 21 & 0.65 [0.61, 0.69] \\
joint fit, wave (ours) & 100 & 1.88 & 73.7 [58.6, 91.5] & 13.0 & 62 & 22 & 0.91 [0.86, 0.95] \\
\midrule
\multicolumn{8}{@{}l}{\emph{least confident 50\% discarded}} \\
conventional DfDD, image energy & 49 & 1.25 & 38.1 [28.6, 50.1] & 13.8 & 64 & 13 & -- \\
one ratio, our curve, image energy & 49 & 1.04 & 34.7 [25.5, 46.6] & 9.4 & 70 & 13 & 0.91 [0.86, 0.95] \\
our joint fit, geometric, image energy & 50 & 0.91 & 29.3 [21.5, 38.5] & 8.8 & 83 & 5 & 0.77 [0.68, 0.87] \\
our joint fit, geometric, from the fit & 50 & 0.83 & 23.8 [20.2, 28.0] & 8.8 & 79 & 9 & 0.62 [0.50, 0.80] \\
our joint fit, calibrated, image energy & 50 & 0.71 & 21.3 [16.7, 27.4] & 7.1 & 84 & 6 & 0.56 [0.46, 0.66] \\
our joint fit, wave, image energy & 48 & 0.82 & 26.7 [19.4, 36.0] & 6.9 & 83 & 6 & 0.70 [0.63, 0.77] \\
joint fit, wave (ours), from the fit & 50 & 0.69 & 16.2 [13.9, 18.9] & 6.7 & 82 & 8 & 0.43 [0.32, 0.56] \\
\midrule
\multicolumn{8}{@{}l}{\emph{least confident 75\% discarded}} \\
conventional DfDD, image energy & 25 & 0.98 & 20.1 [16.0, 24.7] & 9.4 & 76 & 7 & -- \\
one ratio, our curve, image energy & 25 & 0.64 & 14.7 [11.6, 18.3] & 5.9 & 85 & 5 & 0.73 [0.68, 0.79] \\
our joint fit, geometric, image energy & 27 & 0.99 & 22.0 [15.0, 30.6] & 8.7 & 83 & 6 & 1.10 [0.82, 1.47] \\
our joint fit, geometric, from the fit & 25 & 0.59 & 14.6 [12.7, 17.0] & 7.3 & 85 & 6 & 0.73 [0.61, 0.90] \\
our joint fit, calibrated, image energy & 27 & 0.79 & 15.8 [11.6, 21.0] & 6.4 & 84 & 6 & 0.79 [0.64, 1.01] \\
our joint fit, wave, image energy & 27 & 0.83 & 17.3 [11.7, 24.3] & 5.9 & 86 & 6 & 0.86 [0.64, 1.15] \\
joint fit, wave (ours), from the fit & 25 & 0.33 & 8.3 [7.3, 9.5] & 4.9 & 94 & 1 & 0.41 [0.36, 0.49] \\
\bottomrule
\end{tabular}}
\end{table}

\begin{table}[htbp]
\centering
\caption{\textbf{The same windows} (post hoc). Conventional DfDD, one ratio with our curve and the joint fit on the windows kept for ours (confidence from the fit), on those kept for conventional DfDD, and on those kept for both: share of all windows [\%], MRE [\%], MAE [m], and the MRE of ours over that of conventional DfDD [95\% interval].}
\label{S-tab:confsame}
\footnotesize
\setlength{\tabcolsep}{4pt}
\begin{tabular}{@{}llccccccc@{}}
\toprule
discarded & windows & share & \multicolumn{3}{c}{MRE: conv., one ratio, ours} & \multicolumn{2}{c}{MAE: conv., ours} & ours over conv. \\
\midrule
50\% & kept for ours & 50.0 & 24.9 & 21.8 & 16.2 & 1.13 & 0.69 & 0.65 [0.60, 0.71] \\
50\% & kept for conventional & 48.8 & 38.1 & 34.7 & 28.2 & 1.25 & 0.83 & 0.74 [0.66, 0.81] \\
50\% & kept for both & 34.0 & 19.5 & 15.8 & 9.7 & 0.96 & 0.45 & 0.50 [0.45, 0.55] \\
75\% & kept for ours & 25.1 & 20.6 & 16.3 & 8.3 & 0.90 & 0.33 & 0.40 [0.36, 0.45] \\
75\% & kept for conventional & 24.9 & 20.1 & 14.7 & 13.7 & 0.98 & 0.61 & 0.68 [0.58, 0.79] \\
75\% & kept for both & 12.6 & 16.8 & 9.7 & 6.7 & 0.82 & 0.29 & 0.40 [0.34, 0.47] \\
\bottomrule
\end{tabular}
\end{table}

\begin{table}[htbp]
\centering
\caption{\textbf{By geometric blur radius} (post hoc), each estimator filtered by its own confidence: share of the windows of the range that is kept [\%], MRE [\%] [95\% interval], median relative depth error [\%] and share off by more than half [\%].}
\label{S-tab:confblur}
\footnotesize
\setlength{\tabcolsep}{4pt}
\begin{tabular}{@{}llcccccccc@{}}
\toprule
discarded & blur [px] & \multicolumn{4}{c}{conventional DfDD: kept, MRE, median, $>$ half} & \multicolumn{4}{c}{joint fit (ours): kept, MRE, median, $>$ half} \\
\midrule
0\% & 0--1.5 & 100 & 26.1 [20.9, 31.2] & 8.2 & 10.4 & 100 & 33.0 [25.1, 41.6] & 7.6 & 11.6 \\
0\% & 1.5--3.5 & 100 & 49.2 [40.4, 58.7] & 19.2 & 17.9 & 100 & 49.0 [38.7, 60.3] & 12.4 & 16.4 \\
0\% & 3.5--9.5 & 100 & 172.7 [130.6, 216.0] & 65.0 & 63.6 & 100 & 141.9 [103.1, 182.4] & 29.2 & 38.5 \\
50\% & 0--1.5 & 55 & 16.3 [13.0, 20.3] & 6.5 & 6.2 & 65 & 9.1 [8.0, 10.4] & 5.3 & 1.5 \\
50\% & 1.5--3.5 & 57 & 26.2 [21.8, 31.4] & 15.1 & 7.8 & 51 & 12.9 [11.0, 15.4] & 6.2 & 4.0 \\
50\% & 3.5--9.5 & 34 & 95.9 [60.9, 133.4] & 33.9 & 34.7 & 33 & 36.3 [30.4, 43.1] & 17.0 & 28.2 \\
75\% & 0--1.5 & 42 & 13.7 [10.7, 17.0] & 5.9 & 4.9 & 23 & 6.2 [5.5, 7.1] & 3.9 & 0.7 \\
75\% & 1.5--3.5 & 26 & 25.1 [19.2, 32.4] & 14.3 & 8.2 & 36 & 7.2 [6.4, 8.3] & 4.9 & 0.4 \\
75\% & 3.5--9.5 & 6 & 47.3 [32.9, 65.0] & 28.4 & 18.9 & 17 & 13.5 [11.2, 16.8] & 7.5 & 1.6 \\
\bottomrule
\end{tabular}
\end{table}

\begin{table}[htbp]
\centering
\caption{\textbf{By image pair, least confident 75\% discarded} (post hoc). Share of the windows of the pair that is kept [\%] and MRE [\%] [95\% interval] of conventional DfDD and of the joint fit; the span of the distance bins in which the mean relative error stays below 10\% [D], for ours, for conventional DfDD, and their paired difference [95\% interval]; the region contiguous from the focus in which it stays below 10\% [m], its ends interpolated between the centers of the bins ($>$, $<$: the windows of the data end first; none: the bin at the focus is above 10\%).}
\label{S-tab:confpair}
\footnotesize
\setlength{\tabcolsep}{4pt}
\resizebox{\textwidth}{!}{%
\begin{tabular}{@{}lccccccc@{}}
\toprule
focus [m] & \multicolumn{2}{c}{conventional: kept, MRE} & \multicolumn{2}{c}{ours: kept, MRE} & span: ours, conv., difference & contiguous, conv. & contiguous, ours \\
\midrule
0.92 & 10 & 22.6 [18.3, 28.8] & 22 & 12.5 [11.0, 14.5] & 0.53, 0.33, +0.20 [+0.07, +0.39] & $<$0.95 to 1.22 & $<$0.95 to 1.38 \\
1.37 & 21 & 11.5 [8.2, 15.7] & 43 & 7.3 [6.2, 8.7] & 0.86, 0.59, +0.26 [+0.20, +0.39] & $<$0.97 to 2.15 & 0.96 to 4.48 \\
1.94 & 31 & 14.0 [7.7, 24.0] & 42 & 6.0 [5.3, 6.9] & 0.86, 0.72, +0.13 [+0.07, +0.33] & $<$0.99 to 3.17 & $<$0.99 to 5.59 \\
2.62 & 33 & 16.8 [12.5, 22.0] & 32 & 8.7 [7.6, 9.9] & 0.72, 0.33, +0.39 [+0.33, +0.53] & 1.84 to 3.99 & none \\
3.54 & 31 & 16.0 [12.2, 20.2] & 21 & 8.7 [6.5, 11.8] & 0.33, 0.20, +0.13 [+0.00, +0.39] & 2.14 to 4.25 & none \\
4.82 & 27 & 19.1 [15.0, 23.9] & 18 & 6.6 [5.6, 8.2] & 0.66, 0.07, +0.59 [+0.46, +0.72] & none & none \\
5.84 & 24 & 41.3 [30.5, 54.6] & 6 & 14.7 [10.8, 21.7] & 0.26, 0.00, +0.26 [+0.13, +0.33] & none & none \\
7.74 & 22 & 24.3 [14.9, 38.8] & 18 & 9.2 [6.6, 13.6] & 0.39, 0.00, +0.39 [+0.26, +0.59] & none & none \\
\midrule
all pairs & & & & & 4.61, 2.24, +2.37 [+2.17, +3.03] & & \\
\bottomrule
\end{tabular}}
\end{table}

\begin{table}[htbp]
\centering
\caption{\textbf{Main-paper Table~\ref{tab:est} near focus, in distance units} (the mean-based columns are post hoc). Textured windows with uniform LiDAR depth, f/16, held out, within $\pm0.35$\,D of focus: MAE [m], MRE [\%] [95\% interval], median relative depth error [\%], share within a factor 1.25 [\%], and the MRE and the MAE over those of ours [95\% interval].}
\label{S-tab:conft1}
\footnotesize
\setlength{\tabcolsep}{3.4pt}
\begin{tabular}{@{}lcccccc@{}}
\toprule
estimator and curve & MAE & MRE & median & $\delta_{1.25}$ & MRE over ours & MAE over ours \\
\midrule
conventional DfDD: one ratio, $\alpha,\beta$ & 1.09 & 21.9 [18.0, 26.3] & 10.2 & 75 & 1.40 [1.29, 1.56] & 1.49 [1.34, 1.66] \\
one ratio, our curve & 0.72 & 15.3 [11.8, 19.2] & 6.1 & 85 & 0.98 [0.95, 1.00] & 0.98 [0.96, 1.00] \\
our joint fit, geometric disc & 0.85 & 19.0 [15.1, 23.6] & 8.2 & 83 & 1.22 [1.16, 1.28] & 1.16 [1.11, 1.22] \\
\quad fitted radius & 0.79 & 17.2 [13.6, 21.3] & 7.2 & 82 & 1.10 [1.07, 1.14] & 1.08 [1.05, 1.11] \\
\quad fitted radius and gain & 0.77 & 16.6 [13.1, 20.6] & 6.6 & 82 & 1.06 [1.03, 1.10] & 1.05 [1.02, 1.09] \\
\quad radius $\times(1-f_b/f_c)$ & 0.73 & 15.7 [12.1, 19.9] & 6.3 & 85 & 1.01 [1.00, 1.02] & 1.00 [0.99, 1.01] \\
joint fit, wave (ours) & 0.73 & 15.6 [12.0, 19.8] & 6.1 & 85 & -- & -- \\
our joint fit, wave, focus offset per pair & 0.69 & 14.6 [11.1, 18.8] & 5.3 & 86 & 0.94 [0.92, 0.95] & 0.94 [0.92, 0.96] \\
our joint fit, calibrated per pair and band & 0.76 & 15.7 [12.5, 19.4] & 6.3 & 83 & 1.01 [0.95, 1.08] & 1.03 [0.96, 1.12] \\
\quad with a gain & 0.71 & 15.1 [11.9, 18.6] & 5.8 & 84 & 0.97 [0.91, 1.02] & 0.97 [0.92, 1.03] \\
spectral ratio~\cite{S-pentland1987,S-morganmar2014}, calibrated & 0.81 & 17.1 [13.9, 20.6] & 7.3 & 81 & 1.09 [1.02, 1.19] & 1.11 [1.02, 1.22] \\
our perceptron & 0.61 & 11.7 [9.0, 14.9] & 5.2 & 87 & 0.75 [0.63, 0.92] & 0.82 [0.71, 0.96] \\
\bottomrule
\end{tabular}
\end{table}

\begin{table}[htbp]
\centering
\caption{\textbf{Main-paper Table~\ref{tab:est} over the 9.5\,px search, with a confidence filter} (post hoc; the same windows). The MRE [\%] with nothing discarded; with the least confident 75\% and 50\% discarded, each estimator by its own confidence (from the fit for joint fits with model curves, from the image energy otherwise), the MRE [95\% interval], its ratio to ours [95\% interval] and, last, the MRE of the row on the windows kept for ours.}
\label{S-tab:conft1b}
\footnotesize
\setlength{\tabcolsep}{4pt}
\resizebox{\textwidth}{!}{%
\begin{tabular}{@{}lccccccc@{}}
\toprule
estimator and curve & all & 75\%: MRE & over ours & on ours' windows & 50\%: MRE & over ours & on ours' windows \\
\midrule
conventional DfDD: one ratio, $\alpha,\beta$ & 44.7 & 18.2 [13.6, 23.6] & 3.03 [2.41, 3.69] & 19.4 & 20.7 [16.7, 25.4] & 2.57 [2.20, 2.96] & 22.6 \\
one ratio, our curve & 40.6 & 11.6 [8.7, 15.3] & 1.94 [1.53, 2.49] & 11.9 & 14.9 [11.8, 18.4] & 1.85 [1.53, 2.23] & 17.7 \\
our joint fit, geometric disc & 38.0 & 10.3 [8.8, 12.1] & 1.71 [1.50, 1.97] & 11.3 & 12.9 [11.2, 15.0] & 1.60 [1.43, 1.80] & 11.7 \\
\quad fitted radius & 47.2 & -- & -- & -- & -- & -- & -- \\
\quad fitted radius and gain & 52.4 & 11.2 [9.2, 13.6] & 1.87 [1.57, 2.23] & 23.3 & 16.2 [13.7, 19.2] & 2.01 [1.78, 2.26] & 31.0 \\
\quad radius $\times(1-f_b/f_c)$ & 32.7 & 6.5 [5.7, 7.5] & 1.08 [1.04, 1.12] & 6.6 & 8.1 [7.1, 9.4] & 1.01 [0.98, 1.04] & 8.2 \\
joint fit, wave (ours) & 32.1 & 6.0 [5.2, 7.1] & -- & -- & 8.1 [6.9, 9.4] & -- & -- \\
our joint fit, wave, focus offset per pair & 32.0 & -- & -- & -- & -- & -- & -- \\
our joint fit, calibrated per pair and band & 27.9 & 12.3 [8.4, 17.2] & 2.05 [1.54, 2.67] & 6.8 & 13.7 [10.2, 18.0] & 1.70 [1.37, 2.08] & 8.6 \\
\quad with a gain & 27.6 & 9.9 [7.2, 13.5] & 1.65 [1.31, 2.08] & 6.2 & 11.8 [9.2, 15.1] & 1.47 [1.24, 1.73] & 8.1 \\
spectral ratio~\cite{S-pentland1987,S-morganmar2014}, calibrated & 28.6 & 15.2 [11.4, 19.7] & 2.54 [2.01, 3.13] & 8.7 & 15.2 [11.7, 19.4] & 1.88 [1.57, 2.27] & 10.3 \\
our perceptron & 15.2 & 11.6 [8.0, 15.8] & 1.93 [1.40, 2.51] & 6.1 & 10.9 [8.2, 14.0] & 1.35 [1.09, 1.63] & 7.4 \\
\bottomrule
\end{tabular}}
\end{table}

\Para[S-sec:conftest]{Confirmation on the test scenes (pre-specified)}
The analysis above was designed on the validation scenes after their results were known. We therefore fixed a
confirmation on the 50 test scenes in a local record before any distance error or any estimate of conventional
DfDD existed there (\texttt{notes/\allowbreak prespec\_zedd\_confidence\_test\_20261114.md}, with the SHA-256 of 138
files, checked before the run read any data and again after it; \texttt{realdata/\allowbreak
zedd\_confidence\_test.py}, \texttt{results/\allowbreak zedd\_confidence\_test.txt}). The test scenes had been used
before, for the coefficient test of \hyperref[S-par:dslrconfirm]{Sec.~\ref{S-sec:zedd}} and for the defocus errors of
both joint fits in \hyperref[S-par:transfer]{depth on the test scenes}, where a gate on the band energy and scene
medians of the residual of the fit had been computed for ours; the record lists this.
\begin{itemize}
  \item \emph{Protocol.} Every estimator is fitted on the 50 validation scenes with LiDAR depth and scored on the 50
  test scenes against depth from focus at f/2.8 over nine images, as in \hyperref[S-par:transfer]{depth on the test
  scenes}; f/16, one image pair, the 9.5\,px search, every window with a valid reference and no selection by
  texture, 1\,973\,298 samples. The reference exists where depth from focus finds the same vertex in the four
  quadrants of a window, on 53\% of the windows, at distances from 0.98 to 7.41\,m.
  \item \emph{Confidence.} For each estimator the candidate is the one chosen on all validation samples at once, the
  residual fraction for ours, the second minimum for the joint fit with geometric curves and the summed Laplacian
  energy for conventional DfDD at a sparsity of 75\%. The threshold is the quantile over the scored test samples,
  which uses no reference depth, so that every estimator discards the same share.
  \item \emph{Rule.} With the least confident 75\% discarded, the statement that the joint fit with wave-optics
  curves has a smaller MRE than (A) conventional DfDD, (B) the same fit with geometric curves, is confirmed if the
  95\% interval of the ratio, other over ours, over 2\,000 resamples of the test scenes lies above 1.05, refuted if
  it lies below 1.05, and undecided otherwise.
  \item \emph{Predictions written with the rule.} An MRE of ours of 9--16\%, and ratios of 1.3--2.2 and 1.1--1.6,
  smaller than on the validation scenes because the reference adds its own error to every estimator.
\end{itemize}
\emph{Outcome} (Table~\ref{S-tab:conftest}). Both statements are confirmed. The ratio is 2.49 [1.97, 3.12] for
conventional DfDD and 1.77 [1.64, 1.92] for the fit with geometric curves; the MRE is 4.9\% [4.5, 5.5] for ours,
12.2\% [9.6, 15.7] and 8.7\% [7.7, 10.0]. None of our three predictions held. The ratios did not shrink, and the
errors are lower than on the validation scenes because the reference covers only the windows and distances at
which depth from focus is reliable. The validation scenes through the same reference and mask give 5.5\%, 14.2\%
and 9.1\% (post hoc), so on one path the two sets of scenes agree. On the windows kept for ours conventional DfDD
errs 2.96 [2.70, 3.19] times ours, and on those kept for conventional DfDD 1.50 [1.28, 1.82] times. The joint fit
has the smallest MRE for each of the eight pairs and in each range of blur; beyond 3.5\,px it keeps 14\% of the
samples with an MRE of 6.9\%, conventional DfDD 6\% with 27.4\%. The span of the distance bins with a mean error
below 10\% is larger for ours in all eight pairs, 5.33\,D against 1.97\,D in total.

Some results do not favor ours. With nothing discarded the MRE of ours is 28.6\% [22.6, 35.8], and within 1.5\,px
of blur conventional DfDD then has the smaller one (0.87 [0.78, 0.96] times ours). On the windows that the fit with
geometric curves keeps, the two joint fits err alike (1.06 [1.01, 1.11]); statement (B) holds for each fit filtered by
its own confidence, and the advantage lies in the windows of which each fit is confident. For the pair focused at
1.94\,m the interval of the ratio for geometric curves includes 1. The samples kept for ours are spread unevenly
over the pairs, from 5\% of those of the pair focused at 5.80\,m to 49\% of those of the pair at 1.36\,m.

\begin{table}[htbp]
\centering
\caption{\textbf{Depth with a confidence on the 50 test scenes} (pre-specified; reference: depth from focus at
f/2.8; f/16, one image pair, 9.5\,px search, 1\,973\,298 samples). Mean relative depth error (MRE) [\%], mean
absolute distance error (MAE) [m], share of distances within a factor 1.25 of the reference [\%] and share off by
more than half [\%], with 95\% intervals over 2\,000 resamples of the scenes, and the MRE over that of ours. The
rule applies to the two ratios at 75\%.}
\label{S-tab:conftest}
\footnotesize
\setlength{\tabcolsep}{4pt}
\begin{tabular}{@{}llccccc@{}}
\toprule
discarded & estimator & MRE & MAE & $\delta_{1.25}$ & $>$ half & MRE over ours \\
\midrule
75\% & conventional DfDD & 12.2 [9.6, 15.7] & 0.380 [0.276, 0.531] & 84.8 & 1.7 & 2.49 [1.97, 3.12] \\
& our joint fit, geometric & 8.7 [7.7, 10.0] & 0.271 [0.222, 0.330] & 94.8 & 1.7 & 1.77 [1.64, 1.92] \\
& joint fit, wave (ours) & 4.9 [4.5, 5.5] & 0.137 [0.113, 0.167] & 98.8 & 0.1 & -- \\
\midrule
50\% & conventional DfDD & 17.7 [15.7, 20.0] & 0.488 [0.399, 0.605] & 70.0 & 3.2 & 2.41 [2.08, 2.76] \\
& our joint fit, geometric & 11.0 [9.7, 12.7] & 0.339 [0.277, 0.416] & 92.3 & 2.7 & 1.50 [1.43, 1.56] \\
& joint fit, wave (ours) & 7.3 [6.5, 8.5] & 0.234 [0.189, 0.292] & 95.0 & 1.5 & -- \\
\midrule
0 & conventional DfDD & 37.9 [31.7, 46.1] & 0.936 [0.792, 1.115] & 59.2 & 16.6 & 1.33 [1.21, 1.47] \\
& our joint fit, geometric & 35.5 [28.4, 43.8] & 0.972 [0.756, 1.239] & 80.2 & 9.8 & 1.24 [1.16, 1.32] \\
& joint fit, wave (ours) & 28.6 [22.6, 35.8] & 0.752 [0.587, 0.950] & 79.6 & 9.8 & -- \\
\bottomrule
\end{tabular}
\end{table}

\Para[S-par:depthold]{The 20\% limits in defocus drawn as distances}
The 20\% limits of the relative defocus error can be drawn as distances for a camera focused at 1\,m
(Fig.~\ref{S-fig:depthold}a). The main paper does not use this drawing; its Fig.~\ref{fig:depth}a draws the bins of
one image pair by the mean error with a confidence filter. No capture is focused at 1\,m; the drawing pools
both sides of focus and the pairs of a step group; ``to infinity'' is the end of the search; and at its far limits
the error in depth is large (49\% at 3.4\,m). The intervals of Table~\ref{S-tab:depthrange}, in depth and per pair,
replace it. Panel c of the figure shows the labeled scenes that each calibration needs
(\hyperref[S-par:training]{Training data}).

\Para{Texture judged at the estimate's own aperture}
At f/2.8--f/5.6 the tables above judge the texture of a window on the f/16 images of its scene, which a camera at
those apertures does not have. Selecting it on the pair's own images instead (\texttt{zedd\_estimate.TEX\_APERTURE},
\texttt{zedd\_learning\_curve.py --tex-own}) changes little: ours errs by 0.0093, 0.0090 and 0.0086\,D at f/5.6, f/4
and f/2.8 against 0.0092, 0.0088 and 0.0082\,D, the paired ratios of the other methods move by at most 0.06, and at
f/2.8 calibrated curves and spectral ratios still err about 20\% less (0.80 and 0.79). Over the 9.5\,px search at
f/5.6 ours errs by 16\% and 12\% at 1.5--3.5 and 3.5--9.5\,px against 15\% and 10\% with the f/16 selection, and its
median relative error stays below 20\% to the end of the search (\texttt{results/zedd\_r10b\_supp.txt}).

\begin{table}[htbp]
\centering
\caption{f/16, one image pair, 9.5\,px search: mean of the binned median relative error [\%] over 1.5--3.5 / 3.5--9.5\,px
of blur. Textured: the windows of main-paper Table~\ref{tab:est}; low energy: their third with the least in-focus
band energy; all: every window with valid LiDAR depth. 9-tap rows with the window mean
removed (\texttt{results/zedd\_r13\_conv.txt}).}
\label{S-tab:range_energy}
\small
\setlength{\tabcolsep}{3pt}
\begin{tabular}{lccc}
\toprule
Method & textured & low energy & all \\
\midrule
joint fit, wave (ours) & 11.3 / 9.0 & 28.7 / 72.1 & 24.1 / 22.9 \\
our joint fit, wave, noise-aware & 11.3 / 9.0 & 27.3 / 66.9 & 20.3 / 17.4 \\
our joint fit, geometric & 16.4 / 8.6 & 23.5 / 40.5 & 22.4 / 16.0 \\
our joint fit, calibrated & 10.8 / 8.3 & 26.9 / 49.8 & 21.6 / 18.4 \\
spectral ratios~\cite{S-pentland1987,S-morganmar2014}, calibrated & 13.9 / 9.3 & 27.3 / 34.1 & 24.2 / 17.8 \\
spectral ratios, our curves and gain & 14.0 / 13.3 & 28.5 / 64.0 & 24.0 / 23.6 \\
9-tap ratio, 1.5\,px, calibrated & 14.2 / 23.7 & 22.2 / 67.6 & 25.5 / 53.5 \\
9-tap ratio, tuned, our curve & 12.1 / 49.6 & 27.7 / 87.0 & 19.7 / 64.0 \\
9-tap ratio, tuned, calibrated & 14.4 / 28.0 & 22.9 / 71.3 & 23.9 / 46.8 \\
\bottomrule
\end{tabular}
\end{table}

\begin{table}[htbp]
\centering
\caption{f/16, all windows with valid LiDAR depth, 9.5\,px search: the windows whose 0.10\,cycles/px Laplacian energy
at f/16 exceeds $k$ times its first percentile in the scene, their share, and the mean of the binned median relative
error [\%] over 1.5--3.5 / 3.5--9.5\,px (\texttt{zedd\_r10\_supp.py}).}
\label{S-tab:texcrit}
\small
\setlength{\tabcolsep}{3pt}
\begin{tabular}{rrccc}
\toprule
$k$ & share & ours & calibrated & spectral ratios \\
\midrule
1 & 99\% & 23.8 / 22.3 & 21.4 / 18.0 & 23.9 / 17.4 \\
3.2 & 77\% & 17.6 / 12.0 & 16.0 / 10.4 & 18.9 / 10.9 \\
10 & 62\% & 14.2 / 10.2 & 13.1 / 8.0 & 16.1 / 8.6 \\
32 & 48\% & 11.8 / 9.4 & 11.1 / 6.5 & 14.0 / 7.0 \\
100 & 32\% & 9.7 / 9.5 & 9.5 / 6.1 & 11.7 / 6.8 \\
316 & 18\% & 9.0 / 8.5 & 8.6 / 5.2 & 10.7 / 5.7 \\
\bottomrule
\end{tabular}
\end{table}

\begin{table*}[htbp]
\centering
\caption{One image pair: mean of the binned median relative error [\%] over intervals of the geometric blur radius
[95\% interval over scenes], and the 20\% limit [px], held-out scenes; ``none'': the binned error never falls below
20\%.}
\label{S-tab:range}
\small
\resizebox{\linewidth}{!}{%
\begin{tabular}{llccccc}
\toprule
Aperture & Method & 1.5--3.5\,px & 3.5--5.6\,px & 5.6--7.5\,px & 7.5--9.5\,px & 20\% limit \\
\midrule
f/16 & one ratio, calibrated curve & 27.3 [21.7, 34.8] & 35.2 [27.8, 45.8] & 43.9 [31.2, 68.0] & 88.2 [49.3, 93.2] & none \\
    & one ratio, our curve & 18.6 [15.1, 23.8] & 47.6 [37.4, 56.9] & 59.7 [43.3, 78.3] & 93.4 [62.9, 97.3] & 2.9 [2.7, 3.1] \\
    & our joint fit, geometric & 16.4 [14.9, 18.2] & 12.8 [11.4, 14.4] & 7.1 [5.6, 11.8] & 6.5 [5.4, 8.9] & 2.0 [1.6, 9.5] \\
    & our joint fit, geometric, 5 gains & 14.7 [13.2, 16.2] & 12.1 [10.7, 13.7] & 9.6 [7.6, 13.8] & 10.4 [9.1, 12.5] & 9.5 [7.0, 9.5] \\
    & joint fit, wave (ours) & 11.3 [9.4, 14.3] & 9.5 [7.2, 13.8] & 9.2 [7.2, 12.7] & 8.3 [6.7, 11.2] & 9.5 [7.3, 9.5] \\
    & our joint fit, calibrated & 10.8 [9.1, 13.2] & 10.8 [8.0, 14.8] & 6.7 [5.3, 12.0] & 8.0 [4.6, 16.3] & 9.5 [7.3, 9.5] \\
    & our perceptron & 7.8 [7.0, 9.1] & 6.4 [5.2, 8.2] & 5.2 [4.0, 8.6] & 5.3 [4.2, 14.3] & 9.5 [8.9, 9.5] \\
\midrule
f/5.6 & one ratio, calibrated curve & 52.9 [43.1, 64.0] & 39.9 [33.4, 49.6] & 53.3 [43.0, 63.2] & 90.9 [74.0, 94.4] & none \\
     & one ratio, our curve & 28.1 [22.7, 35.9] & 48.8 [40.9, 59.7] & 79.5 [69.5, 86.8] & 95.0 [91.6, 96.5] & none \\
     & our joint fit, geometric & 15.7 [13.0, 19.5] & 11.5 [10.0, 13.7] & 11.1 [9.6, 13.3] & 7.4 [6.3, 8.9] & 9.5 [2.4, 9.5] \\
     & joint fit, wave (ours) & 14.9 [12.3, 18.6] & 11.2 [9.3, 14.2] & 11.0 [9.2, 13.8] & 8.4 [7.2, 10.0] & 9.5 [2.6, 9.5] \\
     & our joint fit, calibrated & 13.3 [11.6, 15.5] & 11.2 [9.2, 14.0] & 9.9 [8.4, 12.0] & 6.4 [5.4, 7.8] & 9.5 [9.5, 9.5] \\
     & our perceptron & 11.9 [10.4, 13.6] & 7.6 [6.6, 8.8] & 6.4 [5.6, 7.3] & 6.7 [5.9, 8.0] & 9.5 [9.5, 9.5] \\
\bottomrule
\end{tabular}}
\end{table*}

\Para[S-par:lowband]{Conventional DfDD on low bands}
The broadband ratio is dominated by its 0.225 cycles/px band, which turns over at a blur radius of about 2\,px. A
conventional estimate tuned for a longer range uses only the low bands. Table~\ref{S-tab:lowband}
(\texttt{realdata/zedd\_lowband.py}) restricts the ratio to the lowest two, three or four bands, each with its own
calibrated curve and the folds of the other estimates. Fewer bands move the turnover away from focus but raise the
error near it, where the low bands carry little signal: with two bands the 20\% criterion holds from 2.4 to 9.5\,px at
f/16 and from 1.8 to 9.5\,px at f/5.6, but the median error within the standard search range is 0.050 and 0.015\,D,
against 0.021 and 0.0092\,D for the sub-band estimate. No fixed subset matches the sub-band estimate both near and far
from focus. With all five bands the binned error at f/16 dips below 20\% only in the bin at 2.1\,px (19.7\%; the runs of
Table~\ref{S-tab:range} search 82 of the 600 candidates and give 21.9\% there and 0.0257\,D, hence no limit).

\begin{table}[htbp]
\centering
\caption{Conventional DfDD with a calibrated curve on subsets of the bands, against ours, the joint fit with
wave-optics curves (last row of each aperture). Error: held-out median $|\hat\psi-\psi|$ [D] within the standard
search range ($|\psi|\le0.35$\,D at f/16, $\le0.08$\,D at f/5.6), overall and closest to focus. Range: blur radius over
which the median relative error stays within 20\% with a search to 9.5\,px, and the 95\% interval of its upper limit.
Last columns: mean relative error [\%] over blur intervals. Focus calibrated on the training scenes of each fold
(\texttt{zedd\_lowband.py --cals}; last rows from the runs of Table~\ref{S-tab:estfull} and from Table~\ref{S-tab:range}).}
\label{S-tab:lowband}
\small
\setlength{\tabcolsep}{4pt}
\begin{tabular}{llccccccc}
\toprule
& & \multicolumn{2}{c}{error [D]} & & \multicolumn{4}{c}{relative error [\%], blur radius [px]} \\
\cmidrule(lr){3-4}\cmidrule(lr){6-9}
Aperture & Bands [cycles/px] & all & nearest & range (20\%) [px] & 1.5--3.5 & 3.5--5.6 & 5.6--7.5 & 7.5--9.5 \\
\midrule
f/16  & all five       & 0.0254 & 0.0188 & 1.8--2.4 [1.8, 2.4] & 26.2 & 36.5 & 44.6 & 88.7 \\
      & 0.045--0.15    & 0.0254 & 0.0245 & 1.2--3.6 [3.0, 4.2] & 15.6 & 25.9 & 29.0 & 67.4 \\
      & 0.045--0.10    & 0.0324 & 0.0349 & 1.8--7.7 [5.9, 8.9] & 12.9 & 12.3 & 14.2 & 27.7 \\
      & 0.045--0.0675  & 0.0503 & 0.0537 & 2.4--9.5 [7.1, 9.5] & 18.5 & 10.1 & 7.0 & 11.8 \\
      & joint fit, wave (ours) & 0.0214 & 0.0177 & 1.2--9.5 [7.3, 9.5] & 11.3 & 9.5 & 9.2 & 8.3 \\
\midrule
f/5.6 & all five       & 0.0134 & 0.0070 & none & 53.7 & 42.0 & 55.8 & 91.7 \\
      & 0.045--0.15    & 0.0109 & 0.0076 & none & 35.6 & 39.7 & 44.2 & 78.5 \\
      & 0.045--0.10    & 0.0116 & 0.0090 & 1.8--4.1 [4.1, 4.7] & 14.1 & 21.7 & 27.4 & 43.6 \\
      & 0.045--0.0675  & 0.0154 & 0.0128 & 1.8--9.5 [8.9, 9.5] & 14.2 & 10.4 & 12.6 & 14.8 \\
      & joint fit, wave (ours) & 0.0092 & 0.0064 & 1.2--9.5 [2.6, 9.5] & 14.9 & 11.2 & 11.0 & 8.4 \\
\bottomrule
\end{tabular}
\end{table}

\Para[S-par:maps]{Depth maps}
Every window with texture gets an estimate from each image pair. Without the ground truth, three pairs are
combined: the pair whose runs are sharpest at the window (largest $S_\times$ at 0.10 cycles/px) and its two neighbors.
An estimate outside its pair's search range is dropped, and the others are averaged in $q$ with weights
$1/s_k(|\hat\psi|)^2$, where $s_k$ is the median absolute error of pair $k$ in six bins of $|\hat\psi|$ on the training
scenes. The baseline is DfF over the nine runs of the same aperture, a log-parabola through the peak run and its
neighbors on the calibrated foci. Tables~\ref{S-tab:maps} and~\ref{S-tab:maps_vergence} give the full results.

\begin{table}[htbp]
\centering
\caption{Depth maps: held-out median $|\hat q-q_{\mathrm{LiDAR}}|$ [D] and median relative depth error over 172\,568
textured windows at 0.7--12.5\,m; focus, gains and error tables from the training scenes of each fold. Every row
but the one-pair row combines three image pairs per window.}
\label{S-tab:maps}
\begin{tabular}{lcccccc}
\toprule
& \multicolumn{2}{c}{f/16} & \multicolumn{2}{c}{f/5.6} & \multicolumn{2}{c}{f/4} \\
\cmidrule(lr){2-3}\cmidrule(lr){4-5}\cmidrule(lr){6-7}
Method & $|\Delta q|$ & $|\Delta z|/z$ & $|\Delta q|$ & $|\Delta z|/z$ & $|\Delta q|$ & $|\Delta z|/z$ \\
\midrule
conventional DfDD, nominal $\alpha,\beta$ & 0.020 & 5.7\% & 0.018 & 5.5\% & 0.025 & 6.7\% \\
one ratio, calibrated curve & 0.011 & 3.1\% & 0.015 & 4.7\% & 0.023 & 7.2\% \\
our joint fit, geometric & 0.012 & 3.3\% & 0.010 & 3.0\% & 0.011 & 3.3\% \\
our joint fit, wave & 0.011 & 3.0\% & 0.0095 & 2.8\% & 0.011 & 3.3\% \\
joint fit, wave (ours), one pair & 0.018 & 4.3\% & 0.010 & 2.8\% & 0.011 & 2.9\% \\
our joint fit, calibrated & 0.011 & 3.0\% & 0.0098 & 3.1\% & 0.012 & 3.8\% \\
DfF, nine runs & 0.018 & 4.5\% & 0.0088 & 2.4\% & 0.0092 & 2.8\% \\
\bottomrule
\end{tabular}
\end{table}

\begin{table}[htbp]
\centering
\caption{Depth maps at f/16: held-out median $|\hat q-q_{\mathrm{LiDAR}}|$ [D] by LiDAR vergence.}
\label{S-tab:maps_vergence}
\begin{tabular}{lcccc}
\toprule
Method & 0.08--0.2\,D & 0.2--0.4\,D & 0.4--0.8\,D & 0.8--1.4\,D \\
\midrule
conventional DfDD, nominal $\alpha,\beta$ & 0.018 & 0.015 & 0.023 & 0.113 \\
one ratio, calibrated curve & 0.012 & 0.010 & 0.011 & 0.044 \\
our joint fit, geometric & 0.010 & 0.0089 & 0.014 & 0.043 \\
our joint fit, wave & 0.010 & 0.0088 & 0.013 & 0.036 \\
joint fit, wave (ours), one pair & 0.021 & 0.018 & 0.016 & 0.030 \\
our joint fit, calibrated & 0.012 & 0.010 & 0.011 & 0.033 \\
DfF, nine runs & 0.025 & 0.021 & 0.014 & 0.026 \\
\bottomrule
\end{tabular}
\end{table}

\begin{figure}[htbp]
\centering
\includegraphics[width=\linewidth]{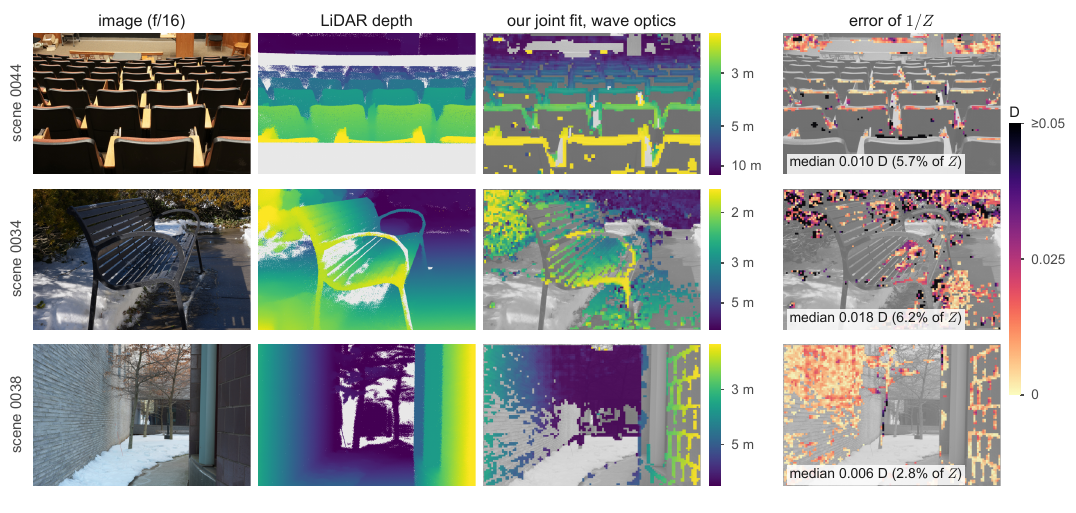}
\caption{\textbf{Depth maps that combine three image pairs per window, f/16.} Columns: the image; LiDAR depth (light
gray: no return); our joint fit (wave-optics curves, held-out fold, three image pairs per window) on the
windows with texture, over a gray copy of the image; and the error of the inverse depth $1/Z$ where the LiDAR depth is
valid, with its median. Depth colors are linear in $1/Z$.}
\label{S-fig:supp_maps}
\end{figure}

\noindent Wave-optics curves beat geometric curves in 31, 24 and 26 of the 50 scenes at f/16, f/5.6 and f/4
(\texttt{results/zedd\_depthmap\_f16\_fold.txt}, \texttt{zedd\_depthmap\_f56\_f4\_fold.txt}). At
f/4--f/5.6 every method fails at 0.8--1.4\,D, with medians of 0.040--0.15\,D: the focus steps there (0.25--0.48\,D) exceed the
working range in defocus of a pair, and the depth of field is too shallow for DfF between runs. Fig.~\ref{S-fig:supp_maps} shows
the combined maps of the three most textured scenes whose depth spans at least a factor of two.

\begin{figure*}[tp]
\centering
\includegraphics[width=\textwidth]{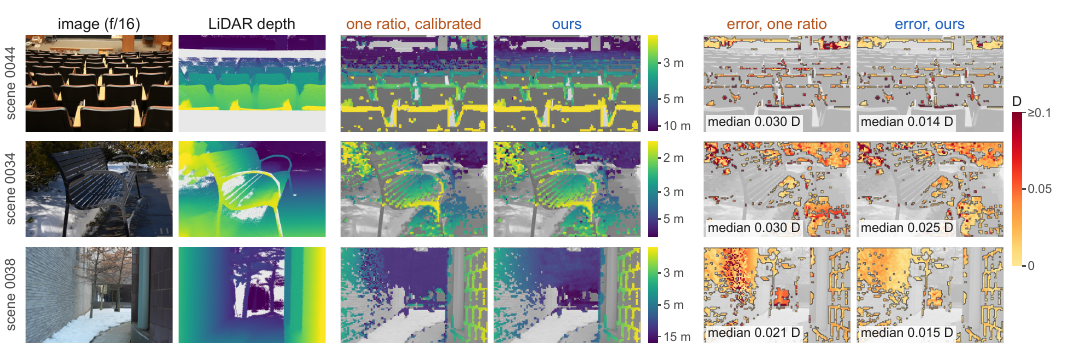}
\caption{\textbf{Depth from one image pair: the two estimates agree near focus and part away from it.} DSLR at f/16,
the pair focused at 1.94\,m, held-out folds, the three most textured scenes whose depth spans at least a factor of two
(main-paper Fig.~\ref{fig:teaser}e shows scene 20 with the pair focused at 7.77\,m, with a confidence filter, and
Fig.~\ref{S-fig:maps_conf} eight more maps).
Columns: image; LiDAR depth (light gray: no return); depth from
one ratio with a calibrated curve and from ours, on windows with texture; error of $1/Z$ with its median, on
a light copy of the image with the evaluated windows outlined; colors
linear in $1/Z$.}
\label{S-fig:maps}
\end{figure*}

\Para[S-par:pairmaps]{Depth maps from one image pair}
A DfDD camera records one image pair, so main-paper Fig.~\ref{fig:teaser}e and Figs.~\ref{S-fig:maps}
and~\ref{S-fig:maps_conf} keep the pairs apart (\texttt{realdata/zedd\_pair\_maps.py}): each of the eight pairs gives
a depth map of every scene, with the search range
widened to a geometric blur radius of 9.5\,px ($\pm1.02$\,D at f/16) as in the working-range analysis, and with the
calibrations, gains and focus distances of the held-out folds (\texttt{--cals results/zedd\_foldcal\_cals.json}).
Table~\ref{S-tab:pairmaps} gives the error against the distance of a window from the focus of the pair. Over the 400 maps
(50 scenes, eight pairs), the sub-band estimate with wave-optics curves is more accurate than conventional DfDD with a
calibrated curve in 85\% of the maps and than conventional DfDD with calibrated $\alpha$ and $\beta$ in 92\%; the
medians of the per-map median errors are 0.027, 0.069 and 0.072\,D. In the 200 maps where more than 30\% of the
windows lie more than 0.3\,D from focus, it is more accurate in 96\% against both (0.044 against 0.165 and 0.184\,D).
Against the sub-band estimate with geometric curves it wins 63\% of the maps (0.036\,D for geometric curves).
Fig.~\ref{S-fig:maps} uses the pair focused at 1.94\,m for every scene; with this pair the sub-band estimate is more accurate than the
calibrated curve in 43 of the 50 scenes, and with the other pairs in 18 to 49 (\texttt{results/zedd\_pair\_maps\_f16\_fold.txt});
by the number of scenes won by the joint fit, this pair ranks sixth of eight against one ratio and first against
geometric curves.

Without a confidence the maps compare three estimators, namely one ratio inverted with the wave-optics curve and one
gain per aperture, which is the estimator of panel c of main-paper Fig.~\ref{fig:teaser} and of the row ``one
ratio, wave'' of main-paper Table~\ref{tab:est}, the joint fit with geometric curves, and ours. The stored maps do
not hold the first, so we
computed it on the same windows, folds, focus calibrations and search grid (\texttt{realdata/zedd\_pair\_maps\_bb.py};
the gains of the five folds are 0.79--0.80, and the one ratio with a calibrated curve, computed again in the same run,
equals the stored maps exactly). Table~\ref{S-tab:pairmaps} gives its error against the distance from focus. Over the
400 maps ours is more accurate than this one ratio in 70\% (medians of the per-map medians 0.027 against 0.040\,D),
and in 93\% of the 200 maps where more than 30\% of the windows lie more than 0.3\,D from focus (0.044 against
0.167\,D); by pair, in 22 to 48 of the 50 scenes, the fewest with the pairs focused at 2.62 and 3.54\,m (22 and 25),
whose windows mostly lie within the range of one ratio. One median over a map therefore mixes two regimes, those of
the windows within and beyond 2.9\,px of blur, the 20\% limit of one
ratio (Fig.~\ref{S-fig:depthold}a). Over all windows of the 400 maps, 39\% of which lie beyond this limit, the
medians are 0.020, 0.026 and 0.020\,D within it and 0.248, 0.054 and 0.049\,D beyond it, for one ratio, the joint fit
with geometric curves and ours; the perceptron errs by 0.019 and 0.031\,D and depth from focus over nine images by
0.019 and 0.018\,D.

Main-paper Fig.~\ref{fig:teaser}e shows one map in the convention of the DfDD prototypes
(\hyperref[S-sec:confidence]{Depth with a confidence}). Its three estimators are conventional DfDD, the joint fit with
geometric curves and ours, each on the windows that it keeps after discarding its least confident 75\% of all
samples, with the confidences and thresholds of Table~\ref{S-tab:confpooled}. The thresholds are those of all scenes, so
an estimator keeps another share of the windows on every map. An error tile carries two numbers, the share of the
estimated windows of the map that its estimator keeps and the mean relative depth error of those windows. The map is
chosen by the rule of the window of main-paper Fig.~\ref{fig:method}, applied to these six numbers. Of the 294 maps
in which each of the three estimators keeps at least 200 windows, the least number behind a bin of main-paper
Fig.~\ref{fig:depth}, it is the one with the smallest sum, over the three estimators and the two numbers, of the
absolute difference between the logarithm of its value and the logarithm of the median over these maps (errors of
13.0\%, 11.1\% and 7.5\%; shares of 25\%, 25\% and 24\%). This is ZEDD scene 20 with the pair focused at 7.77\,m, from
its held-out fold, with errors of 13.9\%, 9.8\% and 7.1\% and shares of 32\%, 21\% and 21\%, a sum of 0.83 against
1.02 for the next map. It is the first also when 100 windows are required, the second when 30 or 500 are, and the
thirteenth when 1\,000 are or when the pooled errors and shares serve as the reference.

The map does not show all that the 400 maps do. Its windows lie 2.5--5.9\,m from the camera (10th--90th percentile),
and 95\% of them lie within 2.9\,px of blur, so it shows the errors near focus and not the range. On it conventional
DfDD keeps more windows than ours, where over the 294 maps ours keeps more in 150; ours has the smaller error in 247
of them against conventional DfDD and in 249 against geometric curves. Pooled over all maps the errors are 20.1\%,
14.6\% and 8.3\%, above those of a typical map, because the maps with large blur carry the mean. Beyond 2.9\,px of
blur, which holds 40\% of the samples, conventional DfDD keeps 8\% of them with a mean error of 43\%, the geometric
fit 15\% with 32\% and ours 19\% with 12\%. On the windows that ours keeps on the map of the figure conventional DfDD
errs by 15.5\%, the geometric fit by 7.8\%, one ratio with our curve by 8.0\% and calibrated curves by
8.7\%. By the errors alone the same rule picks scene 26 with the pair focused at 2.62\,m (13.0\%, 10.5\% and 7.2\%),
on which conventional DfDD keeps 8\% of the windows and ours 27\%; the tiles would show these shares as typical, so
the rule includes them (\texttt{realdata/\allowbreak zedd\_pair\_maps\_conf.py}, \texttt{results/\allowbreak
zedd\_pair\_maps\_conf\_f16.txt}).

\Para[S-par:mapsconf]{More maps with a confidence filter}
Figure~\ref{S-fig:maps_conf} on page~\pageref{S-fig:maps_conf} shows eight more maps in the convention of main-paper
Fig.~\ref{fig:teaser}e, with the depth of conventional DfDD beside that of ours, and the table below gives the
numbers of the eight maps. The upper five rows of the figure continue the rule of main-paper
Fig.~\ref{fig:teaser}e. Ordered by the sum defined above, the 294 maps begin with the map of that figure, and
these rows are the maps at ranks 3, 4, 5, 6 and 8, with sums of 1.08--1.33. One condition was added for this figure
when the order was already known, that a scene appears once. It passes over rank 2, which is scene 20 again with the
pair focused at 1.94\,m, and rank 7, which is scene 21 again with the pair focused at 1.36\,m; on both ours has the
smaller error, 8.1\% against 9.7\% and 6.9\% against 10.9\%.

\begin{center}
\small
\setlength{\tabcolsep}{4.5pt}
\begin{tabular}{@{}ccccccccccc@{}}
\toprule
 & & & & \multicolumn{3}{c}{share kept [\%]} & \multicolumn{3}{c}{mean error [\%]} & ours on the \\
\cmidrule(lr){5-7}\cmidrule(lr){8-10}
part & scene & focus [m] & rank & conv. & geom. & ours & conv. & geom. & ours & windows of conv. \\
\midrule
a & 21 & 3.54 & 3 & 27 & 32 & 29 & 12.6 & 7.9 & 6.4 & 5.6 \\
a & 17 & 2.62 & 4 & 20 & 27 & 23 & 16.7 & 12.3 & 11.2 & 9.3 \\
a & 25 & 4.85 & 5 & 37 & 21 & 25 & 12.9 & 13.3 & 5.6 & 6.6 \\
a & 7 & 1.36 & 6 & 19 & 40 & 26 & 10.6 & 9.8 & 8.6 & 9.6 \\
a & 27 & 1.36 & 8 & 20 & 27 & 51 & 12.9 & 10.8 & 9.4 & 8.9 \\
\midrule
b & 12 & 1.36 & 293 & 3 & 9 & 93 & 1.2 & 4.9 & 3.9 & 1.4 \\
b & 9 & 3.54 & 234 & 14 & 17 & 3 & 10.4 & 11.1 & 29.7 & 7.0 \\
b & 44 & 3.54 & 267 & 76 & 43 & 25 & 48.8 & 29.3 & 47.3 & 69.1 \\
\bottomrule
\end{tabular}
\end{center}

\noindent In the table conv.\ is conventional DfDD and geom.\ our joint fit with geometric curves, which enters the
rule and is not drawn; the last column is the mean error of ours on the windows that conventional DfDD keeps. On
each of the five maps of part (a) ours has the smallest error of the three, and on the windows that conventional
DfDD itself keeps it has a smaller error than conventional DfDD. In their errors the five maps are typical, with
medians of 12.9\%, 10.8\% and 8.6\% in the order of the table against 13.0\%, 11.1\% and 7.5\% over the 294 maps. In
their shares they favor ours more than is typical. Ours keeps more windows than conventional DfDD on four of the
five maps, and on 150 of the 294.

The lower three rows are the maps least favorable to ours among the 294. On 47 of the 294 ours has a larger error
than conventional DfDD, on 144 it keeps fewer windows, and on 22 both. The ratio of the two errors, ours over
conventional DfDD, has a median of 0.61; it is above 1.5 on 14 maps and above 2 on 4 of them. The error of ours is
above 20\% on 17 maps and that of conventional DfDD on 100. Two of the three measures were fixed before the maps
were searched for them, the largest ratio and the largest error of ours.

The largest ratio, 3.2 on scene 12 with the pair focused at 1.36\,m, is a loss in the error alone. Conventional DfDD
keeps 3\% of the windows of this map, 221 of its 235 within 1\,px of blur, and ours 93\%. On the windows that
conventional DfDD keeps ours errs by 1.4\% against 1.2\%, and on those that ours keeps conventional DfDD errs by
22.7\% against 3.9\%. We therefore added a third row after this outcome was known, the largest ratio among the 22
maps on which ours loses in both numbers, 2.9 on scene 9 with the pair focused at 3.54\,m. This loss is a failure of
the confidence of our fit at the focus. Of the windows of the map, 96\% lie within 0.5\,px of blur. There
conventional DfDD keeps 14\% of them with an error of 6.4\% and ours 2\% with 29.0\%, although ours errs by 7.0\% on
the windows that conventional DfDD keeps on the map. The failure is not particular to this scene. Over all scenes,
in the bin of distance at the focus of the pair focused at 3.54\,m, ours keeps 2\% of the windows with an error of
19\% and conventional DfDD 46\% with 5\% (the bins of \hyperref[S-sec:confidence]{Depth with a confidence}). The
largest error of ours, 47.3\% on scene 44 with the same pair, is a second failure of this confidence, and of that of
conventional DfDD. A quarter of the windows of the map lie beyond 6\,m, nearly all of them about 12\,m from the
camera, at 0.08\,D. Ours keeps 52\% of the windows beyond 6\,m and 15\% of the nearer ones. Its kept estimates
beyond 6\,m lie too far, 48\% of them at the clip of 0.04\,D, which is 25\,m, and err by 76\%; nearer it errs by
14.4\%. Conventional DfDD keeps 82\% and 73\% and errs by 110\% and 25.8\%.

The figure is drawn by \texttt{make\_supp\_maps\_conf.py}, whose tiles for the map of main-paper
Fig.~\ref{fig:teaser}e equal those of that figure. The rule, written down before the least favorable maps were
searched for and before any map of the figure was drawn, and the row added afterward are on record
(\texttt{selection\_rule.md}, \texttt{select\_maps.py}, \texttt{selection.txt}, \texttt{extra\_numbers.py}).

\Para[S-par:methodfig]{\texorpdfstring{The window of main-paper Fig.~\ref{fig:method}}{The window of the method figure}}
Main-paper Fig.~\ref{fig:method} follows one window of ZEDD scene val\_0038, the third row of Fig.~\ref{S-fig:maps},
at f/16 in its held-out fold, with the estimator, gain, focus calibration and search to 9.5\,px of main-paper
Table~\ref{tab:est} (\texttt{realdata/\allowbreak fig\_method\_data.py}, \texttt{results/\allowbreak
fig\_method\_data.npz}; its joint-fit estimates agree with the stored depth map to $6\times10^{-8}$\,D). The two pairs
are pair 4 (runs 4 and 5, focus 2.61\,m, step 0.113\,D) and pair 1 (runs 1 and 2, 0.91\,m, 0.500\,D), the 2.62 and
0.92\,m pairs of the text at the
focus and step calibrated in this fold. Panel (a) uses the first because the window lies 0.14\,D from its focus, where
the five band ratios differ clearly; both estimates err by about 8\% there, and the panel is not evidence that the
joint fit is more accurate. The window was chosen by a fixed rule among the 556 windows of the brick wall (centers in
rows 300--1500 and columns 200--1500) that enter the samples of main-paper Table~\ref{tab:est} in both pairs: the
smallest sum, over the joint fit and one ratio on both pairs, of the absolute difference between the logarithm of the
relative error and the logarithm of its median over the 556 windows (medians 7.9\% and 7.3\% on the near pair, 7.3\%
and 46.3\% on the far pair). It is centered at row 672, column 512, at a LiDAR depth of 4.05\,m (0.247\,D). On the near
pair it lies 0.136\,D from focus (1.3\,px of blur), where the joint fit gives 0.125\,D and one ratio 0.126\,D; on the
far pair 0.856\,D (8.0\,px), where they give 0.789 and 0.453\,D. The curves carry the gain of the fold, 0.82 for the
five bands and 0.80 for one ratio, whose band is the sum of the five. The band ratios, converted to diopters by
main-paper Eq.~\eqref{eq:psiapp}, are 0.096, 0.077, 0.078, 0.067 and 0.060\,D from the coarsest to the finest band on
the near pair (0.062\,D for their sum) and 0.53, 0.44, 0.18, 0.016 and 0.012\,D on the far pair (0.071\,D). On the far
pair the more blurred exposure is past the first zero of the clear-disc OTF in the four finer bands; their gray regions
in panel (b) begin where the defocus plus half the step reaches that zero, at 0.80, 0.49, 0.28 and 0.14\,D (1.29\,D
for the coarsest band). The photographs of each panel are the two exposures of the pair, at blur radii of 1.7 and
0.6\,px and of 10.3 and 5.6\,px from the LiDAR depth and the focus of each run as calibrated in this fold. The defocus
of a pair is measured from its own midpoint, the zero crossing of its Laplacian ratio (\hyperref[S-par:focuscal]{focus
calibration}), which for pair 4 lies 0.011\,D closer to the camera than the mean of the foci of its two runs (1.3
against 1.2\,px of blur). The five photographs at the top of main-paper
Fig.~\ref{fig:overlap}b show this window in five of the nine f/16 exposures of the scene, at blur radii of 0, 1.7,
3.3, 5.6 and 10.3\,px (\texttt{realdata/\allowbreak fig\_depth\_data.py}, \texttt{results/\allowbreak
fig\_depth\_data.npz}), above their five bands through the filters of the estimator (\texttt{fig\_theory\_data.py},
\texttt{results/\allowbreak fig\_theory\_data.npz}). Fig.~\ref{S-fig:depthold}a places four of them where a
camera focused at 1\,m would blur them as much: 10.3\,px at 0.47\,m, the focused one at 1\,m, 3.3\,px at 1.55\,m and
5.6\,px at 2.54\,m. The two pictures under the axis of Fig.~\ref{S-fig:evidence}b show the window in focus,
in the 0.225 and 0.0675\,cycles/px bands.

\begin{figure*}[tp]
\centering
\includegraphics[width=\textwidth]{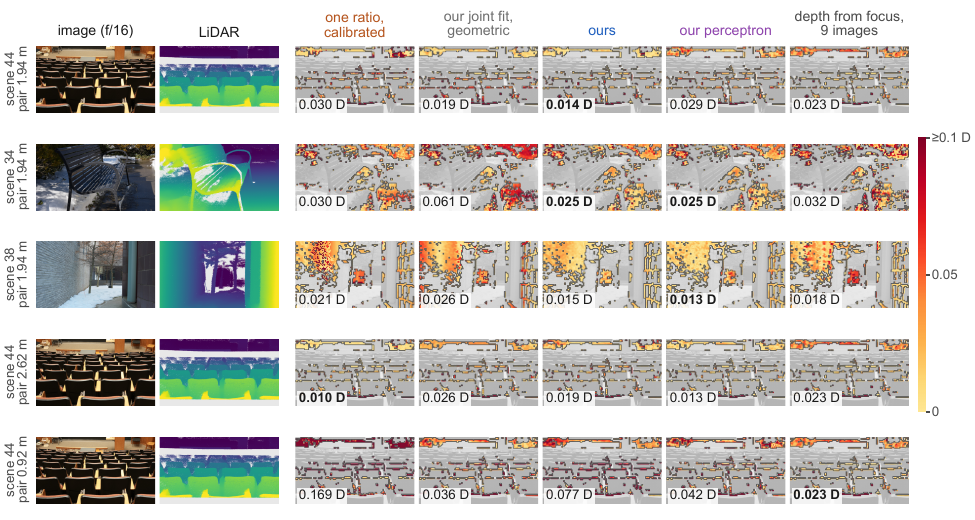}
\caption{\textbf{One image pair, more estimators.} DSLR at f/16, held-out folds. Rows 1--3: the scenes of
Fig.~\ref{S-fig:maps} (ZEDD 44, 34 and 38) with the pair focused at
1.94\,m, the pair on which ours does best against depth from focus; row 4: scene 44 with the pair focused at 2.62\,m,
on which ours loses to one ratio in 32 of 50 scenes and beats depth from focus in 10, this scene among them; row 5:
scene 44 with the pair focused at 0.92\,m, on which depth from focus beats ours in all 50 scenes, and in this row
geometric curves (0.036\,D) and the perceptron (0.042\,D) also beat ours (0.077\,D). Depth from focus uses all nine
images of a scene, so its map is the same in rows 1, 4 and 5. The rows are not typical. Over all 400 maps (50 scenes,
eight pairs) ours loses to the perceptron in 62\% and to depth from focus in 66\%; with the 1.94\,m pair it beats them in 23 and 33 of 50
scenes, with the 0.92\,m pair in 6 and 0. Errors of $1/Z$ against LiDAR with their medians (the lowest of each row in
bold), on the textured windows (outlined): one ratio with a calibrated curve, the joint fit with geometric curves,
ours, the perceptron on the same band statistics, and depth from focus over the nine f/16 focus settings of the scene,
which is not a one-pair method.}
\label{S-fig:maps_more}
\end{figure*}

\Para[S-par:mapsmore]{More estimators on the same maps}
Figure~\ref{S-fig:maps_more} adds the perceptron of main-paper Table~\ref{tab:est}, trained per fold on the band
statistics of the training scenes, and depth from focus over the nine f/16 images of the scene (peak of the
0.10\,cycles/px energy, a parabola through three focus settings; \texttt{zedd\_pair\_maps.py --more},
\texttt{results/zedd\_pair\_maps\_f16\_fold\_more.txt}), on the same windows. Over all 400 maps ours is more accurate than
the perceptron in 38\% (medians of the map medians 0.0273 against 0.0235\,D) and than depth from focus, which uses
nine images, in 34\% (0.0203\,D); in the maps with more than 30\% of their windows beyond 0.3\,D, in 27\% and 16\%. The
2.62\,m pair, whose f/16 focus is off (Errors per pair above), is the pair on which ours does worst against one ratio:
it beats one ratio with a calibrated curve in 18 of 50 scenes, the perceptron in 6 and depth from focus in 10, against
43, 23 and 33 with the 1.94\,m pair. Against the other two methods ours does as badly or worse on the 0.92\,m pair
(perceptron 6, depth from focus 0 of 50 scenes) and against depth from focus also on the 5.80\,m pair (1 of 50). The
first three rows of Fig.~\ref{S-fig:maps_more} show the scenes of Fig.~\ref{S-fig:maps} with the 1.94\,m pair: ours errs by
0.014, 0.025 and 0.015\,D, the perceptron by 0.029, 0.025 and 0.013\,D, depth from focus by 0.023, 0.032 and
0.018\,D, one ratio by 0.030, 0.030 and 0.021\,D and geometric curves by 0.019, 0.061 and 0.026\,D. The fourth row
shows the 2.62\,m pair on scene 44, where ours errs by 0.019\,D against
0.010\,D for one ratio and 0.013\,D for the perceptron; both of these calibrate each pair on the training scenes and so
absorb its focus error. The fifth row shows the 0.92\,m pair on the same scene, a pair on which depth from focus beats
ours in all 50 scenes: ours errs by 0.077\,D, depth from focus by 0.023\,D, the perceptron by 0.042\,D, geometric
curves by 0.036\,D and one ratio by 0.169\,D. With this pair the windows of the scene lie 0.55--1.02\,D beyond focus,
where our estimates of the pair fall short (Errors per pair above).

\begin{table}[htbp]
\centering
\caption{Depth from one image pair at f/16: held-out median $|\hat q-q_{\mathrm{LiDAR}}|$ [D] of textured windows by
their distance $|q-m_k|$ from the focus of the pair, over all scenes and pairs. One ratio: conventional DfDD with calibrated
$\alpha$ and $\beta$ (conv.), a curve calibrated per pair, and our curve with one gain
(\texttt{zedd\_pair\_maps\_bb.py}, \texttt{results/zedd\_pair\_maps\_f16\_fold\_bb.txt}).}
\label{S-tab:pairmaps}
\begin{tabular}{lcccccr}
\toprule
& \multicolumn{3}{c}{one ratio} & \multicolumn{2}{c}{our joint fit} & \\
\cmidrule(lr){2-4}\cmidrule(lr){5-6}
$|q-m_k|$ & conv.\ $\alpha,\beta$ & calib.\ curve & our curve & geometric & wave (ours) & windows \\
\midrule
0--0.1\,D & 0.023 & 0.021 & 0.018 & 0.018 & 0.018 & 328\,737 \\
0.1--0.2\,D & 0.039 & 0.030 & 0.020 & 0.030 & 0.020 & 262\,320 \\
0.2--0.4\,D & 0.090 & 0.076 & 0.040 & 0.043 & 0.029 & 398\,475 \\
0.4--0.6\,D & 0.218 & 0.183 & 0.228 & 0.058 & 0.045 & 224\,316 \\
0.6--1.0\,D & 0.474 & 0.429 & 0.505 & 0.050 & 0.063 & 158\,757 \\
\bottomrule
\end{tabular}
\end{table}

\Para[S-par:learned]{Physical depth and learned depth}
FOSSA~\cite{S-zuo2026} predicts dense depth from five f/2.8 images of a ZEDD focal stack without training on ZEDD; its
released weights reproduce the authors' errors (Table~\ref{S-tab:learned}, first row). Given the images that our
estimator uses, FOSSA errs on the windows with texture by 0.11--0.14\,D from one f/16 pair, against 0.027\,D for our
one-pair estimate, and by 0.041--0.047\,D from all nine f/16 images, against 0.011\,D from three pairs per window
(Table~\ref{S-tab:sameinput}). These inputs lie outside FOSSA's training, which relies on strong defocus (from its own
f/2.8 images it reaches 0.017--0.020\,D), so the gap reflects its training data rather than a limit of learning. We
correct each FOSSA map by one affine map of the inverse depth per scene~\cite{S-ranftl2022}, fitted with a Huber loss to
our f/16 estimates from three pairs per window on the textured windows, an image-only mask, with the focus of every
fold calibrated on its training scenes. Over all LiDAR pixels the abs.\ rel.\ falls from 0.101 to 0.074 for ViT-S and
from 0.095 to 0.060 for ViT-B (Table~\ref{S-tab:learned}), and outside the anchor windows from 0.110 to 0.082 and from
0.092 to 0.062. With anchors from depth from focus over the same nine images the error is higher: the ratio of the
scene means with our anchors to those with depth-from-focus anchors is 0.86 [0.78, 0.96] and 0.82 [0.74, 0.93].
Wave-optics and geometric curves give equal anchors (ratios 1.01 and 1.00, 95\% intervals within 0.96--1.04), so the
improvement comes from the joint fit (\texttt{realdata/fossa\_fusion\_table.py}, \texttt{results/fossa\_fusion\_table\_f16.0\_fold.txt}).

\begin{table*}[htbp]
\centering
\caption{\textbf{Physical estimates correct learned depth.} FOSSA~\cite{S-zuo2026} on the 50 ZEDD validation scenes,
corrected by one affine map of the inverse depth per scene fitted to the f/16 anchors (three pairs per window, held-out
folds); errors over all LiDAR pixels, and the ratio of the scene-mean abs.\ rel.\ to that of FOSSA without correction
[95\% interval over 4\,000 resamples of the scenes].}
\label{S-tab:learned}
\small
\begin{tabular}{lcccccc}
\toprule
& \multicolumn{3}{c}{FOSSA ViT-S} & \multicolumn{3}{c}{FOSSA ViT-B} \\
\cmidrule(lr){2-4}\cmidrule(lr){5-7}
Anchors & abs.\ rel. & $\delta_{1.25}$ & ratio & abs.\ rel. & $\delta_{1.25}$ & ratio \\
\midrule
none & 0.101 & 0.888 & -- & 0.095 & 0.926 & -- \\
depth from focus & 0.086 & 0.910 & 0.84 [0.75, 0.94] & 0.073 & 0.934 & 0.76 [0.64, 0.89] \\
one ratio, calibrated curve & 0.080 & 0.944 & 0.79 [0.66, 0.95] & 0.069 & 0.964 & 0.72 [0.58, 0.88] \\
our joint fit, geometric & 0.073 & 0.945 & 0.71 [0.63, 0.81] & 0.060 & 0.965 & 0.63 [0.54, 0.71] \\
joint fit, wave (ours) & 0.074 & 0.946 & 0.72 [0.63, 0.82] & 0.060 & 0.967 & 0.63 [0.54, 0.72] \\
\midrule
LiDAR (oracle) & 0.063 & 0.957 & 0.62 [0.54, 0.71] & 0.049 & 0.973 & 0.51 [0.44, 0.59] \\
\bottomrule
\end{tabular}
\end{table*}

\Para{FOSSA given the same images}
We ran FOSSA with its released weights and preprocessing on the two f/16 images of each of the eight pairs, on all nine
f/16 images and on five of them (runs 1, 3, 5, 7, 9), and averaged its inverse depth over our windows
(Table~\ref{S-tab:sameinput}). FOSSA was trained on five-image stacks with stronger defocus, so these inputs lie outside
its training setting. The one-pair maps are evaluated on every window with texture and uniform LiDAR depth, without a
restriction to the working range in defocus; with all windows of valid LiDAR depth, uniform or not, the medians rise by 5--11\% for
every method and the ranking is unchanged. Near the focus of a pair ($|q-m_k|<0.1$\,D) our one-pair estimate errs by
0.018\,D, close to FOSSA on its own f/2.8 protocol (0.015--0.018\,D); 0.7--1.1\,D from it, by 0.066\,D, against 0.24--0.42\,D
for FOSSA on the same two images and 0.63\,D for conventional DfDD with a calibrated curve
(\texttt{results/fossa\_same\_input\_fold.txt}).

\Para{Calibrated focus distances for FOSSA}
FOSSA takes the focus distance of every image as an input, and the protocol of its authors supplies the lens readings,
which differ from the LiDAR frame by up to 0.12\,D (Table~\ref{S-tab:calib}). Given instead the calibrated focus
distances $1/q^*_k$ that our estimator uses (\texttt{realdata/fossa\_zedd\_infer.py --foci}), FOSSA changes little. On
its own f/2.8 protocol the abs.\ rel.\ over all LiDAR pixels is 0.099 and 0.098 for ViT-S and ViT-B (0.101 and 0.095
with the lens readings) and the median $|dq|$ on our windows 0.016 and 0.020\,D; from one f/16 pair it errs by 0.148
and 0.117\,D, from the nine f/16 images by 0.048 and 0.043\,D, and from five f/16 images by 0.047 and 0.047\,D. Its
errors change by $-6$ to $+11\%$ against those of Table~\ref{S-tab:sameinput}. The anchors of Table~\ref{S-tab:learned}
lower its abs.\ rel.\ by 27\% (ratio of scene means 0.73 [0.62, 0.86]) and 40\% (0.60 [0.51, 0.70]), and the ratio to
the error with anchors from depth from focus is 0.86 [0.78, 0.96] and 0.82 [0.74, 0.93]
(\texttt{results/fossa\_same\_input\_calfd\_fold.txt}, \texttt{fossa\_fusion\_table\_f16.0\_fossa\_calfd\_fold.txt}).

\begin{table}[htbp]
\centering
\caption{Physical estimates and FOSSA given the same images, on the same windows with texture: median
$|1/\hat z-1/z|$ [D], $\delta_{1.25}$, and the number of scene maps in which ours has the lower median error. Physical
estimates on held-out folds, with the focus calibrated on the training scenes.}
\label{S-tab:sameinput}
\small
\setlength{\tabcolsep}{3.5pt}
\begin{tabular}{llccc}
\toprule
Images & Method & median $|dq|$ & $\delta_{1.25}$ & ours better \\
\midrule
one f/16 pair & joint fit, wave (ours) & \textbf{0.027} & \textbf{0.78} & -- \\
& one ratio, calibrated curve & 0.060 & 0.53 & -- \\
& FOSSA ViT-S / ViT-B & 0.142 / 0.113 & 0.27 / 0.34 & 353 / 330 of 400 \\
\midrule
nine f/16 images & our joint fit, wave, three pairs per window & \textbf{0.011} & \textbf{0.96} & -- \\
& one ratio, calibrated curve, three pairs & 0.011 & 0.95 & -- \\
& depth from focus & 0.018 & 0.90 & -- \\
& FOSSA ViT-S / ViT-B & 0.047 / 0.041 & 0.74 / 0.79 & 48 / 45 of 50 \\
& FOSSA ViT-S / ViT-B, five f/16 images & 0.045 / 0.043 & 0.75 / 0.79 & -- \\
\midrule
five f/2.8 images & FOSSA ViT-S / ViT-B (own protocol) & 0.017 / 0.019 & 0.94 / 0.95 & 35 / 36 of 50 \\
\bottomrule
\end{tabular}
\end{table}

\Para[S-par:correction]{Correction of learned depth}
{\emergencystretch=3em
The script \texttt{fossa\_zedd\_infer.py} runs FOSSA~\cite{S-zuo2026} with its released code, and
\texttt{fossa\_fusion.py} computes the correction of Table~\ref{S-tab:learned}. With f/5.6 anchors (three pairs per window), the
ratio of the scene-mean abs.\ rel.\ to that of FOSSA is 0.79 [0.67, 0.95] for ViT-S and 0.70 [0.58, 0.84] for ViT-B (95\% intervals)
with sub-band anchors, against 0.75 and 0.65 with depth from focus. At f/4 the sub-band ratios are 0.86
[0.70, 1.07] and 0.77 [0.61, 0.95], against 0.78 and 0.69 for depth from focus; broadband anchors with a calibrated
curve do not help there (1.01 and 0.94). At these apertures our anchors are thus no better than those from depth from
focus (ratio 1.06--1.11, every interval containing 1; \texttt{results/fossa\_fusion\_table\_f5.6\_fold.txt},
\texttt{fossa\_fusion\_table\_f4.0\_fold.txt}). Wave-optics and geometric curves give nearly the same anchors at every
aperture (ratios 1.00--1.03). With anchors from one f/16 image pair (\texttt{fossa\_fusion\_pair.py}, held-out folds), the
correction raises the error on average over the eight pairs (mean ratio 1.30--1.33 with wave-optics curves and
1.17--1.19 with geometric curves). Only the pair focused at 1.94\,m lowers it with both curves (0.945--0.993 with
wave-optics and 0.861--0.897 with geometric curves), and with geometric curves also the pair at 4.85\,m
(0.954--0.983). Sub-band anchors still give 1.7 to 4.7 times lower error than broadband anchors with calibrated
$\alpha$ and $\beta$. Keeping only anchors with an estimated defocus below 0.3\,D makes the correction worse (mean
ratio 4.0--4.7), because these anchors span a narrow range of depth (\texttt{results/fossa\_fusion\_pair\_fold.txt},
\texttt{fossa\_fusion\_pair\_psi03\_fold.txt}). A smooth residual field added to the affine map (normalized
Gaussian convolution of the anchor residuals, $\sigma=60$--240 LiDAR pixels) changes the f/16 abs.\ rel.\ by $-0.002$
to $+0.003$ (\texttt{results/fossa\_fusion\_f16\_fold.txt}), so we use the affine map alone.\par}

\Para{Caveats}
The JPEGs carry unknown in-camera processing. A linear filter shared by the two images of an image pair largely
cancels in the per-band statistics; nonlinear processing may not. The controls at f/4 and f/5.6 follow the models
within 6\%, above them in 10 of 12 cases (Table~\ref{S-tab:controls}; Fig.~\ref{S-fig:evidence}c shows the
0.225\,cycles/px band; at f/2.8 in the narrowest window the
data lie 13--25\% above them, inside wide intervals), the simulated
chain of noise, compression and resampling leaves the small-step f/16 test intact (Table~\ref{S-tab:pipesim}), and so
does a chain with color sampling, edge-directed demosaicing and contrast-dependent denoising, up to a denoiser that
smooths the highest band more than the camera (Table~\ref{S-tab:nonlin}). The in-focus diffraction loss is 0.4--0.7
of the clear-disc value; the sharpening we simulated explains it only at the high bands. Other caveats are LiDAR
registration at depth edges (hence the depth-uniform windows) and one lens.

\clearpage
\section{A DfDD prototype: three iris diameters}
\label{S-sec:cod}

The cameras of the two sections above were not built for DfDD. The authors of the coupled-differentiation
prototype~\cite{S-luo2025cod} have released its data, and we fixed a test of main-paper Lemma~\ref{lem:disc} on them
before computing any spatial-frequency statistic of one of its images. The record,
\texttt{notes/\allowbreak preregistration\_\allowbreak cod.md}, dates from 2026-11-13 and comes with the SHA-256 of
33 files. The record is local, so we call the test
pre-specified and not registered (\hyperref[S-par:localrecords]{Sec.~\ref{S-sec:pixel}}). By its rule the outcome is
undecided; we report it in full. An amendment, fixed before the second
released file was analyzed, repeats the test there with three corrections and adds depth from one image pair
(\hyperref[S-par:codamend]{end of this section}). On that file the cutoff scale is undecided again, and the estimator
calibrated on the first file has a median defocus error of 4.7\% and 6.1\% in the two bins of blur
(Table~\ref{S-tab:coddepth}). The main paper reports both in Sec.~\ref{sec:proto}, with the slopes of
Table~\ref{S-tab:cod} in its Fig.~\ref{fig:proto}b (\hyperref[S-par:codfig]{below}).

\Para{Data}
A printed texture on a motorized slide is imaged through a motorized iris of diameter 3, 5 and 7\,mm and a tunable
lens at three optical powers 0.06\,D apart (1.14, 1.20 and 1.26\,D, added to a fixed lens of 8.9\,D), on a
monochrome sensor with $5.86\,\mu$m pixels. With the sensor distance of 110\,mm in the released code the apertures
are f/36.7, f/22 and f/15.7. The file we use holds 97 distances from 0.45 to 1.41\,m in steps of 10\,mm, each with
the nine combinations of diameter and optical power as crops of $400\times400$ pixels. Our six bands, 0.03 to
0.225\,cycles/px, lie at 0.10--0.77 of the nominal cutoff at 3\,mm, at 0.06--0.46 at 5\,mm and at 0.04--0.33 at 7\,mm.
The three diameters move the cutoff by a factor of 2.3 with the lens, the sensor, the target and the processing
unchanged, and at f/36.7 a simple lens is limited by diffraction, so a fall caused by aberrations or by the sensor
would not follow the diameter. Before the record we had read the structure of the files, the stored settings and
intensity ranges, the released code and two contact sheets of 16 images each. Beyond 1.15\,m the edge of the target
enters the crop, so the test uses the 66 distances up to 1.10\,m. A second released file, with 97 distances from 0.91
to 1.87\,m and a coarser print that fills the crop at every distance, was left by the record to an amendment
(\hyperref[S-par:codamend]{end of this section}).

\Para{Statistic and rule}
For an iris diameter and a pair of optical powers $\delta$ apart, each band gives at every distance the apparent
defocus of main-paper Eq.~\eqref{eq:psiapp}, with $r_b$ of Eq.~\eqref{eq:bw_ratio} summed over the whole crop
after subtraction of a noise floor. The floor of each sum is that sum for white noise at the level measured in the
same image at $|\vf|\ge0.64$\,cycles/px, the corners of the spectrum. The slope $\kappa_b$ of the apparent defocus
against $1/Z$ is taken over the distances at which the geometric blur phase $2\pi f_bRs|\psi|$ stays below 2.4 for
both images, at most 0.12\,D from the focus of the pair. A cell (diameter, pair, band) is used if this window holds
at least five distances, two or more on each side of focus, if the band energy is at least three times its floor and
if the slope is positive. The model slopes come from the same regression on the apparent defocus of a clear disc
whose cutoff is $\eta$ times the nominal $2R/(\lambda s)$ (exact overlap integral, the seven wavelengths of the DSLR
experiment with a mean of 544\,nm, the pixel aperture, a texture spectrum $f^{-2}$ and the finite step), and of
geometric optics and a Gaussian pupil. For each diameter a gain and the scale $\eta$ minimize the squared log
residuals of the cells; the gain absorbs the errors of $R$, $s$ and $\delta$. The joint fit has one $\eta$ and one
gain per diameter. The intervals come from 2\,000 resamples of the distances in blocks of three neighbors.
\begin{itemize}
  \item \emph{Predictions.} At 3\,mm and $\eta=1$ the slopes in the bands 0.03 to 0.10\,cycles/px are 0.78, 0.69,
  0.56 and 0.40 of $R^2/4$, where geometric optics gives 1.02--1.05 and a Gaussian pupil 0.93--1.00.
  \item \emph{Rule A, the cutoff scale.} The joint $\eta$ of the two pairs 0.06\,D apart is confirmed if its 95\%
  interval lies inside 0.80--1.25, refuted if the interval lies entirely outside 0.67--1.50, and undecided otherwise.
  \item \emph{Rule B, the same scale at two apertures.} Confirmed if the intervals of $\eta$ at 3\,mm and at 5\,mm
  both lie inside 0.80--1.25, refuted if either lies entirely outside 0.67--1.50, and undecided otherwise. The
  7\,mm diameter, whose eligible bands reach 0.15 of the cutoff, is reported beside them.
  \item \emph{Tolerance.} It covers what the release does not document, namely the effective wavelength of the source
  on a monochrome sensor (520--600\,nm moves $\eta$ from 1.05 to 0.91), the sensor distance, the iris diameter
  (0.1--0.2\,mm, 7\% at 3\,mm) and the texture spectrum. Geometric optics and a Gaussian pupil have no cutoff and put $\eta$ at the
  end of our table, 2.5.
  \item \emph{Synthetic checks.} The same code on synthetic images of the 97 distances with a known pupil, the signal
  levels of the real images, shot noise and quantization gives $\eta=0.98$--1.05 and ``confirmed'' for renderings in
  wave optics with texture spectra $f^{-1}$ to $f^{-3}$, with changes of brightness and magnification between the
  optical powers and with a displaced focus; 1.05 [0.90, 1.31] and ``undecided'' with three times the noise; 0.853
  [0.829, 0.878] for a rendering with $\eta=0.85$; and 2.5 and ``refuted'' for geometric optics and for a Gaussian
  pupil. The focus rule was replaced once on synthetic pairs before the record.
\end{itemize}

\Para[S-par:codoutcome]{Outcome}
No step failed and nothing was changed after the record (\texttt{results/\allowbreak cod\_test\_A.txt}; 24 cells, 28
distances; the 33 frozen files unchanged after the run). \emph{Both outcomes are undecided.} The joint scale is
1.199 [1.134, 1.273], whose upper end passes 1.25. At 3\,mm the scale is 1.243 [1.164, 1.330] and at 5\,mm 1.075
[0.984, 1.141], inside the tolerance; at 7\,mm it is 1.100 [0.995, 1.332]. Table~\ref{S-tab:cod} gives the slopes. At
3\,mm they fall from 0.61--0.64 to 0.37--0.38 between 0.10 and 0.34 of the cutoff, where wave optics with the nominal
cutoff predicts 0.78 to 0.40 and geometric optics 1.02--1.06 at the distances used. The pre-specified secondary
outcomes follow.
\begin{itemize}
  \item \emph{Residuals.} After the best gain, the rms log residual at 3\,mm is 0.017 for wave optics with a free
  scale, 0.062 with the nominal cutoff, 0.200 for geometric optics and 0.164 for a Gaussian pupil. At 5\,mm the four
  values are 0.016, 0.019, 0.111 and 0.100, and at 7\,mm 0.016, 0.018, 0.069 and 0.067. A constant coefficient misses
  the slopes more the smaller the aperture.
  \item \emph{Gains.} 0.750, 0.736 and 0.724 at 3, 5 and 7\,mm. The level of the slopes is thus 26\% below the nominal
  $(Rs)^2\delta$, and the three radii are in their nominal ratio to within 2\%.
  \item \emph{Variants.} The four quadrants of the crop give a joint scale of 1.177, 1.093, 1.243 and 1.265, the
  pair 0.12\,D apart 1.232 (9 cells) and all three pairs 1.193 (33 cells). Without the noise floor the joint scale is
  1.031 [0.988, 1.081].
\end{itemize}

\begin{table}[htbp]
\centering
\caption{\textbf{A DfDD prototype: the slope of the apparent defocus falls with frequency at three iris diameters.}
Slopes of the apparent defocus against $1/Z$ in units of $R^2/4$, for the two pairs of optical powers
0.06\,D apart (24 cells), with the band energy relative to its noise floor and the number of distances; and the
model slopes of a clear disc in wave optics with the nominal cutoff ($\eta=1$) and in geometric optics. The fitted
gains are 0.750, 0.736 and 0.724 (\texttt{results/cod\_test\_A.txt}).}
\label{S-tab:cod}
\footnotesize
\setlength{\tabcolsep}{5pt}
\begin{tabular}{lccccccc}
\toprule
iris & band [cycles/px] & $f/f_c$ & distances & energy / floor & measured & wave, $\eta=1$ & geometric \\
\midrule
3\,mm (f/36.7) & 0.030 & 0.10 & 23 / 20 & 591 / 712 & 0.639 / 0.609 & 0.78 & 1.02 \\
 & 0.045 & 0.15 & 23 / 20 & 366 / 349 & 0.560 / 0.547 & 0.69 & 1.03 \\
 & 0.0675 & 0.23 & 23 / 20 & 66 / 56 & 0.476 / 0.482 & 0.56 & 1.06 \\
 & 0.100 & 0.34 & 20 / 17 & 8.6 / 8.4 & 0.379 / 0.371 & 0.40 & 1.04 \\
\midrule
5\,mm (f/22) & 0.030 & 0.06 & 23 / 20 & 1729 / 2019 & 0.670 / 0.665 & 0.89 & 1.04 \\
 & 0.045 & 0.09 & 23 / 20 & 1102 / 1013 & 0.614 / 0.608 & 0.84 & 1.06 \\
 & 0.0675 & 0.14 & 16 / 15 & 211 / 174 & 0.549 / 0.553 & 0.73 & 1.02 \\
 & 0.100 & 0.20 & 9 / 8 & 36 / 33 & 0.460 / 0.450 & 0.60 & 0.96 \\
\midrule
7\,mm (f/15.7) & 0.030 & 0.04 & 22 / 19 & 4029 / 4668 & 0.691 / 0.697 & 0.94 & 1.05 \\
 & 0.045 & 0.07 & 19 / 16 & 2739 / 2642 & 0.643 / 0.641 & 0.89 & 1.03 \\
 & 0.0675 & 0.10 & 10 / 9 & 520 / 507 & 0.599 / 0.588 & 0.78 & 0.97 / 0.96 \\
 & 0.100 & 0.15 & 5 / 5 & 87 / 88 & 0.523 / 0.508 & 0.70 / 0.69 & 0.96 / 0.93 \\
\bottomrule
\end{tabular}
\end{table}

\Para[S-par:codposthoc]{Why the scale comes out high (post hoc)}
Nothing in this paragraph was fixed in advance. The scale at 3\,mm rests on the 0.10\,cycles/px band, whose energy is
8.6 times the pre-specified floor, so it depends on that floor. Without the floor the joint scale is 1.03, with it
1.20.
\begin{itemize}
  \item \emph{The noise is not white.} At the distances nearest the focus of a pair, its two images show the same
  scene at the same defocus. Half the spectrum of their difference is 0.50 of the corner level below
  0.12\,cycles/px and rises to 1.05 at 0.35--0.50\,cycles/px, so the corners, where the floor was measured, hold twice
  the noise of the bands. A pattern fixed on the sensor adds 0.03--0.10 of the corner level in the bands from
  0.0675\,cycles/px upward at 3\,mm. The pre-specified floor was therefore about 1.7 times too high in the band
  that decides. With the floor measured in the bands the joint scale is 1.114 [1.019, 1.210], and the scales at 3 and
  5\,mm are 1.123 [1.015, 1.238] and 1.036 [0.955, 1.093], all inside 0.80--1.25. The rms log residual at 3\,mm is
  then 0.017 for wave optics, 0.320 for geometric optics and 0.297 for a Gaussian pupil (9 cells).
  \item \emph{The camera sharpens.} The camera of the prototype is a Grasshopper3 GS3-U3-23S6M~\cite{S-luo2025cod},
  and the release does not give its settings. Its technical reference~\cite{S-ptgrey2015} describes a sharpness filter
  computed on $3\times3$ pixels and enabled by default. The noise spectrum of the 3\,mm images follows the transfer
  function of such a filter, the image plus $k$ times its difference from the $3\times3$ mean, with $k=0.51$, to
  3\% rms over 13 rings from 0.02 to 0.64\,cycles/px, where white noise misses by 43\%
  (\texttt{realdata/\allowbreak cod\_posthoc5.py}). The filter multiplies the noise power by 2.14 in the corners and
  by 1.01--1.16 in the four bands used at 3\,mm. A linear filter enters the test only through its variation within a
  band and through the noise floor. Synthetic images of a clear disc with $\eta=1$ passed through this filter give,
  with the pre-specified floor, a joint scale of 1.027 [0.988, 1.052] for the texture spectrum $f^{-2}$, 1.071
  [1.024, 1.103] for $f^{-2.4}$, where the 0.10\,cycles/px band has 8.4 times the energy of its floor as in the real
  images, and 1.180 [1.053, 1.266] for $f^{-3}$; without a floor they give 0.93--0.95. In a camera that sharpens, the
  pre-specified floor thus raises the scale by 3--18\%, and no floor lowers it by 5--7\%.
  \item \emph{The response is linear.} At the three diameters the noise variance in the corners is proportional to
  the image mean above a pedestal of 5.6 counts, within 2\%, as shot noise is on a linear scale, and we take the
  stored values as linear in the light. The same reference enables a gamma of 1.25 by default; added to the
  synthetic images it changes the joint scale by less than 0.005.
  \item \emph{The nominal geometry.} Between neighboring slide positions the image shrinks by 1.006\% per 10\,mm at a
  nominal distance of 0.945\,m, where 1.058\% is expected, so the center of perspective lies 48\,mm farther from the
  target than the nominal distance. With the distances corrected, the focus moves by 0.0652\,D for one step of the
  optical power, 1.086 times the nominal 0.06\,D. With both corrections and the measured floor the joint scale is
  1.059 [0.982, 1.156]. For nominal iris radii the gains then put the sensor distance at 96.6\,mm [95.3, 97.9],
  against the 110\,mm of the released code. The code gives the fixed lens as 8.9\,D; a stock lens of 100\,mm focal
  length, with the tunable lens at 1.17--1.23\,D and the foci at the corrected distances, would have its image at
  98.3\,mm, within 2\% of this estimate, and would raise the nominal cutoff by 1.12. The product of $\eta$ and the
  square root of the gain does not contain the sensor distance; it is 0.930 [0.873, 1.001], against 1 for nominal
  radii and a mean wavelength of 544\,nm, and a mean wavelength of 585\,nm would make it 1. A sensor distance of
  96.6\,mm in place of 110\,mm lowers the gain to $(96.6/110)^2=0.77$ and raises the cutoff by $110/96.6=1.14$,
  inside the intervals of both pre-specified tests of the cutoff, 1.20 [1.13, 1.27] and 1.16 [1.07, 1.36].
  \item \emph{The image means.} Above the pedestal the image means of the three diameters are in the ratios 1 : 2.37
  : 3.71 and the noise variances in the ratios 1 : 2.46 : 3.77, where the nominal areas are 1 : 2.78 : 5.44. The
  modulation of the images follows the areas more closely. In focus the amplitudes of the 0.045\,cycles/px band are
  in the ratios 1 : 2.77 : 5.60, where clear discs of the nominal diameters at equal exposure give 1 : 3.01 : 6.10,
  and the gains of the slopes follow the nominal radii to 2\%. A shorter exposure or a smaller aperture would lower
  the mean and the modulation alike. The images of the two smaller openings therefore carry light that forms no
  image, at least 6 and 11 counts if the 7\,mm images carry none, which is 34\% and 25\% of their image-forming
  light. The paper of the prototype reports a smooth background light that changes with the iris and not with the
  optical power~\cite{S-luo2025cod}. Such light has no energy in the bands of the test; it adds photon noise, which the
  measured floor contains (\texttt{realdata/\allowbreak cod\_posthoc6.py}).
\end{itemize}
\Para[S-par:codamend]{The second file: an amendment fixed before its analysis}
The record left the second released file to an amendment. We wrote it after the analysis above and before computing
any spatial-frequency statistic of an image of that file
(\texttt{notes/\allowbreak preregistration\_\allowbreak cod\_\allowbreak amendment1.md}, with the SHA-256 of 147
files; a local record like the first). Before it we had read the stored settings of the file, its image means and
one contact sheet. The file lies almost wholly on the far side of the focus and reaches 0.46\,m farther than the
first.
\begin{itemize}
  \item \emph{Corrections, fixed from the first file.} The floor of each band sum is multiplied by the level of
  white noise passed through the camera's filter ($k=0.512$) in that band relative to its level in the corners,
  0.475 to 0.807 for the six bands. The distances receive an offset measured on the second file from the change of
  image scale between neighboring positions, the one use of its images before the band sums. The stored optical
  powers are multiplied by 1.086. With the three corrections the first file gives a joint scale of 1.038
  [0.973, 1.135], a post hoc number.
  \item \emph{Part 1, the cutoff scale.} The test of the record on the corrected sums, with rules A and B and their
  tolerances. Two settings follow from the range of the file. Distances up to 1.20\,m are used, because farther out
  the apparent defocus of a band reverses its sign a second time and the focus rule would take that for a focus.
  The fit uses the pair of the two lower optical powers, because the other pair has its focus before the first
  distance of the file.
  \item \emph{Part 2, depth from one image pair.} Every distance, diameter and pair of optical powers 0.06\,D apart
  is a sample of two images, 582 in all. The estimator of main-paper Eq.~\eqref{eq:bw_estimate} fits the defocus
  to the apparent defocus of the bands, weighted by their Laplacian energies, with the through-focus curves of a
  clear disc in wave optics, within 9.5\,px of blur. For each diameter it uses the bands that have an eligible cell
  in the first file (0.03 to 0.15\,cycles/px at 3 and 5\,mm, 0.03 to 0.10 at 7\,mm), the cutoff scale 1.038 and
  one gain, all from the first file. The comparisons are the same fit with the curves of geometric optics and of a
  Gaussian pupil, a single ratio over the same bands against its curve in wave optics or in geometric optics, and
  conventional DfDD (the linear depth equation, main-paper Eq.~\eqref{eq:depth_intro}), each with its own gain per diameter from the first file. The true defocus is $1/Z$
  minus the focus of the first file, shifted by one number for the second file, the median difference between the
  foci of the two files where the second has a valid one. Errors are relative defocus errors in two bins of
  blur, 1.5--3.5 and 3.5--9.5\,px, as in Table~\ref{S-tab:estfull}, with intervals from 2\,000 resamples of the distances in blocks of three.
  \item \emph{Rules of Part 2.} D1, wave optics against geometric optics, is confirmed if the ratio $R_1$ of the
  median error with geometric curves to that with wave curves over 1.5--9.5\,px has its interval above 1, and
  refuted if below. D2, the bands kept apart against one ratio, applies the same rule to the ratio $R_2$ of the
  single ratio to the joint fit, both in wave optics. D3 is confirmed if the median error of the joint fit with
  wave curves has its upper end below 10\% in both bins, and refuted if a lower end is above 10\%.
  \item \emph{Checks before the freeze.} On the first file cut to the distances of the second, the corrected test
  gives 0.998 [0.916, 1.238] and the estimator 2.6\% and 4.3\% in the two bins, in sample. Synthetic files with the
  camera's filter, displaced distances and a wrong power step pass both parts when rendered in wave optics (scale
  1.03 and 1.01 for two textures) and fail Part 1 and D1 when rendered in geometric optics (scale at the end of the
  table, $R_1=0.38$ [0.24, 0.59]). The amendment lists what was changed on the first file before the freeze,
  namely the bands of the fit, the focus rule, the range of D1 and D2 and the 10\% of D3.
\end{itemize}

\Para[S-par:codamendout]{Outcome on the second file}
No step failed and nothing was changed after the freeze (\texttt{results/\allowbreak cod\_amend1/\allowbreak B/};
the 147 frozen files unchanged after the run). The offset of the second file is 54.8\,mm, against 48.4\,mm for the
first.

\emph{Part 1 is undecided, as on the first file.} The joint scale is 1.164 [1.072, 1.360] from 12 cells and 20
distances. At 3\,mm it is 1.195 [1.080, 1.487], at 5\,mm 1.070 [0.910, 1.182], inside the tolerance, and at 7\,mm
1.069 with an interval that reaches the end of the table. The pair of the two higher powers found no focus, as
expected. We had predicted a scale near 1.0. The corrections lower it from 1.361 [1.191, 1.959], the value of the
record's procedure on this file, and do not bring it there. The pre-specified secondary outcomes follow.
\begin{itemize}
  \item \emph{Slopes and residuals.} At 3\,mm the slopes fall from 0.60 to 0.35 between 0.10 and 0.34 of the cutoff
  (wave optics with the nominal cutoff 0.78 to 0.40, geometric optics 1.01--1.05). After the best gain the rms log
  residual is 0.019, 0.016 and 0.018 at the three diameters for wave optics with a free scale, 0.205, 0.108 and
  0.067 for geometric optics and 0.175, 0.096 and 0.061 for a Gaussian pupil. The gains are 0.751, 0.727 and
  0.740.
  \item \emph{Variants.} The pair 0.12\,D apart gives 1.044 (9 cells), all three pairs 1.093 (21 cells) and the
  four quadrants 1.435, 1.207, 1.070 and 1.076. Without the floor correction the scale is 1.460 [1.273, 2.377],
  with the nominal distances 1.174 and with the nominal step 1.072.
\end{itemize}

\emph{Part 2: D1, D2 and D3 are confirmed.} Table~\ref{S-tab:coddepth} gives the errors. With wave curves the joint
fit has a median defocus error of 4.7\% [3.6, 7.1] and 6.1\% [4.6, 7.3] in the two bins, against 2.8\% and 3.7\% in
sample on the first file. Geometric curves give 11.0\% and 8.4\%, with $R_1=1.57$ [1.19, 2.02]; a single ratio
gives 13.2\% and 11.2\%, with $R_2=1.93$ [1.62, 2.63]; conventional DfDD gives 37\% and 20\%. The
pre-specified secondary outcomes follow.
\begin{itemize}
  \item \emph{By bin, diameter and pair.} $R_1$ is 2.37 [1.57, 4.09] below 3.5\,px and 1.39 [1.05, 1.86] beyond,
  $R_2$ 2.83 [1.59, 3.37] and 1.85 [1.57, 2.45]. The joint fit with wave curves has 8.0\% and 7.1\% at 3\,mm,
  4.5\% and 5.1\% at 5\,mm and 3.7\% and 2.8\% at 7\,mm; 3.5\% and 6.0\% for the pair of the two lower powers and
  7.7\% and 6.3\% for the other. A cutoff scale for each diameter changes its error by a factor 0.99 [0.88, 1.10].
  \item \emph{Near focus and in distance.} Below 1.5\,px of blur (29 samples) the median absolute error is 1.9\,mD
  with wave curves, 6.0\,mD with geometric curves and 2.7\,mD for the single ratio. The median distance error of
  the joint fit with wave curves is 3.2\,mm and 16\,mm in the two bins, at distances up to 1.5\,m.
  \item \emph{The focus.} The one number taken from the second file is the shift of the focus, $-0.0062$\,D. With
  the focus of the first file unshifted the joint fit with wave curves has 16.6\% [13.0, 20.0] and 9.0\%
  [7.8, 10.4], and D3 would be refuted. With the focus of each pair taken from the second file where it is valid,
  it has 7.2\% and 7.1\%.
  \item \emph{The single ratio refitted.} With its gains refitted on the second file itself the single ratio has
  6.8\% and 7.8\%, and $R_2$ is 1.25 [1.05, 1.79]. The print of the second file differs from that of the first,
  and the gain of a single ratio depends on the texture.
\end{itemize}
Three observations were not fixed in advance. First, the shift of the focus is what the difference between the two
offsets, 6.4\,mm, does to $1/Z$ at 1\,m. In the coordinates of the released code the focus of the pair lies at 0.958,
0.957 and 0.961\,m in the first file and at 0.960, 0.959 and 0.961\,m in the second. With the offset of the first
file and its foci unshifted, so that nothing of Part 2 comes from the second file, the joint fit with wave curves has
5.2\% [4.3, 7.1] and 6.9\% [5.0, 8.2], with $R_1=1.48$ [1.14, 1.81] and $R_2=1.89$ [1.57, 2.45]. Second, the offset
of the first file measured over the distances of the second is 47.7\,mm, so the two offsets differ by 6--7\,mm for
a reason we have not found. Third, the cutoff scale of the second file is 0.993 [0.924, 1.110] without any noise
floor, where synthetic images lose 5--7\% from leaving the floor out. On both files the scale thus depends on the
floor of the 0.10\,cycles/px band at 3\,mm. With the corrected distances and step it is 0.88 and 0.99 without a
floor, 1.04 and 1.16 with the corrected floor and 1.08 and 1.46 with the floor of the record, on the first and the
second file. The correction fitted on the noise of the first file leaves this dependence in place.

\begin{table}[htbp]
\centering
\caption{\textbf{A DfDD prototype: depth from one image pair on the second file, calibrated on the first.} Median
relative defocus error in percent over the samples of each bin of geometric blur radius, pooled over the three
iris diameters and the two pairs of optical powers, with 95\% intervals from 2\,000 resamples of the distances in
blocks of three. The joint fits and the single ratios use the same bands. The cutoff scale, the gains and the foci
come from the first file; the second file contributes one shift of the focus. The last row is a pre-specified
secondary outcome (\texttt{results/\allowbreak cod\_amend1/\allowbreak B/\allowbreak part2\_eval.txt}).}
\label{S-tab:coddepth}
\footnotesize
\setlength{\tabcolsep}{5pt}
\begin{tabular}{lccc}
\toprule
estimator & 1.5--3.5\,px (40 samples) & 3.5--9.5\,px (125) & 1.5--9.5\,px (165) \\
\midrule
joint fit, wave optics (ours) & 4.7 [3.6, 7.1] & 6.1 [4.6, 7.3] & 5.8 [4.6, 7.1] \\
our joint fit, wave optics, a cutoff scale per iris & 4.8 [3.6, 5.7] & 6.3 [4.4, 7.2] & 5.8 [4.5, 6.9] \\
our joint fit, geometric optics & 11.0 [9.6, 18.5] & 8.4 [6.7, 10.5] & 9.2 [7.1, 10.7] \\
our joint fit, Gaussian pupil & 12.3 [9.1, 16.8] & 14.6 [11.6, 17.7] & 13.8 [11.5, 16.7] \\
one ratio, our curve & 13.2 [7.8, 15.1] & 11.2 [9.8, 13.9] & 11.3 [10.1, 14.4] \\
one ratio, geometric optics & 17.2 [6.8, 28.0] & 9.0 [7.0, 10.7] & 9.9 [7.8, 13.5] \\
conventional DfDD, linear equation & 37.3 [18.7, 39.6] & 19.8 [11.6, 39.0] & 25.1 [14.4, 38.2] \\
\midrule
one ratio, our curve, gains refitted on this file & 6.8 [4.2, 9.2] & 7.8 [5.6, 11.5] & 7.3 [6.0, 10.4] \\
\bottomrule
\end{tabular}
\end{table}

\Para[S-par:codfig]{The slopes against the closed form (main-paper Fig.~\ref{fig:proto}b)}
The pre-specified test fits a cutoff scale. Main-paper Fig.~\ref{fig:proto}b shows the same slopes with nothing
fitted but one gain per iris. Each slope of Table~\ref{S-tab:cod}, and of the pair of the second file, is divided by the
gain that the clear disc with the nominal cutoff gives its iris (0.845, 0.749 and 0.740 on the first file, 0.823,
0.740 and 0.748 on the second) and set against the closed form of main-paper Lemma~\ref{lem:disc} at the nominal $f/f_c$
of the band (Table~\ref{S-tab:codcells}; post hoc tabulation, \texttt{realdata/\allowbreak cod\_figure\_numbers.py},
which reproduces the scales, gains and intervals of the two tests). The 24 cells of the first file deviate from the
closed form by 3.9\% rms and the 12 of the second by 3.6\%; the largest deviations, $-10\%$ at the lowest band and
$+7\%$ at the highest, are those of the 3\,mm iris, whose fall is shallower than predicted, which is what the cutoff
scale of 1.2 expresses. The model slopes of the test, which include the band shape, the pixel aperture and the finite
step, lie within 4\% of the closed form.

Two statements need no gain and no model of the cutoff. Between the 0.03 and the 0.10\,cycles/px band the slope falls
to 0.59 and 0.61 of its value at 3\,mm, 0.69 and 0.68 at 5\,mm and 0.76 and 0.73 at 7\,mm (the two pairs of the first
file; 0.59, 0.68 and 0.74 on the second), where clear-disc curves with the nominal cutoff predict 0.52, 0.67 and
0.73--0.74, geometric curves 1.01--1.03, 0.92--0.93 and 0.88--0.91, and those of a Gaussian pupil 0.93, 0.88--0.89 and
0.87--0.90. And cells of different irises whose nominal $f/f_c$ differ by less than 12\%, 16 pairs of cells on the
first file and 8 on the second, have slopes over gain in ratios of 0.97--1.07 and 0.96--1.09, 3.1\% and 4.6\% rms
from 1, although their bands differ by a factor of 1.5 to 2.25 in cycles/px. A fall set by the sensor, the print or
the processing would follow the frequency on the sensor; this one follows the cutoff of the iris.

\begin{table}[htbp]
\centering
\caption{\textbf{The slopes of main-paper Fig.~\ref{fig:proto}b.} Slope of the apparent defocus divided by the gain
of its iris diameter at the nominal cutoff, with 95\% intervals from the block bootstrap of the tests, for the two
pairs of the first file and the pair of the second; and the closed form of main-paper Lemma~\ref{lem:disc} at
the nominal $f/f_c$ (\texttt{results/\allowbreak cod\_figure\_numbers.txt}).}
\label{S-tab:codcells}
\footnotesize
\setlength{\tabcolsep}{5pt}
\begin{tabular}{@{}ccccccc@{}}
\toprule
iris [mm] & band [cycles/px] & $f/f_c$ & first file, lower pair & first file, upper pair & second file & Lemma 2 \\
\midrule
3 & 0.03 & 0.102 & 0.756 [0.733, 0.783] & 0.721 [0.660, 0.770] & 0.723 [0.673, 0.744] & 0.796 \\
3 & 0.045 & 0.153 & 0.662 [0.641, 0.691] & 0.647 [0.631, 0.666] & 0.670 [0.634, 0.714] & 0.704 \\
3 & 0.0675 & 0.230 & 0.564 [0.549, 0.584] & 0.571 [0.546, 0.590] & 0.581 [0.546, 0.619] & 0.578 \\
3 & 0.1 & 0.341 & 0.448 [0.430, 0.459] & 0.439 [0.420, 0.454] & 0.426 [0.407, 0.462] & 0.419 \\
\midrule
5 & 0.03 & 0.061 & 0.895 [0.869, 0.914] & 0.888 [0.877, 0.905] & 0.881 [0.853, 0.930] & 0.875 \\
5 & 0.045 & 0.092 & 0.820 [0.799, 0.835] & 0.812 [0.799, 0.822] & 0.811 [0.795, 0.831] & 0.816 \\
5 & 0.0675 & 0.138 & 0.733 [0.714, 0.747] & 0.739 [0.727, 0.750] & 0.743 [0.731, 0.763] & 0.731 \\
5 & 0.1 & 0.204 & 0.615 [0.600, 0.637] & 0.601 [0.578, 0.621] & 0.600 [0.577, 0.615] & 0.619 \\
\midrule
7 & 0.03 & 0.044 & 0.934 [0.916, 0.955] & 0.942 [0.924, 0.959] & 0.926 [0.909, 0.942] & 0.910 \\
7 & 0.045 & 0.066 & 0.869 [0.856, 0.875] & 0.867 [0.853, 0.882] & 0.864 [0.854, 0.877] & 0.866 \\
7 & 0.0675 & 0.099 & 0.810 [0.799, 0.819] & 0.794 [0.775, 0.807] & 0.803 [0.780, 0.821] & 0.803 \\
7 & 0.1 & 0.146 & 0.707 [0.688, 0.721] & 0.686 [0.666, 0.703] & 0.687 [0.651, 0.711] & 0.717 \\
\bottomrule
\end{tabular}
\end{table}

\Para[S-par:coddist]{Depth on the second file in units of distance}
Table~\ref{S-tab:coddepth} gives relative defocus errors, the statistic of the amendment. In distance, with
$\hat Z=1/(q_0+\hat\psi)$ for the focus $q_0$ of the pair, the median error of the joint fit with wave curves is
8.5\,mm [3.7, 18.2] over the 194 image pairs within 9.5\,px of geometric blur radius, 0.75\% of the distance, which
runs from 0.96 to 1.53\,m there; conventional DfDD errs by 27.7\,mm [18.2, 43.0], 2.45\%
(Table~\ref{S-tab:coddist}; post hoc tabulation of the saved estimates, \texttt{realdata/\allowbreak
cod\_depth\_units.py}). Within 1.5\,px of blur the median errors are 1.8\,mm and 4.8\,mm.

The prototype papers give a working range, the distances at which the mean depth error stays below 5\% or 10\% of
the depth. On this file the range of every estimator is bounded by the search of the amendment, which ends at
9.5\,px of blur, that is at 1.40--1.54\,m for the 3\,mm iris, 1.18--1.28\,m at 5\,mm and 1.10--1.19\,m at 7\,mm for
the two pairs; beyond it the estimates stay at the end of the search. With the mean taken over five neighboring
slide positions and the two pairs, the joint fit stays below 5\% from the first distance of the file, 0.96\,m, to
1.49, 1.26 and 1.20\,m at the three irises and below 2\% to 1.20, 1.19 and 1.14\,m;
conventional DfDD stays below 5\% to 1.39, 1.11 and 1.10\,m and below 2\% to 1.10, 1.02 and 1.02\,m. A search
beyond 9.5\,px was not part of the amendment and has not been run.

\begin{table}[htbp]
\centering
\caption{\textbf{A DfDD prototype: distance errors on the second file} (post hoc tabulation of the estimates of
Table~\ref{S-tab:coddepth}). Median distance error [mm] with its 95\% interval over 2\,000 resamples of the positions in
blocks of three, and median relative distance error [\%], pooled over the three iris diameters and the two pairs of
optical powers (\texttt{results/\allowbreak cod\_depth\_units\_B.txt}).}
\label{S-tab:coddist}
\footnotesize
\setlength{\tabcolsep}{5pt}
\begin{tabular}{@{}lccc@{}}
\toprule
estimator & 1.5--3.5\,px (40 samples) & 3.5--9.5\,px (125) & within 9.5\,px (194) \\
\midrule
joint fit, wave optics (ours) & 3.2 [2.5, 5.7] / 0.31 & 16.3 [10.1, 25.3] / 1.32 & 8.5 [3.7, 18.2] / 0.75 \\
our joint fit, geometric optics & 5.6 [4.1, 18.5] / 0.53 & 18.1 [12.2, 30.7] / 1.46 & 12.4 [8.6, 22.2] / 1.06 \\
our joint fit, Gaussian pupil & 7.1 [4.6, 17.4] / 0.68 & 30.9 [25.0, 40.4] / 2.70 & 19.9 [11.7, 32.3] / 1.79 \\
one ratio, our curve & 7.3 [6.0, 9.2] / 0.70 & 24.8 [18.5, 57.3] / 2.11 & 12.6 [8.1, 32.2] / 1.18 \\
one ratio, geometric optics & 10.6 [6.6, 12.7] / 1.03 & 22.1 [12.9, 43.5] / 1.71 & 12.6 [8.6, 23.5] / 1.13 \\
conventional DfDD, linear equation & 18.8 [15.3, 24.4] / 1.82 & 46.9 [39.0, 55.8] / 4.08 & 27.7 [18.2, 43.0] / 2.45 \\
\bottomrule
\end{tabular}
\end{table}

\emph{Mean errors.} Table~\ref{S-tab:codmae} gives the same estimates as mean absolute and mean relative distance errors, the measures of the prototype papers (\texttt{realdata/\allowbreak cod\_mae\_mre.py}; post hoc). Within 9.5\,px the joint fit with wave-optics curves has an MAE of 14.8\,mm [10.2, 22.8]; the same fit with geometric curves errs 1.47 [1.25, 1.70] times as much and conventional DfDD 2.59 [1.99, 3.35] times.

\begin{table}[htbp]
\centering
\caption{\textbf{The prototype in distance units} (post hoc tabulation of the estimates of the amendment; second file, calibrated on the first). Mean absolute distance error (MAE) [mm] and mean relative distance error (MRE) [\%] with 95\% intervals from 2\,000 resamples of the positions in blocks of three, and their ratios to the joint fit with wave-optics curves, for the image pairs within 9.5\,px of geometric blur radius (194 pairs, 0.96--1.53\,m) and for those at 1.5--9.5\,px (165 pairs).}
\label{S-tab:codmae}
\footnotesize
\setlength{\tabcolsep}{4pt}
\begin{tabular}{@{}lcccc@{}}
\toprule
estimator & MAE & MRE & MAE over ours & MRE over ours \\
\midrule
\multicolumn{5}{@{}l}{\emph{0--9.5\,px}} \\
joint fit, wave optics (ours) & 14.8 [10.2, 22.8] & 1.18 [0.84, 1.76] & -- & -- \\
our joint fit, wave optics, cutoff per iris & 15.0 [10.2, 23.2] & 1.19 [0.84, 1.78] & 1.01 [0.95, 1.08] & 1.01 [0.95, 1.07] \\
our joint fit, geometric optics & 21.7 [15.3, 32.9] & 1.74 [1.29, 2.53] & 1.47 [1.25, 1.70] & 1.48 [1.27, 1.72] \\
our joint fit, Gaussian pupil & 29.9 [22.1, 43.9] & 2.42 [1.86, 3.39] & 2.02 [1.72, 2.36] & 2.06 [1.75, 2.40] \\
one ratio, our curve & 32.7 [21.6, 52.7] & 2.53 [1.75, 3.93] & 2.21 [1.85, 2.67] & 2.15 [1.82, 2.57] \\
one ratio, geometric optics & 31.5 [21.1, 48.0] & 2.56 [1.76, 3.81] & 2.13 [1.72, 2.53] & 2.17 [1.77, 2.56] \\
conventional DfDD, linear equation & 38.4 [30.9, 49.9] & 3.21 [2.63, 4.03] & 2.59 [1.99, 3.35] & 2.72 [2.06, 3.53] \\
\midrule
\multicolumn{5}{@{}l}{\emph{1.5--9.5\,px}} \\
joint fit, wave optics (ours) & 16.9 [12.0, 24.6] & 1.34 [0.96, 1.88] & -- & -- \\
our joint fit, wave optics, cutoff per iris & 17.2 [11.9, 25.0] & 1.35 [0.96, 1.90] & 1.01 [0.95, 1.08] & 1.01 [0.95, 1.07] \\
our joint fit, geometric optics & 24.4 [17.4, 35.3] & 1.94 [1.44, 2.70] & 1.44 [1.22, 1.67] & 1.45 [1.24, 1.69] \\
our joint fit, Gaussian pupil & 34.0 [25.9, 46.7] & 2.73 [2.17, 3.60] & 2.01 [1.71, 2.35] & 2.05 [1.74, 2.39] \\
one ratio, our curve & 37.8 [25.2, 57.5] & 2.91 [2.03, 4.26] & 2.23 [1.86, 2.71] & 2.18 [1.84, 2.62] \\
one ratio, geometric optics & 36.1 [24.8, 51.1] & 2.91 [2.03, 4.07] & 2.13 [1.70, 2.54] & 2.18 [1.76, 2.57] \\
conventional DfDD, linear equation & 44.2 [38.0, 52.1] & 3.67 [3.23, 4.19] & 2.61 [2.00, 3.41] & 2.75 [2.07, 3.61] \\
\bottomrule
\end{tabular}
\end{table}

\Para[S-sec:codcontrols]{An effective aperture and small windows (pre-specified controls)}
The depth comparison of the amendment leaves two questions open. A geometric disc of another radius might do as
well as the wave-optics curves, as on the DSLR (\hyperref[S-sec:attrctl]{attribution controls}), and the estimates
above use one whole crop of $400\times400$ pixels, where DfDD cameras report windows of 5 to 71 pixels. We fixed
both controls in a local record before any of their estimates was computed (\texttt{notes/\allowbreak
prespec\_cod\_controls\_20261114.md}, with the SHA-256 of 72 files, written after the dry checks and a rehearsal on
synthetic images and checked after the runs; \texttt{realdata/\allowbreak cod\_controls\_A.py},
\texttt{cod\_controls\_B.py}, \texttt{results/\allowbreak cod\_controls/}). Both runs first reproduce the saved
estimates, gains and foci of the amendment, with differences of zero.

\emph{Control A, an effective aperture.} Protocol and endpoint are those of the amendment, that is the calibration
of the first file, one focus shift from the second, and the median relative defocus error over 1.5--9.5\,px on the
second file with its block bootstrap. Three families of the joint fit are fitted per iris on the first file. G-rg
is a geometric disc with a fitted radius and gain, G-r the same with the radius alone and a gain of 1, and S-g a
geometric disc whose radius in band $b$ is $1-f_b/f_c$ times nominal, the factor of Stokseth's
approximation~\cite{S-stokseth1969}, with a fitted gain. The gain is the least-squares gain of the amendment, the
geometric mean over the eligible cells of the measured slope over the slope of the model, and the radius minimizes
the same sum of squared logarithmic residuals, on a search from 0.50 to 1.20 times nominal that a fine step may
pass by 0.04. The statement that the advantage of wave-optics curves on the prototype is specific to wave optics,
not an effective aperture, is confirmed if the 95\% interval of the error of G-rg over ours lies above 1.05, refuted
if it lies below 1.05, and undecided otherwise. We predicted a radius of 0.6--0.8, a ratio of 1.1--1.5 and S-g
within 10\% of ours.

\emph{Outcome} (Table~\ref{S-tab:codctlA}). By the rule the statement is confirmed, since G-rg errs 3.03 [2.49, 3.77]
times ours. The rival of the rule is weak, however. The criterion of the slopes has no minimum inside the search,
the radius of G-rg stops at its end, 1.24 times nominal at all three irises, and G-rg then errs twice as much as the
nominal geometric curves (17.7\% against 9.2\%). A rehearsal on synthetic images had shown this before the run, and
the record says so. We do not rest the claim on this number. The radius alone, 0.64, 0.73 and 0.78 times nominal,
gives 1.55 [1.03, 2.51], which the same margin leaves undecided. Two descriptive variants that choose the radius by
the depth error on the first file give 1.66 [0.94, 2.59] with a gain and 1.71 [1.25, 2.33] without. In the mean
absolute distance error within 9.5\,px no disc of a single radius, nominal or fitted, reaches ours (1.47--2.76
times, every interval above 1.25). For the fitted radii the gap in the endpoint comes from the 3\,mm iris (2.2--2.7
times ours); at 5 and 7\,mm a fitted radius is level with ours within intervals that span a factor of five or more.
S-g equals ours (1.00 [0.86, 1.15], and 0.94 [0.85, 1.03] in distance). On the prototype, as on the DSLR, the
depth advantage belongs to an aperture that shrinks with frequency as the diffraction cutoff sets it, and not to
the finer shape of the wave-optics curves.

\begin{table}[htbp]
\centering
\caption{\textbf{Control A on the prototype, an effective aperture} (pre-specified; second file, calibrated on the
first). Fitted radius in units of the nominal one and gain at 3, 5 and 7\,mm; median relative defocus error over
1.5--9.5\,px [\%], the endpoint of the amendment, and its ratio to ours; mean absolute distance error within
9.5\,px [mm] and its ratio to ours; 95\% intervals from 2\,000 resamples of the positions in blocks of three. The
rule applies to G-rg. The last two rows are descriptive and choose the radius by the depth error on the first
file. $^{\ast}$The end of the search.}
\label{S-tab:codctlA}
\footnotesize
\setlength{\tabcolsep}{3pt}
\resizebox{\textwidth}{!}{%
\begin{tabular}{@{}lcccccc@{}}
\toprule
curves of our joint fit & radius & gain & defocus error & ratio to ours & MAE & ratio to ours \\
\midrule
wave optics, nominal radius (ours) & 1 & 0.832, 0.745, 0.751 & 5.85 [4.57, 7.09] & -- & 14.8 [10.2, 22.8] & -- \\
geometric, nominal radius & 1 & 0.412, 0.536, 0.623 & 9.19 [7.15, 10.67] & 1.57 [1.19, 2.02] & 21.7 [15.3, 32.9] & 1.47 [1.25, 1.70] \\
G-rg, geometric, fitted radius and gain & 1.24, 1.24, 1.24$^{\ast}$ & 0.277, 0.372, 0.440 & 17.74 [14.63, 20.75] & 3.03 [2.49, 3.77] & 40.9 [27.9, 64.7] & 2.76 [2.39, 3.23] \\
G-r, geometric, fitted radius, gain 1 & 0.64, 0.73, 0.78 & 1 & 9.04 [6.56, 13.35] & 1.55 [1.03, 2.51] & 24.7 [18.4, 35.3] & 1.67 [1.37, 1.99] \\
S-g, geometric, radius $\times(1-f_b/f_c)$ & -- & 0.795, 0.720, 0.736 & 5.86 [4.72, 6.45] & 1.00 [0.86, 1.15] & 13.9 [9.8, 20.9] & 0.94 [0.85, 1.03] \\
\midrule
geometric, radius by the depth error, fitted gain & 1.14, 0.78, 0.81 & 0.322, 0.861, 0.923 & 9.71 [5.49, 15.11] & 1.66 [0.94, 2.59] & 29.0 [18.1, 48.4] & 1.96 [1.52, 2.43] \\
geometric, radius by the depth error, gain 1 & 0.55, 0.74, 0.77 & 1 & 9.99 [6.46, 14.19] & 1.71 [1.25, 2.33] & 34.8 [22.3, 56.4] & 2.35 [2.00, 2.70] \\
\bottomrule
\end{tabular}}
\end{table}

\emph{Control B, windows of $64\times64$ pixels.} The band statistics are summed over 16 windows of $64\times64$
pixels on the central $256\times256$ pixels of the crop, without the taper of the whole-crop pipeline; on the whole
crop and with the taper the window routine returns the saved band sums. Gains, foci and the constant of conventional
DfDD are fitted again on the first file from these windows and differ from the whole-crop values in the third
digit. The second file is scored on the 194 image pairs within 9.5\,px, 3\,104 windows. For X the joint fit with
geometric curves and for X conventional DfDD, the statement that the joint fit with wave-optics curves has
the smaller mean absolute distance error is confirmed if the interval of MAE(X) over MAE(ours) lies above 1.05,
refuted if it lies below, and undecided otherwise.

\begin{figure}[htbp]
\centering
\includegraphics[width=\textwidth]{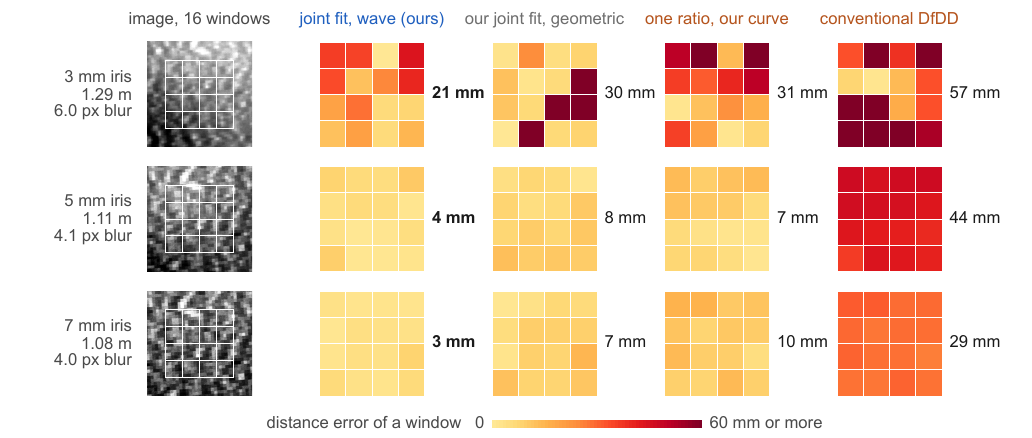}
\caption{\textbf{The print through windows of $64\times64$ pixels} (control B). Each row shows one iris, with one
image of its example pair and the 16 windows, and the absolute distance error of every window for four estimators;
the number beside a map is the mean over its 16 windows. By a rule in the record the example of an iris is, among
its image pairs within 9.5\,px, the pair whose four mean errors lie closest to the medians over those pairs.}
\label{S-fig:protomaps}
\end{figure}

\begin{figure}[p]
\centering
\includegraphics[width=\textwidth]{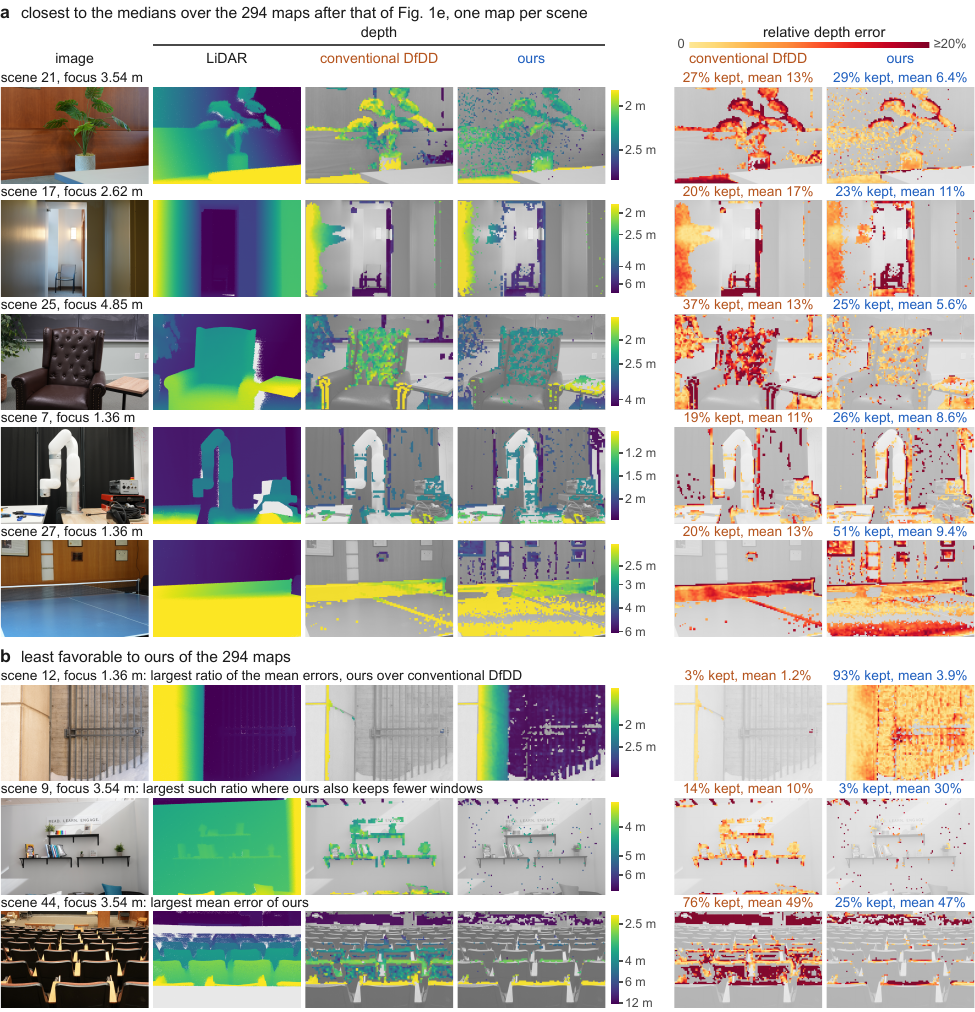}
\caption{\textbf{Depth from one image pair with a confidence filter: eight more maps.} DSLR at f/16, as in main-paper
Fig.~\ref{fig:teaser}e. Each estimator is drawn on the windows that it keeps under a threshold set on the other
four folds to discard their least confident 75\%; over an error tile stand the share kept of the estimated
windows of the map and their mean relative depth error. The 294 maps are those of 400 (50 scenes, eight pairs) on
which these two and our joint fit with geometric curves each keep at least 200 windows. (a)~The five next after the
map of main-paper Fig.~\ref{fig:teaser}e by its rule, one per scene. (b)~The least favorable to ours by the
measure over each row, the second chosen after the first was seen; over the 294 ours has the larger error on 47 and
keeps fewer windows on 144. Gray marks no kept estimate or no LiDAR return. Depth colors are linear in $1/Z$ on a
scale per row; ``focus'' is the focus distance of the pair
(\hyperref[S-par:mapsconf]{More maps with a confidence filter}).}
\label{S-fig:maps_conf}
\end{figure}

\emph{Outcome} (Table~\ref{S-tab:codctlB}, Fig.~\ref{S-fig:protomaps}). Both statements are confirmed, with 1.41
[1.28, 1.55] and 3.06 [2.71, 3.48]. The MAE of ours is 17.3\,mm [12.6, 25.6], against 14.8\,mm for the whole crop,
and 26.9, 8.5 and 3.7\,mm at the three irises. We had predicted that the errors would grow by a factor of 2 to 5;
they grow by 12--38\%. With the least confident 75\% of the windows discarded, by a confidence chosen on the first
file (the curvature of the cost for the joint fits, the Laplacian energy for the other two), the MAE is 4.4\,mm
[3.4, 8.4] for ours, 6.4\,mm with geometric curves, 6.1\,mm for one ratio and 19.3\,mm for conventional DfDD;
the interval of one ratio over ours includes 1. One threshold serves the three irises, and of the windows kept for
ours 7\%, 59\% and 34\% come from 3, 5 and 7\,mm. By the mean criterion of the prototype papers the far limit of
the working range at 10\% and at 5\%, pooled over the irises, lies at 1.37 and 1.24\,m for ours, at 1.35 and
1.23\,m with geometric curves and at 1.24 and 1.13\,m for conventional DfDD. The near limit is the first distance
of the file, 0.96\,m, for every estimator, and the search of the amendment, which ends at 9.5\,px, bounds every
range as above.

\begin{table}[htbp]
\centering
\caption{\textbf{Control B on the prototype, windows of $64\times64$ pixels} (pre-specified; second file, 194 image
pairs within 9.5\,px, 3\,104 windows). Mean absolute distance error (MAE) over all windows [mm] and its ratio to
ours, the endpoint; mean relative distance error (MRE) [\%]; the MAE of the windows over that of the whole crop;
and the MAE with the least confident 75\% of the windows discarded, with its ratio to ours. 95\% intervals from
2\,000 resamples of the positions in blocks of three. The rule applies to the ratios of the second and fourth row.}
\label{S-tab:codctlB}
\footnotesize
\setlength{\tabcolsep}{3pt}
\resizebox{\textwidth}{!}{%
\begin{tabular}{@{}lcccccc@{}}
\toprule
estimator & MAE & ratio to ours & MRE & over whole crop & MAE, 75\% discarded & ratio to ours \\
\midrule
joint fit, wave optics (ours) & 17.3 [12.6, 25.6] & -- & 1.39 [1.04, 1.98] & 1.17 [1.10, 1.25] & 4.4 [3.4, 8.4] & -- \\
our joint fit, geometric optics & 24.4 [17.7, 36.4] & 1.41 [1.28, 1.55] & 1.96 [1.50, 2.80] & 1.12 [1.04, 1.25] & 6.4 [5.4, 10.5] & 1.45 [1.03, 1.77] \\
one ratio, our curve & 38.2 [25.6, 61.5] & 2.21 [1.86, 2.63] & 2.96 [2.06, 4.59] & 1.17 [1.09, 1.26] & 6.1 [5.0, 8.8] & 1.37 [0.85, 1.71] \\
conventional DfDD, linear equation & 52.9 [40.8, 73.2] & 3.06 [2.71, 3.48] & 4.30 [3.43, 5.68] & 1.38 [1.22, 1.60] & 19.3 [14.5, 31.1] & 4.36 [2.78, 5.52] \\
\bottomrule
\end{tabular}}
\end{table}

The release does not document the spectrum of the illumination, the point from which the distances were measured or
the settings of the camera. By the rules
fixed in advance the cutoff scale is undecided on both files. Apart from the rules, the pre-specified statistics show
a slope that falls with frequency at three apertures of one DfDD prototype, follows the clear disc with one gain and
one cutoff scale per aperture to 2\% rms, and departs from a constant coefficient by 7--20\% rms on the first file
and by 6--21\% on the second. On the second file the estimator of the main paper, calibrated on the first, has a
median defocus error of 5.8\% over 1.5--9.5\,px of blur, against 9.2\% with geometric curves and 11.3\% for a single
ratio.

\clearpage
\section{The posterior of the joint fit}
\label{S-sec:posterior}

This section defines the posterior of main-paper Sec.~\ref{sec:bw} and gives the numbers behind the row ``read
from the posterior'' of main-paper Table~\ref{tab:est}. All of it is exploratory: the 50 validation scenes of
ZEDD, five folds over scenes, every constant fixed on the training scenes of a fold. A second, independent
implementation, which calibrates on all scenes, agrees with the numbers quoted here within 0.15 points, with two
exceptions: for the minimum of the fit with the foci at f/16 it gives 0.996 in the first column of
Table~\ref{S-tab:posterior}, where the pair with a step of 0.02\,D decides, and on the windows that the posterior
keeps it finds no difference between the two read-outs.

\paragraph{Camera noise}
The images are divided by their local mean before the bands are formed, so photon noise adds to the Laplacian energy
of band $b$ a floor that is inversely proportional to the mean linear luma $\mu$ of the window, $N_b=\alpha_b/\mu$.
The five constants $\alpha_b$ are properties of the camera. We measure them without depth: of the f/2.8 images
of a window that are not saturated we take the three whose energy at 0.045\,cycles/px is lowest, in which the
window is blurred away, and $\alpha_b$ is the median of their band energy times $\mu$ over the training scenes of
each fold. Windows chosen by a LiDAR defocus of 30\,px or more give constants within 7\% of these; between scenes
those vary by a factor of 0.86--1.23 at 0.225\,cycles/px and of 0.92--1.07 at 0.10\,cycles/px (quartiles). The
share of noise in the weight
of a band is $\eta_b=N_b/\omega_b$, clipped to $[0,1]$.

\paragraph{Likelihood and posterior}
Noise in the denominator of the band ratio shrinks its mean to $\gamma(1-\eta_b)g_b(\psi)$ and leaves a variance
proportional to $(1-\eta_b/2)/\omega_b$. With residuals of unknown scale $\sigma$, the likelihood of the $n$ band
ratios is $\sigma^{-n}\exp[-C(\psi)/2\sigma^2]$ with the cost $C$ of main-paper Sec.~\ref{sec:bw}. Real windows
do not obey a variance model (the texture is not Gaussian, the gain and the foci have errors), so $\sigma$ is
integrated out under the prior $\dd\sigma^2/\sigma^2$, which gives $p(\psi\,|\,r)\propto\pi(\psi)\,C(\psi)^{-n/2}$.
A band that is pure noise adds a constant to $C$: it does not vote for a defocus, but it sets the scale against
which the fit of the other bands is judged. The prior $\pi$ is uniform in diopters on the range of the search, cut
at 0.04\,D (25\,m, the limit at which estimates are also clipped when they are scored).

\paragraph{Estimate and expected error}
Let the posterior have weights $p_j$ on a grid of vergences $q_j$. The expected relative distance error of an
estimate $\hat q$ is $\sum_jp_j\,|q_j/\hat q-1|$; it is smallest when $\hat q$ is the median of the $q_j$ under the
weights $p_jq_j$. For a narrow posterior $N(q_0,s^2)$ this gives $\hat q=q_0+s^2/q_0$ and an expected error of
$\sqrt{2/\pi}\,s/q_0$; for a flat posterior on $[a,b]$ it gives $\hat q=\sqrt{(a^2+b^2)/2}$. The minimum of the
cost of the main paper is, in the linear range of one band, the mean of a posterior whose prior is centered on the
midpoint of the pair with a width of $C_b/(\gamma\kappa_b\sqrt{M_b})$, 46\,mD for the band at 0.10\,cycles/px and a
step of 0.25\,D, an order of magnitude narrower than the spread of the defocus of real windows.

\paragraph{Foci at f/16}
The calibration of the main paper takes the nine foci from f/2.8--f/5.6. Stopped down to f/16 the lens focuses
6.8\,mD farther [5.3, 7.7]. We measure the foci at f/16 without depth, on the training scenes: a window whose
main-lobe energies are equal in two runs lies midway in diopters between their foci. The first run focuses at 1.26\,D, not at
the 1.33\,D of the calibration; the focus distances stored in the metadata of the dataset and the motor positions
in the file names, fitted to runs 2--9 and extrapolated, give 1.23--1.26\,D.

\begin{table}[ht]
\centering
\caption{\textbf{Read-outs of the joint fit on the DSLR at f/16} (exploratory; samples and columns of main-paper
Table~\ref{tab:est}). Ratio to the minimum of the fit [95\% interval over scenes].}
\label{S-tab:posterior}
\setlength{\tabcolsep}{4pt}
\begin{tabular}{lcc}
\toprule
& MRE within $\pm0.35$\,D & 75\% discarded \\
\midrule
minimum of the fit, calibrated foci (main paper) & 15.6\% & 6.0\% \\
minimum of the fit, foci at f/16 & 0.964 [0.946, 0.980] & 0.915 [0.876, 0.956] \\
posterior, foci at f/16 & 0.688 [0.617, 0.771] & 0.675 [0.603, 0.750] \\
\quad image pairs 1--6 & 0.823 [0.785, 0.863] & \\
\quad image pairs 7 and 8 & 0.551 [0.444, 0.693] & \\
posterior, calibrated foci & 0.717 & 0.776 [0.709, 0.843] \\
minimum of the fit, discarded by the expected error & & 0.773 [0.706, 0.841] \\
\bottomrule
\end{tabular}
\end{table}

\paragraph{Results}
Table~\ref{S-tab:posterior}. Within 0.35\,D of focus the MRE falls from 15.6\% to 10.8\%. The improvement is largest in the
two farthest pairs, whose steps are small and whose windows are noisy; 0.8\% of the minima of the fit lie at the
25\,m limit and supply 2.6 of the 15.6 points, and the posterior has no such estimate. On the windows that the
posterior keeps, the two read-outs differ by 4\% (0.959 [0.904, 1.006]), so most of the improvement in the last column
comes from the confidence. At f/5.6 the posterior changes the first column by a factor of 0.999 and the last by
0.895. The expected error ranks the windows correctly but is too small as it stands: 73.1\% of the windows are
expected to err by less than 10\%, and their MRE is 7.1\%, against 12.0\% for as many windows kept by the residual
(the most confident tenth: expected 0.4\%, measured 3.4\%). A calibrated posterior has four more constants, fitted
on the training scenes against the LiDAR: an uncertainty of the origin of the defocus axis (18--20\,mD at f/16),
an uncertainty of the gain, the exponent $n$, and a share of the prior mixed in. Its central 50, 68, 90 and 95\%
intervals contain the LiDAR vergence in 52, 70, 90 and 95\% of the held-out windows; 53.9\% of the windows are
then expected to err by less than 10\%, and their MRE is 5.3\%, against 8.4\% for as many windows kept by the
residual (ratio 0.624 [0.552, 0.694]).

\section{Turnovers and sign changes on the DSLR}
\label{S-sec:signs}

A gain, a tone curve or a noise floor rescales a through-focus curve; none of them moves the defocus at which the
curve turns over, or the defocus at which a band has equal energy in two images. This section gives the two tests
of main-paper Sec.~\ref{sec:pixel} that rest on these positions.

\paragraph{Where the two models differ}
In Stokseth's approximation the first zero of the defocused OTF at frequency $f$ lies at a geometric blur radius
\begin{equation}
  b_0=\frac{0.61}{f\,(1-f/f_c)}=\frac{0.61}{f}+\frac{0.61}{f_c-f},
\end{equation}
where a pillbox has $0.61/f$. With the exact polychromatic OTF of the main paper the first minimum of the energy of
a narrow band at 0.12--0.26\,cycles/px lies 0.87\,px beyond the geometric one at f/16 (0.81--0.95\,px), 0.10\,px
beyond Stokseth's value; the radius of the Airy disc is 1.28\,px. Plotted against $1/f$ the minima of the
clear disc lie on a line with an intercept of 1.07\,px, those of the geometric disc on a line through the origin.
A geometric disc can only change the slope of this line through its radius, so it needs a radius that falls with
frequency.

\paragraph{Turnover ratio (pre-specified, confirmed)}
In geometric optics the curve of a band depends only on the product of the blur radius and the frequency, so the
defocus at which two bands fall to half their peak is in the ratio of the frequencies for a disc of any radius;
wave optics adds a factor that depends on $f/f_c$. For the bands at 0.225 and 0.15\,cycles/px, on the windows of
the test, a geometric disc of radius 0.60--1.00 gives 0.670--0.690 and a clear disc of radius 0.85--1.05 gives
0.738--0.777. The rule, fixed before the test scenes were read through this path, was: geometric optics is excluded
if the lower end of the 95\% interval lies above the larger of 0.690 and the largest geometric value on the test
windows, wave optics if the upper end lies below the smaller of 0.740 and the smallest clear-disc value there.
On the 50
test scenes, with depth from focus at f/2.8 as the reference and the shift of that reference (0.032, measured on
the validation scenes) subtracted, the ratio is 0.719 [0.697, 0.757]: geometric optics is excluded by a margin of
0.007 and wave optics is not. One radius fitted to the turnovers of three bands, with no noise subtracted, gives
$0.99R$ and $\chi^2=18$ for a clear disc and $\chi^2=331$ for a geometric one.

\paragraph{Sign changes without depth (pre-specified, undecided)}
Windows whose main-lobe energies at 0.128--0.190\,cycles/px are equal in run 2 and in run 3, 4 or 5 lie midway
in diopters between the two foci, at 0.718, 0.633 and 0.575\,D; on the validation scenes the LiDAR vergence of these windows
lies $-8.5$, $+0.2$ and $-10.2$\,mD from the three midpoints. Their blur in a pair of far-focused runs is then
known from the foci alone, $\beta=4.0$--$5.5$\,px for the 16 pairs of runs 5--9 with a step below 1.5\,px. Over 48
narrow bands (0.02--0.45\,cycles/px), the share of these windows in which the nearer-focused image of the far pair
has the larger energy falls through one half at a frequency $f^\ast$, which gives one point $(f^\ast,\beta)$ of the
locus of sign changes; the radius of a clear disc or of a geometric disc that puts its sign change there follows
from model tables. Which image has more energy does not change under any monotonic processing of the energies.

\begin{table}[ht]
\centering
\caption{\textbf{Sign changes of narrow bands without depth.} Radii in units of the nominal one; slopes against
$\ln f^\ast$; 95\% intervals over 2\,000 resamples of the scenes. Criteria fixed before the test scenes were
read: C1 to C3 in this order.}
\label{S-tab:signs}
\setlength{\tabcolsep}{4pt}
\resizebox{\linewidth}{!}{%
\begin{tabular}{lccl}
\toprule
& validation scenes & 50 test scenes & criterion \\
\midrule
clear disc, median radius & 0.907 [0.900, 0.913] & 0.920 [0.912, 0.923] & inside 0.88--0.94: met \\
clear disc, slope & $-0.017$ [$-0.041$, $+0.023$] & $-0.047$ [$-0.062$, $-0.021$] & inside $\pm0.06$: not met \\
geometric disc, slope & $-0.231$ [$-0.273$, $-0.191$] & $-0.252$ [$-0.272$, $-0.224$] & below $-0.10$: met \\
points with a sign change & 16 of 16 & 14 of 16 & at least 8 \\
windows of the three rungs & 2193, 3960, 1740 & 8267, 5336, 2037 & \\
median clear-disc radius by rung & 0.906, 0.911, 0.900 & 0.926, 0.910, 0.903 & \\
scatter of the clear-disc radius & 0.0054 & 0.0098 & \\
scatter of the geometric radius & 0.0345 & 0.0374 & \\
intercept of $\beta$ against $1/f^\ast$ [px] & 1.72 [1.40, 1.89] & 1.73 [1.64, 1.88] & \\
\bottomrule
\end{tabular}}
\end{table}

Table~\ref{S-tab:signs}. The test was run once. By its rule the three criteria must all be met for a confirmation,
and the interval of the slope of the clear-disc radius leaves its band by 0.002, so the verdict is undecided. On
the test scenes the radius that a geometric disc needs falls from 0.76 at 0.148\,cycles/px to 0.63 at
0.243\,cycles/px; the radius of the clear disc lies between 0.900 and 0.935 and its slope is a fifth of the
geometric one. The three rungs occupy different frequencies, and a rung that is 10\,mD from its midpoint moves
$\beta$ by 0.093\,px and the radius by 2\%, which is the size of the difference between the rungs. A second
implementation, written from the record alone, reproduces the radius, the two slopes and the intercept of the
validation column within 0.001; its middle rung has one window more. With LiDAR
depth in the place of the rungs, the sign changes of seven narrow bands on the validation scenes give a
clear-disc radius of 0.923 [0.916, 0.931] that varies by 0.3\% between the bands, where the radius of a geometric
disc varies by 2.4\%.

\paragraph{The radius and the gain}
On the validation scenes, with LiDAR depth and with the camera noise of Sec.~\ref{S-sec:posterior} subtracted from
the band energies (exploratory), one clear disc of radius
$0.92R$ fits the turnovers of the bands at 0.10, 0.15 and 0.225\,cycles/px at f/16, and its curves then need a
gain of 1.01 where the curves of the nominal radius need 0.82; at f/5.6 the radius is 1.00 and the gain 0.97. The
in-focus loss of contrast between f/5.6 and f/16 would need a disc of $1.25R$, so the pupil is not identified.

\section{Conditions and limits of the results}
\label{S-sec:conditions}

The main paper states its limits in Sec.~\ref{sec:discussion}. This section gives them in full.

\paragraph{The working range}
The range needs the confidence. With nothing discarded the mean error of the minimum of the joint fit stays below
10\% over no wider a span than that of conventional DfDD, as 38\% of its estimates beyond 3.5\,px are off by more
than half (Table~\ref{S-tab:confblur}). With half of the windows discarded and a criterion of 5\%, conventional DfDD
keeps the wider span in the three nearest pairs. With the least confident 75\% discarded and a criterion of 10\%,
a range contiguous from the focus exists for three of the eight pairs for ours and for five for conventional
DfDD; ours has none at 2.6 and 3.5\,m, where its mean error in the bin at the focus is 11\% and 19\%
(Table~\ref{S-tab:confpair}). The range comes from the low bands. In the median defocus error without a confidence
filter it does not come from diffraction, since geometric curves and calibrated spectral ratios reach it too
(Table~\ref{S-tab:estfull}); with the filter they err 1.7 and 2.5 times ours (Table~\ref{S-tab:conft1b}). With nothing discarded ours collapses onto the pair
midpoint beyond 3.5\,px in three of the 50 test scenes (Sec.~\ref{S-sec:zedd}).

\paragraph{Learned and calibrated estimators}
The perceptron has no confidence. Near focus it errs 25\% less than the minimum of our fit, and as much as ours
without the pair index (1.06 [0.81, 1.53]). With nothing discarded it errs half as much as the minimum of our fit
on textured windows, 15\% against 32\%, and on the windows that our confidence keeps as much, 6.1\% against 6.0\%.
Over all 400 one-pair maps without a confidence ours loses to the perceptron in 62\% and to depth from focus over
nine images in 66\% (Figs.~\ref{S-fig:maps},~\ref{S-fig:maps_more}). At f/2.8 calibrated curves and spectral ratios
err 14--32\% less than ours, so the range is stated for f/16. FOSSA~\cite{S-zuo2026}, the dataset's network, is more
accurate on its own five f/2.8 images than our one-pair estimate, and it is dense (Table~\ref{S-tab:sameinput}).

\paragraph{Prototype}
A test of the cutoff itself, pre-specified before any band statistic was computed, is undecided on both files;
the fitted scale moves from 0.88 to 1.46 with the noise floor assumed for the weakest band (Sec.~\ref{S-sec:cod}). In
the first of the two controls a geometric disc fitted per iris ran its radius to the bound of the search; with
only the radius fitted it errs 1.55 times ours [1.03, 2.51], and with Stokseth's scaling as ours
(Sec.~\ref{S-sec:codcontrols}). Without the focus shift from the second file the fit errs by 16.6\% at
1.5--3.5\,px. The coupled differentiation of~\cite{S-luo2025cod} reaches a longer range than any estimator here,
which change the focus only.

\paragraph{Phone and DSLR}
The confirmation on the phone counted five sweeps twice, which does not change the outcome
(Sec.~\ref{S-sec:pixelconfirm}); the fall it confirms at 95 cycles/mm is far steeper than for a clear disc, and we
cannot name the origin of this frequency dependence. On the validation scenes of the DSLR a test on all eight
pairs at f/16, fixed before we ran it, is undecided; the five pairs with small steps, a subset we chose after
seeing these scenes, show the fall of main-paper Lemma~\ref{lem:disc}. The confirmation that we then
pre-specified on the 50 test scenes, with depth from focus at f/2.8 as the reference, refuted the result: at
0.225 cycles/px the data give 0.82 [0.73, 0.94], against 0.62 [0.58, 0.68] on the validation scenes through the
same path. Subtracting a noise floor estimated per scene and reweighting lowers the simulated biases of the fit,
though not within our preset bound, and changes the DSLR error of its minimum by $+2\%$ on textured windows
(Table~\ref{S-tab:baselines}); the posterior of Sec.~\ref{S-sec:posterior} uses camera constants instead.

\section{Further tests fixed in advance on the DSLR}
\label{S-sec:furthertests}

Three more tests had their rules fixed before the test scenes were read for them (Table~\ref{S-tab:protocol}). Their
subjects are not results of the main paper, so they are reported here.

\paragraph{Two constants per image pair (confirmed)}
\label{S-par:twocalib}
Near focus the residual of the band ratios of a pair has two directions that the theory names: a shift of the
midpoint of the pair, nearly the same in every band, and a difference of contrast between its two images, whose
effect on the band ratio scales as $1/(f_b^2\Delta\psi)$. Two constants per pair, fitted on the validation scenes,
were tested on the 50 test scenes against depth from focus at f/2.8. The statistic is the relative defocus error
beyond 3.5\,px of blur (the median in bins of the defocus, averaged over the bins) of the calibrated fit over that
of ours, and the rule asked for a 95\% interval below 1. The ratio is 0.782 [0.657, 0.891]: the error falls from
7.7\% to 6.1\%, and the median defocus error within 0.35\,D from 20.6 to 16.3\,mD. The contrast term alone gives
0.756 [0.696, 0.842] beyond 3.5\,px and leaves the error near focus unchanged (1.009). With both constants the
improvement beyond 3.5\,px comes from the pairs with large steps (0.715 [0.584, 0.849]); for the pairs with small steps
the error does not improve (1.022 [0.964, 1.117]). The estimators of the main paper do not use these constants.

\paragraph{Astigmatism of the lens (both criteria met)}
\label{S-par:astig}
A grating whose gradient makes the angle $x$ with the radius of the field focuses at $\bar q+(\Delta/2)\cos2x$,
where $\Delta$ is the difference between the tangential and the sagittal focus, so the vergence at which a window
is sharpest depends on the orientation of its texture. We measure $\Delta$ without depth. The gradient energy of a
window is split into the part along the radius of the field and the part across it, each part is fitted with the
through-focus profile over the nine images of the stack, and the two foci differ by $\kappa\Delta$ with
$\kappa=(m_2-a^2)/(1-a^2)$, $a=\langle\cos2x\rangle$ and $m_2=\langle\cos^22x\rangle$ over the energy of the
window. On the validation scenes $\Delta$ rises to 21\,mD at 0.7 of the half image height, falls back to 11.5\,mD
at 1.0 and reaches 59\,mD in the corners; f/16 and f/5.6 agree. The test on the 50 test scenes, which have no
depth, had two criteria. The profile of the validation scenes predicts the difference of the two foci on the test
scenes with a slope of 1.03 [0.99, 1.05] (band 0.85--1.15), and between windows 64\,px apart the estimates of the
main paper follow the predicted shift $a\Delta/2$ with a slope of 0.73 [0.66, 0.79] (band 0.50--1.10). The second
slope depends on how the orientation index treats windows of weak texture and is not a response of the estimator;
where the noise is negligible it is 0.93--0.96 on the validation scenes. Compared inside one image pair and
calibrated against the LiDAR, the difference of the two foci is 1.07 [1.03, 1.11] times the profile. Subtracting
the predicted shift lowers the MRE near focus at f/5.6 by 9\% and by 1--5\% at f/16 (validation scenes).

\paragraph{Noise floor from camera constants (fixed, not run)}
\label{S-par:camerafloor}
With the noise floor of the band energies set to the camera constants of Sec.~\ref{S-sec:posterior} and nothing else
changed, the share of estimates beyond 3.5\,px that are off by more than half falls from 19.7\% to 10.9\% on the
validation scenes (ratio 0.555 [0.448, 0.678]). A test of this ratio on the test scenes was fixed on 15 Nov.\ and
has not been run.

\clearpage
\section{Code and reproduction}
\label{S-sec:repro}

All results come from public data. The scripts that produce them (Python with numpy, scipy and Pillow; matplotlib for the
figures) will be released with the paper, together with the records of Sec.~\ref{S-sec:protocol}. Where this document
cites a second implementation (Secs.~\ref{S-sec:posterior} and~\ref{S-sec:signs}), both implementations were written
with the LLM assistant named in the usage statement of the main paper, in separate sessions, each with its own
scripts. The scripts fall into five groups.
\begin{itemize}
  \item \emph{Theory checks}: closed-form and quadrature OTFs, the Gaussian-pupil constraint, the tests of exact diffusion, the COD
  tables and the sub-band simulation.
  \item \emph{Phone data}: streaming extraction of the needed members (7.4\,GB of 108\,GB), per-capture processing
  (registration, focus-step gauge, frame quality, flat field, DfF, band sums), pooled and within-point statistics,
  synthetic sweeps, aliasing model, pupil fits and depth estimation.
  \item \emph{DSLR data}: range-request download (4.3\,GB), per-scene window sums (about 8\,min per scene and core),
  focus calibration, band-ratio tests, in-focus ratios, the errors-in-variables bracket, single-pair estimation, the
  attribution controls, the working range in defocus and in distance, the confidence of the fit with its
  confirmation on the test scenes, and depth maps.
  \item \emph{Learned depth}: inference with the released FOSSA code and weights (BSD-3 license, PyTorch) under the
  authors' ZEDD protocol~\cite{S-zuo2026}, and the per-scene affine correction.
  \item \emph{Prototype data} (Sec.~\ref{S-sec:cod}): band energies and noise floors, the slopes of the apparent
  defocus, the model tables, the synthetic checks, the post hoc diagnostics, and for the second file the
  corrections, the calibration on the first file, the depth estimators and the decision script; the tabulations of
  the slopes against the closed form and of the distance errors, and the two controls with an effective aperture
  and with windows of $64\times64$ pixels.
\end{itemize}
The records of the tests fixed in advance, with their hash lists, are among these files.

\end{document}